\documentclass[%
 reprint,
 amsmath,amssymb,
 aps,
rmp,
]{revtex4-2}

\usepackage{graphicx}
\usepackage{dcolumn}
\usepackage{bm}
\usepackage{hyperref}
\hypersetup{
    colorlinks=true,
    urlcolor= blue,
    citecolor=blue,
linkcolor= blue}
\usepackage{array}
\newcolumntype{P}[1]{>{\centering\arraybackslash}p{#1}}

\newcommand{\am}[1]{{\color{black} #1}}
\newcommand{\amgreen}[1]{{\color{black} #1}}
\newcommand{\eq}[1]{\begin{equation} #1 \end{equation}}

\newcommand{\addCXL}[1]{{\color{black} #1}}
\newcommand{\czchang}[1]{{\color{black} #1}}

\begin{document}

\preprint{APS/123-QED}

\title{Colloquium: Quantum anomalous Hall effect}

\author{Cui-Zu Chang}
\email{cxc955@psu.edu}
\affiliation{Department of Physics, The Pennsylvania State University, University Park, Pennsylvania 16802, USA}%

\author{Chao-Xing Liu}
\email{cxl56@psu.edu}
\affiliation{Department of Physics, The Pennsylvania State University, University Park, Pennsylvania 16802, USA }%

\author{Allan H. MacDonald}
\email{macd@physics.utexas.edu}
\affiliation{Department of Physics, The University of Texas at Austin, Austin, Texas 78712, USA}%

\date{\today}

\begin{abstract}
The quantum Hall (QH) effect, quantized Hall resistance combined with zero longitudinal resistance,
is the characteristic experimental fingerprint of Chern insulators - topologically non-trivial states of two-dimensional matter 
with broken time-reversal symmetry. In Chern insulators, non-trivial bulk band topology 
is expressed by chiral states that carry current along sample edges without dissipation.
The quantum anomalous Hall (QAH) effect refers to QH effects that occur in the absence of external 
magnetic fields due to spontaneously broken time-reversal symmetry.  The QAH effect has now 
been realized in four different classes of two-dimensional materials: (\textit{i}) thin films of magnetically (Cr- and/or V-) doped
topological \am{insulators in the} (Bi,Sb)$_2$Te$_3$ family, (\textit{ii}) thin films of the intrinsic magnetic topological 
insulator MnBi$_2$Te$_4$, (\textit{iii}) moir\'e materials formed from graphene, and
({\textit{iv}) moir\'e materials formed from transition metal dichalcogenides.
In this Article, we review the physical mechanisms responsible for each class of QAH insulator, 
highlighting both differences and commonalities, and comment on potential applications of the QAH effect. 
\begin{description}
\item[DOI]
10.1103/RevModPhys.xx.xxxxxx
\end{description}
}
\end{abstract}

\maketitle

\tableofcontents

%
%
%
%

\section{Introduction}
\label{Sec:intro}


The quantum Hall (QH) effect refers to quantization of the Hall conductance of a material at integer multiples of $e^2/h$, a value that is 
dependent only on two fundamental physical constants, the electron charge $e$ and the Planck constant $h$.
\am{Precision} QH effect measurements \am{currently} achieve deviations from
exact quantization of 1 part in 10$^{10}$ \cite{schopfer2007testing}, an accuracy that allows the QH effect to play an essential role in metrology \cite{jeckelmann2001quantum,poirier2009resistance}.  
The QH effect was first observed \cite{klitzing1980new} 
in semiconductor quantum well two-dimensional (2D) electron gas systems in  
strong perpendicular magnetic fields that cause 
electrons to execute closed cyclotron orbits with quantized kinetic energy.
The quantum mechanical problem of \am{independent} electrons in a magnetic field was first addressed by 
Lev Landau \cite{landau1930diamagnetismus,landau2013quantum},
and the set of orbitals with a discrete allowed kinetic energy is known as a Landau level.  

Although early explanations of the QH effect were based on the specific properties of Landau levels, 
one could ask if the phenomenon of precise quantization of Hall resistance could also occur, at least in principle,
in the absence of a magnetic field. An affirmative answer to this question was supplied early in the 
development of the theory of the QH effect.
In 1982, Thouless, Kohmoto, Nightingale, and \czchang {den} Nijs (TKNN) \cite{thouless1982quantized} derived a formula that 
related the Hall conductivity of a 2D crystalline insulator to
the integral of the Bloch state momentum-space 
Berry curvature \cite{xiao2010berry} over the finite area toroidal momentum space (the Brillouin zone) of a \am{2D} crystal.
For each band, the integral is always an integer multiple of $2 \pi$, and the integer is
the Chern number - a topological index that classifies how the Bloch wavefunction depends on 
crystal momentum.  It is truly remarkable that this abstract mathematical quantity is observable experimentally
simply by performing a very standard transport measurement.  

Since momentum space is finite only for electrons in a periodic potential, the TKNN formula 
does not apply directly to the Landau levels of electron gas systems.
In the TKNN paper, the formula was applied to a single-band square-lattice tight-binding model with an external magnetic field. Commensurability \cite{macdonald1983landau} between the crystal unit cell area and 
the area that encloses one quantum of magnetic flux then leads to fractal energy spectra often referred to as 
Hofstadter butterflies \cite{hofstadter1976energy}.  Initially, the possibility  
that Chern numbers could in principle be non-zero \am{even} in the absence of a magnetic field
was not emphasized. This property was first explicitly highlighted in an important 1988 paper \cite{haldane1988model} by Duncan Haldane, in which 
he constructed a toy model of electrons on a honeycomb lattice \czchang{with zero spatially averaged perpendicular field whose bands nevertheless
have non-zero Chern numbers.}
Crucially, the bands of the model, like those of any magnetic crystal, do break time-reversal (TR) symmetry.

A band insulator that exhibits non-zero integer Hall conductance but 
preserves lattice translation symmetry is called a Chern insulator, or sometimes a Hofstadter-Chern insulator if a magnetic field is present.
The terms Chern insulator and quantum anomalous Hall (QAH) insulator are almost identical in usage, although the
term QAH insulator 
is sometimes used to refer to the zero magnetic field subset of Chern insulators.
In this Article, we will adopt this terminology.  The adjective anomalous for a QH effect without 
a magnetic field is adopted from the terminology used to describe the Hall effects of three-dimensional(3D) magnetic 
conductors in the absence of a magnetic field, which are generically non-zero as \am{already} 
discovered by Edwin Hall \cite{hall1880new} in the nineteenth century.  

The underlying microscopic mechanism of Hall's bulk anomalous Hall (AH) effect was controversial for more than a century. 
The modern Berry phase theory of the AH effect \cite{nagaosa2010anomalous,xiao2010berry} in magnetic 
conductors, and interest in the role of momentum-space Berry phases in transport more generally, 
can be traced in part to the TKNN paper.  \am{It is now clear that the intrinsic Berry phase contribution to the AH conductivity,
proportional to the Chern number summed over occupied bands and 
in the case of 2D insulators the only contribution to the Hall conductivity, is often also important in metals.}  
In the metal case, the integral of the Berry curvature, which is taken over occupied electronic states only, 
is not the only contribution to the Hall effect, and is not quantized.

The TKNN formula tells us that the Hall conductance in 2D magnetic insulators and semiconductors 
is always quantized.  Unfortunately, the quantized value is almost always zero \am{in the absence of a magnetic field}.
For several decades, little progress was made toward experimental realization of the QAH effect \cite{onoda2003quantized,nagaosa2010anomalous}. 
Prospects improved around 2006 with the discovery of 
topological insulators (TIs)
 \cite{kane2005quantum,bernevig2006quantum1,bernevig2006quantum,konig2007quantum,fu2007topological,fu2007topologicala,roy2009topological,moore2007topological,xia2009observation,zhang2009topological,chen2009experimental},
 which were to provide a natural platform.
A TI possesses a topologically non-trivial electronic band structure protected by 
TR symmetry \cite{hasan2010colloquium,qi2011topological}, and has topologically protected boundary modes. Because 2D TIs have a quantized spin Hall effect, opposite sign Hall effects in opposite spin sectors, 
it was natural to consider \cite{qi2006topological} what could be achieved by adding magnetism to TIs.
The first theoretical proposal was to dope Mn ions into HgTe/CdTe quantum wells, a 2D TI system \cite{liu2008quantum}. 
Unfortunately, HgTe remains paramagnetic when Mn-doped and does not break TR symmetry. 
A weak magnetic field ($\sim$0.07 T) is therefore necessary to drive the system into a quantized state \cite{budewitz2017quantum}. 
Soon after, a related strategy was proposed \cite{qi2008topological,yu2010quantized}, namely
introducing magnetic dopants (such as Cr or Fe) into the 3D TI (Bi/Sb)$_2$(Se/Te)$_3$ [Fig. \ref{fig2:CrTI}(a)]. This approach led
to successful realization of the QAH effect in Cr-doped (Bi,Sb)$_2$Te$_3$ films in 2013 \cite{chang2013experimental}, and in 
V-doped (Bi,Sb)$_2$Te$_3$ films in 2015 \cite{chang2015high}. 
Since then, the QAH effect and the physical properties of Cr- or V- doped (Bi,Sb)$_2$Te$_3$ have been 
extensively studied by experimental groups worldwide. 

Despite impressive progress \cite{chang2013experimental,chang2015high,mogi2015magnetic,ou2018enhancing}, 
the magnetic doping approach has the clear disadvantage that it inevitably degrades the sample quality and thus limits the critical temperature of the QAH state. It is therefore desirable to realize the QAH effect in materials with intrinsic magnetism. 
This goal was recently achieved in manually exfoliated MnBi$_2$Te$_4$ flakes \cite{deng2020quantum}. \am{Separately, the QAH effect has now also been achieved \cite{sharpe2019emergent,serlin2020intrinsic,chen2020ferromagnetism} in moir\'e superlattice systems \cite{andrei2021marvels} that do not contain magnetic elements, including twisted bilayer graphene (TBG) \cite{sharpe2019emergent,serlin2020intrinsic}, ABC trilayer graphene  on 
hexagonal boron nitride (\textit{h}-BN) \cite{chen2020ferromagnetism}, and 
transition metal dichalcogenide (TMD) heterobilayers \cite{li2021quantum}. In moir\'e superlattices}, the AH effect is driven not by local-moment 
spin magnetism in combination with spin-orbit coupling (SOC), 
but instead by unusual purely orbital magnetic states.

Early developments in the theory of the QAH effect, and the work responsible for the effect's
first experimental realization in magnetically doped TIs 
have both been discussed in a number of review papers \cite{chang2016quantum,he2018topological,weng2015quantum,tokura2019magnetic,liu2016quantum}.
The aim of this Article is to provide a detailed description of the QAH effects that have been realized 
in magnetically doped TIs, the intrinsic magnetic TI MnBi$_2$Te$_4$, and moiré materials (Sec. \ref{Sec:II}).   
We seek to identify important features that are unique to each class of 
systems (Sec. \ref{Sec:MTI} to Sec. \ref{Sec:tbg}), and also
to identify commonalities.  
We will also discuss potential applications of the QAH effect in quantum information devices, spintronics, 
and metrology and provide an outlook on future research directions (Sec. \ref{SectionVfutures}). 

\section{Physical Mechanisms of the QAH Effect}
\label{Sec:II}
\subsection{Dirac models and Berry curvature}
\label{Sec:II_Nontrivial}

\subsubsection{The two-dimensional Dirac equation}
\label{Sec:II1berry}
A common feature of all the QAH systems that are established as of this writing, magnetically doped TI films, films of the intrinsic magnetic TI MnBi$_2$Te$_4$, magic-angle TBG, \am{ABC trilayer graphene on \textit{h}-BN,} and TMD moirés, is \am{adiabatic connection to a limit in which} 
the band states close to the Fermi level can be described by 2D 
massive Dirac equations.  
We therefore use this model as a springboard for our discussion.
The 2D massive Dirac Hamiltonian is
\eq{\label{eq:Dirac_1}
H_{\rm{D}}=\hbar v_{{\rm{D}},x}k_x \sigma_x+\hbar v_{{\rm{D}},y}k_y \sigma_y+m\sigma_z
}
where $\sigma_{x,y,z}$ are Pauli matrices, 
${\bf k}$ is the electron momentum, \addCXL{ $v_{{\rm{D}},x(y)}$ is the Dirac velocity along the {\it x} ({\it y}) direction}, 
and $m$ is a mass parameter that characterizes the energy gap. The 2D Dirac model is not periodic in momentum and is therefore not a crystal Hamiltonian.
When applied to crystalline electronic degrees of freedom, it is intended to apply only in small isolated portions 
of the Brillouin zone (BZ) with large Berry curvatures, and the zero-of-momentum is chosen to be at the center of that region.
The two-level degrees of freedom that the Pauli matrices 
in $H_D$ act on are defined by the two bands that cross at ${\bf k}=0$
\am{when $m=0$}. 


\begin{figure}[hbt!]
    \centering
    \includegraphics[width=3.5in]{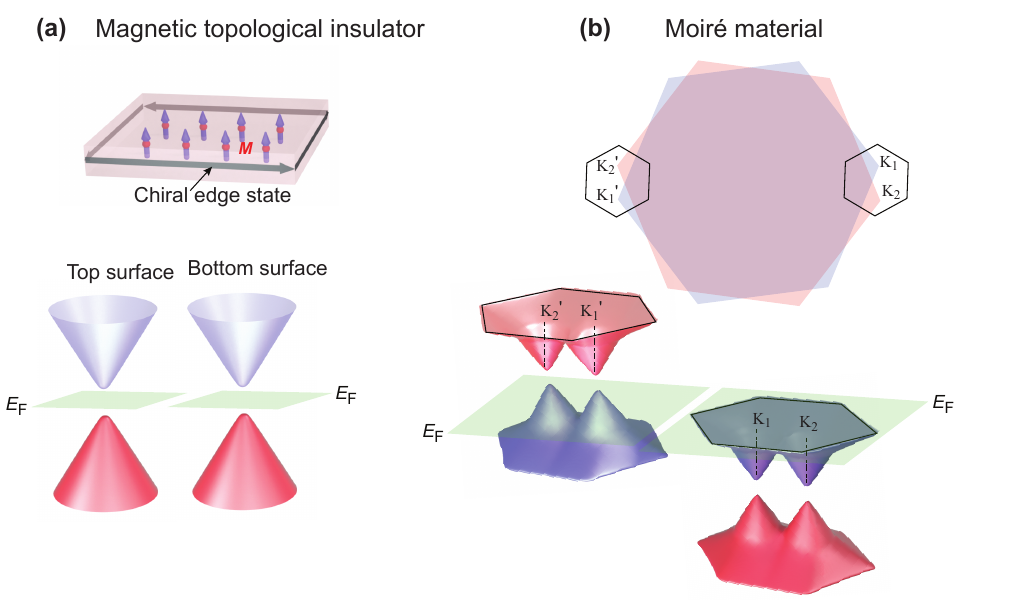}
    \caption{QAH effects in magnetic TI and moir\'e superlattice systems.  
    In both cases gapped Dirac cones contribute $\sigma_{xy}$ = $\pm$ \textit{e}$^2$/2\textit{h} to the Hall conductance.
    The sign of the contribution is distinguished by the color (red or blue) of the lower energy occupied band.
    The Fermi level (green \czchang{plane}) must be in a gap for the Hall effect to be quantized.
    (a) Schematic of a magnetic TI thin film. The TR symmetry of a TI is broken by the appearance of ferromagnetism in the sample. 
    The total Hall conductance of a magnetic TI film is $\sigma_{xy}$ = (1/2+1/2) \textit{e}$^2$/\textit{h}= \textit{e}$^2$/\textit{h}  
    when the two surfaces have parallel magnetization directions.  (b) The BZ and moir\'e BZ of a triangular lattice 
    moir\'e material.  The moir\'e minibands located near the $K$ and $K'$ valleys are related by TR.  
    A QAH effect occurs when the two 2D Dirac cones in the same valley contribute half-quantized Hall conductances 
    of the same sign and the moir\'e miniband filling factors in the two valleys differ. \czchang {The green planes in (a) and (b) are the Fermi surfaces.}
    }
    \label{fig1:overview}
\end{figure}

The electronic momentum-space Berry curvature in a crystal is defined in terms of the 
dependence of the Bloch wavefunctions on momentum:
\eq{\label{eq:BerryCurvature}
\Omega_{xy,s}({\bf k})=i(\langle \partial_{k_x}u_s|\partial_{k_y}u_s\rangle-\langle \partial_{k_y}u_s|\partial_{k_x}u_s\rangle),
}
where $s$ is a band index, and $u_{s,{\bf k}}$ is a Bloch state eigen-wavefunction. 
The famous TKNN formula states that the Hall conductivity $\sigma_{xy}$ in a band insulator is related to the Chern number $C$ by
\eq{\label{eq:HallConductivity}\sigma_{xy}=\frac{e^2}{h}C,}
\am{where} the Chern number
\eq{\label{eq:ChernNumber}C=\frac{1}{2\pi}\int_{\rm BZ} d^2k \sum_{s\in {\rm{occ}}} \Omega_{xy,s}({\bf k})} 
is an integer. Here the sum is over occupied bands and
${\rm BZ}$ specifies that the integral if over the entire BZ.

Chern numbers are particularly easy to evaluate in two-band models since the Berry curvature is then proportional \cite{auerbach2012interacting} 
to the area enclosed on the two-level Bloch sphere per area enclosed in momentum space.
It follows that the Berry curvature of the 2D Dirac Hamiltonian (Eq. \ref{eq:Dirac_1}) is
\eq{\Omega_{xy,s}=-\frac{s m \hbar^2 v_{{\rm D},x}v_{{\rm D},y}}{2\left[(\hbar v_{{\rm D},x} k_x)^2+(\hbar v_{{\rm D},y} k_y)^2+m^2\right]^{3/2}},}
where $s=\pm $ is a band label that specifies the sign of 2D Dirac band energy, 
and $E_s=s\sqrt{(\hbar v_{{\rm D},x} k)^2+(\hbar v_{{\rm D},y} k_y)^2+m^2}$. 
From this expression, one can see that the Berry curvature in the 2D Dirac equation is concentrated within a 
momentum-space area proportional to \am{$(m/\hbar v_{\rm D})^2$
(where $v_{\rm D} \sim v_{{\rm D},x}\sim v_{{\rm D},y}$)}, 
and that it decays as ${|\bf k|}^{-3}$ for large ${|\bf k|}$. 
The Hall conductivity $\sigma_{xy}$ cannot be 
directly determined from the 2D Dirac model since the former requires an integral over the 
whole BZ [Eqs. (\ref{eq:HallConductivity}) and (\ref{eq:ChernNumber})], including \am{momenta} outside the Dirac model's range of 
validity. This effective model approach is nevertheless 
useful in understanding the topological properties of realistic materials that have a small value of $m$.
\am{In that case} we can choose a momentum space cut-off $\Lambda$ that is simultaneously large compared to 
$m/\hbar v_{\rm D}$, and therefore large enough to capture nearly all the Berry curvature integral,
and small compared to the size of BZ, and therefore within the range of validity of the 2D Dirac model.
In this limit, the Hall conductivity in Eq. (\ref{eq:HallConductivity}) $\sigma_{xy} \to \textit{sign}\left(v_{{\rm D},x}v_{{\rm D},y}m\right)(e^2/2h)$.
Each 2D Dirac Hamiltonian therefore contributes  $\pm (e^2/2h)$ to the Hall conductivity; the band states cover half the Block sphere 
starting from a pole position at $\bf{k}=0$, circling the sphere when $\bf{k}$ executes a circle and moving toward the equator as 
$|\bf{k}| \to \infty$. The sign of the Hall contribution is determined by the relative signs of $v_{{\rm D},x}$, $v_{{\rm D},y}$ and $m$.

\addCXL{The TKNN paper tells us that the total Hall conductivity in any 2D crystal must be an integer multiple of $e^2/h$.  
When level crossings in a 2D crystal
are weakly gapped by TR symmetry breaking, so that
almost all the Berry curvature contributions are associated
with 2D Dirac fermions, 
the integer can be determined simply by summing half-quantized contributions of variable sign.  It follows that 
the number of local Dirac Hamiltonian [Eq. (\ref{eq:Dirac_1})] 
across the BZ must be even.   
This property can be viewed as a 2D version \cite{semenoff1984condensed,fradkin1986physical,haldane1988model,fang2019new} of 
the fermion-doubling theorem\cite{nielsen1981absence1,nielsen1981absence2} 
for Weyl points in 3D crystals with TR symmetry.}

\begin{figure}[hbt!]
    \centering
    \includegraphics[width=3.5in]{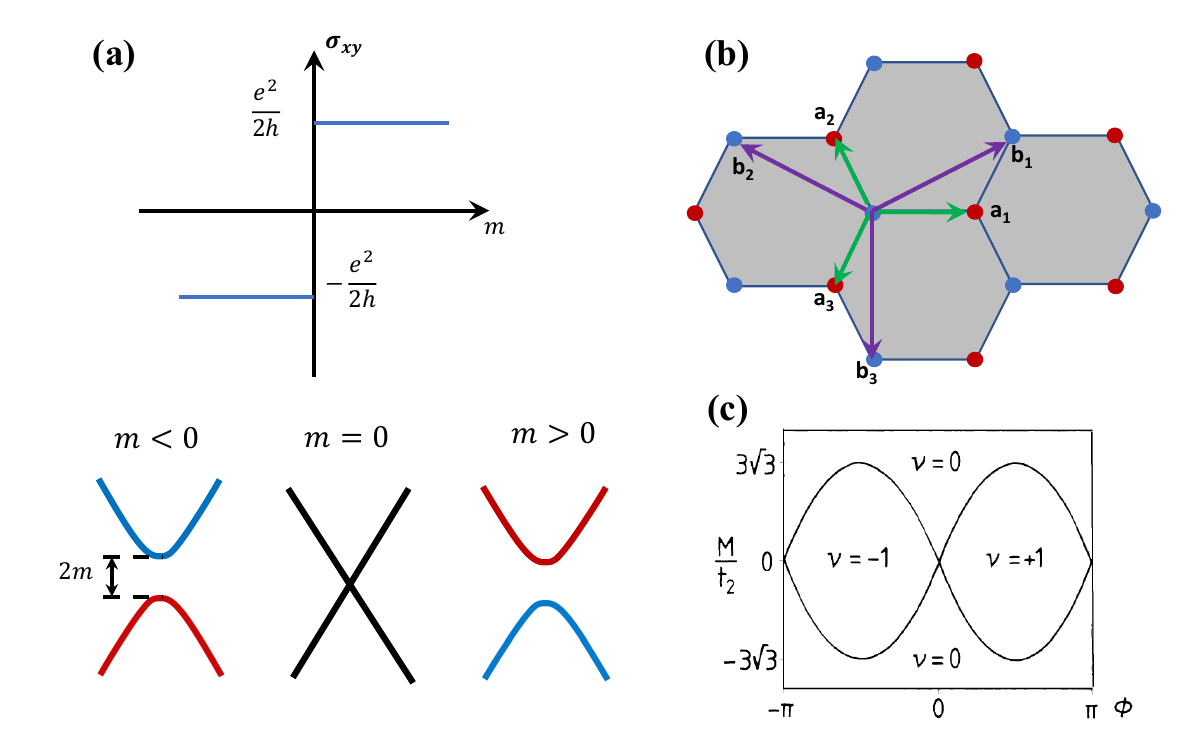}
    \caption{The Haldane model for the QAH effect. (a) The Hall conductivity $\sigma_{xy}$ as a function of the mass $m$ in 2D Dirac equation. 
    When $m$ changes from a negative value to a positive value, $\sigma_{xy}$ jumps from $-\frac{e^2}{2h}$ to $\frac{e^2}{2h}$. 
    (b) The honeycomb lattice of the Haldane model. Here ${\bf a}_i$ and ${\bf b}_i$ respectively label the NN and NNN translation vectors. 
    (c) The phase diagram of the Haldane model.$\nu$ is the Chern number. From \onlinecite{haldane1988model}. }
    \label{fig1:Haldane}
\end{figure}

\addCXL{The relationship of 2D Dirac fermions to realistic QAH systems,
summarized in Fig.\ref{fig1:overview} (see next two subsections).  Here we consider a toy model, the Haldane model - which was the first theoretical model for the QAH effect,}
was constructed by adjusting the electronic structure of graphene \cite{haldane1988model} 
in a way that was mainly of academic interest at the time.
The low-energy electronic structure of graphene \cite{castro2009electronic} can be described using 
a single $\pi$-band tight-binding model on a honeycomb lattice with the nearest 
neighbor (NN) hopping between A and B sub-lattices.  This model yields gapless Dirac cones centered on the $K$ and $K'$
BZ corners.  In order to demonstrate that the QAH effect was a realistic experimental possibility, Haldane added artificial 
complex next-nearest neighbor (NNN) 
intra-sublattice (A to A and B to B) hopping terms,
illustrated in Fig. \ref{fig1:Haldane}(b), 
to the graphene tight-binding model.  The complex hopping terms 
break TR symmetry and can be viewed as being induced by a magnetic field that has the periodicity of the lattice and 
a vanishing spatial average.  When both NN and NNN hopping are included, the modified tight-binding model Hamiltonian,
\begin{widetext}
\eq{ 
H({\bf k})=t_1\sum_{i\in {\rm{NN}}}\left(\cos({\bf k\cdot a}_i)\sigma_x+\sin({\bf k\cdot a}_i)\sigma_y\right)+2t_2 \cos\phi \sum_{i\in \rm{NNN}} \cos({\bf k\cdot b}_i) +\left(M-2t_2\sin\phi \sum_{i\in \rm{NNN}}\sin({\bf k\cdot b}_i)\right)\sigma_z,
\label{eq:HaldaneHamiltonianExplicit}}
\end{widetext}
preserves the original translational symmetry.  In Eq.~\ref{eq:HaldaneHamiltonianExplicit}, 
$\sigma_{x,y,z}$ are three Pauli matrices for the A and B sublattices, $t_1$ is the NN hopping parameter,  
$t_2$ and $\phi$ are the amplitude and phase of the NNN hopping parameter, 
$M$ accounts for a possible on-site energy difference between A and B sublattices, 
and ${\bf a}_i$ and ${\bf b}_i$ are the lattice vectors between the NN and NNN sites, respectively. In the limit $t_2, M\ll t_1$, the low-energy physics will be completely determined by the 2D Dirac fermions located at K and K'. We then can expand the \czchang{tight binding} Hamiltonian around K and K' as
\eq{ 
H_{\bf K (K')}= \mp \frac{3t_1}{2}(\kappa_x\sigma_x\pm \kappa_y\sigma_y)+(M\pm 2t_2\cos\phi)\sigma_z,
}
where a constant term can be absorbed into the zero of energy.
Comparing with Eq.(\ref{eq:Dirac_1}), this Hamiltonian can be recognized as a massive Dirac model with velocities
$\hbar v_{{\rm D},x}=\pm \frac{3}{2}t_1$ and $\hbar v_{{\rm D},y}=-\frac{3}{2}t_1$, and mass $m=M\pm 2t_2\cos\phi$. 
Therefore, in the limit $M,t_2\ll t_1$, the Hall conductance $\sigma_{xy}$ can be directly read from the relative signs of $v_{D,x(y)}$ 
and $m$. Since the sign of $v_{{\rm D},x}v_{{\rm D},y}$ is opposite for the Dirac cones at ${\bf K}$ and ${\bf K}'$, 
the contribution to the Hall conductance $\sigma_{xy}$ of the Dirac cones at ${\bf K}$ and ${\bf K}'$ will share the same sign when 
$t_2$ dominates, but will be opposite if \am{the}
$M$ term is dominant. A phase transition [Fig. \ref{fig1:Haldane}(a)] occurs at $M=\pm 2t_2\cos\phi$. 
(The $t_2=0$, $M\ne0$ version of this model is realized by graphene on aligned \textit{h}-BN, as we have discussed previously.) The model has a QAH state with $\pm e^2/h$ Hall conductivity when $|2t_2\cos\phi|>M$, as shown in Fig. \ref{fig1:Haldane}(c). 

The Haldane model can describe some TMD moir\'e homobilayers.  When the two valence band tops in a TMD bilayer are close in energy, 
interlayer hybridization becomes important 
and the one-band Hubbard model \cite{wu2018hubbard} of typical heterobilayers must be decorated with a \am{two-level} layer 
degree-of-freedom that acts much like the sublattice degree-of-freedom in the Haldane model.  The orbitals in 
each layer can be approximated by triangular lattice Hubbard models, with the lattice points located at the position where the 
metal atoms of one layer are directly above or below chalcogen atoms of the other layer.  In bilayers, the two
triangular lattice models combine to form a honeycomb lattice, and the one-band tight-binding model 
generalizes to a $2 \times 2$ tight-binding model.  Band crossings at the moir\'e BZ corners
are avoided because the site-diagonal terms in the Hamiltonian favor one layer over the other.
When opposite layers are favored at opposite moir\'e BZ corners, as happens in the heterobilayer case \cite{wu2019topological,pan2022topological}, 
the bilayer is described by a moir\'e superlattice length scale version of the Haldane model 
and the valley-projected bands carry non-zero Chern numbers.  

\subsubsection{Magnetic TI films}
\label{Sec:magnetic}
The discussion in this subsection applies to both thin films of both magnetically doped TIs and the intrinsic magnetic TI MnBi$_2$Te$_4$.
\addCXL{We direct readers who are not familiar with the concept of TIs to 
Refs.~\onlinecite{qi2011topological,hasan2010colloquium}.}
In the prototype (Bi/Sb)$_2$(Se/Te)$_3$ TI family, 
the form \addCXL{for one topological surface state (TSS)} is particularly simple and described by the effective Hamiltonian 
\eq{\label{eq:HTSS}
H_{\rm{sur}}=\hbar v_{\rm{D}} (k_x\sigma_y-k_y\sigma_x),
}
where the Pauli matrices $\sigma_i$ act on electron spin, \addCXL{or more strictly speaking a pseudospin formed by the ${\bf k}=0$ Kramers doublet. }
This effective Hamiltonian $H_{\rm{sur}}$ coincides with Eq. (\ref{eq:Dirac_1}) with $m \to 0$, after performing the spin rotation $\sigma_y\rightarrow \sigma_x$ and $\sigma_x\rightarrow -\sigma_y$. 
Note that $H_{\rm{sur}}$ has the same form as the SOC terms in the Rashba
Hamiltonian \cite{bychkov1984properties} that describes 2D electron gases in semiconductor quantum wells 
and the surface states of certain heavy metals \cite{koroteev2004strong,krupin2005rashba}. 
This similarity reflects the common SOC origin of the spin-splitting. 
Different from ordinary Rashba bands, one branch of TSSs merges with the bulk conduction band at 
large momentum while the other branch merges with the valence band. 
When the Fermi energy is located in the bulk gap of a TI, there is only a single 
spin-momentum locked helical Fermi surface for one TSS. 
Ordinary Rashba bands have $k^2$-terms in their Hamiltonians that dominate at large momenta
so that there are always two Fermi surfaces with opposite helicities. 

The gapless nature of the TSSs ($m=0$ in the 2D Dirac equation) 
is protected by TR symmetry and Kramers' theorem, which in solid states 
that an eigenstate at the momentum ${\bf k}$ has the same energy as an orthogonal time-reversed eigenstate  
at the momentum $-{\bf k}$.  
The energy spectrum of $H_{\rm{sur}}$ clearly satisfies the requirements of Kramers' theorem. 
At TR-invariant momenta, Kramers' theorem requires double degeneracies, $E_{s}(0)=E_{-s}(0)$.  
Thus the band touching point, called the Dirac point, between two branches of $H_{\rm{sur}}$ is at a TR-invariant 
momentum. Inducing magnetization in a TI produces exchange coupling between the ordered magnetic moments and 
the TSS spins, and introduces an additional non-zero mass term $H_{\rm{ex}}=m\sigma_z$ 
in $H_{\rm{sur}}$ when the magnetization has an out-of-plane component.
According to the discussion in Sec. \ref{Sec:II1berry}, the Dirac fermions at each surface are described by a 2D Dirac Hamiltonians $H_{\rm{sur}}+H_{\rm{ex}}$ that together gives rise to a contribution of $\pm e^2/2h$ to the Hall conductivity. 
This property was referred to as a “surface half QH effect" \cite{qi2011topological,chu2011surface,qi2008topological}.
\addCXL{The TSSs exist for both the top and bottom surfaces for a TI film,} and since both Dirac velocities have opposite signs on opposite surfaces, 
the two 2D Dirac cones contribute Hall conductivities of the same sign when their masses have the same sign,
{\it i.e.} when the magnetizations are parallel near two surfaces.
Consequently, one expects a total Hall conductivity $\pm e^2/h$ for a TI film with a uniform magnetization, as shown in Fig. \ref{fig1:overview}(a), where the $\pm$ sign depends on the magnetization direction.  Since the QAH effect requires only 
surface magnetization, the role of the bulk magnetic order is simply to lock two surface magnetizations into a 
parallel configuration.


\subsubsection{Moir\'e materials}
\label{Sec:moirematerial}
After the first realization of the QAH effect in Cr-doped (Bi,Sb)$_2$Te$_3$ \cite{chang2013experimental}, the search for new 
QAH materials focused on compounds with heavy elements that can host both strong SOC and local magnetic moments.
Unexpectedly, the QAH effect was however next discovered  \cite{sharpe2019emergent,serlin2020intrinsic}  
in a system composed entirely of carbon atoms, light elements with very weak SOC, and no magnetic moments, twisted bilayer graphene (TBG).
The TBG case represents a new route to realize the QAH effect in which emergent translational symmetries,
orbital magnetic moments, and Coulomb interactions all 
play essential roles.
Related QAH effects have now been oberved 
in \czchang {ABC} trilayer graphene  on \textit{h}-BN \cite{chen2019topological} and TMD bilayer 
moir\'e superlattices\cite{li2021quantum}.

Moir\'e patterns are formed in few-layer van der Waals thin films by overlaying layers that have 
different lattice constants or different orientations\cite{andrei2020graphene,andrei2021marvels}.  
When the host 2D materials are 
semiconductors or semimetals and the period of the moir\'e pattern is long compared to the lattice constants of the host 2D crystals, 
the moir\'e pattern gives rise to low-energy effective Hamiltonians with 
an emergent translational symmetry that has the periodicity of the moir\'e pattern - {\it i.e.} to 
artificial crystals (moir\'e materials) with lattice constants on the $\sim$10 nm scale.
TBG and group VI TMD bilayers are the prototypical moir\'e materials.
In graphene and most TMDs, the host materials have triangular Bravais lattices and 
low-energy states have crystal momenta in valleys centered on one of two inequivalent BZ-corner momenta [$K,K'$ in Fig. \ref{fig1:overview}(b)]
that are related to each other by TR symmetry.  
The valley flavor is a good quantum 
number in the emergent moir\'e band Hamiltonian, and the valley projected Hamiltonians are not in general TR invariant.
It follows that TBG and TMD moir\'e bands can have non-zero Chern numbers.

The low-energy valley-projected 
Hamiltonian of an isolated graphene layer is the massless Dirac model of Eq.~\ref{eq:Dirac_1}.
The mass $m$ can be generated \cite{giovannetti2007substrate,quhe2012tunable,yankowitz2012emergence,san2014spontaneous}
by interaction with an adjacent layer of \textit{h}-BN - the material that is now nearly universally
employed to isolate graphene multilayer systems chemically.
A finite Dirac mass $m$ can also be generated spontaneously 
by electron-electron interactions \cite{min2008pseudospin,xue2018time,bultinck2020mechanism} when the Dirac velocities are sufficiently low.
Because of its emergent moir\'e bands, gaps can appear in the spectrum not only at neutrality but also at electron densities 
that are multiples of $4/A_{\rm M}$, where $A_{\rm M}$ is the moir\'e pattern unit cell area and the factor of four accounts for 
spin and valley degeneracy. 
In electrically neutral twisted bilayers, the separate half QH effects 
of the two layers add.
TBG therefore normally has a quantum valley Hall effect \cite{xiao2007valley,andrei2020graphene,liu2021orbital,liu2019quantum,martin2008topological,li2018valley,li2016gate}, 
opposite QH effect contributions from opposite valleys
at neutrality.
The quantum valley Hall effect is promoted to a true charge QAH effect when the
the Fermi level lies in a Dirac gap for an odd number of spin/valley flavors 
as depicted in Fig. \ref{fig1:overview}(b). 
The origin of the QAH effect is therefore spontaneous valley polarization, 
a purely orbital type of TR symmetry breaking that also leads to magnetization with a dominantly 
orbital nature \cite{zhu2020voltage,liu2021magnetic}. 

In TMD heterobilayers, electrons and holes are often strongly confined to one layer or the other.
They can then be described \cite{wu2018hubbard} by single-band Hubbard models with an approximately
parabolic kinetic energy operator and an external potential with moir\'e periodicity. Since the Hubbard model is TR invariant, the valley projected bands do
not carry non-zero Chern numbers when this approximation is valid. The situation changes in homobilayers and heterobilayers, in which gate voltages have been adjusted to favor strong hybridization between 
layers \cite{wu2019topological,pan2022topological,li2021quantum}.  
Under these circumstances, the valley projected 
bands have a layer degree-of-freedom that acts as a pseudospin and allows each valley to provide 
a physical realization of the Haldane model, which is 
the first theoretical model of the QAH effect and is discussed below.
The QAH effect was recently discovered in a TMD heterobilayer system \cite{li2021quantum} in which the layer degree of freedom is activated in
by applying large gate-controlled transverse displacement fields.

\subsection{Physical mechanisms of magnetism}
\label{Sec:Magnetism}

\subsubsection{Magnetically doped TIs}

Magnetically doped TIs can be viewed as a special type of diluted magnetic semiconductor
\cite{sato2010first,jungwirth2006theory,dietl2014dilute,zener1953exchange,van1953models} with strong SOC. 
A qualitative understanding of the magnetic properties of these systems can be obtained by 
separating the whole system into two subsystems, 
topologically non-trivial band electrons and local magnetic moments from magnetic doping. 
The two subsystems are coupled to each other, through interactions that are also (confusingly) typically
referred to as exchange interactions and have the form 
$H_{\rm ex}=J_{\rm ex}\sum_i{\bf S}_i\cdot{\bf s}({\bf R}_i)$ where ${\bf s}({\bf r})$ is the electron spin operator at 
position ${\bf r}$, 
and ${\bf S}_i$ is the spin of the $i$-th magnetic impurity.
Here $J_{\rm ex}$ is the exchange coupling parameter.  The microscopic mechanism of this coupling will be discussed below.

Provided that the magnetic moments are uniformly distributed on characteristic electronic 
length scales, $\sum_i{\bf S}_i \cdot{\bf s}({\bf R}_i)$ can be replaced by 
$n_{\rm M} \int d{\bf R} \, {\bf S}({\bf R}) \cdot s(\bf{R})$, where $n_{\rm M}$ is the local moment density.
When this interaction is treated in a mean-field approximation which can be traced to an early paper 
by Zener \cite{zener1951interaction}, the band spins see a mean magnetic 
field (in enegy units) $ -J_{\rm ex} n_{\rm M} \langle {\bf S} \rangle$ and the local moments see a magnetic field 
${\bf H} = -J_{\rm ex} \langle {\bf s} \rangle$, where $\langle {\bf S}\rangle$ and $ \langle {\bf s} \rangle$ are the thermal averages
of the local moment spins and the band electron spin-densities.  Given the effective magnetic field $\langle {\bf S}\rangle$
can be calculated from the text-book formula for polarization of an isolated local moment in a thermal bath,
$\langle {\bf S} \rangle = S \hat{{\bf H}} B_S(S|{\bf H}|/k_BT)$, where $B_s(x)$ is the Brillouin function.
We can eliminate $\langle {\bf s} \rangle$ from these mean-field equations by assuming that the band electron spin-response is 
linear, {\it i.e.} that $\langle {\bf s}\rangle=J_{\rm ex}N_{\rm M} \chi_{s}\langle {\bf S}\rangle$, where $\chi_{s}$ is electron spin susceptibility.
The mean-field critical temperature is determined by solving
self-consistently for ${\bf S}$ and yields a critical temperature proportional to $\chi_{s}$.  
This mean-field theory is accurate when \cite{jungwirth2006theory} the momentum-dependent band spin-susceptibility is strongly peaked 
near zero-wave vector uniform response.  

In the Zener picture, the ferromagnetic critical temperature is proportional to the square of the exchange coupling 
$J_{\rm ex}$ and the first power of the electron spin susceptibility $\chi_s$. 
For transition metal local moments, the exchange coupling $J_{ex}$ originates from hybridization between
metal \textit{d}-electrons and the \textit{p} orbitals of the host compounds \cite{zener1951interaction}.
Strong hybridization between the \textit{p}-orbitals of Te atoms and the \textit{d}-orbitals of transition metal magnetic 
impurities has been demonstrated experimentally in Cr- or V-doped (Bi,Sb)$_2$Te$_3$ \cite{tcakaev2020comparing}. 
When the band system is metallic, the electron spin susceptibility is usually dominated by the Fermi level Pauli response
which induces Ruderman-Kittel-Kasuya-Yosida (RKKY) type \cite{ruderman1954indirect,kasuya1956theory,yosida1957magnetic}) of interactions 
between the local moments.  However, the QAH effect requires the band system to be insulating.  
Because the electron spin-operator has large inter-band matrix elements in systems 
with strong SOC, the electron spin susceptibility $\chi_s$ can be non-zero and large when gaps are small.
This mechanism for induced coupling between local moments is known as the van Vleck mechanism \cite{van1932theory} (or Bloembergen-Rowland mechanism \cite{bloembergen1955nuclear}). Therefore, the van Vleck mechanism can give rise to ferromagnetism in insulating magnetically doped TIs \cite{yu2010quantized,li2015experimental}, 
and in this way allows the realization of the QAH effect.

The situation in real materials is, as usual, more complex.  The magnetic properties of magnetically doped TIs are 
sensitive to chemical composition, the types, and locations of magnetic impurities and other unintended defects, 
the energies of the \textit{d}-orbitals of the magnetic impurities, and even external gate voltages.
Motivated by the observation that insulating ferromagnetism survives in density functional theory (DFT) calculations even when SOC is set to zero, it has been suggested \cite{kim2017ordering} that \am{the segregation of magnetic dopants on the group III sites of (Bi/Sb)$_2$(Se/Te)$_3$ 
plays a crucial role in favoring magnetic order.} 
Furthermore, a finite carrier concentration is normally present in the (Bi/Sb)$_2$(Se/Te)$_3$ TIs
and, along with carriers in the TSSs, can induce a ferromagnetic RKKY interaction on surfaces \cite{liu2009magnetic}. 
Moreover, \textit{d}-orbitals may appear inside the bulk gap of TIs for several compounds, e.g. V-doped (Bi,Sb)$_2$Te$_3$ \cite{peixoto2016impurity}, and form impurity bands near $E_F$\cite{zhang2018electronic}. 
This can lead either to a ferromagnetic superexchange mechanism when $E_F$ lies in the gap 
between the lower and upper crystal-field-split majority-spin bands, similar to the case of (Ga,V)As, or to a
Zener's double exchange mechanism when a finite \textit{d}-orbital density of states appears at $E_F$, 
similar to the case of (Ga,Mn)N \cite{sato2010first,jungwirth2006theory}. 
In addition, the spatial randomness of magnetic impurities may require a magnetic 
percolation picture. Experiments have revealed super-paramagnetic dynamics, in which magnetic impurities are 
clustered to form magnetic islands with a size of tens of nanometers
and the magnetic island behaves nearly independently, in several types of magnetically doped TI compounds \cite{lachman2015visualization,chang2014chemical,grauer2015coincidence,liu2016large}.

\subsubsection{Intrinsic magnetic TIs}

MnBi$_2$Te$_4$ is a stoichiometric compound with Mn local moments that form strong bonds with neighboring Te atoms 
located at the center of weakly coupled septuple layer.
Both the higher density of Mn local moments and the absence of intended randomness 
offer potential advantages over magnetically doped TIs.
The Mn local moments have valence $+2$ by losing two $4s$ electrons,
and the remaining five $3d$ electrons give rise to an isotropic fully spin-polarized $5 \mu_B$ local moment. 
Since the majority-spin \textit{d}-levels of Mn are fully filled and the minority-spin \textit{d}-levels are completely unoccupied, 
the super-exchange mechanism plays an essential role in MnBi$_2$Te$_4$ \cite{otrokov2017highly,li2019intrinsic,sliwa2021superexchange}. 
For the super-exchange mechanism \cite{anderson1950antiferromagnetism},
virtual electron transfer between two cations through an intermediate non-magnetic anion 
couples the magnetic moments of the cations. 

The super-exchange interaction can be ferromagnetic or antiferromagnetic, 
depending on the relative angle between two cation-anion bonds, according to the Goodenough-Kanamori rules 
\cite{goodenough1955theory,kanamori1959superexchange}. 
In MnBi$_2$Te$_4$, two neighboring Mn atoms in one septuple layer are coupled through the adjacent Te atoms 
and the angle of two Mn-Te bonds is around $86^{\circ}$, close to $90^{\circ}$. 
Thus, the intra-layer super-exchange between two Mn atoms is ferromagnetic. 
In contrast, the coupling between two Mn magnetic moments in two adjacent septuple layers is mediated by several intermediate Te or Bi atoms and is antiferromagnetic type since all the cation-anion bonds are roughly along the \textit{z}-direction. 
Therefore, MnBi$_2$Te$_4$ has intrinsic A-type antiferromagnetism, namely intralayer ferromagnetism and interlayer anti-ferromagnetism. MnBi$_2$Te$_4$ is only one member of a large family of compounds Mn(Bi/Sb)$_{2n}$(Se/Te)$_{3n+1}$, in which the Mn(Bi/Sb)$_2$(Se/Te)$_4$ and (Bi/Sb)$_{2(n-1)}$(Se/Te)$_{3(n-1)}$ layers alternate and form a superlattice structure. 
More properties of the Mn(Bi/Sb)$_{2n}$(Se/Te)$_{3n+1}$ family of materials are discussed in Sec.\ref{sec:MBTproperties}.

\subsubsection{Magnetism in moiré materials}

In graphene and TMD multilayer materials, the carriers are often concentrated near the two inequivalent  
triangular lattice BZ corners, referred to as valleys, that transform into each other under TR symmetry.
For the moir\'e material, the QAH states that are known as of this writing, all have valley projected moir\'e bands 
with broken TR symmetry and non-zero Chern numbers. 
QAH states then appear when opposite valleys, have 
different total filling factors.  The mechanism that favors ground states with spontaneous valley polarization 
is itinerant electron exchange interactions, the same interactions that are responsible for ferromagnetism in many metals.
Because of the long-range nature of Coulomb interactions, itinerant electron exchange is stronger between states that are 
close together in momentum space, {\it i.e.} in the same valley.  Maximizing the magnitude of 
attractive itinerant electron exchange interactions competes with minimizing band energy, which has lesser importance 
when bands are narrow.  

Occupying states in momentum space in a way that violates TR symmetry allows broken TR symmetry observables that depend only on orbital degrees of freedom to be non-zero even when SOC is weak. 
The AH effect is an example of such a physical quantity, and can be grouped in this 
sense with orbital magnetization \cite{thonhauser2011theory,xiao2010berry,aryasetiawan2019modern} (see below and Sec.\ref{Sec:orbital}) and magneto-optical observables like the Kerr and Faraday effects.
The origin of the AH effect and the QAH effect in moir\'e materials is quite distinct from its origin in systems like magnetic TIs that rely on SOC combined with spin-ferromagnetism.

The evaluation of orbital magnetization in periodic crystals is more subtle than the evaluation of spin magnetism because 
the electron position operator is ill-defined for extended Bloch 
wavefunctions \cite{thonhauser2011theory,xiao2010berry,aryasetiawan2019modern}. As in the modern theory of charge polarization, 
orbital magnetization can be related to Bloch state Berry phases that are conveniently evaluated using tight-binding models based on  
maximally localized Wannier functions. \cite{thonhauser2005orbital,ceresoli2006orbital,bianco2013orbital}.
The connection between orbital magnetization and Bloch state Berry phases can be established by examining  
semi-classical wave-packet dynamics \cite{xiao2005berry,xiao2010berry} or using linear-response theory \cite{shi2007quantum}. 

\begin{figure*}[hbt!]
    \centering
    \includegraphics[width=7 in]{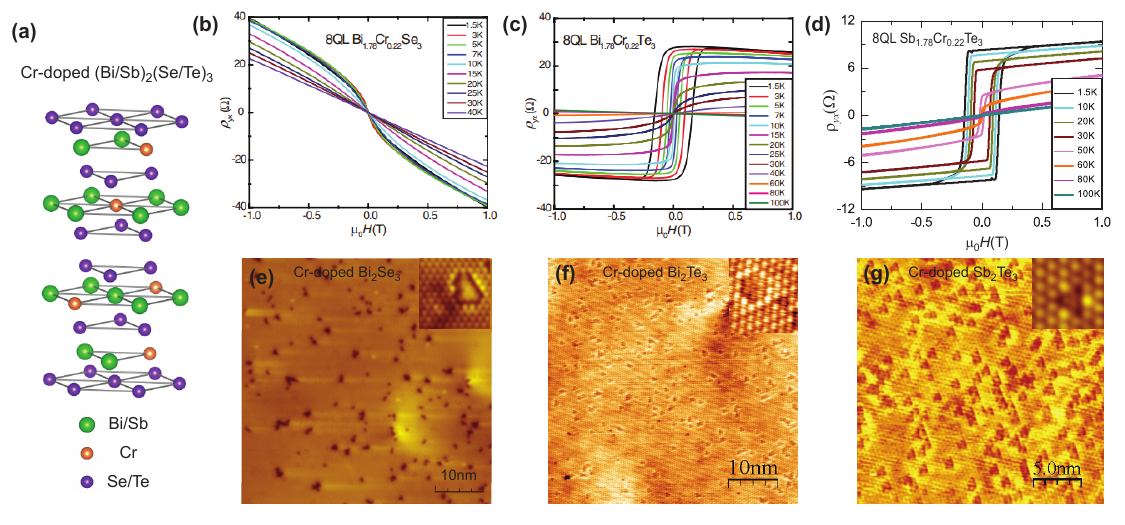} 
    \caption{Magnetically doped TI films. (a) Schematic structure of Cr-doped (Bi/Sb)$_2$(Se/Te)$_3$.(b-d) Hall traces of 8 QL Bi$_{1.78}$Cr$_{0.22}$Se$_3$ (b), Bi$_{1.78}$Cr$_{0.22}$Te$_3$ (c), and Sb$_{1.78}$Cr$_{0.22}$Te$_3$ (d) films. (e-g) STM images of Cr-doped Bi$_2$Se$_3$ (e), Bi$_2$Te$_3$ (f), and Sb$_2$Te$_3$ (g).  From \onlinecite{chang2014chemical,zhang2013topology,chang2013thin}.}
    \label{fig2:CrTI}
\end{figure*}

The general expression for the orbital magnetization ${\bf M}_{\rm orb}$ of Bloch bands 
includes two contributions: a local circulation contribution ${\bf M}_{\rm orb,L}$ that comes from atomic orbital moments and an itinerant
contribution ${\bf M}_{\rm orb,I}$ that originates from the free current flowing at the boundary of the finite system \cite{thonhauser2005orbital}.
\am {A different decomposition of the orbital magnetization into distinct partial contributions, one from self-rotation of the wave packet and the other due to center-of-mass motion, arises naturally in the semi-classical wave-packet formalism\cite{xiao2010berry}.}
The itinerant contribution ${\bf M}_{\rm orb,I}$ is of particular interest since it is directly 
related to the Hall conductivity. In Chern insulators \cite{zhu2020voltage,ceresoli2006orbital}, 
the itinerant contribution ${\bf M}_{\rm orb,I}$ is produced by currents that circulate around the sample boundary
and are related to the quantized Hall conductivity and the associated band Chern numbers $C$ via the relation,
\begin{equation}
\label{eq:edgemagnetism}
   \frac{dM_{\rm orb,I}}{d\mu}=\frac{Ce}{h},
\end{equation}
where $\mu$ is the chemical potential. In fact, most experimental measurements of the QH effect can
be viewed \cite{macdonald1995proceedings} as direct measurements of the left-hand side of Eq.~\ref{eq:edgemagnetism}.

\section{Magnetically Doped TIs}\label{Sec:MTI}
\subsection{Material properties}\label{Sec:Magneticallydoped}

Crystals in the (Bi/Sb)$_2$(Se/Te)$_3$ family share the same rhombohedral crystal structure with the space group $D_{\rm 3d}^5$ ($R\bar{3}m$) with five atoms in one layer, as shown in Fig. \ref{fig2:CrTI}(a). These five-atom layers arranged along the \textit{z}-direction are known as quintuple layers (QLs). Each QL consists of 3 Se/Te layers and 2 Bi/Sb layers. 
Among the four stoichiometric members, Bi$_2$Se$_3$, Bi$_2$Te$_3$, and Sb$_2$Te$_3$ are topologically nontrivial, whereas Sb$_2$Se$_3$ is a trivial insulator due to its weaker SOC. 
In TI films, the TSSs at the top and bottom surfaces hybridize with each other and open a gap
that shrinks exponentially with film thickness.
The gaps become unobservably small at around 6 QLs in Bi$_2$Se$_3$ \cite{zhang2010crossover,chang2011growth}, 2 QLs for Bi$_2$Te$_3$ \cite{li2010intrinsic}, and 4 QLs for Sb$_2$Te$_3$ \cite{wang2010atomically,jiang2012landau}. 
\czchang{The ferromagnetic order can be introduced in TI} via three different routes: 
(\textit{i}) dilute magnetic doping of a TI, 
i.e. introducing magnetic ions into TI materials; (\textit{ii}) discovery of single-crystalline materials that host coexisting topological and magnetic states intrinsically; and (\textit{iii}) fabrication of TI/ferromagnetic insulator heterostructures. 
Approach (\textit{i}) was the first to be successful, while the approach (\textit{ii}) has also been achieved in thin films exfoliated from the bulk intrinsic magnetic TI, MnBi$_2$Te$_4$ \cite{deng2020quantum} (See Sec. \ref{Sec:MnBi2Te4}).
Despite significant efforts, the approach (\textit{iii}) has still not been fully successful. 
A large AH effect and even the QAH effect has been claimed to be observed in MBE-grown Cr$_2$Ge$_2$Te$_6$/(Bi,Sb)$_2$Te$_3$/Cr$_2$Ge$_2$Te$_6$ \cite{mogi2019large} and (Zn,Cr)Te/(Bi,Sb)$_2$Te$_3$/(Zn,Cr)Te \cite{watanabe2019quantum} sandwiches. 
We noticed that the ferromagnetic layers in both systems include the element Cr. When the MBE growth of these sandwich samples was 
performed at high temperatures, Cr diffusion inevitably occurs between the middle (Bi,Sb)$_2$Te$_3$ and the Cr-based ferromagnetic surface layers. 
The QAH effect realized in MBE-grown (Zn,Cr)Te/(Bi,Sb)$_2$Te$_3$/(Zn,Cr)Te sandwich heterostructures might represent success of approach (\textit{i}) [i.e. Cr-doped (Bi,Sb)$_2$Te$_3$]
instead of approach (\textit{iii}). This section focuses on the approach (\textit{i}).

Transition metal element doping is a convenient approach to induce long-range ferromagnetic order in TIs [Fig. \ref{fig2:CrTI}(a)] \cite{chien2005growth,zhou2005thin,zhou2006thin,hor2010development}.  The ordering 
mechanism is analogous to that in conventional dilute magnetic semiconductors\cite{ohno1999properties,ohno2000electric,sato2010first,jungwirth2006theory,dietl2014dilute}, 
as discussed in Sec. \ref{Sec:Magnetism}. In Cr-doped(Bi/Sb)$_2$(Se/Te)$_3$, 
scanning tunneling microscopy (STM) studies have demonstrated that Cr atoms usually substitute for Bi/Sb atoms [Fig. \ref{fig2:CrTI}(a)] \cite{chang2014chemical,chang2013thin,lee2015imaging,jiang2015mass,zhang2018electronic}. 
Below we compare the properties of Cr-doped Bi$_2$Se$_3$, Bi$_2$Te$_3$, and Sb$_2$Te$_3$ and explain why one is preferred 
as the parent material for the realization of the QAH effect. It turns out that the magnetic properties in these three TI compounds 
are highly sensitive to inhomogeneity of the Cr impurities. In Cr-doped Bi$_2$Se$_3$ thin films, no long-range ferromagnetic order has been observed down to \textit{T} =1.5 K [Fig. \ref{fig2:CrTI}(b)] \cite{zhang2013topology,chang2013thin}. The nonlinear Hall traces 
at low temperatures reflect the presence of weakly coupled superparamagnetic multimers formed by 
aggregated substitutional Cr atoms in Bi$_2$Se$_3$ matrices [Fig. \ref{fig2:CrTI}(e)]. 
The appearance of a large gap-opening at the Dirac point in Cr- and Fe-doped Bi$_2$Se$_3$ might also reflect substitutional Cr or Fe aggregations\cite{chen2010massive,chang2014chemical}. The absence of long-range ferromagnetic order due to inhomogeneity rules out 
Cr-doped Bi$_2$Se$_3$ as the candidate for the realization of the QAH effect. In contrast, both Cr-doped Bi$_2$Te$_3$ and Cr-doped Sb$_2$Te$_3$ exhibit pronounced hysteresis loops [Figs. \ref{fig2:CrTI}(c) and \ref{fig2:CrTI}(d)], demonstrating robust long-range ferromagnetic order in both materials\cite{zhang2013topology,chang2016field}. Compared to Cr-doped Bi$_2$Te$_3$, Cr-doped Sb$_2$Te$_3$ has a much better ferromagnetic order as seen by a more square hysteresis loop at low temperatures [Fig. \ref{fig2:CrTI}(d)]. STM measurements find that three Cr atom aggregations are common in Bi$_2$Te$_3$, and a more uniform Cr atom distribution in Sb$_2$Te$_3$ [Figs. \ref{fig2:CrTI}(f) and \ref{fig2:CrTI}(g)]\cite{chang2013thin}.  Overall Cr doping homogeneity is better in Cr-doped Sb$_2$Te$_3$ films. 
Doping homogeneity is the key ingredient for the long-range ferromagnetic order, as well as for the quality of the QAH state in magnetically doped TIs discussed below. Therefore, both transport and STM results suggest Cr-doped Sb$_2$Te$_3$ as a promising parent material for the realization of the QAH effect\cite{chang2013experimental,chang2013thin,chang2016field}.
  

\subsection{Realization of the QAH effect in magnetically doped TIs}
\subsubsection{QAH effect in Cr-doped TI}\label{Sec:crqah}


\begin{figure}[hbt!]
    \centering
    \includegraphics[width=3.5in]{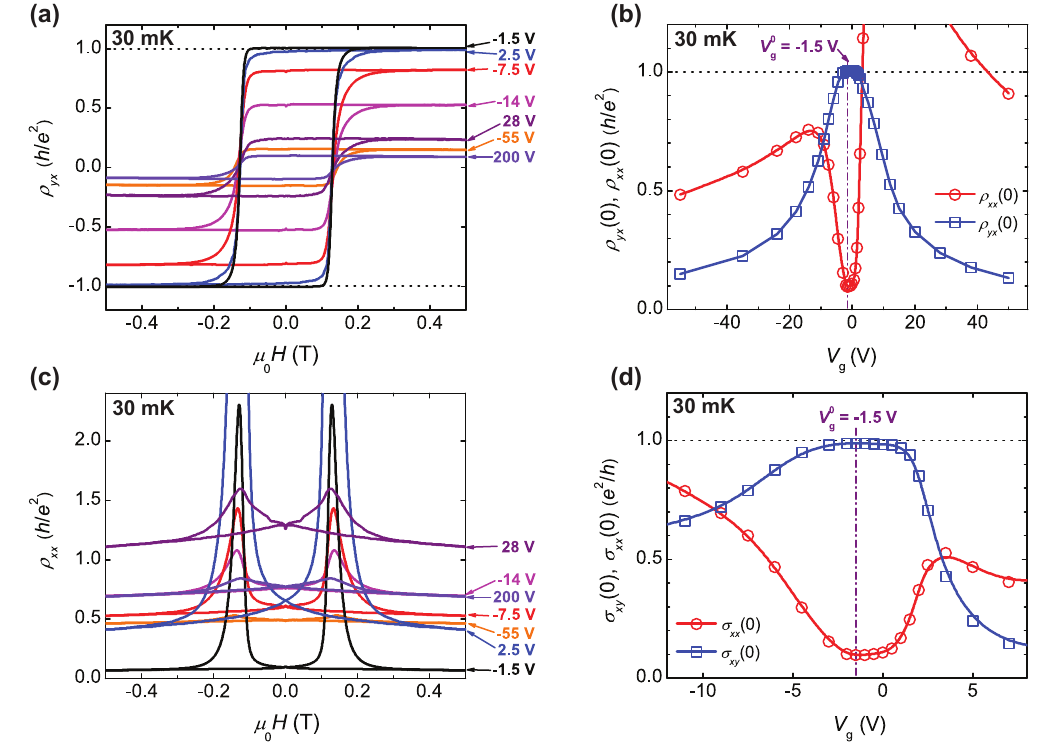} 
    \caption{Experimental realization of the QAH effect in 5 QL Cr-doped (Bi,Sb)$_2$Te$_3$ films. (a, c) Magnetic field $\mu_0 H$ dependence of the Hall resistance $\rho_{yx}$ (a) and the longitudinal resistance $\rho_{xx}$ (c) at different $V_{g}$s. (b) Gate dependence of the zero magnetic field Hall resistance  $\rho_{yx}$(0)  and the zero magnetic field longitudinal resistance $\rho_{xx}$(0). (d) Gate dependence of the zero magnetic field Hall resistance  $\sigma_{xy}$(0)  and the zero magnetic field longitudinal resistance $\sigma_{xx}$(0). From \onlinecite{chang2013experimental}.}
    \label{fig3:Cr}
\end{figure}

Cr-doped Bi$_2$Te$_3$ and Sb$_2$Te$_3$ films both have large carrier densities because of common defects that act as 
donors or acceptors.  The QAH effect could be achieved only after a systematic and careful exploration of
the relationship between unintended doping and stochiometry \cite{chang2013experimental,chang2013thin}. 
In Cr-doped Sb$_2$Te$_3$, doping Bi can reduce the high hole carrier density\cite{chang2016field,chang2013thin}. 
The QAH effect was first realized in 5 QL Cr$_{0.15}$(Bi,Sb)$_{1.85}$Te$_3$ thin films on heat-treated SrTiO$_3$(111) on October 9, 2012 \cite{chang2013experimental}. Figures. \ref{fig3:Cr}(a) and \ref{fig3:Cr}(b) show the magnetic field $\mu_0 H$ dependence of $\rho_{yx}$ and $\rho_{xx}$, respectively, measured at $T \sim 30$ mK for various bottom gate biases $V_{\rm g}$. 
The shape and the coercive fields of the AH hysteresis loops are nearly 
independent of $V_{\rm g}$, indicating a carrier-independent ferromagnetic order \cite{chang2013thin,chang2013experimental}. 
The zero magnetic field AH resistance $\rho_{yx}(0)$ changes dramatically with $V_{\rm g}$, showing a maximum value of $h/e^2$ ($\sim25.8$ k$\Omega$) at the charge neutral point $V_{\rm g}=V_{\rm g}^0$. The $\mu_0H$ dependence of $\rho_{xx}$ exhibits the typical hysteretic shape 
commonly observed in conventional ferromagnetic materials.

The most important observation, that the Hall resistance exhibits a distinct plateau with the quantized value $h/e^2$ at zero magnetic field, is shown in Fig. \ref{fig3:Cr}(b). Concomitant with the $\rho_{yx}$ plateaus, the zero magnetic field longitudinal resistance $\rho_{xx}(0)$ shows a dip, reaching a value of $0.098 h/e^2$, yielding a Hall angle $\theta$ of $\sim$84.4$^\circ$. For comparison with theory, we convert $\rho_{yx}(0)$ and $\rho_{xx}(0)$ into the sheet conductances $\sigma_{xy}(0)$ and $\sigma_{xx}(0)$ [Fig. \ref{fig3:Cr} (d)]. 
$\sigma_{xy}(0)$ exhibits a notable plateau at $0.987 e^2/h$, while $\sigma_{xx}(0)$ exhibits a dip down to $0.096 e^2/h$. These results mark the first experimental realization of the QAH effect. Note that the deviation of $\sigma_{xy}(0)$ from $e^2/h$ and nonzero $\sigma_{xx}(0)$ can be attributed to residual dissipative channels in the sample, which are expected to vanish completely at zero temperature or under a high magnetic field. By applying a magnetic field, \onlinecite{chang2013experimental} did see near perfect quantization of $\rho_{yx}$ 
and vanishing $\rho_{xx}$.
The observation of a QAH effect in MBE-fabricated Cr-doped TI thin films 
was relatively quickly reproduced by many research groups \cite{checkelsky2014trajectory,kou2014scale,bestwick2015precise,lachman2015visualization,kandala2015giant,liu2016large,kawamura2017current},
using film thickness from 5 to 10 QLs on various substrates including SrTiO$_3$(111), InP(111), and GaAs(111). 
We will review the physical properties of the QAH state in Cr-doped TI system in Sec. \ref{Sec:PhysicalQAH}. 

\begin{figure}[hbt!]
    \centering
    \includegraphics[width=3.5in]{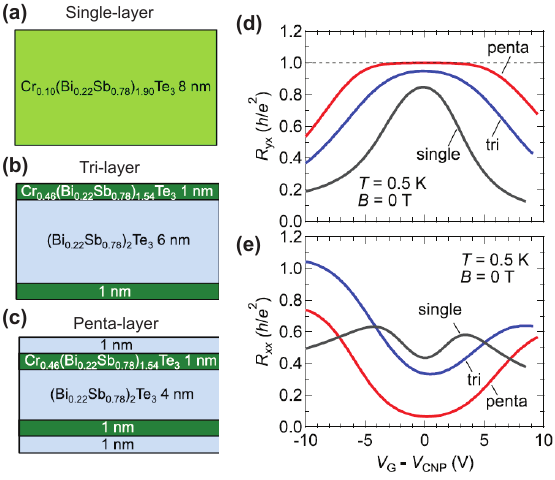} 
    \caption{Higher temperature QAH effects in TI heterostructures with modulation magnetic doping. (a-c) Schematics for 
    different Cr-doping geometries: uniformly-doped (a), Cr-doped TI tri-layer (b), and penta-layer (c).  (d, e) Gate dependence of the Hall resistance $R_{yx}$ (d) and the longitudinal resistance $R_{xx}$ (e) of the three samples shown in (a-c) measured at \textit{T} =0.5 K. From \onlinecite{mogi2015magnetic}.}
    \label{fig4:modulation}
\end{figure}

Despite great effort, the critical temperature of the QAH state in Cr-doped TI films [{Fig. \ref{fig4:modulation}}(a)] is still
limited to $\sim 2$ K \czchang{, an order of magnitude lower than the value of their Curie temperature \textit{T}$_C$}\cite{chang2013experimental,checkelsky2014trajectory,kou2014scale,bestwick2015precise,lachman2015visualization,kandala2015giant,liu2016large,kawamura2017current,feng2016thickness}. \czchang{This difference is \addCXL{presumably} due to the following two effects: (\textit{i})In magnetically doped TI films, the magnetic dopants may induce spatial fluctuations of the chemical potential and the exchange coupling across the film, and thus make the effective excitation gap much smaller than the average magnetic exchange gap \cite{chang2016observation}\amgreen{, enhancing the role of 
thermal fluctuations \cite{lin2022spin,lin2022influence}}(\textit{ii}) The energy of the Dirac point of the surface states in (Bi,Sb)$_2$Te$_3$ is close to or even buried into the bulk valence band maximum along the  direction \cite{chang2015zero,wang2018direct,li2016origin}.}

\czchang{Different} metrics have been used in the literature to identify
the critical temperature of QAH insulators. In this Article, we define the critical temperature of the QAH state as that at which the $\rho_{yx}(0)$/$\rho_{xx}(0)$ ratio is greater than 1. The low critical temperature not only impedes quantitative studies of the QAH state,  examination of the scaling behaviors of the plateau transitions \cite{checkelsky2014trajectory,kou2014scale}, and current-induced QAH breakdown \cite{fox2018part,kawamura2017current},
but also limits potential applications for electronic and spintronic devices with low power consumption.

\onlinecite{mogi2015magnetic} found that magnetic Cr modulation doping
(Bi,Sb)$_2$Te$_3$ to form tri- and penta-layer heterostructures can enhance the critical temperature of the QAH effect. 
By introducing two $1$ nm thick heavily-Cr-doped (Bi,Sb)$_2$Te$_3$ layers near the top and bottom surfaces of a $6$ nm nonmagnetic (Bi,Sb)$_2$Te$_3$ layer to form tri- and penta-layer heterostructures  [Figs. \ref{fig4:modulation}(a-c)], a better QAH state with a higher critical temperature could be achieved [Figs. \ref{fig4:modulation}(d) and \ref{fig4:modulation}(e)]. At $T=0.5$ K and the charge neutral point ($V_{\rm g}=V_{\rm g}^0$), the 
zero-magnetic-field Hall resistance $\rho_{yx}(0)$ is  $\sim0.85 h/e^2$, $\sim 0.95h/e^2$,  $\sim 1 h/e^2$ for single-, tri- and penta-layer heterostructure samples respectively. The monotonic decrease of the zero magnetic field longitudinal resistance $\rho_{xx}(0)$ further confirms that a better QAH state is achieved. In these modulation-doped tri- and penta-layer heterostructure samples, the fluctuation of the magnetic exchange gap on the sample surfaces is apparently suppressed and thus the effective magnetic exchange gap becomes larger, leading to a critical temperature $\sim$10 K \cite{mogi2015magnetic}. We note that the heavily Cr-doped (Bi,Sb)$_2$Te$_3$ layers used in tri- and penta-layer samples may drive the systems to a trivial ferromagnetic insulator state
due to reduction of SOC as a result of the heavy Cr doping in (Bi,Sb)$_2$Te$_3$ \cite{zhang2013topology,zhao2020tuning}, which will be further discussed in Sec. \ref{Sec:highqah}.

\onlinecite{okada2016terahertz} performed Terahertz Faraday and Kerr rotation measurements on 
penta-layer QAH samples and found that the Faraday and Kerr rotation angles quantitatively agree with the estimates
based on electrical transport measurements. QAH trilayer samples have also been used in 
scaling studies of the plateau to plateau transitions \cite{kawamura2020current,kawamura2018topological}. 
More recently, motivated by this modulation doing method, \onlinecite{jiang2020concurrence} 
fabricated 3 QL Cr-doped (Bi,Sb)$_2$Te$_3$ /5 QL (Bi,Sb)$_2$Te$_3$/3 QL Cr-doped (Bi,Sb)$_2$Te$_3$ sandwich structures and observed the concurrence of the topological Hall effect and the QAH effect through electric field gating. 
This concurrence indicates an interplay between the chiral edge states and chiral spin textures in magnetic TI heterostructures. 

\subsubsection{QAH effect in V-doped TI}\label{Sec:vqah} 
\begin{figure}[hbt!]
    \centering
    \includegraphics[width=3.5in]{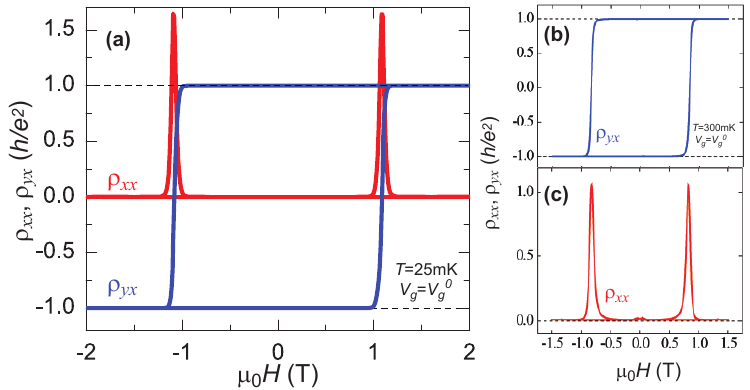}  
    \caption{QAH effect in V-doped (Bi,Sb)$_2$Te$_3$ films and Cr/V co-doped (Bi,Sb)$_2$Te$_3$ films. (a) Magnetic field dependence of the Hall resistance $\rho_{yx}$ (blue) and the longitudinal resistance $\rho_{xx}$ (red) of 4 QL V-doped (Bi,Sb)$_2$Te$_3$ films at the charge neutral point $V_{\rm g} =V_{\rm g}^0$ and temperature $T=$25 mK. (b,c) Magnetic field dependence of $\rho_{yx}$ (b) and $\rho_{xx}$ (c) of 5 QL Cr/V co-doped (Bi,Sb)$_2$Te$_3$ films at the charge neutral point $V_{\rm g} =V_{\rm g}^0$ and $T=$300 mK. From \onlinecite{chang2015high,ou2018enhancing}.}
    \label{fig5:vqah}
\end{figure}

The first-principles calculations by \onlinecite{yu2010quantized} suggested that 
QAH insulators should occur in thin films of TIs with Cr or Fe doping, whereas
the QAH state was expected to be absent for Ti or V doping because of \textit{d}-electron impurity bands at the Fermi surface.
\onlinecite{chang2015high} found that among all 3\textit{d }transition metal elements \cite{chien2005growth,zhou2005thin,zhou2006thin,hor2010development}, V-doped Sb$_2$Te$_3$ exhibits the most robust ferromagnetic order. Following a strategy similar to that employed earlier for Cr-doped QAH samples\cite{chang2013experimental}, 
a high-precision QAH effect was observed in a 4 QL V-doped (Bi,Sb)$_2$Te$_3$ film [Fig. \ref{fig5:vqah}(a)]\cite{chang2015high,chang2015zero}. 
The QAH effect in V-doped TI system was later replicated by 
other research groups \cite{grauer2015coincidence,grauer2017scaling,ou2018enhancing,lippertz2022current}. 

\czchang{For both Cr- and V-doped (Bi,Sb)$_2$Te$_3$ films, the Curie temperature \textit{T}$_C$ strongly depends on the magnetic doping concentration \textit{x}. The maximum value of \textit{T}$_C$ can be greater than 100 K in heavily Cr- and/or V-doped Sb$_2$Te$_3$ films \cite{chang2015high,zhou2006thin,zhou2005thin}. However, heavy Cr- and/or V-doping greatly reduces the SOC of the magnetically doped TI and drives it into a trivial insulator state\cite{zhang2013topology,chang2014chemical,zhao2020tuning,zhao2022zero}.Therefore, an optimal Cr- and/or V-doping concentration \textit{x} is required. The values of magnetic doping concentration \textit{x} are usually different in different groups worldwide, but the \textit{T}$_C$ value of samples that exhibit the QAH effect is usually in a range of 15$\sim$30 K.} \czchang{Compared to Cr-doped Sb$_2$Te$_3$, the} two main advantages of V-doped Sb$_2$Te$_3$ are: (\textit{i}) higher Curie temperature \textit{T}$_C$ - almost double that of Cr-doped Sb$_2$Te$_3$ at the same doping level $x$; and (\textit{ii}) larger coercive field $H_c$ - about an order larger than that of the Cr-doped Sb$_2$Te$_3$ with the same doping level at a fixed temperature \cite{chang2015high}.

Figure \ref{fig5:vqah}(a) shows a nearly ideal QAH state in a 4 QL V$_{0.11}$(Bi,Sb)$_{1.89}$Te$_3$/SrTiO$_3$(111) film. At the charge neutral point $V_{\rm g}=V_{\rm g}^0$, the zero magnetic field Hall resistance $\rho_{yx}(0)$ is $\sim1.00019\pm 0.00069 h/e^2$ ($25.8178\pm 0.0177   k\Omega$), while the zero magnetic field longitudinal resistance $\rho_{xx}(0)$ is as low as $\sim 0.00013\pm 0.00007 h/e^2$ ($\sim 3.35\pm 1.76 \Omega$) measured at $T=25$mK. The ratio $\rho_{yx}(0)/\rho_{xx}(0)$ corresponds to an AH angle $\sim 89.993\pm0.004^\circ$. The corresponding Hall conductance at zero magnetic field $\sigma_{yx}(0)$ is $\sim0.9998\pm 0.0006 e^2/h$ and the corresponding longitudinal conductance $\sigma_{xx}(0)$ is $\sim0.00013\pm 0.00007 e^2/h$ \cite{chang2015high}. The critical temperature of the QAH state in the V-doped TI system is around 1 K. Measuring 
the gate and temperature dependence of local and non-local magnetoresistance, \onlinecite{chang2015zero} established the presence of 
dissipationless chiral edge transport in this system in the absence of a magnetic field. 
By tuning the chemical potential, \onlinecite{chang2016observation} also observed a quantum phase transition from a QAH insulator to an Anderson insulator. 
The critical scaling behavior of this topological quantum phase transition will be discussed in Sec.\ref{Sec:moreqah}.

\onlinecite{grauer2015coincidence} synthesized 10 QL V-doped (Bi,Sb)$_2$Te$_3$ films on hydrogen passivated Si(111) substrates and observed a coincidence of super-paramagnetism and perfect quantization. The scaling behaviors of uniformly V-doped (Bi,Sb)$_2$Te$_3$ films of various thicknesses on Si(111) and InP(111) substrates are studied and the QAH insulators are classified as either 2D or 3D \cite{grauer2017scaling,fijalkowski2021any}. The different origins of the $(\sigma_{xy}, \sigma_{xx}) = (0, 0)$ point in scaling plots of individual magnetically doped QAH sample and magnetic TI sandwich samples will be discussed in Sec.\ref{Sec:MTIaxion}.

As noted above, the QAH effect has been demonstrated in both Cr- and V-doped (Bi,Sb)$_2$Te$_3$ films.
\onlinecite{ou2018enhancing} synthesized 5 QL Cr and V co-doped (Bi,Sb)$_2$Te$_3$ films with various Cr/V ratios and found a significant increase in the critical temperature of the QAH effect for Cr/(Cr+V) $\sim$ 0.16. It is also around 10 K. For the optimal Cr to V ratio,
a co-doped (Bi,Sb)$_2$Te$_3$ sample yielded $\rho_{yx}(0) \sim h/e^2$ to
within experimental uncertainty and $\rho_{xx}(0) \sim 0.009 h/e^2$
at $V_{\rm g} = V_{\rm g}^0$ and $T = 300$mK [Figs. \ref{fig5:vqah}(b) and \ref{fig5:vqah}(c)]. 
At \textit{T} = 1.5 K, $\rho_{yx}(0) \sim 0.97 h/e^2$ and $\rho_{xx}(0)\sim 0.19 h/e^2$. Such a quantization level has only been observed at \textit{T} $\sim$ 100 mK in either individual Cr-doped or V-doped QAH samples\cite{chang2013experimental,chang2015high,checkelsky2014trajectory,kou2014scale}. The enhancement of the critical temperature of the QAH effect in  Cr and V co-doped QAH samples is attributed to an improvement of the homogeneity of the ferromagnetic order and modulation of the sample band structure \cite{ou2018enhancing}.

A follow-up study combined magnetic force microscopy with electrical transport to
reveal typical ferromagnetic domain behaviors in the Cr and V co-doped QAH films, 
in contrast to the much weaker magnetic signals observed in either Cr-doped or V-doped QAH films \cite{wang2018direct}, 
possibly due to role of local super-para-magnetism \cite{lachman2015visualization,grauer2015coincidence}. 
In addition, by studying a series of  Cr and V co-doped and individual Cr-doped QAH samples,  \onlinecite{liu2020distinct} found that the ground state of all samples can be categorized as either a QAH insulator or as an AH insulator. 
In the low-disorder limit, a universal quantum longitudinal resistance $\rho_{xx}=h/e^2$ is observed at the coercive field $H_c$ of the QAH samples. Modulation doping (Sec. \ref{Sec:crqah}) \cite{mogi2015magnetic} and magnetic co-doping \cite{ou2018enhancing} are now established as techniques to enhance the critical temperature of the QAH effect in magnetically doped TI systems. 
The QAH critical temperatures of 5$\sim$10 K are regularly obtained.

\subsection{Physical properties of the QAH states in magnetically doped TIs}\label{Sec:PhysicalQAH}
\am{Following} the realization of the QAH effect in a Cr-doped TI system \cite{chang2013experimental}, close synergy between theory and experiment has greatly advanced our understanding of the QAH state. 
The QAH effect and the QH effect share the same topological properties, but their physical origins are different. 
The QH effect arises from the formation of Landau levels under high magnetic fields, and its realization relies on carrier mobility \cite{klitzing1980new} that is high enough to weaken mixing between different Landau levels,
{\it i.e.} to make $\omega_c \tau \gg 1$ large where $\tau$ is the Bloch state lifetime, $\omega_c$ is the cyclotron frequency and $\hbar \omega_c$ is the Landau level separation which has a typical value $\sim 1$ meV.
In contrast, the QAH effect \am{in magnetically doped TIs} is a consequence of an interplay between strong SOC and 
magnetic exchange interactions \cite{chang2020marriage}. The QAH effect can appear in materials with very low carrier mobility \cite{chang2013experimental,chang2015high} since it requires only that $\Delta \tau \gg \hbar$, where 
$\Delta$ is the topologically non-trivial band gap.  
The energy scale $\Delta$ is set by 
SOC and exchange interactions in magnetically doped TIs, \am{and is} normally on the order of hundreds of meV. 
The different physical origins of the QAH and QH states motivates studies of the
physical properties of QAH states in magnetically doped TIs. 
This subsection will review a variety of physical properties associated with QAH states, 
including in particular evidence for the coexistence of chiral and non-chiral edge states, the 
appearance of zero Hall conductance plateaus, and their relation 
to axion insulator states and the topological magneto-electric (TME) effect, 
global phase diagrams, scaling behaviors, and current-induced breakdown
\am{properties}. 

\subsubsection{Chiral and non-chiral edge states}

The QH effect and the QAH effect are both characterized by electric currents that 
flow along sample edges in equilibrium thanks to topologically protected chiral edge states. 
In both cases, the microscopic physics of the edge states can be complex.  We focus here on the 
peculiarities of the edge state systems associated with QAH state in magnetically doped TIs.
As the mobility of magnetically doped TIs is several orders smaller than that in 
typical QH systems, disorder scattering is usually much stronger in the QAH system.

The energy of the Dirac point of TSSs in parent (Bi,Sb)$_2$Te$_3$ compounds is close to the valence band maximum \cite{chang2015zero,wang2018direct,li2016origin}.  It follows that
disordered bulk transport likely coexists with edge transport in magnetically doped devices. 
When the magnetization is perpendicular to the thin film plane,
the TSSs remain gapless on the side surfaces of TI films and give rise to an edge state system that 
contains a number of channels proportional to the film thickness \cite{wang2013anomalous}. 
Both the residual bulk carriers and the multi-channel edge state systems play a role in 
understanding the experimental fact that the QAH effect in magnetically doped TIs sometimes shows a 
residual $\rho_{xx}$ that is several $k\Omega$, while $\rho_{yx}$ is nearly quantized.


The presence of multi-channel side walls was experimentally confirmed by performing the non-local transport 
measurements on both uniformly 10 QL Cr-doped (Bi, Sb)$_2$Te$_3$ \cite{kou2014scale} and uniformly 4 QL V-doped  (Bi, Sb)$_2$Te$_3$ \cite{chang2015zero} Hall bar devices.  In the absence of disorder, wavevector along the side-wall is a good quantum number 
and the edge-system can be separated into a single chiral channel and a set of non-chiral pairs.
$\rho_{yx}$ quantization is perfect when bulk transport is localized near the side walls, and the entire edge state system is 
in local equilibrium. \onlinecite{kou2014scale} showed that the non-local resistance
is hysteretic, with both low-resistance and high-resistance states appearing during the magnetization reversal process,  due to the interplay between edge and bulk channels. \onlinecite{chang2015zero} further showed asymmetric behavior of 
local and non-local transport when the carrier type is tuned from electrons to holes through an external gate voltage, implying the dominant dissipative channels vary from quasi-helical edge states to bulk states. Dissipation can occur because of transport across the bulk, or because
the current-carrying multi-channel edge system is not in local equilibrium.
Recently, \onlinecite{wang2020demonstration} performed more systematic temperature and magnetic field dependent transport measurements on a 3 QL Cr-doped (Bi,Sb)$_2$Te$_3$/5 QL (Bi,Sb)$_2$Te$_3$/3 QL Cr-doped (Bi,Sb)$_2$Te$_3$ sandwich sample and 
concluded that in thick QAH samples the dominant dissipation mechanism switches between edge and bulk states in different magnetic-field regimes.

Another unique feature of magneto-transport in QAH materials
is that the magnetization direction in a magnetic domain is locked not only to the Hall conductance but also to the propagation direction (i.e.clockwise or counterclockwise) of the chiral edge current. Chiral currents flow not only at sample edges but also along domain walls with opposite 
magnetization orientations. This property makes electrical probes sensitive to the motion of magnetic domain walls.
\onlinecite{yasuda2017quantized} utilized the magnetic force microscope to control the magnetization direction 
of a QAH sample with modulated magnetic doping and concluded that the observed non-local transport 
is consistent with chiral edge conduction along the walls in a written domain pattern. 
We note that direct imaging of magnetic domain walls is absent in these non-local transport measurements. \onlinecite{rosen2017chiral} partially covered the QAH films with a bulk niobium superconductor and demonstrated chiral edge conduction along
the magnetic domain walls via Meissner–Ochsenfeld effect. Microwave impedance microscopy (MIM) can directly image the distribution of chiral edge currents at the boundary and at magnetic domain walls in a  QAH sample \cite{allen2019visualization}. 
In addition, chiral edge states can also carry spin polarization \cite{zhang2016electrically,cheraghchi2020spin} or form spin texture \cite{wu2014topological}, which implies that the chiral edge states can be utilized to control magnetic domain walls and drive a domain-wall motion \cite{kim2019electrically,upadhyaya2016domain}. The interplay between dissipationless chiral modes and magnetic domain walls in QAH insulators
points to potential applications in spintronics if higher critical temperatures can be achieved.

\subsubsection{Zero Hall conductance plateau and the axion insulator state}\label{Sec:MTIaxion}
In uniformly doped thin films, the magnetization of a QAH sample 
reverses its sign at the coercive field $H_c$,  
and correspondingly the Hall conductance $\sigma_{xy}$ will vary from $e^2/h$ to $-e^2/h$ with a change of the Chern number \textit{C} by 2. 
This scenario does not, however, exhaust the available phenomenology. 
In some devices, the system undergoes two separated topological phase transitions with each varying \textit{C} by 1, 
and consequently, an intermediate zero Hall conductance plateau (ZHCP) with \textit{C} = 0 emerges in the magnetization reversal process.

\begin{figure*}[hbt!]
    \centering
    \includegraphics[width=7in]{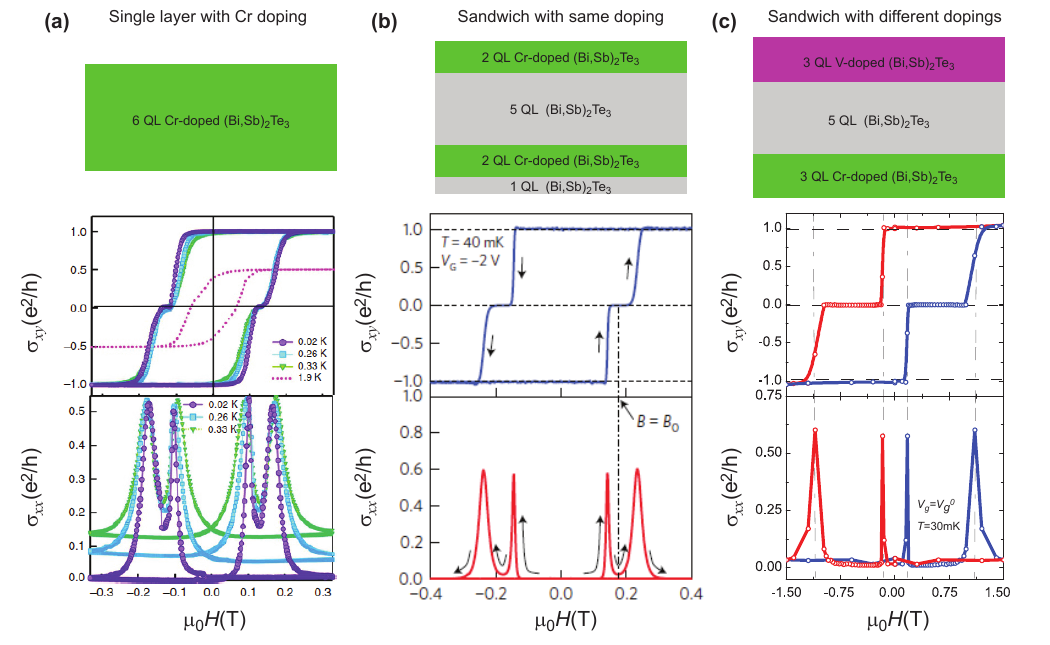} 
    \caption{Zero Hall conductance plateaus and axion insulator states in magnetic TI films/heterostructures. 
    (a-c) Top: schematics of a uniformly Cr-doped TI, a magnetic TI sandwich with the asymmetric Cr doping near top and 
    bottom surfaces, and a magnetic TI sandwich with magnetic doping with Cr and V. 
    Bottom: magnetic field $\mu_0 H$ dependence of the Hall conductance $\sigma_{xy}$ and the longitudinal conductance $\sigma_{xx}$ of 6 QL Cr-doped (Bi,Sb)$_2$Te$_3$ films (a), a 2QL Cr-doped(Bi,Sb)$_2$Te$_3$/5QL  (Bi,Sb)$_2$Te$_3$/2QL Cr-doped(Bi,Sb)$_2$Te$_3$/1QL  (Bi,Sb)$_2$Te$_3$ heterostructure (b), a 3QL V-doped(Bi,Sb)$_2$Te$_3$/5QL  (Bi,Sb)$_2$Te$_3$/3QL Cr-doped(Bi,Sb)$_2$Te$_3$ heterostructure (c). From \onlinecite{kou2015metal,mogi2017magnetic,xiao2018realization}.}
    \label{fig7:zero}
\end{figure*}

\addCXL{Two scenarios can be identified for ZHCP formation in magnetic TI films. The first scenario assumes that the magnetization is 
uniform across the device so that the magnetic gaps, denoted as $\Delta_{\rm M}$ below, for the TSSs on top and bottom surfaces are the same. A finite hybridization (denoted as $m_0$) between two TSSs is considered, and when the magnetic domains in the sample are well-aligned in the same direction, the magnetic gap $\Delta_{\rm M}$ is dominant so that the system is in the QAH state with the Chern number $C = \Delta_{\rm M}/|\Delta_{\rm M}| = \pm 1$, depending on the sign of $\Delta_{\rm M}$. During the magnetization reversal process, the number of upward and downward oriented magnetic domains changes, and the spatially averaged magnetic gap continuously varies from positive to negative. This scenario applies when the typical domain size is small compared to the characteristic electronic length scale $\hbar v_{\rm D}/\Delta_M$, allowing the non-uniform magnetization to be replaced by its spatial average. The topological phase transition between $C =0$ and $|C| \ne 0$ states takes place at the gap closings when $|\Delta_0|=|m_0|$. When the hybridization gap is dominant \czchang {($|\Delta_0|<|m_0|$)}, the Chern number becomes $C = 0$, leading to a ZHCP. This scenario is more
likely to apply in thinner TI films with a large hybridization between two TSSs \cite{feng2015observation,kou2015metal}. Since the hybridization between two TSSs 
\am{declines exponentially with increasing TI film thickness and samples always have some disorder, this picture becomes untenable beyond a 
relatively small numbers of layers.}  On the other hand, with 
increasing film thickness the \am{magnetism in the two surface regions 
is more weakly coupled, and modulation doping makes it possible to design devices in which the two coercive fields differ}. 
This leads to an alternative scenario for the ZHCP. 
When the magnetic gap for top TSS has the opposite sign to that of the bottom TSS, 
the two TSSs give the opposite contributions to the Hall conductivity and this leads to a ZHCP \cite{wang2014universal,mogi2017tailoring,xiao2018realization}.
}
\am{\addCXL{When ZHCP states are observed in magnetic sandwiches,
both sidewall and bulk quasiparticles must be absent
at the Fermi level, or if present must be localized.} 
ZHCP states therefore 
can be viewed as a special subclass of normal 2D insulators,
or, in the case of where localized states are present, as 
Anderson insulators.  
However they still have broken TR symmetry and, as we discuss below,
are expected to exhibit properties that are related to the TME effect, a characteristic of bulk 3D TIs.}

Soon after the theoretical predictions \cite{wang2014universal}, two experimental groups separately reported ZHCP observations in 5 QL and 6 QL uniformly Cr-doped (Bi,Sb)$_2$Te$_3$ films [Fig. \ref{fig7:zero}(a)] \cite{feng2015observation,kou2015metal}.  
Given \am{the uniform magnetization}, the observed ZHCPs are likely to have their origin from the hybridization gap. Experimental identifications of ZHCP states can be somewhat ambiguous because measurements 
are always performed at finite temperatures
where insulating behaviors are not fully developed.
Examinations of the Hall conductivity and resistivity, 
related by $\sigma_{xy}=\rho_{yx}/(\rho_{xx}^2+\rho_{yx}^2)$ can give different impressions.
$\sigma_{xy}$ is small whenever $\rho_{xx}$ at $H_c$ is much greater than the quantized $\rho_{yx}$. 
This can result in the appearance of a zero $\sigma_{xy}$ plateau but no zero $\rho_{yx}$ plateau at $H_c$. 
Indeed, pronounced ZHCPs are found in disordered 5 QL Cr-doped (Bi,Sb)$_2$Te$_3$ films \cite{feng2015observation}, 
but are usually absent in magnetically 
doped QAH samples with higher quality \cite{chang2015high}. 
The appearance of ZHCP plateaus also depends strongly on MBE growth conditions.
For 6 QL Cr-doped (Bi,Sb)$_2$Te$_3$ films from the same group, 
some samples show the ZHCP [Fig. \ref{fig7:zero}(a)] \cite{kou2014scale}, while others that have
relatively small $\rho_{xx}$ at $H_c$ do not  \cite{rosen2017chiral}. 
For the hybridization gap induced ZHCP, \onlinecite{haim2019quantum} pointed out that strong magnetic 
disorder can drive the system into a new topological state supporting helical edge modes that are protected by crystalline 
reflection symmetry, instead of TR symmetry. This phenomenon is dubbed the “quantum anomalous parity Hall effect", but not yet supported by  
experimental evidence. 



\am{In the literature, \addCXL{the observation of ZHCPs in 2D magnetic TI thin films has always been connected to the concept of axion insulator states}, a moniker motivated by the TME effect - 
a charactertistic property of 3D TIs \cite{wilczek1987two,qi2008topological,essin2009magnetoelectric,wang2015quantized,morimoto2015topological,pournaghavi2021nonlocal}. In the field theory description of 3D TIs, a topological $\theta$ term $\theta e^2 {\bf E\cdot B}/2\pi h$ is added into the ordinary Maxwell electromagnetic Lagrangian \cite{qi2008topological,sekine2021axion}, 
yielding axion electrodynamics \cite{wilczek1987two}.
Here ${\bf E}$ and ${\bf B}$ are the conventional electric and magnetic fields inside an insulator, and $\theta$ is a pseudo-scalar. 
It is normally argued that $\theta$ can differ from 
$-\theta$ only by a multiple of $ 2 \pi$ when TR symmetry or inversion symmetry is present in the bulk.
It follows that $\theta$ is always a multiple of $\pi$,
an even multiple (normally 0) for a normal 3D insulator 
and an odd multiple (normally $\pm 1$) 
for a 3D TI \cite{qi2008topological,essin2009magnetoelectric}.
Breaking TR and inversion symmetry in the bulk
can generally lead to an arbitrary value of $\theta$. 
The term {\it axion} is adopted because the mathematical structure of the electrodynamics of TIs with gapped surface states is 
similar to that of the hypothetical particles christened by Frank Wilczek 
to solve the strong charge conjugation-parity problem in particle physics \cite{peccei1977cp}. 
In a field theory description, 
the parameter $\theta$ in the axion effective action continuously varies from the value $\pi$ in the TI to $0$ in the vacuum 
across the surface region. 
It turns out that this $\pi$ change of $\theta$ implies 
the surface half QH effect, described in Sec.\ref{Sec:magnetic}. 
Therefore, the quantized value $e^2/2h$ of the surface $\sigma_{xy}$, as well as the quantized magneto-electric response of bulk TIs, 
are equivalent to the quantized $\theta$ value of the axion term in 
electromagnetic Lagrangian of 3D bulk TIs \cite{sekine2021axion}.

\addCXL{The definition of the term axion insulators has been somewhat inconsistent in the literature. Generally, an axion insulator is viewed as a 3D TI with a non-zero quantized $\theta$ parameter in the bulk and the surface states gapped either by surface magnetism or bulk magnetism \cite{essin2009magnetoelectric,turner2012quantized,varnava2018surfaces}. In this terminology, axion insulators that allow for the observation of the TME effect have not only insulating bulks, but also insulating surfaces and hinges, a requirement that is satisfied
by all-in or all-out magnetization configurations \cite{varnava2018surfaces}. 
A TI heterostructure in which a thick TI film has an undoped interior sandwiched by magnetically doped regions near the film surface, 
the sample configuration in Fig. \ref{fig7:zero}(c), with anti-parallel magnetization orientations on the two surfaces belongs to this category. 
When the TI film is thick enough that the surface regions are electrically isolated, this system will support half-quantized surface Hall conductances of opposite signs.}
The orbital magnetizations associated with the corresponding equilibrium 
orbital currents then do not cancel perfectly tn the presence of a weak electric field applied across the film. 
The net magnetization per cross-sectional area is 
\begin{equation}
M_{\rm orb} = \frac{e}{2h} \times {eEt},
\label{eq:TME}
\end{equation}
where $t$ is the film thickness, the first factor on the right-hand side is the surface state Hall conductance from
Eq.~\ref{eq:edgemagnetism}, and the second is the electric potential drop across the film.  
The electric field creates a parallel bulk magnetization per unit volume
with a universal coefficient of proportionality that is quantized in units of $e^2/2h$.
\addCXL{This property known as the TME effect, is equivalent to
having $\theta=\pi$ in the electromagnetic Lagrangian and to 
half quantized surface Hall conductivities.  The TME is therefore 
activated by surface magnetizations whose projections onto the surface normal never change sign.}
For thin film geometries, the necessary condition is that the top and bottom surface magnetizations are 
opposite in orientation, so that the chiral edge channel is absent in Fig.~\ref{fig7:zero}(c).

Experimental confirmation of the TME effect is challenging.  The following 
three conditions are required: (\textit{i}) the TI film is in the 3D regime so that TSSs at opposite surfaces are decoupled; 
(\textit{ii}) all surfaces are gapped, with the chemical potential lying within both gaps; (\textit{iii}) the interior of the TI must 
preserves either TR symmetry or certain crystalline symmetries (e.g. inversion) to maintain the $\theta=\pi$ constraint. 
Given that the TME is equivalent to a magnetic-surface 
localized half-quantized Hall conductance, it is reasonable to 
argue \cite{morimoto2015topological,wang2015quantized,mogi2017magnetic,xiao2018realization,mogi2017tailoring}
that \addCXL{the second scenario of the ZHCP discussed above, }
in concert with observation of the QAH effect, provides experimental evidences of the TME effect. 
A jump of the total 2D Hall conductance by $\pm e^2/h$ due to reversal 
of magnetization at one surface, must mean that the contribution of 
that surface to the Hall conductance is $\pm e^2/2h$.
\onlinecite{mogi2017magnetic} fabricated asymmetric (Bi,Sb)$_2$Te$_3$ sandwich heterostructures with surface Cr doping. 
The observation of a zero $\sigma_{xy}$ plateau [Fig. \ref{fig7:zero}(b)] was interpreted as evidence of a TME effect 
associated with antiparallel alignment of the top and bottom Cr-doped (Bi,Sb)$_2$Te$_3$ magnetized layers. 
Subsequent magnetic domain imaging measurements on the same sample failed, however, to find evidence of uniform antiparallel magnetization alignment at any external magnetic field \cite{lachman2017observation,allen2019visualization},
making it less convincing that the observed ZHCP is fully explained by opposite magnetic exchange gaps.  
Soon after, \onlinecite{xiao2018realization} and \onlinecite{mogi2017tailoring} replaced the top Cr-doped TI layer with a V-doped layer to form a V-doped TI/TI/Cr-doped TI sandwich heterostructures [Fig. \ref{fig7:zero}(c)]. 
The significant difference in $H_c$ between the Cr- and V-doped TI systems then 
leads to a broad ZHCP, as shown in Fig. \ref{fig7:zero}(c). Magnetic force microscopy images in this system reveal two separate magnetization reversals, as expected from the coercive field difference of the two magnetic layers \cite{xiao2018realization}. This observation suggests that the observed ZHCP has a different origin from that in the uniformly magnetically doped QAH samples, and is a result of opposite sign half-quantized $\sigma_{xy}$ in the two surfaces regions.}
\addCXL{An alternative approach to probe the TME effect is through magneto-optical measurements
\cite{wu2016quantized,dziom2017observation,okada2016terahertz}, but these
have the disadvantage of probing the response properties of topological matter at finite frequencies, whereas the TME refers to a characteristic static response.}


\begin{figure}[hbt!]
    \centering
    \includegraphics[width=3.5in]{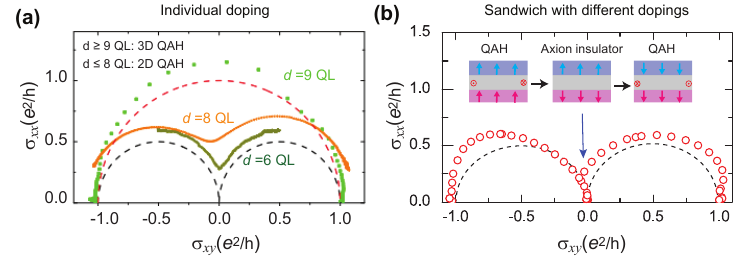} 
    \caption{Scaling behaviors of the QAH effect in uniformly magnetically doped TI films, and in sandwiches with V and Cr 
    surface doping. (a) $(\sigma_{xy}, \sigma_{xx})$ flow diagram of uniformly V-doped (Bi,Sb)$_2$Te$_3$ with a  thicknesses of 9 QL (green), 8 QL (orange), and 6 QL(dark yellow). (b) $(\sigma_{xy}, \sigma_{xx})$ flow diagram 
    of a sample with an axion insulator doping profile:  3QL V-doped (Bi,Sb)$_2$Te$_3$/5QL (Bi,Sb)$_2$Te$_3$/3QL Cr-doped (Bi,Sb)$_2$Te$_3$). Two semicircles of radius $0.5e^2/h$ centered at $(0.5e^2/h,0)$ and $(-0.5e^2/h,0)$ are plotted as black dashed lines in (a) and (b)
    for comparison. A semicircle of the radius $e^2/h$ centered at $(0, 0)$ is shown as a red dashed lines in (a). From \onlinecite{grauer2017scaling,xiao2018realization}.}
    \label{fig8:scaling}
\end{figure}

\addCXL{The essential feature of axion insulator is the surface half QH effect, but metallic modes may exist on certain surfaces or hinges of a finite sample in a more broad definition once its bulk $\theta$ value is quantized. }
This has led to theoretical proposals of high-order topology in axion
insulators \cite{xu2019higher,zhang2020mobius,wieder2018axion,yue2019symmetry,varnava2018surfaces,chen2021using,varnava2021controllable}, in which metallic modes exist at the hinges of finite samples. The critical behaviors for the transition from a 3D axion insulator with high-order topology 
to a trivial insulator have also been theoretically studied \cite{song2021delocalization,li2021critical}. Experimentally, the one semicircle behavior of the $(\sigma_{xx},\sigma_{xy})$ flow diagram observed in thick QAH samples \cite{grauer2017scaling,fijalkowski2021any} is consistent with the scaling behavior of a surface half QH effect \cite{nomura2011surface} if one assumes that the phase transitions on the two surfaces occur simultaneously. 


\subsubsection{Scaling behavior of plateau transitions}\label{Sec:ScalingPlateau}
The scaling behavior of plateau transitions in the space of temperature and 
Landau level filling factor $\nu$ has been extensively studied in conventional 
QH systems \cite{girvin1987quantum,huckestein1995scaling}.
In QAH systems, the external magnetic field plays the role of $\nu$, and the plateau 
behavior is complicated by the role of the magnetic domain structure
near the coercive field $H_c$. \onlinecite{wang2014universal}
used a Chalker-Coddington-type network model based on an 
intuitive picture of random magnetic domain switching and chiral edge state
propagation along magnetic domain walls to discuss the scaling properties of transitions between QAH plateaus. 
Phase diagrams and scaling behaviors have also been addressed by performing numerical simulations of
disordered QAH systems \cite{onoda2003quantized,nomura2011surface,qiao2016anderson,chang2016observation,chen2019effects}. 
The possible presence of ZHCP states discussed in Sec. \ref{Sec:MTIaxion} strongly influences 
the scaling behavior of the plateau transitions.  In the absence of a ZHCP, the Chern number change on the plateau 
transition is 2, and this case must be described by a Chalker-Coddington model with two channels \cite{lee1994unified,xiong2001metallic,chen2019effects}. 


\begin{table*}[hbt!]
\caption{QAH insulator scaling study summary. QL is the sample quintuple layer number, Cr-TI specifies 
Cr-doped (Bi,Sb)$_2$Te$_3$, and V-TI specifies V-doped (Bi,Sb)$_2$Te$_3$.  
$\kappa$ characterizes the temperature-dependence of the plateau transition widths.}
    \label{table1:scaling}
    \centering
    \begin{tabular}{|c|P{5cm}|c|c|c|}\hline
   QAH-related phase transition & Sample & Extracted $\kappa$ & $T$ range for scaling & References  \\\hline
    Plateau to plateau & 6 QL Cr-TI & 0.22 & 20 $\sim$ 100 mK &	\onlinecite{kou2015metal} \\\hline
    QAH to Anderson insulator &  4 QL V-TI &	0.62 $\pm$ 0.03 & 25 $\sim$ 500 mK & \onlinecite{chang2016observation} \\\hline
    QAH to trivial insulator & 2 QL Cr-TI/2QL TI/2QL Cr-TI/1QL TI  &	0.61 $\pm$ 0.01 & 60 $\sim$ 800 mK & \onlinecite{kawamura2018topological}\\\hline
    QAH to AH insulator &  5QL Cr-TI	& 0.31 $\pm$ 0.01 &	100 $\sim$ 800 mK &	\onlinecite{liu2020distinct} \\\hline
    QAH to axion insulator & 3QL V-TI/5QL TI/3QL Cr-TI & 0.38 $\pm$ 0.02 & 45 $\sim$ 100 mK & \onlinecite{wu2020scaling}\\\hline
    \end{tabular}
\end{table*}

Experimental efforts have explored global phase diagrams and scaling 
behaviors of QAH effects in a variety of different magnetically doped TI
heterostructures \cite{checkelsky2014trajectory,chang2016observation,kawamura2018topological,liu2020distinct,wu2020scaling,kou2015metal,grauer2017scaling}. 
\onlinecite{checkelsky2014trajectory} found the first evidence of quantum criticality 
by examining $(\sigma_{xy},\sigma_{xx})$ as a function of temperature and electrostatic doping 
near neutrality, concluding that the delocalization behavior of QAH states 
can be described in terms of the renormalization group flow of the integer QH effect. 
The global $(\sigma_{xy},\sigma_{xx})$ phase diagram 
depends qualitatively on whether or not a ZHCP is present.
When a ZHCP is absent the $(\sigma_{xy},\sigma_{xx})$ flow diagram can be approximated by a 
semi-circle with radius $e^2/h$ centered at $(0, 0)$; when the ZHCP is present
it is approximated by two semi-circles with radius $0.5e^2/h$ centered at $(0.5e^2/h,0)$ and $(-0.5e^2/h,0)$. 
Two semi-circle $(\sigma_{xy},\sigma_{xx})$ behavior was first observed by \onlinecite{kou2015metal}. 
Soon after, \onlinecite{grauer2017scaling} found a systematic crossover from two semi-circle to one \am{semi-circle} behavior
with increasing thickness of V-doped (Bi,Sb)$_2$Te$_3$ films,
and interpreted this behavior as a dimensional crossover from a 2D QAH state to a 3D \czchang {TI} state [Fig.\ref{fig8:scaling}(a)]. It was suggested that the critical thickness of this crossover 
is $8\sim 9$ QL for TI films with uniform V doping \cite{fijalkowski2021any}. 
The thickness dependence strongly relies on the quality of magnetically doped QAH thin films,
however, \am{and has additional} sensitivity to the typical magnetic domain size and the magnetic dopant chemical environment.

The physical picture of dimensional crossover, which relies on magnetization reversals at 
both top and bottom surfaces that occur simultaneously, can be relevant only for uniformly doped systems.
It cannot be applied to V-doped TI/TI/Cr-doped TI sandwich structures 
since the middle TI layer acts like a magnetic buffer layer that
decouples the top and bottom magnetic TI layers. When the number of non-magnetic interior TI layers is increased, 
the exchange coupling between top and bottom magnetic TI layers is quickly reduced 
and leads to a broader ZHCP \cite{xiao2018realization,mogi2017tailoring}. As discussed in Sec. \ref{Sec:MTIaxion}, independent 
surface magnetization reversal processes are observed in V-doped TI/TI/Cr-doped TI heterostructure,
but not in uniformly magnetically doped QAH samples. The scaling behavior of the V-doped TI/TI/Cr-doped TI 
samples always shows
two semi-circle behavior, as shown in Fig. \ref{fig8:scaling}b, independent of thicknesses.

\onlinecite{fijalkowski2021any} recently observed \am{an unexpected}
ZHCP in a symmetric magnetic TI thin film
sandwich structure and ascribed the plateau to hybridization between
surfaces. They suggested on this basis that the ZHCP observed in intentionally
asymmetric V-doped TI/TI/Cr-doped TI sandwich heterostructures
\cite{xiao2018realization,mogi2017tailoring} might also originate from a hybridization gap, rather than from
the formation of the axion insulator state.  
\addCXL{This interpretation requires the reduction of the exchange coupling between the local moments and the Dirac cone electrons, allowing inter-surface hybridization to dominate at small external magnetic fields. }
An alternative explanation for the appearance of two peaks in 
nominally symmetric samples is that asymmetry was induced unintentionally during the photo lithography process.
We favor this latter interpretation since Zeeman coupling at the coercive field scale 
has a small direct influence on electronic structure. Future experiments that more fully
reveal trends {\it vs.} film thickness, on which hybridization depends exponentially, should be 
able to clearly distinguish between these scenarios.

\begin{figure}[hbt!]
    \centering
    \includegraphics[width=3.5in]{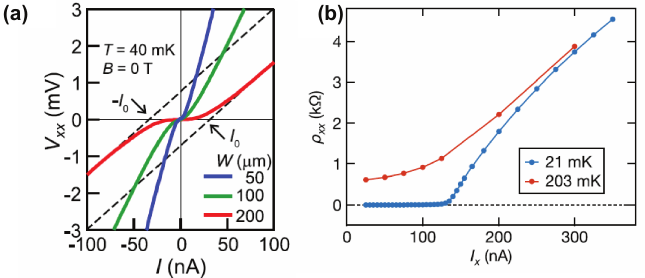} 
    \caption{Current-induced breakdown of the QAH effect.  (a) Bias current dependence of the voltage $V_{xx}$ for 
    three Hall bars with different widths W = 50 $\mu m$ (blue), 100 $\mu m$ (green),  and 200 $\mu m$ (red). The QAH samples used here are 9 QL Cr-doped (Bi,Sb)$_2$Te$_3$ films. (b) The bias current $I_x$ dependence of the longitudinal resistance $\rho_{xx}$ of a 6 QL Cr-doped (Bi,Sb)$_2$Te$_3$ film.  All curves are measured at the charge neutral point $V_{\rm g} =V_{\rm g}^0$. From \onlinecite{kawamura2017current,fox2018part}.}
    \label{fig6:breakdown}
\end{figure}

The critical exponents $\kappa$ of the QAH plateau transitions extracted from the temperature 
dependence of the plateau transition widths are summarized in Table \ref{table1:scaling}. 
$\kappa$ varies over a broad range from 0.22 to 0.62. This might be because the critical temperature 
of the QAH state in the magnetically doped TI films/heterostructures is limited to a few Kelvin, limiting the temperature range
available to identify the plateau-width power law.
Studies on the scaling behaviors of the plateau transitions have to be performed at  
temperatures that are well below the critical temperature of the QH/QAH state \cite{wei1986localization,wei1988experiments,huckestein1995scaling}. As a comparison, the critical temperature of the QH state in (In, Ga)As/InP heterostructure is $\sim$50 K \cite{wei1986localization}, and the scaling behaviors were studied in the temperature range 0.1 K $<$ \textit{T} $<$ 4.2 K \cite{wei1988experiments}. 
In spite of these uncertainties, the determination of the critical exponents still 
provides valuable information. It is expected that the plateau transition should be identical to that 
of QH systems when the magnetization reversals on the two surfaces occur separately. 
In sandwich samples, \onlinecite{wu2020scaling} demonstrated that the QAH to axion insulator phase transition indeed shares
scaling universality with the standard plateau transition in the QH effect. 
However, if the magnetization reversal at the top and bottom surfaces occurs simultaneously, 
the plateau transition should be described by a  Chalker-Coddington model with two channels \cite{lee1994unified,xiong2001metallic,chen2019effects}. 
Other features of the magnetization reversal process have also been explored experimentally.  
\onlinecite{lachman2015visualization} imaged the magnetic structure of a 7 QL Cr-doped (Bi,Sb)$_2$Te$_3$/SrTiO$_3$(111) in a QAH state 
and found a superparamagnetic state with weakly interacting magnetic domains. 
\onlinecite{liu2016large} observed large jumps in the Hall and longitudinal resistances during magnetic transition between 
QAH states in a 10 QL Cr-doped (Bi,Sb)$_2$Te$_3$/SrTiO$_3$(111) sample, and attributed these large jumps to 
quantum tunneling between magnetic domains. These experiments imply that the magnetic dynamics during the reversal is 
complex and likely requires a theoretical description that is more sophisticated than simple percolation theory. 
In addition, \onlinecite{kandala2015giant} performed angular magneto-transport measurements
on a 10 QL Cr-doped (Bi,Sb)$_2$Te$_3$/SrTiO$_3$(111) sample
by rotating the magnetic field and found that a giant anisotropic magnetoresistance is 
induced by a magnetic-field-tilt driven crossover from chiral edge transport to diffusive transport.

\subsubsection{Current breakdown}\label{Sec:moreqah}

The stability of the QAH state against excitation currents
has also been studied experimentally in the magnetically doped TI system. \onlinecite{kawamura2017current} claimed that
some of their observations are similar to those in QH states.
For example, they found that the critical current for QAH breakdown is roughly proportional 
to the width of the Hall bar [Fig. \ref{fig6:breakdown}(a)].  However, some behavior differs qualitatively 
from QH state breakdown phenomenology.  In particular, there is no sudden increase in dissipation at the breakdown current, which they attributed to a variable-range-hopping bulk transport mechanism at low temperatures. 
\onlinecite{fox2018part} observed current-induced QAH breakdown at \textit{T}$\sim$ 21 mK 
and attributed this phenomenon to runaway electron heating in bulk current flow [Fig. \ref{fig6:breakdown}(b)]. \onlinecite{lippertz2022current} recently investigated the QAH breakdown through local and nonlocal measurements and found that the QAH breakdown is absent in nonlocal regions, indicating that the transverse electric field might be responsible for the QAH breakdown. Because the QAH state in magnetically doped TIs has a strong temperature dependence 
and a small effective exchange gap ($\sim$ 100 $\mu$eV),  QAH breakdown occurs at a significantly smaller current in TI QAH devices (tens to hundreds of nA) \cite{kawamura2017current,fox2018part,lippertz2022current} than that in conventional QH samples (a few tens of $\mu$A) \cite{mensz1989high,kawaji1996breakdown,singh2009nonequilibrium,alexander2012high,connolly2012unraveling}. Further studies that address the breakdown in higher temperature QAH insulators are desirable.


\subsection{High-Chern-number QAH effect}\label{Sec:highqah}
High-Chern-number QAH effects were proposed theoretically for Cr-doped Bi$_2$(Se,Te)$_3$  films \cite{wang2013quantum},
and for films of the topological crystalline insulator SnTe  when magnetically doped \cite{fang2014large}. 
The high-Chern-number QAH effect in the former system appears when two or more pairs of inverted sub-bands are induced by strong exchange fields \cite{jiang2012quantum,wang2013quantum}, while the high-Chern-number QAH effect conjectured in the magnetically doped SnTe system derives from 
the presence of multiple TSSs \cite{fang2014large}. In practice, realization of high-Chern-number QAH states is unlikely in 
Cr-doped Bi$_2$(Se,Te)$_3$ films, given the complex magnetic configurations suggested by non-square
Hall conductivity hysteresis loops and the possible relevance of metallic phases \cite{zhang2013topology,chang2014chemical}. 
The proposed high-Chern-number QAH state in magnetically doped SnTe is also challenging because of the absence of ferromagnetism, and because the multiple Dirac points are normally located at different energies. 
This property makes it difficult to have a fully gapped surface in the SnTe system \cite{fang2014large,wang2018chromium}. 
In addition to these two systems, high-Chern-number QAH states can 
also be realized in magnetic TI-based multilayer structures with alternating $C = 1$ QAH and normal insulator layers \cite{burkov2011weyl}. 
The thickness of the normal insulator layer modulates the coupling between two $C = 1$ QAH layers and thus tunes the Chern numbers of QAH insulators. 

\begin{figure*}[hbt!]
    \centering
    \includegraphics[width=7in]{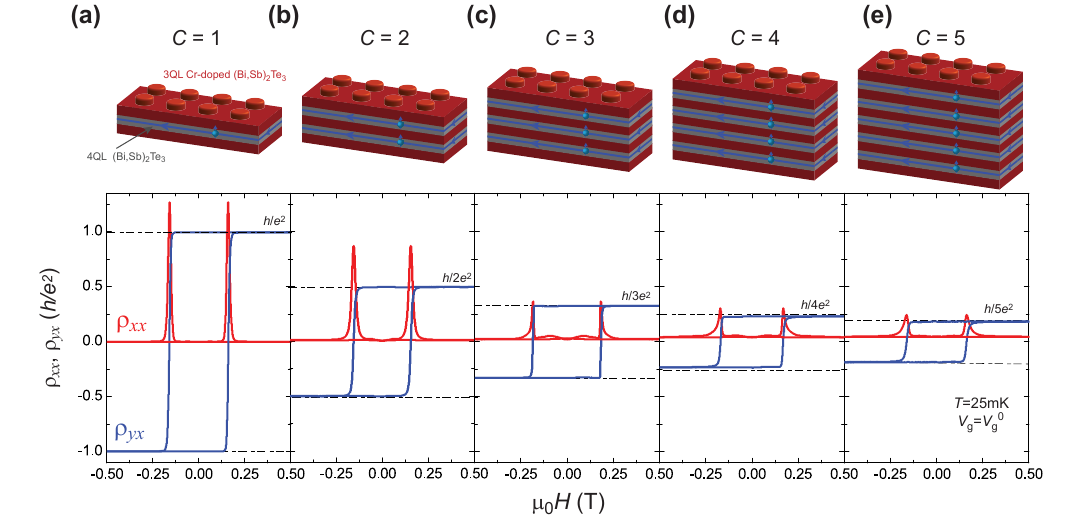}  
    \caption{High-Chern-number QAH effect in magnetic-TI/TI multilayer structures. (a-e), Top: schematic (Legos) multilayer structures for the QAH effect with Chern numbers from \textit{C} = 1 to \textit{C} = 5. The red and gray Legos correspond to 3 QL Cr-doped (Bi,Sb)$_2$Te$_3$ and 4 QL (Bi,Sb)$_2$Te$_3$, respectively. Bottom: magnetic field $\mu_0 H$ dependence of the longitudinal resistance $\rho_{xx}$ (red curve) and the Hall resistance $\rho_{yx}$ (blue curve) measured at the charge neutral point $V_{\rm g} = V_{\rm g}^0$ and \textit{T} =25 mK. From \onlinecite{zhao2020tuning}.}
    \label{fig9:high}
\end{figure*}

\onlinecite{jiang2018quantum} initiated a search for high-Chern-number QAH states
in Cr-doped (Bi,Sb)$_2$Te$_3$/CdSe multilayers.
However, the CdSe layers had a wurtzite structure.  The difference in structure compared to the tetradymite 
structure of magnetic TIs inevitably leads to stacking faults in Cr-doped (Bi,Sb)$_2$Te$_3$/CdSe multilayer samples. 
These defects can induce a large longitudinal resistance and also make the Hall resistances exceed the
quantized value. Recently, \onlinecite{zhao2020tuning} grew heavily Cr-doped TI/TI multilayers with symmetric structures, specifically [3QL Cr-doped (Bi, Sb)$_2$Te$_3$/4QL (Bi, Sb)$_2$Te$_3$]$_m$/3QL Cr-doped (Bi, Sb)$_2$Te$_3$ multilayer structures, where $m$ is an integer reflecting the number of bilayer periods. Well-quantized high-Chern-number QAH effects with Chern numbers $C$ of 1 to 5 were observed [Figs. \ref{fig9:high}(a) to \ref{fig9:high}(e)]. 
It is thought that the bulk band gap is no longer inverted in the heavily Cr-doped TI layer because heavy Cr doping greatly reduces SOC. 
However, the heavy doping can break the TR symmetry of the undoped TI layers, and allows for the $C = 1$ QAH effects
in each undoped TI layer. Heavily Cr-doped-TI/TI multilayer samples behave like several $C = 1$ QAH insulators in parallel.
The Chern number $C$ of the QAH insulators is determined by the number $m$ of undoped TI layers. 
In the same sample configuration, \onlinecite{zhao2020tuning} also showed that the Chern number of the QAH insulators can be tuned by varying either the magnetic doping concentration or the thickness of the interior magnetic TI layers. 
\addCXL{Increasing magnetic TI/TI layer numbers is expected to further raise the Chern number, potentially leading to the 3D QAH state with chiral surface states\cite{halperin1987possible,wang20173d,bernevig2007theory,lu20193d}.  } 
The realization of QAH insulators with tunable Chern numbers facilitates potential applications of dissipationless chiral edge currents in energy-efficient electronic devices and opens opportunities for developing multi-channel quantum computing and higher-capacity chiral circuit interconnects.

\section{MnBi$_2$Te$_4$: an Intrinsic Magnetic TI}\label{Sec:MnBi2Te4}
\subsection{Material properties}\label{sec:MBTproperties}
\begin{figure}[hbt!]
    \centering
    \includegraphics[width=3.5in]{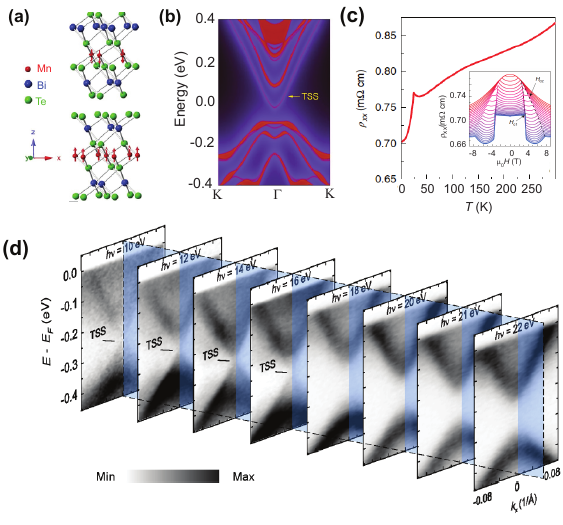}  
    \caption{Properties of MnBi$_2$Te$_4$ single crystals.(a) Crystal structure of MnBi$_2$Te$_4$. 1 SL of MnBi$_2$Te$_4$ is composed of Te-Bi-Te-Mn-Te-Bi-Te. The red arrows indicate the magnetic moments on the Mn layer. (b) Calculated electronic band structure of MnBi$_2$Te$_4$. (c) Temperature dependence of the longitudinal resistivity $\rho_{xx}$. The hump feature occurs at the Néel temperature \textit{T}$_N$ of MnBi$_2$Te$_4$ single crystals.Inset: magnetic field $\mu_0 H$ dependence of $\rho_{xx}$ at different temperatures. (d) ARPES band structures of MnBi$_2$Te$_4$ single crystals measured at different photon energies. From \onlinecite{chen2019intrinsic,chen2019topological,zhang2019topological,liu2020robust}.}
    \label{fig11:MBTcrystal}
\end{figure}

MnBi$_2$Te$_4$, is a tetradymite antiferromagnetic semiconductor, and was recently theoretically predicted \cite{otrokov2017highly,li2019intrinsic,zhang2019topological} and experimentally demonstrated \cite{gong2019experimental,otrokov2019prediction} to be an intrinsic magnetic TI. 
Like the (Bi/Sb)$_2$(Se/Te)$_3$ compounds discussed in Sec.\ref{Sec:MTI}, 
MnBi$_2$Te$_4$ crystals have a rhombohedral structure with the space group $D_{3d}^5$ ($R\bar{3}m$). In the MnBi$_2$Te$_4$ case the building blocks are seven atom thick (Te-Bi-Te-Mn-Te-Bi-Te) septuple layers (SLs). The lattice structure of MnBi$_2$Te$_4$ can be viewed as intercalating a MnTe bilayer into a Bi$_2$Te$_3$ QL [Fig. \ref{fig11:MBTcrystal}(a)]. As noted in Sec.\ref{Sec:Magnetism}, MnBi$_2$Te$_4$ is one member of the Mn(Bi/Sb)$_{2n}$(Se/Te)$_{3n+1}$ family of compounds, 
which includes MnBi$_4$Te$_7$ \cite{li2019dirac,wu2019natural,yan2020type,hu2020van,vidal2019topological,shi2019magnetic,xu2020persistent,ding2020crystal,klimovskikh2020tunable}, MnBi$_6$Te$_{10}$ \cite{li2019dirac,shi2019magnetic,yan2020type,jo2020intrinsic,tian2020magnetic,klimovskikh2020tunable}, MnBi$_8$Te$_{13}$ \cite{hu2020realization},  MnSb$_2$Te$_4$ \cite{yan2019evolution,liu2021site,chen2020ferromagnetism,ge2021direct,murakami2019realization,yan2021site,wimmer2021mn}, MnSb$_4$Te$_7$ \cite{huan2021multiple}, Mn(Bi,Sb)$_2$Te$_4$ \cite{chen2019intrinsic,yan2019evolution,lee2021evidence,chen2020ferromagnetism,jiang2021quantum}, and MnBi$_2$Se$_4$ \cite{zhu2021synthesis}.
Mn(Bi/Sb)$_{2n}$(Se/Te)$_{3n+1}$ can be viewed as (\textit{n}-1) (Bi/Sb)$_2$(Se/Te)$_3$ QLs stacked on 
a Mn(Bi/Sb)$_2$(Se/Te)$_4$ parent to form one unit cell.

As noted in Sec.\ref{Sec:Magnetism}, the magnetic properties of MnBi$_2$Te$_4$ stem primarily from the presence of 
the Mn 3\textit{d} local moments. The Mn$^{2+}$ ions are located at SL centers and carry $5 \mu_{\rm B}$ magnetic moments.
They have strong ferromagnetic exchange interactions within each SL, and weaker 
antiferromagnet interactions between adjacent layers [Fig. \ref{fig11:MBTcrystal}(a)], and therefore 
have A-type antiferromagnetic order with a Néel temperature \textit{T}$_N$ of $\sim$25 K [Fig.\ref{fig11:MBTcrystal} (c)] \cite{lee2019spin,otrokov2019prediction,yan2019crystal,liu2020robust}. As discussed in Sec. \ref{Sec:Magnetism}. 
the intralayer ferromagnetism (interlayer antiferromagnetism) is induced by ferromagnetic superexchange (antiferromagnetic super-superexchange), 
and captured by DFT calculations \cite{otrokov2017highly,li2019intrinsic,zhang2019topological}.
Magnetic properties in these compounds are strongly influenced by impurity states;
for some synthesis conditions, MnSb$_2$Te$_4$ can exhibit ferromagentism with \textit{T}$_{\rm C}$ of 25$ \sim$50 K \cite{liu2021site,ge2021direct,murakami2019realization,wimmer2021mn,yan2021site}. 

MnBi$_{2n}$Te$_{3n+1}$ compounds have $n$ QLs of Bi$_2$Te$_3$ between each MnBi$_2$Te$_4$ SL.  
The strength of the antiferromagnetic coupling between neighboring Mn layers therefore decreases strongly with 
$n$, and can become ferromagnetic. 
For \textit{n} = 2 and 3 (i.e. MnBi$_4$Te$_7$ and MnBi$_6$Te$_{10}$), the compounds still show the A-type antiferromagnetim, while for \textit{n} $\geq$ 4, the compounds become ferromagnetic \cite {wu2019natural,yan2020type,klimovskikh2020tunable}.
The Néel temperature \textit{T}$_N$ values of MnBi$_4$Te$_7$ and MnBi$_6$Te$_{10}$ are $\sim$13 K and $\sim$11 K, respectively \cite{yan2020type,shi2019magnetic,tian2020magnetic}, while the Curie temperature \textit{T}$_C$ of MnBi$_8$Te$_{13}$ is $\sim$ 10.5 K \cite{hu2020realization}. 


The electronic band structures and topological properties of MnBi$_2$Te$_4$ 
and its relatives depend qualitatively on magnetic configurations, and these can be 
altered by relatively weak magnetic fields because of the weak interlayer exchange interactions.
For example, bulk MnBi$_2$Te$_4$ is a magnetic TI in its ground antiferromagnetic configuration,
with the topological nontrivial surface states formed by 
Bi and Te \textit{p} orbital bands that are inverted and have strong SOC [Fig. \ref{fig11:MBTcrystal}(b)] \cite{li2019intrinsic,zhang2019topological,otrokov2019prediction}.
In the aligned spin configuration induced by relatively modest magnetic fields, 
it becomes a type-II Weyl semimetal with a single pair of Weyl points \cite{li2019intrinsic,zhang2019topological,lei2020magnetized}. 

Although the electronic bands of these compounds are theoretically expected to be relatively simple,
early experimental studies have painted a more confusing picture.
In early ARPES measurements, the TSSs of MnBi$_2$Te$_4$ bulk crystals were found to have surface gaps greater 
than 50 meV \cite{lee2019spin,otrokov2019prediction,vidal2019surface}. 
However, the observed gap had a weak temperature dependence and could be detected even above $T_N$, 
suggesting that it was not related to magnetic order in MnBi$_2$Te$_4$. 
Soon after, several groups performed high-resolution synchrotron/laser ARPES measurements with photon energy $h\nu$ $\leq$ 16 eV 
and observed gapless TSSs which cannot be resolved for $h\nu$ $>$ 16 eV [Fig.\ref{fig11:MBTcrystal} (d)] \cite{chen2019topological,hao2019gapless,li2019dirac,swatek2020gapless,nevola2020coexistence,chen2019intrinsic,yan2021origins}. 
This finding suggests that the gap observed in prior studies \cite{lee2019spin,otrokov2019prediction,vidal2019surface}
could be a bulk gap. The fact that the gapless TSSs are observed only for $h\nu$ $\leq$ 16 eV 
might be a result of matrix element and cross-section effects in ARPES measurements \cite{chen2019topological}. The observation of the gapless TSSs is also independent of magnetic ordering in MnBi$_2$Te$_4$. The formation of the gapless TSSs in MnBi$_2$Te$_4$ bulk crystals has been attributed to the existence of multiple antiferromagnetic domains \cite{chen2019topological}, to unexpected types of magnetic order, {\it e.g.} in-plane \textit{A}-type antiferromagentism
or \textit{G}-type antiferromagnetism) \cite{swatek2020gapless,hao2019gapless}, and to 
weak hybridization between the localized magnetic moments and the surfaces of MnBi$_2$Te$_4$ \cite{li2019dirac}. 
Further study is needed to clarify the origin of the gapless TSSs observed in some ARPES studies of
MnBi$_2$Te$_4$ bulk crystals.

For MnBi$_{2n}$Te$_{3n+1}$ with \textit{n} $\geq$ 2, the ARPES results are also controversial. \onlinecite{hu2020universal,xu2020persistent} observed gapless TSSs for all terminations of MnBi$_4$Te$_7$ and MnBi$_6$Te$_{10}$, whereas \onlinecite{jo2020intrinsic} observed a temeprature-dependent gap opening on MnBi$_2$Te$_4$ with SL termination,
but gapless surface states with Bi$_2$Te$_3$ QL terminations. Separately, \onlinecite{gordon2019strongly,ma2020hybridization,hu2020realization,hu2020van,wu2020distinct,yan2021delicate} observed a gapped TSS 
in MnBi$_6$Te$_{10}$ and MnBi$_8$Te$_{13}$ on with 1QL Bi$_2$Te$_3$/1SL MnBi$_2$Te$_4$ termination, 
which might be interpreted in terms of of surface-bulk band hybridization, 
but gappless TSSs for other terminations.  The topological character of MnSb$_2$Te$_4$ is still under 
debate \cite{chen2019intrinsic,wimmer2021mn}. More recently, MnSb$_4$Te$_7$ was demonstrated to be an intrinsic antiferromagnetic TI with \textit{T}$_N$ $\sim$ 13.5 K, and to possess multiple topological phases, including \addCXL{antiferromagnetic TI phase, ferromagnetic inversion-symmetry-protected axion insulator phase, and ferromagnetic Weyl semimetal phase with multiple Weyl nodes, depending on magnetic configurations and doping levels \cite{huan2021multiple}}.  

In addition to direct magnetic and ARPES measurements, transport, STM, and magnetic force microscopy measurements have also been 
performed on MnBi$_2$Te$_4$ bulk crystals. Negative magnetoresistance below and above \textit{T}$_N$, the emergence of a canted antiferromagnetic state between the antiferromagnetic and spin-aligned states, and nonlinear Hall traces in the canted antiferromagnetic state, have all been reported in transport studies \cite{lee2019spin,cui2019transport,li2020antiferromagnetic,chen2019intrinsic}. 
Low-temperature STM measurements have found large densities (3 $\sim$ 5\%) of Mn/Bi antisites, 
which might affect intralayer and interlayer magnetic coupling of MnBi$_2$Te$_4$  \cite{yan2019crystal,liang2020mapping,huang2020native,yuan2020electronic}. 
Very recently, antiferromagnetic domains as well as the \textit{A}-type antiferromagnetic order were imaged in MnBi$_2$Te$_4$ and Mn(Bi,Sb)$_2$Te$_4$ bulk crystals by magnetic force microscopy \cite{sass2020magnetic,sass2020robust}. 
More focused reviews of the MnBi$_2$Te$_4$ family of compounds can be found in \onlinecite{ning2020recent,wang2021intrinsic,zhao2021routes,sekine2021axion}.

\subsection{MnBi$_2$Te$_4$ thin films}


The \textit{A}-type antiferromagnetic order has a substantial influence on the topological properties of MnBi$_2$Te$_4$ thin films. Because of the ferromagnetism within each SL, the surface states develop a magnetic exchange gap driven by 
surface magnetism.  The signs of the half-quantized Hall conductance $\sigma_{xy}$ contributed by the top and bottom
surfaces are determined by the direction of magnetic order at that surface.
For perfect A-type order, the surface states on top and bottom SLs will respond to 
\addCXL{parallel} magnetizations when the number $N$ of septuple layers is odd and to \addCXL{antiparallel} magnetizations 
when $N$ is even.  The total Hall conductance is therefore expected to be quantized in the absence of a 
magnetic field at $\pm e^2/h$ when $N$ is odd and the 
films are thick enough to weaken hybridization between the top and bottom surfaces.  Even $N$ films have no Hall conductance. 
\am{Like the ZHCP states of magnetically doped sandwich 
structures discussed previously, even $N$ MnBi$_2$Te$_4$
\addCXL{thick} films are referred to as axion insulators in the literature  \cite{otrokov2019unique,li2019intrinsic,zhang2019topological,liu2020robust},
a practice that we will follow. 
The axion electrodynamics field theory model applies only to 
even $N$ MnBi$_2$Te$_4$ thin films.   
Films with an odd number of SLs will yield QAH states.}
It has been predicted that an in-plane magnetic field can drive MnBi$_2$Te$_4$ 
into a higher-order M$\ddot{o}$bius insulator with surface “M$\ddot{o}$bius fermions” and chiral hinge states \cite{zhang2020mobius}. 


Atomically thin MnBi$_2$Te$_4$ films can be formed either by MBE growth \cite{gong2019experimental,lapano2020adsorption,tai2021polarity,bac2022topological,rienks2019large,kagerer2020molecular,zhu2020investigating,trang2021crossover,zhao2021even} or by manual exfoliation from bulk crystals \cite{deng2020quantum,ge2020high,liu2020robust,liu2021magnetic,ovchinnikov2021intertwined,ying2022experimental,gao2021layer}. MBE growth of MnBi$_2$Te$_4$ thin films have been achieved by alternate deposition of Bi$_2$Te$_3$ and MnTe layers \cite{gong2019experimental,bac2022topological,trang2021crossover,zhao2021even}, co-evaporation of Mn, Bi, and Te elements  \cite{zhu2020investigating,lapano2020adsorption,rienks2019large,tai2021polarity} and co-deposition of MnTe and Bi$_2$Te$_3$ sources \cite{kagerer2020molecular}. Due to unusual growth dynamics, Bi$_2$Te$_3$, MnTe, and Mn-doped Bi$_2$Te$_3$
inevitably coexist with the dominant MnBi$_2$Te$_4$ phase in all these MBE-grown materials.
MBE growth of MnBi$_2$Te$_4$ follows a SL-by-SL growth mode, 
which is similar to the QL-by-QL growth mode of the well-studied Bi$_2$Te$_3$ family TI \cite{zhang2010crossover,li2010intrinsic}. 
For this reason, it is difficult to achieve uniform thickness MnBi$_2$Te$_4$ thin films
over millimeter size ranges.  Due to this property, the Hall resistances measured in 
MBE-grown MnBi$_2$Te$_4$ thin films \cite{gong2019experimental,tai2021polarity,lapano2020adsorption,bac2022topological,zhao2021even}
are normally small.  Realization of the QAH effect requires uniform thickness across entire devices, 
which is challenging for MBE-grown samples. 

\begin{figure}[hbt!]
    \centering
    \includegraphics[width=3.5in]{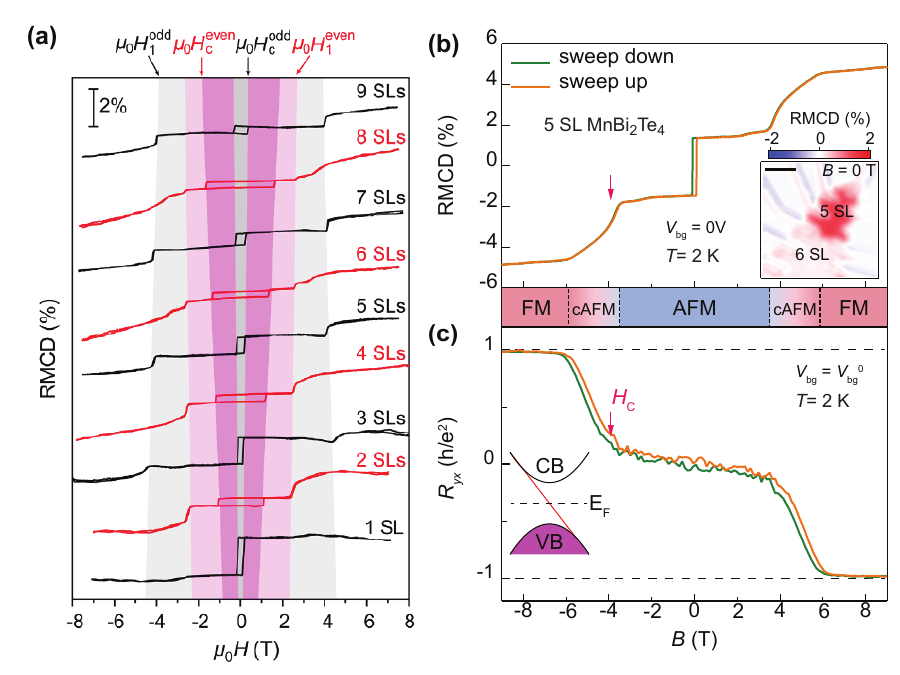}  
    \caption{Relationship between RMCD and Hall data. (a) RMCD measurements on MnBi$_2$Te$_4$ flakes (from 1 SL to 9 SLs) at \textit{T} =1.6 K. The shaded areas highlight the thickness dependence of the low-field spin-flip and spin-flop phase transitions in odd- and even-number SL samples. (b,c) Magnetic field $\mu_0 H$ dependence of RMCD signal and the Hall resistance $R_{yx}$ of a 5SL MnBi$_2$Te$_4$ device. The RMCD and transport characteristics in (b) and (c) are measured in
    the same sample at $V_{\rm g} = 0$, and at the charge neutral point $V_{\rm g} = V_{\rm g}^0$,  respectively. 
    From \onlinecite{yang2021odd,ovchinnikov2021intertwined}.}
    \label{fig12:rmcd}
\end{figure}

\begin{figure}[hbt!]
    \centering
    \includegraphics[width=3.5in]{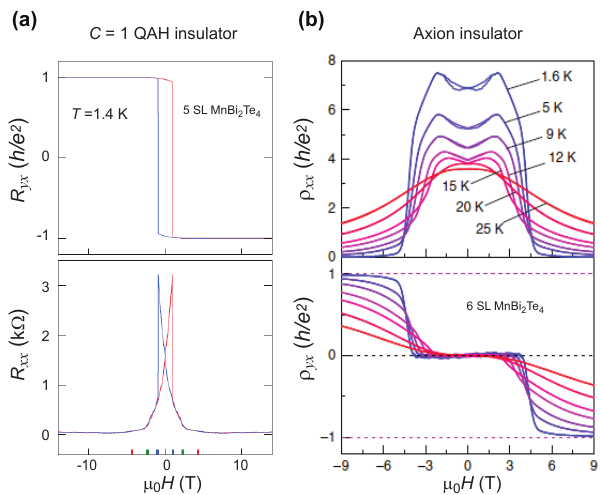}  
    \caption{Realization of the \textit{C}= 1 QAH insulator and a zero Hall plateau in exfoliated MnBi$_2$Te$_4$ flakes. 
    (a) \textit{C} = 1 QAH insulator observed in 5 SL MnBi$_2$Te$_4$. (b) Zero Hall plateau, 
    taken as evidence for an axion insulator state, observed in 6 SL MnBi$_2$Te$_4$. From \onlinecite{deng2020quantum,liu2020robust}.}
    \label{fig12:QAHMBT}
\end{figure}

Mechanically-exfoliated MnBi$_2$Te$_4$ thin flakes are typically small and can be formed into uniform $\mu$m scale devices.
Moreover, exfoliated flakes can be flexibly combined with other 2D materials to 
fabricate gated devices in which unintentional doping can be compensated.
Indeed, both QAH and axion insulator states have been claimed to be realized in the absence of a magnetic field \cite{deng2020quantum,liu2020robust}.
Magnetic fields influence electronic structure directly, but most strongly via their impact on magnetic 
configurations.  Both antiferromagnetic and spin-aligned Chern insulator states have been realized 
under high magnetic fields in exfoliated MnBi$_2$Te$_4$ flakes [Figs.\ref{fig12:rmcd}(b),\ref{fig12:rmcd}(c),and \ref{fig12:QAHMBT}] \cite{deng2020quantum,ge2020high,liu2020robust,liu2021magnetic,ovchinnikov2021intertwined,ying2022experimental,gao2021layer}, 
and will be discussed in Sec.\ref{Sec:mbtqah}. 

We now discuss the SL-number $N$ dependence of the magnetic properties of exfoliated MnBi$_2$Te$_4$ flakes.
Since even- and odd-SL MnBi$_2$Te$_4$ films respectively have compensated and uncompensated magnetic moments, 
their total magnetizations should exhibit distinct magnetic-field dependences. 
Recently, two independent studies reported an even-odd effect 
in the magnetism of thin MnBi$_2$Te$_4$ flakes thin flakes using reflection magnetic circular dichroism (RMCD) measurements [Fig. \ref{fig12:rmcd}(a)] \cite{ovchinnikov2021intertwined,yang2021odd}. 
Specifically: (\textit{i}) odd SL flakes have pronounced hysteresis loops and 
significant remanent RMCD signals at weak magnetic fields that reflect uncompensated magnetic moments.
On the other hand, even SL samples have vanishingly small RMCD signals and hysteresis loops, consistent with their zero net magnetization. 
(\textit{ii}) the critical magnetic field for the spin-flop transition between antiferromagnetic and canted antiferromagnetic states [labeled by $H_1$ in Fig. \ref{fig12:rmcd}(a)],
which occurs at intermediate field strengths, is $\sim$2.6 T for even SL and $\sim$4.2 T for odd SL MnBi$_2$Te$_4$ samples. The larger $H_1$ for odd SL MnBi$_2$Te$_4$ compared to the even ones can be understood in terms of the Zeeman energy of coupling to 
uncompensated magnetization in odd SL samples \cite{ovchinnikov2021intertwined,yang2021odd}. This behavior was also observed in transport measurements on both exfoliated and MBE-grown MnBi$_2$Te$_4$ films \cite{chen2019intrinsic,zhao2021even}. 

\subsection{Chern insulators, axion insulators, and the QAH effect}\label{Sec:mbtqah}



\onlinecite{deng2020quantum} realized the \textit{C} = 1 QAH effect in a 5 SL exfoliated MnBi$_2$Te$_4$ device soon after it was predicted theoretically. Figure \ref{fig12:QAHMBT}(a) shows the magnetic field $\mu_0H$ dependence of the Hall resistance $R_{yx}$ and the longitudinal resistance $R_{xx}$. At \textit{T} = 1.4 K, $R_{yx}\sim 0.97 h/e^2$ and $R_{xx} \sim 0.061 h/e^2$ at zero magnetic field. 
Hall quantization accuracy can be improved by applying an intermediate strength 
external magnetic field to stabilize an aligned spin magnetic configurations in which 
 $R_{yx}\sim0.998 h/e^2$ at $\mu_0H$ $\sim$ 2.5 T. 
 \addCXL{To date, the \textit{C} = 1 QAH effect at zero magnetic field in exfoliated MnBi$_2$Te$_4$ devices has been realized by this group.} 
 However, the \textit{C} = 1 Chern insulator state in
 spin-aligned MnBi$_2$Te$_4$ under magnetic fields has been realized repeatedly by several different groups in devices with both even and odd numbers of SLs \cite{ge2020high,liu2021magnetic,liu2020robust,ovchinnikov2021intertwined,deng2020quantum,ying2022experimental,gao2021layer}. 
 The \textit{C} = 1 Chern insulator state with $R_{yx}\ge 0.904 h/e^2$ in exfoliated MnBi$_2$Te$_4$ devices
 is found to survive even up to $\sim$45 K under high magnetic fields, higher than $T_\textit{N}$ $\sim$25 K \cite{ge2020high}.

\begin{figure}[hbt!]
    \centering
    \includegraphics[width=3.5in]{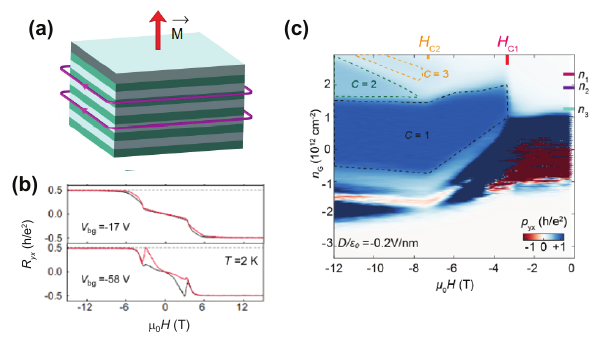}  
    \caption{ \textit{C}=2 Chern insulator state in MnBi$_2$Te$_4$ devices. (a) Schematics for the \textit{C }= 2 Chern insulator state with two chiral edge states propagating along the edges of the sample. (b) Magnetic field $\mu_0 H$ dependence of the Hall resistance $R_{yx}$ of a 10 SL MnBi$_2$Te$_4$ device at fixed bottom gate voltages $V_{bg} = -17V$ and -58V. (c) $\rho_{yx}$ as a function of magnetic field $\mu_0 H$ and gate injected carrier density $n_G$ at fixed electric field $D/\varepsilon_0$ = -0.2 V/nm in a 7 SL MnBi$_2$Te$_4$ device. Black, green, and yellow dashed lines enclose \textit{C} = 1, 2, 3 states, respectively. The marked densities are $(n_1,n_2,n_3) = (2.23,1.87,1.29) \times 10^{12} cm^{-2}$. From \onlinecite{ge2020high, cai2022electric}. } \label{fig13:high}
\end{figure}


\onlinecite{liu2020robust} measured axion insulator characteristics in an exfoliated MnBi$_2$Te$_4$ device with a nominal
6 SL thickness and the ground antiferromagnetic spin configuration.  Figure \ref{fig12:QAHMBT}(b) shows the $\mu_0 H$ dependence of $\rho_{xx}$ and $\rho_{yx}$ at $V_{\rm g} =V_{\rm g}^0$. At the base temperature \textit{T }= 1.6 K, $\rho_{yx}$ is found to be zero over the magnetic field range -3.5 T $<\mu_0 H <$ 3.5 T, concomitant with a large $\rho_{xx}$ of $\sim 7 h/e^2$. 
In this work, the authors considered the large $\rho_{xx}$ combined with vanishing $\rho_{yx}$ as the key. 
However, \onlinecite{ovchinnikov2021intertwined} performed combined electrical transport and RMCD measurements and found that the RMCD and AH hysteresis loops are not synchronized in their devices, whether even or odd SL. 
For a 5 SL MnBi$_2$Te$_4$ device, they found that both $\rho_{yx}$ and its hysteresis loop 
nearly vanish over the magnetic field range -3.8 T $<\mu_0 H <$ 3.8 T, in sharp contrast to RMCD
characteristics which showed pronounced hysteresis [Figs. \ref{fig12:rmcd}(b) and \ref{fig12:rmcd}(c)]. 
The nearly vanishing $R_{yx}$ and the large $R_{xx}$, which reaches $\sim 7h/e^2$, in 
this nominal 5 SL MnBi$_2$Te$_4$ device, are both similar to the behavior of the nominally 6 SL MnBi$_2$Te$_4$
device shown in Fig. \ref{fig12:QAHMBT}(b) \cite{liu2020robust}. 
Since the large $\rho_{xx}$ combined with vanishing $\rho_{yx}$ are also signatures of a trivial 2D insulator,
\onlinecite{ovchinnikov2021intertwined} speculated that the antiferromagnetic state of 5 SL MnBi$_2$Te$_4$ 
devices might be a trivial magnetic insulator.  Indeed the only practical distinction between 
axion insulators and ordinary insulators in \am{zero magnetic field transport measurements} is the tendency of the former to have 
very small side-wall energy gaps that likely limit the resistivity of  high-quality devices \am{in which bulk transport is negligible.}    


In addition to axion insulator and \textit{C} =1 Chern insulator states , 
\textit{C} = 2 Chern insulator states have also been observed in the magnetic-field induced
spin-aligned states of exfoliated MnBi$_2$Te$_4$ devices with 9 SL and 10 SL thicknesses [Figs. \ref{fig13:high}(a) and \ref{fig13:high}(b)]. \onlinecite{ge2020high} attributed the observation of the \textit{C} = 2 Chern insulator in thick MnBi$_2$Te$_4$ devices to an interplay 
between quantum confinement effects and the Weyl semimetal physics of MnBi$_2$Te$_4$ in spin-aligned high magnetic field configurations \cite{li2019intrinsic,zhang2019topological}. In the limit of thick films that ratio of the Chern number $C$ to the septuble layer 
number $N$ is expected \cite{lei2020magnetized} to approach the Hall conductivity per layer of the bulk semimetal in quantum
$e^2/h$ units.  
\onlinecite{cai2022electric} recently fabricated the MnBi$_2$Te$_4$ devices with dual gates and mapped the $\rho_{yx}$ of a 
7 SL MnBi$_2$Te$_4$ device as a function of both $\mu_0 H$ and injected carrier density $n_G$ at a fixed 
displacement field $D/\varepsilon_0 = -0.2 V/nm$. In addition to the usual \textit{C} = 1 plateaus in both canted 
antiferromagnetic and ferromagnetic regimes, \textit{C} = 2 and 3 states, signaled by $\rho_{yx}\sim h/Ce^2$, appear in spin-aligned 
states at higher carrier densities $n_G$. Unlike the \textit{C} = 1 phase, the \textit{C}= 2 and 3 phases are
stablized by strong magnetic fields and are centered on densities that are proportional to field $\mu_0 H$ [Fig. \ref{fig13:high}(c)], which suggests that they are related to the formation of Landau levels \cite{cai2022electric,ge2020high,deng2020quantum,li2021coexistence}.


More recently, \onlinecite{gao2021layer} observed that an AH effect is induced in an even SL number ($N=6$) MnBi$_2$Te$_4$ device by a gate electric field. \czchang{This effect is referred to as the layer Hall effect, \addCXL{ in which the Hall current responses are layer dependent and can cancel each other in two opposite layers due to the combined symmetry of inversion and TR. Applying a gate voltage breaks inversion and thus leads to the emergence of a large, layer-polarized AH effect. Reversing gate voltages can change the sign of AH resistance.} 
} \czchang{In this work, \onlinecite{gao2021layer}} interpreted their results in terms of electric field-induced polarization and 
layered-locked Berry curvatures. We note that the interpretations of transport observations like 
these can be reinforced by RMCD data, as discussed above \cite{yang2021odd,ovchinnikov2021intertwined}. The studied MnBi$_2$Te$_4$ device has a 
quantized AH effect in the field-induced spin-aligned state 
when the chemical potential is tuned to the charge neutral point.
The layer Hall effect, on the other hand, is observed when the Fermi level lies outside the gap and is not quantized.
A gate-induced sign reversal of the AH effect near-zero magnetic field was systematically studied in 
exfoliated MnBi$_2$Te$_4$ devices by \onlinecite{zhang2019experimental}.  
In odd SL number $N$ MnBi$_2$Te$_4$ thin films, the QAH effect 
is expected \cite{lei2021gate} to break down after gate electric fields exceed a breakdown
field $\sim 10$ meV/nm.


\section{Moir\'e Materials}\label{Sec:tbg} 
The large lattice constants \am{of the moir\'e} material 
systems introduced in Section ~\ref{Sec:moirematerial} are 
important because they allow the number of electrons per 
period to be varied by up to around ten purely with electrical gates, effectively moving 
through the periodic table of these artificial materials without doping.
In many cases\cite{andrei2021marvels}, 
moir\'e materials can be tuned into regimes 
in which correlations are strong and broken symmetries are common.  
The broken symmetry states that have been realized include 
superconductors \cite{cao2018unconventional,yankowitz2019tuning,lu2019superconductors,balents2020superconductivity}, Mott insulators \cite{cao2018correlated}, 
Wigner crystals \cite{xu2020correlated,li2021imaging} and - 
of particular interest to this review - unusual orbital 
ferromagnets with Chern insulator ground states \cite{sharpe2019emergent,serlin2020intrinsic,chen2020tunable,chen2020ferromagnetism,
polshyn2020electrical,sharpe2021evidence,li2021quantum,chen2020tunable,tschirhart2021imaging}.  In the following, we explain the QAH effects seen in graphene and  
TMD moir\'e materials, 
which are alike in that they rely on spontaneous valley polarization, but 
distinct in that their non-trivial topologies are related respectively to sublattice and 
layer degrees of freedom.

\subsection{Strong correlations in twisted bilayer graphene}
\begin{figure}[hbt!]
    \centering
    \includegraphics[width=3.5in]{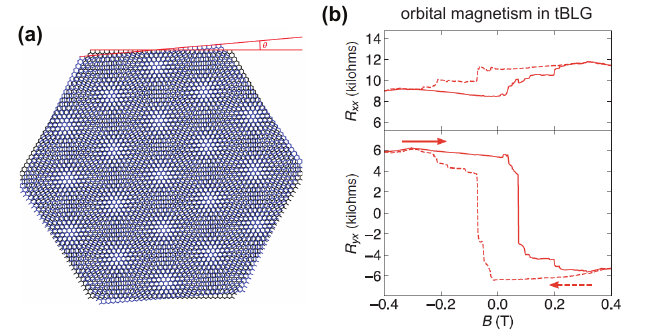}  
    \caption{Orbital magnetism in twisted bilayer graphene systems. (a) Schematic of twisted bilayer graphene. The relative orientation 
    angle $\theta$ determines the periodicity of the moiré superllattice. (b) Magnetic field $\mu_0 H$ dependence of the longitudinal resistance $R_{xx}$ and the Hall resistance $R_{yx}$ with $n/n_s = 3/4$ and $D/D_0=-0.62$V/nm at \textit{T} =30 mK, demonstrating the hysteretic AH effect resulting from emergent orbital magnetic order. From \onlinecite{sharpe2019emergent}. }
    \label{fig14:orbital}
\end{figure}

\am{Because of its superlattice periodicity twisted bilayer graphene (TBG) [Fig.\ref{fig14:orbital}(a)], the first moir\'e material, has low energy emergent
Bloch bands \cite{dos2007graphene,li2010observation} in a BZ defined by the moir\'e superlattice.  
\onlinecite{bistritzer2011moire} predicted that when the rotation angle 
was adjusted to a magic-angle $\sim$1.1°, electronic velocities would slow markedly, 
converting bilayer graphene from a 2D Fermi liquid with moderate interaction strengths
to a strongly correlated electron system.  
Progress in accurate twist angle control \cite{kim2017tunable,ribeiro2018twistable}, \onlinecite{cao2018correlated,cao2018unconventional} eventually enabled this prediction to 
be tested.  In 2018 it was discovered \cite{cao2018correlated,cao2018unconventional} 
that magic-angle TBG has a rich phase diagram as a 
function of twist angle and moir\'e band filling factor $\nu$
with correlated insulators 
common at integer values of $\nu$, and superconducting domes common over intervals of non-integer $\nu$.  
Thd discovery was confirmed 
and the rich properties of magic-angle TBG and related multilayer systems 
were more fully revealed by subsequent experiments \cite{yankowitz2019tuning,lu2019superconductors}.
Some of this ongoing work is reviewed in \onlinecite{andrei2020graphene,liu2021orbital}. 
The discussion below focuses on the unusual 
ferromagnetic Chern insulator states that are responsible for the QAH effect.}

\subsection{Orbital TR symmetry breaking}\label{Sec:orbital}

In TBG, spin and valley degrees of freedom combine to 
provide four flavors for the superlattice Bloch states.    The Hamiltonian is spin-independent,
and the two valleys (located at opposite corners of the atomic-scale BZ)
are exchanged by TR. It follows that every moir\'e Bloch state is 
four-fold degenerate.  Because there are two flat bands centered on 
neutrality for each flavor, the flat bands are partially occupied for moir\'e band 
filling factors $ \nu \in (-4,4)$.  The broken symmetry states that are common in
magic-angle TBG have conventional \addCXL{particle-hole density matrix
order parameters in the spin/valley flavor
space}, and in most cases do not break translational symmetry.  
They can therefore be viewed as 
generalized ferromagnets.  As we have discussed previously, the AH effects 
that have been observed in moir\'e materials result from the order that is characterized by a finite valley polarization and not,
as in conventional ferromagnets and magnetic TIs, from SOC combined with spontaneous spin-polarization.  
Magic-angle TBG is quite exceptional in that 
SOC does not play an essential role in its AH effect. 
Finite valley polarization also leads to finite values of orbital magnetization, as well as 
Faraday and Kerr responses.
The QAH effect in magic-angle TBG is most common when
the moiré band filling factor is an odd integer, but this 
condition is neither sufficient nor necessary as we explain below.   

\subsection{QAH effect at odd moir\'e band filling factors}
\onlinecite{sharpe2019emergent} aligned a TBG  moiré superlattice with 
a \textit{h}-BN cladding layer and found evidence of emergent ferromagnetism with a 
giant AH effect at $\nu=3$ [Fig. \ref{fig14:orbital}(b)].  Over a narrow range of carrier density \textit{n} near 
$\nu=3$, the magnetic field dependence of the longitudinal resistance $R_{xx}$ and the Hall resistance $R_{yx}$ show hysteretic behavior suggestive of collective behavior. 
At zero magnetic field, a large $R_{yx} \approx \pm 6 k\Omega$ remains after 
magnetic field cycling, with the sign depending on the direction of the magnetic field sweep. 
This observation is clear evidence for orbital ferromagnetic order in TBG. 
Non-local transport measurements suggest that the TBG system might be a Chern insulator. 
The authors also found that the Hall effect of the TBG sample could be reversed by 
applying a very small external current. More recently, anomalous 
Hall effects were also realized in twisted monolayer-bilayer graphene \cite{polshyn2020electrical,chen2021electrically}, in ABC trilayer graphene on \textit{h}-BN, and in 
TBG on WSe$_2$ \cite{lin2022spin}. 

\begin{figure}[hbt!]
    \centering
    \includegraphics[width=3.5in]{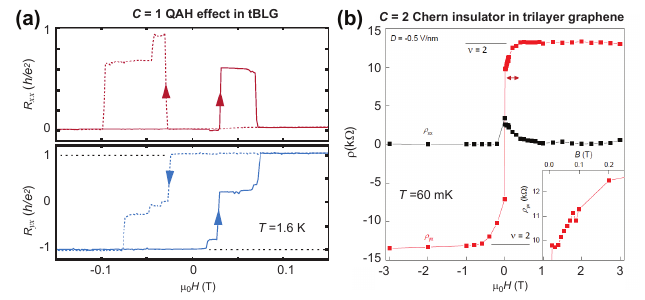}  
    \caption{\textit{C}=1 QAH effect in TBG  and \textit{C}=2 Chern insulator state in trilayer graphene. (a) Magnetic field $\mu_0 H$ dependence of the longitudinal resistance $R_{xx}$ and the Hall resistance $R_{yx}$ in TBG measured at $n = 2.37\times 10^{12} cm^{-2}$. (b) Magnetic field $\mu_0 H$ dependence of the longitudinal resistance $\rho_{xx}$ and the Hall resistance $\rho_{yx}$ in a rhombohedral trilayer graphene-\textit{h}-BN moiré superlattice. A \textit{C} =2 Chern insulator state is realized under a finite external magnetic field. From \onlinecite{serlin2020intrinsic, chen2020tunable}.
    }
    \label{fig15:TBG}
\end{figure}
Soon after the discovery of the AH effect, which signals spin/valley order that breaks TR symmetry,  
\onlinecite{serlin2020intrinsic} optimized the device fabrication process and indeed realized the \textit{C} = 1 QAH effect in TBG [Fig.\ref{fig15:TBG}(a)]. At zero magnetic field and \textit{T} =1.6 K, the Hall resistance $R_{yx}$ approaches $h/e^2$ in a narrow range of density near 
moir\'e band filling filling $\nu=3$, concomitant with a deep minimum in the longitudinal resistance $R_{xx}$.  The Curie temperature for flavor ordering $T_{\rm{C}}\sim$ 7.5 K. 
Accurate $R_{yx}$ quantization can survive up to $T\sim$3 K. 
So far, this experiment has not been reproduced by other groups, presumably due to the difficulties in the fabrication of high-quality TBG/{\it h}-BN devices. There is experimental evidence that some
insulating states at odd values of $\nu$ Chern insulators and some are not \cite{lu2019superconductors}. 

Several theoretical papers \cite{po2018origin,seo2019ferromagnetic,kang2019strong,xie2020nature,bultinck2020mechanism}
had raised the theoretical possibility of a QAH effect in magic-angle TBG due to spin/valley 
ferromagnetism, some prior to the experimental observation.
These papers are compatible with the experimental observation that although the QAH effect is 
sometimes observed at odd integer filling factors, it often is not observed.
First, the flavor splitting produced by scattering off the 
order must be larger than the width of the single-particle flat 
bands in order to eliminate the Fermi surfaces present for $\nu \in (-4,4) $ in the normal state.
The ratio of the flavor-splitting of the bands to the band
width depends \cite{xie2020nature} strongly on twist angle, 
and on the proximity of electrodes used for 
gating, which tend to screen \cite{liu2021tuning} electron-electron interactions.
Second, the Chern numbers of the occupied bands must have a non-zero sum.
\onlinecite{bultinck2020mechanism} emphasized that a non-zero Chern number in 
an individual valley can originate from the coupling between the graphene device 
and the encapsulating \textit{h}-BN dielectric, which breaks the $C_2T$ symmetry that protects the \addCXL{2D}
Dirac points \cite{ahn2017unconventional}.  $C_2T$ symmetry can also break spontaneously.

In general, we expect the flat band Chern numbers to be $\pm 1$
whenever the $C_2T$ symmetry is violated. The expectation is justified 
provided that there are no band closings between the large twist angle
limit \cite{macdonald2012pseudospin} where layers are weakly coupled and the smaller 
twist angles at which the bands become flat.  In the large twist angle 
limit, the isolated layer half-quantized Hall conductivities add to 
yield Chern number one.
Level crossings do sometimes occur \cite{xie2020nature}, depending on the details of the band and 
interaction model, and this may explain the occasional observation of 
odd filling-factor insulating states that appear not to have a QAH effect.
QAH states with higher Chern numbers have been 
studied in a number of twisted multi-layer systems \cite{zhang2019nearly,liu2019quantum}, whose properties 
are sensitive to gate-applied vertical displacement fields. 

\onlinecite{liu2019pseudo} suggested a pseudo-magnetic field picture to understand 
the non-zero Chern number in each valley and the resulting QAH effect in TBG.
Given the similarity between the flat bands in TBG and Landau levels, fractional Chern insulator phases have also been discussed theoretically \cite{ledwith2020fractional,repellin2020chern}. 
\onlinecite{he2020giant} proposed a giant orbital magnetoelectric effect in TBG, which can explain the extremely small 
current that can switch the magnetization in TBG \cite{sharpe2019emergent}.
The band basis with non-zero Chern number in each valley, sometimes called “Chern band basis”, 
is also valuable for understanding other topological band property, e.g. fragile topology \cite{ahn2019failure,po2019faithful,song2019all,song2021twisted}, 
and other correlated phases in twisted bilayer graphene \cite{lian2021twisted,bernevig2021twisted,liu2021theories}. 


The QAH effect also occurs in graphene multilayer moir\'e superlattices when the 
moir\'e bands are flat, sometimes with systematics that are different from those of the bilayer case.
\onlinecite{chen2020tunable} observed hysterisis and a \textit{C}= 2 Chern insulator state at 
band filling $\nu=1$ in an ABC trilayer graphene/\textit{h}-BN moiré superlattice [Fig. \ref{fig15:TBG}(b)]. At \textit{T}= 60 mK, the Hall resistance $\rho_{yx}$ is quantized to 13.0$\pm$ 0.2 k$\Omega$ for magnetic fields larger than $\sim$0.4 T, demonstrating the \textit{C} = 2 Chern insulator state.  
The Hall resistance drops to $\rho_{yx} \sim$  8 k$\Omega$, much smaller than the quantized 
value at $\mu_0 H$ $\sim$ 0 T, suggesting that the Chern insulator state is stabilized by a 
magnetic field.  Furthermore, when the carrier density \textit{n} and the displacement field \textit{D}
are varied, $\rho_{xx}$ does not show a dip at the same point that
$\rho_{yx}$ has a peak.  This observation might be related to domain walls separating states 
with different Chern numbers; more studies are needed to clarify its physical origin. 
In addition to the QAH states observed at zero magnetic field, 
TBG Chern insulator phases at the finite magnetic field have been found and studied 
by many different groups \cite{nuckolls2020strongly,saito2021hofstadter,wu2021chern}.
\addCXL{We notice that the Streda formula \cite{streda1982theory}, which relates the Chern number to the derivative of 2D electron density at which the charge gap occurs with respect to the applied magnetic field, provides a useful approach to demonstrate the appearance of Chern insulator phases. See for example ~\onlinecite{choi2021correlation,nuckolls2020strongly}, who use scanning tunneling microscopy to 
identify the gap densities, or \onlinecite{pierce2021unconventional,xie2021fractional} who use 
local compressibility measurements.  The Streda formula approach has been less useful in magnetically doped \czchang {TI} because, we believe, disorder closes or substantially reduced the thermodynamic gaps.}
In many cases, Chern insulators appear at surprisingly weak magnetic fields, suggesting the presence  
competing many-body ground states with different Chern numbers \cite{zhang2012chiral}.  

We emphasize that TBG does not possess any local magnetic moments. The magnetization of TBG is therefore always dominated by its
orbital contribution, which varies \cite{zhu2020voltage} with chemical potential across 
the bulk band gap because of the role of topologically 
protected chiral edge states.  The relationship of Eq. (\ref{eq:edgemagnetism}),
which applies inside the gaps and is equivalent to  
Hall conductivity quantization \cite{macdonald1995proceedings}, implies that there is a 
jump of orbital magnetization $\Delta M_{orb}=Ce E_g/h$ 
occurs when the Fermi energy is tuned across the energy gap of QAH state. 
In magnetically doped TIs, this jump is almost negligible compared to the spin magnetization of the local moments, which is weakly sensitive to doping. 
The jump of orbital magnetization induced by tuning Fermi energy can allow for magnetic order to be reversed electrically in a weak magnetic field, 
as demonstrated experimentally by \onlinecite{polshyn2020electrical}. 


\subsection{QAH effect in AB-stacked MoTe$_2$/WSe$_2$ heterobilayers} 

\begin{figure}[hbt!]
    \centering
    \includegraphics[width=3.5in]{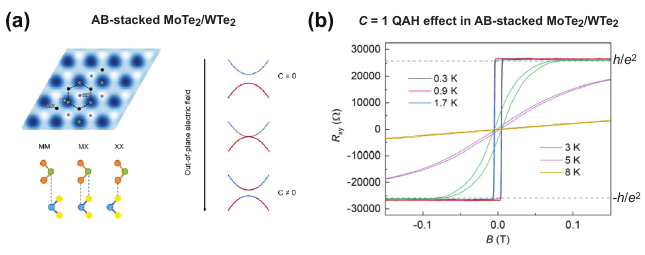}  
    \caption{\textit{C}=1 QAH effect in an AB-stacked MoTe$_2$/WSe$_2$ heterobilayer. (a) Left: triangular moiré superlattice with high-symmetry stacking sites MM, MX and XX (M = Mo/W; X = Se/Te). Right: schematic of out-of-plane electric field-induced band inversion in AB-stacked MoTe$_2$/WSe$_2$ heterobilayer. (b) Magnetic field $\mu_0 H$ dependence of the Hall resistance $R_{yx}$ in an AB-stacked MoTe$_2$/WSe$_2$ heterobilayer with the filling factor \textit{v} = 1 at different temperatures. A \textit{C} =1 QAH insulator state is realized. From \onlinecite{li2021quantum}.
    }
    \label{fig18:TwistedTMD}
\end{figure}

The QAH effect was also realized recently in AB stacked MoTe$_2$/WSe$_2$ moiré heterobilayers. The two semiconducting TMD layers form a triangular moiré superlattice with three high-symmetry staking sites: metal on metal (MM), metal on chalcogen (MX), and
chalcogen on chalcogen (XX) (M = Mo, W; X = Te, Se), as shown in Fig.\ref{fig18:TwistedTMD}(a). 
Since the lattice mismatch between MoTe$_2$ and WSe$_2$ is $\sim$ 7\%, the period of the moiré lattice is
relatively short $\sim$ 5 nm and the moiré density is $\sim$ 5$\times$10$^{12}$ cm$^{-2}$ \cite{li2021continuous}. 
Application of an out-of-plane electric field can control the bandwidth and 
change the band topology by intertwining moiré bands associated with different layers or with different orbital characters within a layer [Fig.\ref{fig18:TwistedTMD}(a)]. At moir\'e band filling 
$\nu = 1$, a state with Hall resistance $R_{yx}$ quantized at $\sim h/e^2$ and vanishing
longitudinal resistance $R_{xx}$ has been observed [Fig.\ref{fig18:TwistedTMD}(b)]. The Hall resistance $\rho_{yx}$ of the QAH effect in magnetically doped TI films \czchang{[Figs. \ref{fig3:Cr} to \ref{fig5:vqah}]} and exfoliated MnBi$_2$Te$_4$ flakes [Fig.\ref{fig12:QAHMBT}] is always less than 
$h/e^2$ when the longitudinal resistance $\rho_{xx}$ is finite. However, for the QAH effect observed in TBG [Fig.\ref{fig15:TBG}(a)] and in AB stacked TMD [Fig.\ref{fig18:TwistedTMD}(b)] bilayers, $\rho_{yx}$ is obviously greater than $h/e^2$ when $\rho_{xx}$ is greater than zero. This difference
might be related to the different types of disorders in the two classes of QAH systems. 

TMD bilayer moir\'e superlattices have valley degrees of freedom in each layer that are locked to spin \czchang{by strong SOC}.
One theoretical scenario \cite{zhang2021spin,pan2022topological} 
for a QAH effect echoes that thought to apply to TBG, namely a combination of valley-projected Chern bands of opposite Chern numbers in the two valleys combined with TR symmetry breaking by spontaneous valley polarization. 
Berry curvatures of opposite sign in opposite valleys respect TR invariance 
and are present in the electronic bands of 2D TMD 
crystals \cite{xiao2012coupled} even without moir\'e modulation.
If this is the operable mechanism, there is still a need to understand how precisely the valley-projected bands develop finite Chern numbers, and the answer to this question is not as obvious as in 
previous cases because of the absence of simple isolated Dirac cones that 
can readily be gapped.  It seems clear experimentally that the QAH effect occurs when the perpendicular
electric field applied between layers is strong enough to bring the valence band tops in the 
WSe$_2$ and MoTe$_2$ layers close to alignment so that they can have avoided crossings that yield large Berry curvatures.  
In the case of AB stacked bilayers, however,
the relationship between spin and valley is opposite in the two layers.  
For this reason, single-particle hybridization between layers is expected to be weak, making it less obvious how valley 
projected Chern bands could develop.  \onlinecite{devakul2022quantum}, \onlinecite{zhang2021spin},\onlinecite{pan2022topological},\onlinecite{xie2022valley}, and \onlinecite{xie2022topological} have separately provided plausible scenarios that rely on specific details of 
the bilayer moir\'e band electronic structure for how Chern bands could emerge.  \am{Very recent studies \cite{tao2022valley} of the magneto-optical properties of these states suggest that they have inter-valley order, apparently invalidating valley polarization mechanisms and favoring those
that invoke \cite{xie2022topological} time-reversal broken 
inter-valley coherent states.}  
The general lesson from all these studies seems to be 
that topologically non-trivial bands are even more common in moir\'e materials than in atomic-scale crystals, in which
bands have a strong tendency to organize according to the 
atomic shells of constituent atoms.  


\section{QAH Effect Research Challenges}
\label{SectionVfutures}


\subsection{Theoretical proposals and challenges} 
In this Article, we have focused our attention on quasi-2D electron systems
in which the QAH effect has already been observed. 
Below we discuss \am{some 
theoretical ideas that have been proposed over the years, but not yet realized in experiment.  
The list of possibilities mentioned below
is far from exhaustive, especially since new QAH mechanisms that are currently completely unanticipated are likely to emerge in the coming years.}

\subsubsection{QAH materials}

As discussed in Sec.\ref{Sec:II}, the QAH effect requires 
both band inversion and spontaneous TR symmetry breaking. 
A natural design strategy for the QAH effect is therefore to catalyze magnetism in materials 
that are already topological. The \am{magnetism promotion strategies} 
that are considered most often are 
(\textit{i}) magnetic doping and (\textit{ii}) the fabrication of  
ferromagnetic (or antiferromagnetic) insulator/topological material heterostructures. 
A typical example of the magnetic doping approach is to design the
AH effect in magnetically doped II-VI or III-V semiconductor compound heterostructures.
\am{Ideally} the host semiconductor should already have a topologically non-trivial band 
structure, as in the case of HgTe quantum wells \cite{liu2008quantum,budewitz2017quantum}).
More flexibility can be achieved by 
integrating two different material compounds to form 
type-II quantum wells in which the conduction band bottom in one material has lower energy than the valence band top in the other material.  
Following this approach, a QAH state has been predicted to occur in magnetically doped InAs/GaSb 
quantum wells \cite{wang2014quantum,liu2008quantumSpin}, and in other similar 
\textit{p}-\textit{n} heterojunctions between II-VI, III-V, and group IV semiconductors \cite{zhang2014quantumSpin}.  
\am{Unfortunately, magnetic doping does not always lead to magnetic order and this approach 
has not yet been successful.}

\addCXL{One possibility to overcome this obstacle is to combine two 2D ferromagnetic materials to form a type-II junction \cite{garrity2014chern,pan2020quantum}.}
An alternative theoretical possibility follows from the observation that
when the hybridization between layers in \textit{p}-\textit{n} heterojunctions is weak compared to interaction strengths, 
rich phase diagrams can emerge  
\cite{budich2014time,pikulin2014interplay,xue2018time,zeng2022plane,zhu2019gate} in which TR symmetry breaking and QAH states appear spontaneously without the complication of magnetic doping. 

Motivated by the Haldane model \cite{haldane1988model} and the Kane-Mele model \cite{kane2005quantum}, a body
of theoretical work focused on compounds with honeycomb type lattice structures and strong SOC. Graphene possesses a typical honeycomb lattice, but has weak SOC and no local 
magnetic moments. It was therefore proposed that QAH states could be induced by 
adding heavy transition elements to graphene as adatoms \cite{qiao2010quantum,zhang2012electrically,qiao2012microscopic,tse2011quantum,li2015theory,deng2017realization,de2018quantum,zhang2018realizing}, or by forming ferromagnetic insulator/graphene heterostructures \cite{zhang2015robust,zhang2019possible,hogl2020quantum,vila2021valley}. 
So far, these strategies have not succeeded, 
presumably because the spatial distribution of adatoms is inevitably disordered and
the hybridization between graphene and 
adjacent materials tends to be weak.  Heavier elements that can also form honeycomb or buckled honeycomb lattice structure, such as \czchang {silicene} \cite{ezawa2012valley,pan2014valley,ezawa2013spin,kaloni2014prediction,zhang2013quantum,qian2019robust,zhang2013abundant}, germanene \cite{hsu2017quantum,zou2020intrinsic,wu2014prediction,zhang2019converting,pham2020designing}, stanene \cite{xu2013large,wu2014prediction,zhang2016quantum,li2019stanene,zhang2016quantumanomalous} and bismuth systems \cite{niu2015functionalized,ji2016giant,jin2015quantum,liu2015valley}, have aslo 
been considered as possible QAH effect hosts.  In addition to these 2D materials containing a single element, honeycomb lattices are also common in more complex materials, 
including organic triphenyl-transition-metal compounds \cite{wang2013quantum}, half-fluorinated GaBi \cite{chen2016prediction}, co-decorated in-triangle adlayers on a Si(111) surface \cite{zhou2017valley}, \addCXL{heavy atom layers on magnetic insulator substrates\cite{garrity2013chern}}, monolayer jacutingaite \cite{luo2021functionalization}, monolayer PtCl$_3$ \cite{you2019two}, monolayer EuO$_2$ \cite{meng2021multiple} and (111) perovskite-type transition-metal oxides \cite{xiao2011interface}.

The approach of inducing magnetism externally has the obvious disadvantage that
magnetic doping introduces disorder and thus degrades sample quality,
while the ferromagnetic (or antiferromagnetic) insulator/topological material heterostructure 
approach has the disadvantage that it relies on magnetic proximity effects that are often weak.
Thus, it is desirable to find topological materials with intrinsic magnetism.
Unfortunately they are rare \cite{huang2017quantum,wu2017high}. 
One approach is to search for magnetic Weyl semimetals,
since the 3D magnetic Weyl semimetal phase is closely related to the 2D QAH state.
In thin films, Weyl nodes are gapped by finite-size effects allowing the QAH effect to occur.
In the limit of thick films, finite-thickness gaps become smaller 
and the total Chern number per layer of the film  
approaches \cite{lei2020magnetized} the 
non-quantized Hall conductivity per layer of the bulk material.
Thus, one may expect the QAH effect in 2D films of magnetic Weyl semimetals with 
broken TR symmetry.  Following this idea, a number of intrinsic magnetic materials have been proposed, including HgCr$_2$Se$_4$ \cite{xu2011chern},  Co$_3$Sn$_2$S$_2$ \cite{morali2019fermi,liu2019magnetic}, Mn$_2$Sn \cite{kuroda2017evidence}, and Co$_2$MnGa \cite{belopolski2019discovery}. The QAH state is expected to emerge in all these systems in the 
quasi-2D thin film limit \cite{xu2011chern,muechler2020emerging}. 

An additional path to achieve the QAH effect is to create  
systems in which interaction effects  are likely to 
spontaneously break TR and induce non-zero Chern number bands \cite{min2008pseudospin,raghu2008topological,nandkishore2010quantum,martin2008itinerant,zhang2011spontaneous}. The QAH states seen in 2D moir\'e super-lattices provide one successful 
example of this route to QAH states \cite{shi2021moire,liu2021theories,wu2020collective,wu2018hubbard,kwan2021exciton}. 
\addCXL{Other 2D quantum spin Hall systems with small Dirac velocities
have also been proposed as possible QAH hosts that do not require 
magnetic dopants\cite{min2008pseudospin,xue2018time,cao2016heavy}. }


\subsubsection{Fractional QAH effect}
In 2D electron gases with partial Landau level
filling can under strong magnetic fields, strong correlations  
can result in gapped many-body states that have
quasi-particles with fractional charge and fractional statistics.
This phenomenon is known as the fractional QH effect \cite{stormer1999fractional}.  It is natural to 
ask \cite{neupert2011fractional,sun2011nearly,tang2011high,klinovaja2015integer} whether 
or not a fractional version of the QAH effect, the fractional QAH effect, can occur in strongly interacting systems. 
The search for materials in which the fractional QAH effect occurs is 
challenged by the lack of simple predictive theoretical tools that account for 
strong correlations in realistic materials.  Even for the weakly correlated materials 
in which DFT calculations are helpful, theory is not always able to account for all important 
complications.  At present, the flat bands induced by moiré superlattices provide the 
most promising direction towards the realization of fractional QAH states \cite{ledwith2020fractional,repellin2020chern}. 




\subsubsection{Higher temperature QAH effect} 

The successful theoretical prediction and subsequent experimental realization of the QAH effect 
has been recognized as a great triumph of our understanding of topological states of matter,
and as a demonstration of our ability to engineer complex topological materials.  
To utilize these states in technology, however,
the low critical temperature at which the QAH effect is observed is a weighty obstacle.
The QAH effects in magnetically doped TIs \cite{chang2013experimental,chang2015high,mogi2015magnetic}, 
in the intrinsic magnetic TI MnBi$_2$Te$_4$ \cite{deng2020quantum}, and in moiré materials \cite{serlin2020intrinsic,li2021quantum}
have comparable and low onset temperatures.  

An important goal of QAH research is to 
greatly enhance the temperature at which the QAH effect is observed. One of 
the following two approaches might ultimately prove successful: (\textit{i}) form high-quality heterostructures between TIs and high Curie temperature ferromagnetic (or high N\'eel temperature antiferromagnetic) insulators \cite{tang2017above,wang2019observation}; (\textit{ii}) discover new moir\'e bilayers \cite{andrei2020graphene} that order with shorter moir\'e periods. The former requires that TI and oxide film growth be successfully combined. The latter approach requires a joint effort, involving both theoretical predictions and materials synthesis. The quest to realize higher temperature QAH states not only promotes new fundamental inquiries,
but could also carry far-reaching implications for topological quantum computation and low-energy cost spintronic devices.

\subsection {Potential applications}
\label{Sec:IVmf}

\begin{figure} [hbt!]
    \centering
    \includegraphics[width=3.5in]{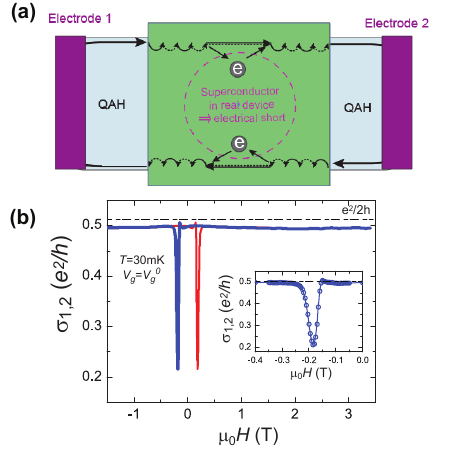}  
    \caption{Absence of the chiral Majorana edge modes in millimeter-size QAH-superconductor devices. (a) A millimeter-size superconductor strip covering a QAH device induces an electrical short that circumvents a chiral MEM signal. (b) Magnetic field $\mu_0 H$ dependence of the two-terminal conductance $\sigma_{1,2}$ of a QAH-superconductor device. $\sigma_{1,2}\sim 0.5e^2/h$ for the entire $\mu_0 H$ range when the magnetization is accurately aligned. No change in $\sigma_{1,2}$ is observed when the Nb strip transitions from the superconducting state to the normal state. From \onlinecite{kayyalha2020absence}.}
    \label{fig16:majorana}
\end{figure}

\subsubsection{Chiral Majorana edge modes}

During the past decades, the search for Majorana modes has been a key activity of 
condensed matter physics \cite{mourik2012signatures,das2012zero,deng2012anomalous,rokhinson2012fractional,nadj2014observation,jeon2017distinguishing,wang2018evidence,kayyalha2020absence}. Majorana modes exhibit non-local quantum exchange 
statistics and are promising as a platform for topological quantum computation \cite{nayak2008non,beenakker2013search,alicea2012new}. 
Majorana modes can appear at the boundary or inside the vortex cores of topological superconductors (TSC). 

\onlinecite{qi2010chiral} predicted that chiral TSC can be realized in 
QAH/superconductor hybrid structure like that illustrated in Fig. \ref{fig16:majorana}(a). 
A chiral TSC has a full pairing gap in the bulk and an odd number $\mathcal{N}$ of chiral 
Majorana edge modes (MEMs). Proximity-induced superconductivity in a QAH insulator 
can give rise to chiral MEMs, appearing at the topological phase transition between trivial and nontrivial phases. In general, a QAH insulator with \textit{C} chiral edge modes 
coupled to an s-wave superconductor is a chiral TSC with $\mathcal{N}$=2\textit{C} chiral MEMs. 
Here \textit{C} is the Chern number of the isolated QAH insulator
and $\mathcal{N}$ also labels the superconductor Chern number. 
When proximitized, the transition from a topologically nontrivial phase of a \textit{C}=1 QAH phase to a \textit{C}=0 topologically trivial (i.e. normal insulator) phase has to go through a chiral TSC phase with $\mathcal{N}=1$.  Based on the quasiparticle properties of this state, \onlinecite{wang2015chiral} 
predicted that a half-quantized two-terminal conductance $\sigma_{12}$ plateau
would appear near the coercive field $H_c$ of a QAH/superconductor hybrid structure and be 
a fingerprint of a $\mathcal{N}=1$ chiral TSC.  As we explain below, however, several theoretical papers 
later raised concerns about this conclusion \cite{chen2017effects,chen2018quasi,huang2018disorder,ji20181,lian2018quantum}.
Separately it was theoretically proposed that 1D nanowires formed from 
QAH insulators in proximity to superconductivity can host Majorana zero modes at the ends of the wire \cite{zeng2018quantum,chen2018quasi,xie2020creating}. 

Experimental studies of QAH/superconductor heterostructure have been controversial. 
The predicted half-quantized $\sigma_{12}$ plateau was reported in a 
millimeter-sized QAH/Nb hybrid structure in 2017 \cite{he2017chiral}.
However, \onlinecite{kayyalha2020absence} recently carried out a systematic Andreev reflection 
spectroscopy study of the contact transparency between the QAH insulator and superconducting Nb 
layers on more than 30 QAH/Nb devices. These authors found that $\sigma_{12}$
is half quantized in QAH-Nb devices even when the magnetization is well aligned and
the coercive field is exceeded [Fig. \ref{fig16:majorana}(b)]. The theoretical calculations in \onlinecite{wang2015chiral} accounted only for quasi-particle transport processes in the hybrid system,
and did not account for collective currents carried by the superconducting 
condensate.  When these are allowed, the superconductor can act as an electrical short 
for the QAH device [Fig. \ref{fig16:majorana}(a)], making transport an ineffective probe of 
quasi-particle properties.  For this reason, the observation of the half-quantized plateau cannot be considered as conclusive evidence for the existence of chiral MEMs in the millimeter-size QAH-superconductor hybrid structures \cite{kayyalha2020absence}.




\subsubsection{Resistance standards} 

As explained in Sec.\ref{Sec:intro}, the QAH effect can be used to create  
a resistance standard that is free from the need for superconducting magnets.  
To make this goal a reality, it is necessary to improve the accuracy of the 
Hall resistance quantization in the QAH effect and understand sources of deviation from perfect 
quantization.  To date, all high precision measurements on the QAH effect have been performed on magnetically doped TI films/heterostructures.  In both Cr- and V-doped QAH samples,  the quantization of the Hall resistance has been already demonstrated at the level of parts-per-million (ppm) uncertainty. \onlinecite{fox2018part} used a cryogenic current comparator system to perform the electrical transport on a 6 QL uniformly Cr-doped (Bi,Sb)$_2$Te$_3$ film and measured quantization of the Hall resistance to within 1 ppm at zero magnetic field. \onlinecite{gotz2018precision} measured a 9 QL V-doped (Bi,Sb)$_2$Te$_3$ film and determined a value of $0.17\pm 0.25$ ppm. \onlinecite{okazaki2020precise} performed high precision transport measurements on a 9 QL modulation-doped QAH sandwich sample and demonstrated 2 ppm accuracy. 
More recently, \onlinecite{okazaki2021quantum} measured the same QAH sandwich device with a permanent magnet and claimed precision of 10 parts per billion (ppb). Note that the uncertainties of the order of $0.1$ ppm $\sim$ 10 ppb still do not reach the requirements for a quantum resistance standard, which requires quantization with uncertainty on the order of 1 ppb \cite{poirier2009resistance,jeckelmann2001quantum}. Therefore, further optimization of QAH samples, whether magnetically doped TI (Sec.\ref{Sec:MTI}),
or MnBi$_2$Te$_4$ (Sec.\ref{Sec:MnBi2Te4}),or moir\'e superlattice (Secs.\ref{Sec:tbg}) systems, will be needed if the QAH effect is to 
achieve a breakthrough in resistance metrology.


\subsubsection{Dissipation free interconnects}

The dissipationless chiral edge current of the QAH state can potentially 
be integrated into contemporary electronic or spintronic technology to develop chiral electronic or chiral spintronic devices. 
Since chiral edge transport eliminates local electrical resistivity, 
leaving only terminal or contact resistances, it was theoretically proposed that 
QAH devices can be used to construct chiral interconnects in computing devices
in which the total resistance is independent of the circuit length \cite{zhang2012chiral}. 
The spin polarization \cite{zhang2016electrically,cheraghchi2020spin} and real-space spin textures \cite{wu2014topological} carried by magnetically doped TI chiral edge states 
might also enable electrical control of magnetic domain wall motion \cite{kim2019electrically,upadhyaya2016domain}. 
Experimentally, the interplay between chiral edge states and magnetic domain walls has been demonstrated \cite{yasuda2017quantized,rosen2017chiral}. 
In TBG moir\'e superlattice devices, electrical switching of magnetic order has been demonstrated \cite{sharpe2019emergent,serlin2020intrinsic,polshyn2020electrical,chen2021electrically,zhu2020voltage,tschirhart2022intrinsic}. \czchang {In} magnetic TI systems with the well-quantized QAH effect,  \czchang {the electrical switching of the edge state chirality is recently realized through spin-orbit torque \cite {yuan2022electrical}}. The realization of all-electrical switching of the QAH edge current chirality can be important for the development of the chiral electronic devices. 
Chiral edge state-based electronic devices could potentially usher in a new era of dissipation-free 
memory and logic devices, and have a transformative impact on the grand challenges facing 
semiconductor electronics.


\begin{acknowledgments}
C.-Z. Chang, C.-X. Liu, and A. H. MacDonald thank R. Bistritzer, 
M. H. W. Chan, Y.-T. Cui, X. Dai, Z. Fang, K. He, T. Hughes, J. Jain, M. Kayyalha, C. Lei, B. Lian,
X. Lin, L. Lv, L. Molenkamp, N. Morales-Duran, J. Moodera, Q. Niu, P. Potasz, X.-L. Qi, N. Samarth, 
C. G. Smith, Y.-Y. Wang, F. Wu, W.-D. Wu, D. Xiao, X.D. Xu, Q.-K. Xue, B.-H. Yan, J.Q. Yan, Q.-M. Yan, F. Zhang, H.-J. Zhang, and S.-C. Zhang 
for long-term collaborations and interactions related to the subject of this review. 
C.-Z. Chang acknowledges the support from DOE grant (DE-SC0019064), ARO grant (W911NF1810198\czchang{, W911NF2210159}), NSF-CAREER award (DMR-1847811), Penn State NSF-MRSEC grant (DMR-2011839), AFOSR grant (FA9550-21-1-0177), Gordon and Betty Moore Foundation’s EPiQS Initiative (GBMF9063 to C.-Z. C.), and Alfred P. Sloan Foundation. C.-X. Liu acknowledges the support from ONR grant (No. N00014-18-1-2793), DOE grant (DE-SC0019064), Penn State NSF-MRSEC grant (DMR-2011839), Princeton NSF-MRSEC (DMR-2011750), and Kaufman grant (KA2018-98553). A. H. MacDonald acknowledges support from the U.S. Department of 
Energy, Office of Science, Basic Energy Sciences, under Awards DE-SC0022106 and DE‐SC0019481, from the Simons Foundation, and from the Army Research Office under Grant Number W911NF-16-1-0472.

\end{acknowledgments}

\bibliographystyle{apsrmp.bst} 
\bibliography{apssamp.bib}

\providecommand{\noopsort}[1]{}\providecommand{\singleletter}[1]{#1}%
\begin{thebibliography}{434}
\expandafter\ifx\csname natexlab\endcsname\relax\def\natexlab#1{#1}\fi
\expandafter\ifx\csname bibnamefont\endcsname\relax
  \def\bibnamefont#1{#1}\fi
\expandafter\ifx\csname bibfnamefont\endcsname\relax
  \def\bibfnamefont#1{#1}\fi
\expandafter\ifx\csname citenamefont\endcsname\relax
  \def\citenamefont#1{#1}\fi
\expandafter\ifx\csname url\endcsname\relax
  \def\url#1{\texttt{#1}}\fi
\expandafter\ifx\csname urlprefix\endcsname\relax\def\urlprefix{URL }\fi
\providecommand{\bibinfo}[2]{#2}
\providecommand{\eprint}[2][]{\url{#2}}

\bibitem[{\citenamefont{Ahn} \emph{et~al.}(2019)\citenamefont{Ahn, Park, and
  Yang}}]{ahn2019failure}
\bibinfo{author}{\bibnamefont{Ahn}, \bibfnamefont{J.}},
  \bibinfo{author}{\bibfnamefont{S.}~\bibnamefont{Park}}, and
  \bibinfo{author}{\bibfnamefont{B.-J.} \bibnamefont{Yang}},
  \bibinfo{year}{2019}, \bibinfo{journal}{Physical Review X}
  \textbf{\bibinfo{volume}{9}}(\bibinfo{number}{2}), \bibinfo{pages}{021013}.

\bibitem[{\citenamefont{Ahn and Yang}(2017)}]{ahn2017unconventional}
\bibinfo{author}{\bibnamefont{Ahn}, \bibfnamefont{J.}}, and
  \bibinfo{author}{\bibfnamefont{B.-J.} \bibnamefont{Yang}},
  \bibinfo{year}{2017}, \bibinfo{journal}{Physical Review Letters}
  \textbf{\bibinfo{volume}{118}}(\bibinfo{number}{15}),
  \bibinfo{pages}{156401}.

\bibitem[{\citenamefont{Alexander-Webber}
  \emph{et~al.}(2012)\citenamefont{Alexander-Webber, Baker, Buckle, Ashley, and
  Nicholas}}]{alexander2012high}
\bibinfo{author}{\bibnamefont{Alexander-Webber}, \bibfnamefont{J.}},
  \bibinfo{author}{\bibfnamefont{A.}~\bibnamefont{Baker}},
  \bibinfo{author}{\bibfnamefont{P.~D.} \bibnamefont{Buckle}},
  \bibinfo{author}{\bibfnamefont{T.}~\bibnamefont{Ashley}}, and
  \bibinfo{author}{\bibfnamefont{R.}~\bibnamefont{Nicholas}},
  \bibinfo{year}{2012}, \bibinfo{journal}{Physical Review B}
  \textbf{\bibinfo{volume}{86}}(\bibinfo{number}{4}), \bibinfo{pages}{045404}.

\bibitem[{\citenamefont{Alicea}(2012)}]{alicea2012new}
\bibinfo{author}{\bibnamefont{Alicea}, \bibfnamefont{J.}},
  \bibinfo{year}{2012}, \bibinfo{journal}{Reports on Progress in Physics}
  \textbf{\bibinfo{volume}{75}}(\bibinfo{number}{7}), \bibinfo{pages}{076501}.

\bibitem[{\citenamefont{Allen} \emph{et~al.}(2019)\citenamefont{Allen, Cui, Ma,
  Mogi, Kawamura, Fulga, Goldhaber-Gordon, Tokura, and
  Shen}}]{allen2019visualization}
\bibinfo{author}{\bibnamefont{Allen}, \bibfnamefont{M.}},
  \bibinfo{author}{\bibfnamefont{Y.}~\bibnamefont{Cui}},
  \bibinfo{author}{\bibfnamefont{E.~Y.} \bibnamefont{Ma}},
  \bibinfo{author}{\bibfnamefont{M.}~\bibnamefont{Mogi}},
  \bibinfo{author}{\bibfnamefont{M.}~\bibnamefont{Kawamura}},
  \bibinfo{author}{\bibfnamefont{I.~C.} \bibnamefont{Fulga}},
  \bibinfo{author}{\bibfnamefont{D.}~\bibnamefont{Goldhaber-Gordon}},
  \bibinfo{author}{\bibfnamefont{Y.}~\bibnamefont{Tokura}}, and
  \bibinfo{author}{\bibfnamefont{Z.-X.} \bibnamefont{Shen}},
  \bibinfo{year}{2019}, \bibinfo{journal}{Proceedings of the National Academy
  of Sciences} \textbf{\bibinfo{volume}{116}}(\bibinfo{number}{29}),
  \bibinfo{pages}{14511}.

\bibitem[{\citenamefont{Anderson}(1950)}]{anderson1950antiferromagnetism}
\bibinfo{author}{\bibnamefont{Anderson}, \bibfnamefont{P.~W.}},
  \bibinfo{year}{1950}, \bibinfo{journal}{Physical Review}
  \textbf{\bibinfo{volume}{79}}(\bibinfo{number}{2}), \bibinfo{pages}{350}.

\bibitem[{\citenamefont{Andrei} \emph{et~al.}(2021)\citenamefont{Andrei,
  Efetov, Jarillo-Herrero, MacDonald, Mak, Senthil, Tutuc, Yazdani, and
  Young}}]{andrei2021marvels}
\bibinfo{author}{\bibnamefont{Andrei}, \bibfnamefont{E.~Y.}},
  \bibinfo{author}{\bibfnamefont{D.~K.} \bibnamefont{Efetov}},
  \bibinfo{author}{\bibfnamefont{P.}~\bibnamefont{Jarillo-Herrero}},
  \bibinfo{author}{\bibfnamefont{A.~H.} \bibnamefont{MacDonald}},
  \bibinfo{author}{\bibfnamefont{K.~F.} \bibnamefont{Mak}},
  \bibinfo{author}{\bibfnamefont{T.}~\bibnamefont{Senthil}},
  \bibinfo{author}{\bibfnamefont{E.}~\bibnamefont{Tutuc}},
  \bibinfo{author}{\bibfnamefont{A.}~\bibnamefont{Yazdani}}, and
  \bibinfo{author}{\bibfnamefont{A.~F.} \bibnamefont{Young}},
  \bibinfo{year}{2021}, \bibinfo{journal}{Nature Reviews Materials}
  \textbf{\bibinfo{volume}{6}}(\bibinfo{number}{3}), \bibinfo{pages}{201}.

\bibitem[{\citenamefont{Andrei and MacDonald}(2020)}]{andrei2020graphene}
\bibinfo{author}{\bibnamefont{Andrei}, \bibfnamefont{E.~Y.}}, and
  \bibinfo{author}{\bibfnamefont{A.~H.} \bibnamefont{MacDonald}},
  \bibinfo{year}{2020}, \bibinfo{journal}{Nature Materials}
  \textbf{\bibinfo{volume}{19}}(\bibinfo{number}{12}), \bibinfo{pages}{1265}.

\bibitem[{\citenamefont{Aryasetiawan and
  Karlsson}(2019)}]{aryasetiawan2019modern}
\bibinfo{author}{\bibnamefont{Aryasetiawan}, \bibfnamefont{F.}}, and
  \bibinfo{author}{\bibfnamefont{K.}~\bibnamefont{Karlsson}},
  \bibinfo{year}{2019}, \bibinfo{journal}{Journal of Physics and Chemistry of
  Solids} \textbf{\bibinfo{volume}{128}}, \bibinfo{pages}{87}.

\bibitem[{\citenamefont{Auerbach}(2012)}]{auerbach2012interacting}
\bibinfo{author}{\bibnamefont{Auerbach}, \bibfnamefont{A.}},
  \bibinfo{year}{2012}, \emph{\bibinfo{title}{Interacting electrons and quantum
  magnetism}} (\bibinfo{publisher}{Springer Science \& Business Media}).

\bibitem[{\citenamefont{Bac} \emph{et~al.}(2022)\citenamefont{Bac, Koller, Lux,
  Wang, Riney, Borisiak, Powers, Zhukovskyi, Orlova, Dobrowolska}
  \emph{et~al.}}]{bac2022topological}
\bibinfo{author}{\bibnamefont{Bac}, \bibfnamefont{S.-K.}},
  \bibinfo{author}{\bibfnamefont{K.}~\bibnamefont{Koller}},
  \bibinfo{author}{\bibfnamefont{F.}~\bibnamefont{Lux}},
  \bibinfo{author}{\bibfnamefont{J.}~\bibnamefont{Wang}},
  \bibinfo{author}{\bibfnamefont{L.}~\bibnamefont{Riney}},
  \bibinfo{author}{\bibfnamefont{K.}~\bibnamefont{Borisiak}},
  \bibinfo{author}{\bibfnamefont{W.}~\bibnamefont{Powers}},
  \bibinfo{author}{\bibfnamefont{M.}~\bibnamefont{Zhukovskyi}},
  \bibinfo{author}{\bibfnamefont{T.}~\bibnamefont{Orlova}},
  \bibinfo{author}{\bibfnamefont{M.}~\bibnamefont{Dobrowolska}}, \emph{et~al.},
  \bibinfo{year}{2022}, \bibinfo{journal}{npj Quantum Materials}
  \textbf{\bibinfo{volume}{7}}(\bibinfo{number}{1}), \bibinfo{pages}{1}.

\bibitem[{\citenamefont{Balents} \emph{et~al.}(2020)\citenamefont{Balents,
  Dean, Efetov, and Young}}]{balents2020superconductivity}
\bibinfo{author}{\bibnamefont{Balents}, \bibfnamefont{L.}},
  \bibinfo{author}{\bibfnamefont{C.~R.} \bibnamefont{Dean}},
  \bibinfo{author}{\bibfnamefont{D.~K.} \bibnamefont{Efetov}}, and
  \bibinfo{author}{\bibfnamefont{A.~F.} \bibnamefont{Young}},
  \bibinfo{year}{2020}, \bibinfo{journal}{Nature Physics}
  \textbf{\bibinfo{volume}{16}}(\bibinfo{number}{7}), \bibinfo{pages}{725}.

\bibitem[{\citenamefont{Beenakker}(2013)}]{beenakker2013search}
\bibinfo{author}{\bibnamefont{Beenakker}, \bibfnamefont{C.}},
  \bibinfo{year}{2013}, \bibinfo{journal}{Annual Review of Condensed Matter
  Physics} \textbf{\bibinfo{volume}{4}}(\bibinfo{number}{1}),
  \bibinfo{pages}{113}.

\bibitem[{\citenamefont{Belopolski}
  \emph{et~al.}(2019)\citenamefont{Belopolski, Manna, Sanchez, Chang, Ernst,
  Yin, Zhang, Cochran, Shumiya, Zheng} \emph{et~al.}}]{belopolski2019discovery}
\bibinfo{author}{\bibnamefont{Belopolski}, \bibfnamefont{I.}},
  \bibinfo{author}{\bibfnamefont{K.}~\bibnamefont{Manna}},
  \bibinfo{author}{\bibfnamefont{D.~S.} \bibnamefont{Sanchez}},
  \bibinfo{author}{\bibfnamefont{G.}~\bibnamefont{Chang}},
  \bibinfo{author}{\bibfnamefont{B.}~\bibnamefont{Ernst}},
  \bibinfo{author}{\bibfnamefont{J.}~\bibnamefont{Yin}},
  \bibinfo{author}{\bibfnamefont{S.~S.} \bibnamefont{Zhang}},
  \bibinfo{author}{\bibfnamefont{T.}~\bibnamefont{Cochran}},
  \bibinfo{author}{\bibfnamefont{N.}~\bibnamefont{Shumiya}},
  \bibinfo{author}{\bibfnamefont{H.}~\bibnamefont{Zheng}}, \emph{et~al.},
  \bibinfo{year}{2019}, \bibinfo{journal}{Science}
  \textbf{\bibinfo{volume}{365}}(\bibinfo{number}{6459}),
  \bibinfo{pages}{1278}.

\bibitem[{\citenamefont{Bernevig} \emph{et~al.}(2007)\citenamefont{Bernevig,
  Hughes, Raghu, and Arovas}}]{bernevig2007theory}
\bibinfo{author}{\bibnamefont{Bernevig}, \bibfnamefont{B.~A.}},
  \bibinfo{author}{\bibfnamefont{T.~L.} \bibnamefont{Hughes}},
  \bibinfo{author}{\bibfnamefont{S.}~\bibnamefont{Raghu}}, and
  \bibinfo{author}{\bibfnamefont{D.~P.} \bibnamefont{Arovas}},
  \bibinfo{year}{2007}, \bibinfo{journal}{Physical review letters}
  \textbf{\bibinfo{volume}{99}}(\bibinfo{number}{14}), \bibinfo{pages}{146804}.

\bibitem[{\citenamefont{Bernevig} \emph{et~al.}(2006)\citenamefont{Bernevig,
  Hughes, and Zhang}}]{bernevig2006quantum}
\bibinfo{author}{\bibnamefont{Bernevig}, \bibfnamefont{B.~A.}},
  \bibinfo{author}{\bibfnamefont{T.~L.} \bibnamefont{Hughes}}, and
  \bibinfo{author}{\bibfnamefont{S.-C.} \bibnamefont{Zhang}},
  \bibinfo{year}{2006}, \bibinfo{journal}{Science}
  \textbf{\bibinfo{volume}{314}}(\bibinfo{number}{5806}),
  \bibinfo{pages}{1757}.

\bibitem[{\citenamefont{Bernevig} \emph{et~al.}(2021)\citenamefont{Bernevig,
  Lian, Cowsik, Xie, Regnault, and Song}}]{bernevig2021twisted}
\bibinfo{author}{\bibnamefont{Bernevig}, \bibfnamefont{B.~A.}},
  \bibinfo{author}{\bibfnamefont{B.}~\bibnamefont{Lian}},
  \bibinfo{author}{\bibfnamefont{A.}~\bibnamefont{Cowsik}},
  \bibinfo{author}{\bibfnamefont{F.}~\bibnamefont{Xie}},
  \bibinfo{author}{\bibfnamefont{N.}~\bibnamefont{Regnault}}, and
  \bibinfo{author}{\bibfnamefont{Z.-D.} \bibnamefont{Song}},
  \bibinfo{year}{2021}, \bibinfo{journal}{Physical Review B}
  \textbf{\bibinfo{volume}{103}}(\bibinfo{number}{20}),
  \bibinfo{pages}{205415}.

\bibitem[{\citenamefont{Bernevig and Zhang}(2006)}]{bernevig2006quantum1}
\bibinfo{author}{\bibnamefont{Bernevig}, \bibfnamefont{B.~A.}}, and
  \bibinfo{author}{\bibfnamefont{S.-C.} \bibnamefont{Zhang}},
  \bibinfo{year}{2006}, \bibinfo{journal}{Physical Review Letters}
  \textbf{\bibinfo{volume}{96}}(\bibinfo{number}{10}), \bibinfo{pages}{106802}.

\bibitem[{\citenamefont{Bestwick} \emph{et~al.}(2015)\citenamefont{Bestwick,
  Fox, Kou, Pan, Wang, and Goldhaber-Gordon}}]{bestwick2015precise}
\bibinfo{author}{\bibnamefont{Bestwick}, \bibfnamefont{A.}},
  \bibinfo{author}{\bibfnamefont{E.}~\bibnamefont{Fox}},
  \bibinfo{author}{\bibfnamefont{X.}~\bibnamefont{Kou}},
  \bibinfo{author}{\bibfnamefont{L.}~\bibnamefont{Pan}},
  \bibinfo{author}{\bibfnamefont{K.~L.} \bibnamefont{Wang}}, and
  \bibinfo{author}{\bibfnamefont{D.}~\bibnamefont{Goldhaber-Gordon}},
  \bibinfo{year}{2015}, \bibinfo{journal}{Physical Review Letters}
  \textbf{\bibinfo{volume}{114}}(\bibinfo{number}{18}),
  \bibinfo{pages}{187201}.

\bibitem[{\citenamefont{Bianco and Resta}(2013)}]{bianco2013orbital}
\bibinfo{author}{\bibnamefont{Bianco}, \bibfnamefont{R.}}, and
  \bibinfo{author}{\bibfnamefont{R.}~\bibnamefont{Resta}},
  \bibinfo{year}{2013}, \bibinfo{journal}{Physical Review Letters}
  \textbf{\bibinfo{volume}{110}}(\bibinfo{number}{8}), \bibinfo{pages}{087202}.

\bibitem[{\citenamefont{Bistritzer and MacDonald}(2011)}]{bistritzer2011moire}
\bibinfo{author}{\bibnamefont{Bistritzer}, \bibfnamefont{R.}}, and
  \bibinfo{author}{\bibfnamefont{A.~H.} \bibnamefont{MacDonald}},
  \bibinfo{year}{2011}, \bibinfo{journal}{Proceedings of the National Academy
  of Sciences} \textbf{\bibinfo{volume}{108}}(\bibinfo{number}{30}),
  \bibinfo{pages}{12233}.

\bibitem[{\citenamefont{Bloembergen and
  Rowland}(1955)}]{bloembergen1955nuclear}
\bibinfo{author}{\bibnamefont{Bloembergen}, \bibfnamefont{N.}}, and
  \bibinfo{author}{\bibfnamefont{T.}~\bibnamefont{Rowland}},
  \bibinfo{year}{1955}, \bibinfo{journal}{Physical Review}
  \textbf{\bibinfo{volume}{97}}(\bibinfo{number}{6}), \bibinfo{pages}{1679}.

\bibitem[{\citenamefont{Budewitz} \emph{et~al.}(2017)\citenamefont{Budewitz,
  Bendias, Leubner, Khouri, Shamim, Wiedmann, Buhmann, and
  Molenkamp}}]{budewitz2017quantum}
\bibinfo{author}{\bibnamefont{Budewitz}, \bibfnamefont{A.}},
  \bibinfo{author}{\bibfnamefont{K.}~\bibnamefont{Bendias}},
  \bibinfo{author}{\bibfnamefont{P.}~\bibnamefont{Leubner}},
  \bibinfo{author}{\bibfnamefont{T.}~\bibnamefont{Khouri}},
  \bibinfo{author}{\bibfnamefont{S.}~\bibnamefont{Shamim}},
  \bibinfo{author}{\bibfnamefont{S.}~\bibnamefont{Wiedmann}},
  \bibinfo{author}{\bibfnamefont{H.}~\bibnamefont{Buhmann}}, and
  \bibinfo{author}{\bibfnamefont{L.}~\bibnamefont{Molenkamp}},
  \bibinfo{year}{2017}, \bibinfo{journal}{arXiv:1706.05789} .

\bibitem[{\citenamefont{Budich} \emph{et~al.}(2014)\citenamefont{Budich,
  Trauzettel, and Michetti}}]{budich2014time}
\bibinfo{author}{\bibnamefont{Budich}, \bibfnamefont{J.~C.}},
  \bibinfo{author}{\bibfnamefont{B.}~\bibnamefont{Trauzettel}}, and
  \bibinfo{author}{\bibfnamefont{P.}~\bibnamefont{Michetti}},
  \bibinfo{year}{2014}, \bibinfo{journal}{Physical Review Letters}
  \textbf{\bibinfo{volume}{112}}(\bibinfo{number}{14}),
  \bibinfo{pages}{146405}.

\bibitem[{\citenamefont{Bultinck} \emph{et~al.}(2020)\citenamefont{Bultinck,
  Chatterjee, and Zaletel}}]{bultinck2020mechanism}
\bibinfo{author}{\bibnamefont{Bultinck}, \bibfnamefont{N.}},
  \bibinfo{author}{\bibfnamefont{S.}~\bibnamefont{Chatterjee}}, and
  \bibinfo{author}{\bibfnamefont{M.~P.} \bibnamefont{Zaletel}},
  \bibinfo{year}{2020}, \bibinfo{journal}{Physical Review Letters}
  \textbf{\bibinfo{volume}{124}}(\bibinfo{number}{16}),
  \bibinfo{pages}{166601}.

\bibitem[{\citenamefont{Burkov and Balents}(2011)}]{burkov2011weyl}
\bibinfo{author}{\bibnamefont{Burkov}, \bibfnamefont{A.}}, and
  \bibinfo{author}{\bibfnamefont{L.}~\bibnamefont{Balents}},
  \bibinfo{year}{2011}, \bibinfo{journal}{Physical Review Letters}
  \textbf{\bibinfo{volume}{107}}(\bibinfo{number}{12}),
  \bibinfo{pages}{127205}.

\bibitem[{\citenamefont{Bychkov}(1984)}]{bychkov1984properties}
\bibinfo{author}{\bibnamefont{Bychkov}, \bibfnamefont{Y.~A.}},
  \bibinfo{year}{1984}, \bibinfo{journal}{JETP Letters}
  \textbf{\bibinfo{volume}{39}}(\bibinfo{number}{2}), \bibinfo{pages}{78}.

\bibitem[{\citenamefont{Cai} \emph{et~al.}(2022)\citenamefont{Cai, Ovchinnikov,
  Fei, He, Song, Lin, Wang, Cobden, Chu, Cui} \emph{et~al.}}]{cai2022electric}
\bibinfo{author}{\bibnamefont{Cai}, \bibfnamefont{J.}},
  \bibinfo{author}{\bibfnamefont{D.}~\bibnamefont{Ovchinnikov}},
  \bibinfo{author}{\bibfnamefont{Z.}~\bibnamefont{Fei}},
  \bibinfo{author}{\bibfnamefont{M.}~\bibnamefont{He}},
  \bibinfo{author}{\bibfnamefont{T.}~\bibnamefont{Song}},
  \bibinfo{author}{\bibfnamefont{Z.}~\bibnamefont{Lin}},
  \bibinfo{author}{\bibfnamefont{C.}~\bibnamefont{Wang}},
  \bibinfo{author}{\bibfnamefont{D.}~\bibnamefont{Cobden}},
  \bibinfo{author}{\bibfnamefont{J.-H.} \bibnamefont{Chu}},
  \bibinfo{author}{\bibfnamefont{Y.-T.} \bibnamefont{Cui}}, \emph{et~al.},
  \bibinfo{year}{2022}, \bibinfo{journal}{Nature communications}
  \textbf{\bibinfo{volume}{13}}(\bibinfo{number}{1}), \bibinfo{pages}{1}.

\bibitem[{\citenamefont{Cao} \emph{et~al.}(2016)\citenamefont{Cao, Zhang, Tang,
  Yang, Sofo, Duan, and Liu}}]{cao2016heavy}
\bibinfo{author}{\bibnamefont{Cao}, \bibfnamefont{W.}},
  \bibinfo{author}{\bibfnamefont{R.-X.} \bibnamefont{Zhang}},
  \bibinfo{author}{\bibfnamefont{P.}~\bibnamefont{Tang}},
  \bibinfo{author}{\bibfnamefont{G.}~\bibnamefont{Yang}},
  \bibinfo{author}{\bibfnamefont{J.}~\bibnamefont{Sofo}},
  \bibinfo{author}{\bibfnamefont{W.}~\bibnamefont{Duan}}, and
  \bibinfo{author}{\bibfnamefont{C.-X.} \bibnamefont{Liu}},
  \bibinfo{year}{2016}, \bibinfo{journal}{2D Materials}
  \textbf{\bibinfo{volume}{3}}(\bibinfo{number}{3}), \bibinfo{pages}{034006}.

\bibitem[{\citenamefont{Cao}
  \emph{et~al.}(2018{\natexlab{a}})\citenamefont{Cao, Fatemi, Demir, Fang,
  Tomarken, Luo, Sanchez-Yamagishi, Watanabe, Taniguchi, Kaxiras}
  \emph{et~al.}}]{cao2018correlated}
\bibinfo{author}{\bibnamefont{Cao}, \bibfnamefont{Y.}},
  \bibinfo{author}{\bibfnamefont{V.}~\bibnamefont{Fatemi}},
  \bibinfo{author}{\bibfnamefont{A.}~\bibnamefont{Demir}},
  \bibinfo{author}{\bibfnamefont{S.}~\bibnamefont{Fang}},
  \bibinfo{author}{\bibfnamefont{S.~L.} \bibnamefont{Tomarken}},
  \bibinfo{author}{\bibfnamefont{J.~Y.} \bibnamefont{Luo}},
  \bibinfo{author}{\bibfnamefont{J.~D.} \bibnamefont{Sanchez-Yamagishi}},
  \bibinfo{author}{\bibfnamefont{K.}~\bibnamefont{Watanabe}},
  \bibinfo{author}{\bibfnamefont{T.}~\bibnamefont{Taniguchi}},
  \bibinfo{author}{\bibfnamefont{E.}~\bibnamefont{Kaxiras}}, \emph{et~al.},
  \bibinfo{year}{2018}{\natexlab{a}}, \bibinfo{journal}{Nature}
  \textbf{\bibinfo{volume}{556}}(\bibinfo{number}{7699}), \bibinfo{pages}{80}.

\bibitem[{\citenamefont{Cao}
  \emph{et~al.}(2018{\natexlab{b}})\citenamefont{Cao, Fatemi, Fang, Watanabe,
  Taniguchi, Kaxiras, and Jarillo-Herrero}}]{cao2018unconventional}
\bibinfo{author}{\bibnamefont{Cao}, \bibfnamefont{Y.}},
  \bibinfo{author}{\bibfnamefont{V.}~\bibnamefont{Fatemi}},
  \bibinfo{author}{\bibfnamefont{S.}~\bibnamefont{Fang}},
  \bibinfo{author}{\bibfnamefont{K.}~\bibnamefont{Watanabe}},
  \bibinfo{author}{\bibfnamefont{T.}~\bibnamefont{Taniguchi}},
  \bibinfo{author}{\bibfnamefont{E.}~\bibnamefont{Kaxiras}}, and
  \bibinfo{author}{\bibfnamefont{P.}~\bibnamefont{Jarillo-Herrero}},
  \bibinfo{year}{2018}{\natexlab{b}}, \bibinfo{journal}{Nature}
  \textbf{\bibinfo{volume}{556}}(\bibinfo{number}{7699}), \bibinfo{pages}{43}.

\bibitem[{\citenamefont{Castro~Neto}
  \emph{et~al.}(2009)\citenamefont{Castro~Neto, Guinea, Peres, Novoselov, and
  Geim}}]{castro2009electronic}
\bibinfo{author}{\bibnamefont{Castro~Neto}, \bibfnamefont{A.~H.}},
  \bibinfo{author}{\bibfnamefont{F.}~\bibnamefont{Guinea}},
  \bibinfo{author}{\bibfnamefont{N.~M.~R.} \bibnamefont{Peres}},
  \bibinfo{author}{\bibfnamefont{K.~S.} \bibnamefont{Novoselov}}, and
  \bibinfo{author}{\bibfnamefont{A.~K.} \bibnamefont{Geim}},
  \bibinfo{year}{2009}, \bibinfo{journal}{Reviews of Modern Physics}
  \textbf{\bibinfo{volume}{81}}(\bibinfo{number}{1}), \bibinfo{pages}{109}.

\bibitem[{\citenamefont{Ceresoli} \emph{et~al.}(2006)\citenamefont{Ceresoli,
  Thonhauser, Vanderbilt, and Resta}}]{ceresoli2006orbital}
\bibinfo{author}{\bibnamefont{Ceresoli}, \bibfnamefont{D.}},
  \bibinfo{author}{\bibfnamefont{T.}~\bibnamefont{Thonhauser}},
  \bibinfo{author}{\bibfnamefont{D.}~\bibnamefont{Vanderbilt}}, and
  \bibinfo{author}{\bibfnamefont{R.}~\bibnamefont{Resta}},
  \bibinfo{year}{2006}, \bibinfo{journal}{Physical Review B}
  \textbf{\bibinfo{volume}{74}}(\bibinfo{number}{2}), \bibinfo{pages}{024408}.

\bibitem[{\citenamefont{Chang}
  \emph{et~al.}(2016{\natexlab{a}})\citenamefont{Chang, Liu, Zhang, Wang, He,
  and Xue}}]{chang2016field}
\bibinfo{author}{\bibnamefont{Chang}, \bibfnamefont{C.}},
  \bibinfo{author}{\bibfnamefont{M.}~\bibnamefont{Liu}},
  \bibinfo{author}{\bibfnamefont{Z.}~\bibnamefont{Zhang}},
  \bibinfo{author}{\bibfnamefont{Y.}~\bibnamefont{Wang}},
  \bibinfo{author}{\bibfnamefont{K.}~\bibnamefont{He}}, and
  \bibinfo{author}{\bibfnamefont{Q.}~\bibnamefont{Xue}},
  \bibinfo{year}{2016}{\natexlab{a}}, \bibinfo{journal}{Science China Physics,
  Mechanics \& Astronomy} \textbf{\bibinfo{volume}{59}}(\bibinfo{number}{3}),
  \bibinfo{pages}{637501}.

\bibitem[{\citenamefont{Chang}(2020)}]{chang2020marriage}
\bibinfo{author}{\bibnamefont{Chang}, \bibfnamefont{C.-Z.}},
  \bibinfo{year}{2020}, \bibinfo{journal}{Nature Materials}
  \textbf{\bibinfo{volume}{19}}(\bibinfo{number}{5}), \bibinfo{pages}{484}.

\bibitem[{\citenamefont{Chang} \emph{et~al.}(2011)\citenamefont{Chang, He,
  Wang, Ma, Liu, Zhang, Chen, Wang, and Xue}}]{chang2011growth}
\bibinfo{author}{\bibnamefont{Chang}, \bibfnamefont{C.-Z.}},
  \bibinfo{author}{\bibfnamefont{K.}~\bibnamefont{He}},
  \bibinfo{author}{\bibfnamefont{L.-L.} \bibnamefont{Wang}},
  \bibinfo{author}{\bibfnamefont{X.-C.} \bibnamefont{Ma}},
  \bibinfo{author}{\bibfnamefont{M.-H.} \bibnamefont{Liu}},
  \bibinfo{author}{\bibfnamefont{Z.-C.} \bibnamefont{Zhang}},
  \bibinfo{author}{\bibfnamefont{X.}~\bibnamefont{Chen}},
  \bibinfo{author}{\bibfnamefont{Y.-Y.} \bibnamefont{Wang}}, and
  \bibinfo{author}{\bibfnamefont{Q.-K.} \bibnamefont{Xue}},
  \bibinfo{year}{2011}, in \emph{\bibinfo{booktitle}{Spin}}
  (\bibinfo{organization}{World Scientific}), volume~\bibinfo{volume}{1}, pp.
  \bibinfo{pages}{21--25}.

\bibitem[{\citenamefont{Chang and Li}(2016)}]{chang2016quantum}
\bibinfo{author}{\bibnamefont{Chang}, \bibfnamefont{C.-Z.}}, and
  \bibinfo{author}{\bibfnamefont{M.}~\bibnamefont{Li}}, \bibinfo{year}{2016},
  \bibinfo{journal}{Journal of Physics: Condensed Matter}
  \textbf{\bibinfo{volume}{28}}(\bibinfo{number}{12}), \bibinfo{pages}{123002}.

\bibitem[{\citenamefont{Chang} \emph{et~al.}(2014)\citenamefont{Chang, Tang,
  Wang, Feng, Li, Zhang, Wang, Wang, Chen, Liu}
  \emph{et~al.}}]{chang2014chemical}
\bibinfo{author}{\bibnamefont{Chang}, \bibfnamefont{C.-Z.}},
  \bibinfo{author}{\bibfnamefont{P.}~\bibnamefont{Tang}},
  \bibinfo{author}{\bibfnamefont{Y.-L.} \bibnamefont{Wang}},
  \bibinfo{author}{\bibfnamefont{X.}~\bibnamefont{Feng}},
  \bibinfo{author}{\bibfnamefont{K.}~\bibnamefont{Li}},
  \bibinfo{author}{\bibfnamefont{Z.}~\bibnamefont{Zhang}},
  \bibinfo{author}{\bibfnamefont{Y.}~\bibnamefont{Wang}},
  \bibinfo{author}{\bibfnamefont{L.-L.} \bibnamefont{Wang}},
  \bibinfo{author}{\bibfnamefont{X.}~\bibnamefont{Chen}},
  \bibinfo{author}{\bibfnamefont{C.}~\bibnamefont{Liu}}, \emph{et~al.},
  \bibinfo{year}{2014}, \bibinfo{journal}{Physical Review Letters}
  \textbf{\bibinfo{volume}{112}}(\bibinfo{number}{5}), \bibinfo{pages}{056801}.

\bibitem[{\citenamefont{Chang}
  \emph{et~al.}(2013{\natexlab{a}})\citenamefont{Chang, Zhang, Feng, Shen,
  Zhang, Guo, Li, Ou, Wei, Wang} \emph{et~al.}}]{chang2013experimental}
\bibinfo{author}{\bibnamefont{Chang}, \bibfnamefont{C.-Z.}},
  \bibinfo{author}{\bibfnamefont{J.}~\bibnamefont{Zhang}},
  \bibinfo{author}{\bibfnamefont{X.}~\bibnamefont{Feng}},
  \bibinfo{author}{\bibfnamefont{J.}~\bibnamefont{Shen}},
  \bibinfo{author}{\bibfnamefont{Z.}~\bibnamefont{Zhang}},
  \bibinfo{author}{\bibfnamefont{M.}~\bibnamefont{Guo}},
  \bibinfo{author}{\bibfnamefont{K.}~\bibnamefont{Li}},
  \bibinfo{author}{\bibfnamefont{Y.}~\bibnamefont{Ou}},
  \bibinfo{author}{\bibfnamefont{P.}~\bibnamefont{Wei}},
  \bibinfo{author}{\bibfnamefont{L.-L.} \bibnamefont{Wang}}, \emph{et~al.},
  \bibinfo{year}{2013}{\natexlab{a}}, \bibinfo{journal}{Science}
  \textbf{\bibinfo{volume}{340}}(\bibinfo{number}{6129}), \bibinfo{pages}{167}.

\bibitem[{\citenamefont{Chang}
  \emph{et~al.}(2013{\natexlab{b}})\citenamefont{Chang, Zhang, Liu, Zhang,
  Feng, Li, Wang, Chen, Dai, Fang} \emph{et~al.}}]{chang2013thin}
\bibinfo{author}{\bibnamefont{Chang}, \bibfnamefont{C.-Z.}},
  \bibinfo{author}{\bibfnamefont{J.}~\bibnamefont{Zhang}},
  \bibinfo{author}{\bibfnamefont{M.}~\bibnamefont{Liu}},
  \bibinfo{author}{\bibfnamefont{Z.}~\bibnamefont{Zhang}},
  \bibinfo{author}{\bibfnamefont{X.}~\bibnamefont{Feng}},
  \bibinfo{author}{\bibfnamefont{K.}~\bibnamefont{Li}},
  \bibinfo{author}{\bibfnamefont{L.-L.} \bibnamefont{Wang}},
  \bibinfo{author}{\bibfnamefont{X.}~\bibnamefont{Chen}},
  \bibinfo{author}{\bibfnamefont{X.}~\bibnamefont{Dai}},
  \bibinfo{author}{\bibfnamefont{Z.}~\bibnamefont{Fang}}, \emph{et~al.},
  \bibinfo{year}{2013}{\natexlab{b}}, \bibinfo{journal}{Advanced Materials}
  \textbf{\bibinfo{volume}{25}}(\bibinfo{number}{7}), \bibinfo{pages}{1065}.

\bibitem[{\citenamefont{Chang}
  \emph{et~al.}(2015{\natexlab{a}})\citenamefont{Chang, Zhao, Kim, Wei, Jain,
  Liu, Chan, and Moodera}}]{chang2015zero}
\bibinfo{author}{\bibnamefont{Chang}, \bibfnamefont{C.-Z.}},
  \bibinfo{author}{\bibfnamefont{W.}~\bibnamefont{Zhao}},
  \bibinfo{author}{\bibfnamefont{D.~Y.} \bibnamefont{Kim}},
  \bibinfo{author}{\bibfnamefont{P.}~\bibnamefont{Wei}},
  \bibinfo{author}{\bibfnamefont{J.~K.} \bibnamefont{Jain}},
  \bibinfo{author}{\bibfnamefont{C.}~\bibnamefont{Liu}},
  \bibinfo{author}{\bibfnamefont{M.~H.} \bibnamefont{Chan}}, and
  \bibinfo{author}{\bibfnamefont{J.~S.} \bibnamefont{Moodera}},
  \bibinfo{year}{2015}{\natexlab{a}}, \bibinfo{journal}{Physical Review
  Letters} \textbf{\bibinfo{volume}{115}}(\bibinfo{number}{5}),
  \bibinfo{pages}{057206}.

\bibitem[{\citenamefont{Chang}
  \emph{et~al.}(2015{\natexlab{b}})\citenamefont{Chang, Zhao, Kim, Zhang,
  Assaf, Heiman, Zhang, Liu, Chan, and Moodera}}]{chang2015high}
\bibinfo{author}{\bibnamefont{Chang}, \bibfnamefont{C.-Z.}},
  \bibinfo{author}{\bibfnamefont{W.}~\bibnamefont{Zhao}},
  \bibinfo{author}{\bibfnamefont{D.~Y.} \bibnamefont{Kim}},
  \bibinfo{author}{\bibfnamefont{H.}~\bibnamefont{Zhang}},
  \bibinfo{author}{\bibfnamefont{B.~A.} \bibnamefont{Assaf}},
  \bibinfo{author}{\bibfnamefont{D.}~\bibnamefont{Heiman}},
  \bibinfo{author}{\bibfnamefont{S.-C.} \bibnamefont{Zhang}},
  \bibinfo{author}{\bibfnamefont{C.}~\bibnamefont{Liu}},
  \bibinfo{author}{\bibfnamefont{M.~H.} \bibnamefont{Chan}}, and
  \bibinfo{author}{\bibfnamefont{J.~S.} \bibnamefont{Moodera}},
  \bibinfo{year}{2015}{\natexlab{b}}, \bibinfo{journal}{Nature Materials}
  \textbf{\bibinfo{volume}{14}}(\bibinfo{number}{5}), \bibinfo{pages}{473}.

\bibitem[{\citenamefont{Chang}
  \emph{et~al.}(2016{\natexlab{b}})\citenamefont{Chang, Zhao, Li, Jain, Liu,
  Moodera, and Chan}}]{chang2016observation}
\bibinfo{author}{\bibnamefont{Chang}, \bibfnamefont{C.-Z.}},
  \bibinfo{author}{\bibfnamefont{W.}~\bibnamefont{Zhao}},
  \bibinfo{author}{\bibfnamefont{J.}~\bibnamefont{Li}},
  \bibinfo{author}{\bibfnamefont{J.}~\bibnamefont{Jain}},
  \bibinfo{author}{\bibfnamefont{C.}~\bibnamefont{Liu}},
  \bibinfo{author}{\bibfnamefont{J.~S.} \bibnamefont{Moodera}}, and
  \bibinfo{author}{\bibfnamefont{M.~H.} \bibnamefont{Chan}},
  \bibinfo{year}{2016}{\natexlab{b}}, \bibinfo{journal}{Physical Review
  Letters} \textbf{\bibinfo{volume}{117}}(\bibinfo{number}{12}),
  \bibinfo{pages}{126802}.

\bibitem[{\citenamefont{Checkelsky}
  \emph{et~al.}(2014)\citenamefont{Checkelsky, Yoshimi, Tsukazaki, Takahashi,
  Kozuka, Falson, Kawasaki, and Tokura}}]{checkelsky2014trajectory}
\bibinfo{author}{\bibnamefont{Checkelsky}, \bibfnamefont{J.}},
  \bibinfo{author}{\bibfnamefont{R.}~\bibnamefont{Yoshimi}},
  \bibinfo{author}{\bibfnamefont{A.}~\bibnamefont{Tsukazaki}},
  \bibinfo{author}{\bibfnamefont{K.}~\bibnamefont{Takahashi}},
  \bibinfo{author}{\bibfnamefont{Y.}~\bibnamefont{Kozuka}},
  \bibinfo{author}{\bibfnamefont{J.}~\bibnamefont{Falson}},
  \bibinfo{author}{\bibfnamefont{M.}~\bibnamefont{Kawasaki}}, and
  \bibinfo{author}{\bibfnamefont{Y.}~\bibnamefont{Tokura}},
  \bibinfo{year}{2014}, \bibinfo{journal}{Nature Physics}
  \textbf{\bibinfo{volume}{10}}(\bibinfo{number}{10}), \bibinfo{pages}{731}.

\bibitem[{\citenamefont{Chen}
  \emph{et~al.}(2019{\natexlab{a}})\citenamefont{Chen, Fei, Zhang, Zhang, Liu,
  Zhang, Wang, Wei, Zhang, Zuo} \emph{et~al.}}]{chen2019intrinsic}
\bibinfo{author}{\bibnamefont{Chen}, \bibfnamefont{B.}},
  \bibinfo{author}{\bibfnamefont{F.}~\bibnamefont{Fei}},
  \bibinfo{author}{\bibfnamefont{D.}~\bibnamefont{Zhang}},
  \bibinfo{author}{\bibfnamefont{B.}~\bibnamefont{Zhang}},
  \bibinfo{author}{\bibfnamefont{W.}~\bibnamefont{Liu}},
  \bibinfo{author}{\bibfnamefont{S.}~\bibnamefont{Zhang}},
  \bibinfo{author}{\bibfnamefont{P.}~\bibnamefont{Wang}},
  \bibinfo{author}{\bibfnamefont{B.}~\bibnamefont{Wei}},
  \bibinfo{author}{\bibfnamefont{Y.}~\bibnamefont{Zhang}},
  \bibinfo{author}{\bibfnamefont{Z.}~\bibnamefont{Zuo}}, \emph{et~al.},
  \bibinfo{year}{2019}{\natexlab{a}}, \bibinfo{journal}{Nature Communications}
  \textbf{\bibinfo{volume}{10}}(\bibinfo{number}{1}), \bibinfo{pages}{1}.

\bibitem[{\citenamefont{Chen} \emph{et~al.}(2017)\citenamefont{Chen, He, Xu,
  and Law}}]{chen2017effects}
\bibinfo{author}{\bibnamefont{Chen}, \bibfnamefont{C.-Z.}},
  \bibinfo{author}{\bibfnamefont{J.~J.} \bibnamefont{He}},
  \bibinfo{author}{\bibfnamefont{D.-H.} \bibnamefont{Xu}}, and
  \bibinfo{author}{\bibfnamefont{K.~T.} \bibnamefont{Law}},
  \bibinfo{year}{2017}, \bibinfo{journal}{Physical Review B}
  \textbf{\bibinfo{volume}{96}}(\bibinfo{number}{4}), \bibinfo{pages}{041118}.

\bibitem[{\citenamefont{Chen}
  \emph{et~al.}(2019{\natexlab{b}})\citenamefont{Chen, Liu, and
  Xie}}]{chen2019effects}
\bibinfo{author}{\bibnamefont{Chen}, \bibfnamefont{C.-Z.}},
  \bibinfo{author}{\bibfnamefont{H.}~\bibnamefont{Liu}}, and
  \bibinfo{author}{\bibfnamefont{X.}~\bibnamefont{Xie}},
  \bibinfo{year}{2019}{\natexlab{b}}, \bibinfo{journal}{Physical Review
  Letters} \textbf{\bibinfo{volume}{122}}(\bibinfo{number}{2}),
  \bibinfo{pages}{026601}.

\bibitem[{\citenamefont{Chen} \emph{et~al.}(2018)\citenamefont{Chen, Xie, Liu,
  Lee, and Law}}]{chen2018quasi}
\bibinfo{author}{\bibnamefont{Chen}, \bibfnamefont{C.-Z.}},
  \bibinfo{author}{\bibfnamefont{Y.-M.} \bibnamefont{Xie}},
  \bibinfo{author}{\bibfnamefont{J.}~\bibnamefont{Liu}},
  \bibinfo{author}{\bibfnamefont{P.~A.} \bibnamefont{Lee}}, and
  \bibinfo{author}{\bibfnamefont{K.~T.} \bibnamefont{Law}},
  \bibinfo{year}{2018}, \bibinfo{journal}{Physical Review B}
  \textbf{\bibinfo{volume}{97}}(\bibinfo{number}{10}), \bibinfo{pages}{104504}.

\bibitem[{\citenamefont{Chen}
  \emph{et~al.}(2020{\natexlab{a}})\citenamefont{Chen, Sharpe, Fox, Zhang,
  Wang, Jiang, Lyu, Li, Watanabe, Taniguchi} \emph{et~al.}}]{chen2020tunable}
\bibinfo{author}{\bibnamefont{Chen}, \bibfnamefont{G.}},
  \bibinfo{author}{\bibfnamefont{A.~L.} \bibnamefont{Sharpe}},
  \bibinfo{author}{\bibfnamefont{E.~J.} \bibnamefont{Fox}},
  \bibinfo{author}{\bibfnamefont{Y.-H.} \bibnamefont{Zhang}},
  \bibinfo{author}{\bibfnamefont{S.}~\bibnamefont{Wang}},
  \bibinfo{author}{\bibfnamefont{L.}~\bibnamefont{Jiang}},
  \bibinfo{author}{\bibfnamefont{B.}~\bibnamefont{Lyu}},
  \bibinfo{author}{\bibfnamefont{H.}~\bibnamefont{Li}},
  \bibinfo{author}{\bibfnamefont{K.}~\bibnamefont{Watanabe}},
  \bibinfo{author}{\bibfnamefont{T.}~\bibnamefont{Taniguchi}}, \emph{et~al.},
  \bibinfo{year}{2020}{\natexlab{a}}, \bibinfo{journal}{Nature}
  \textbf{\bibinfo{volume}{579}}(\bibinfo{number}{7797}), \bibinfo{pages}{56}.

\bibitem[{\citenamefont{Chen}
  \emph{et~al.}(2021{\natexlab{a}})\citenamefont{Chen, Li, Sun, Liu, Zhao, Lu,
  and Xie}}]{chen2021using}
\bibinfo{author}{\bibnamefont{Chen}, \bibfnamefont{R.}},
  \bibinfo{author}{\bibfnamefont{S.}~\bibnamefont{Li}},
  \bibinfo{author}{\bibfnamefont{H.-P.} \bibnamefont{Sun}},
  \bibinfo{author}{\bibfnamefont{Q.}~\bibnamefont{Liu}},
  \bibinfo{author}{\bibfnamefont{Y.}~\bibnamefont{Zhao}},
  \bibinfo{author}{\bibfnamefont{H.-Z.} \bibnamefont{Lu}}, and
  \bibinfo{author}{\bibfnamefont{X.}~\bibnamefont{Xie}},
  \bibinfo{year}{2021}{\natexlab{a}}, \bibinfo{journal}{Physical Review B}
  \textbf{\bibinfo{volume}{103}}(\bibinfo{number}{24}),
  \bibinfo{pages}{L241409}.

\bibitem[{\citenamefont{Chen}
  \emph{et~al.}(2021{\natexlab{b}})\citenamefont{Chen, He, Zhang, Hsieh, Fei,
  Watanabe, Taniguchi, Cobden, Xu, Dean} \emph{et~al.}}]{chen2021electrically}
\bibinfo{author}{\bibnamefont{Chen}, \bibfnamefont{S.}},
  \bibinfo{author}{\bibfnamefont{M.}~\bibnamefont{He}},
  \bibinfo{author}{\bibfnamefont{Y.-H.} \bibnamefont{Zhang}},
  \bibinfo{author}{\bibfnamefont{V.}~\bibnamefont{Hsieh}},
  \bibinfo{author}{\bibfnamefont{Z.}~\bibnamefont{Fei}},
  \bibinfo{author}{\bibfnamefont{K.}~\bibnamefont{Watanabe}},
  \bibinfo{author}{\bibfnamefont{T.}~\bibnamefont{Taniguchi}},
  \bibinfo{author}{\bibfnamefont{D.~H.} \bibnamefont{Cobden}},
  \bibinfo{author}{\bibfnamefont{X.}~\bibnamefont{Xu}},
  \bibinfo{author}{\bibfnamefont{C.~R.} \bibnamefont{Dean}}, \emph{et~al.},
  \bibinfo{year}{2021}{\natexlab{b}}, \bibinfo{journal}{Nature Physics}
  \textbf{\bibinfo{volume}{17}}(\bibinfo{number}{3}), \bibinfo{pages}{374}.

\bibitem[{\citenamefont{Chen} \emph{et~al.}(2016)\citenamefont{Chen, Huang,
  Crisostomo, Hsu, Chuang, Lin, and Bansil}}]{chen2016prediction}
\bibinfo{author}{\bibnamefont{Chen}, \bibfnamefont{S.-P.}},
  \bibinfo{author}{\bibfnamefont{Z.-Q.} \bibnamefont{Huang}},
  \bibinfo{author}{\bibfnamefont{C.~P.} \bibnamefont{Crisostomo}},
  \bibinfo{author}{\bibfnamefont{C.-H.} \bibnamefont{Hsu}},
  \bibinfo{author}{\bibfnamefont{F.-C.} \bibnamefont{Chuang}},
  \bibinfo{author}{\bibfnamefont{H.}~\bibnamefont{Lin}}, and
  \bibinfo{author}{\bibfnamefont{A.}~\bibnamefont{Bansil}},
  \bibinfo{year}{2016}, \bibinfo{journal}{Scientific Reports}
  \textbf{\bibinfo{volume}{6}}(\bibinfo{number}{1}), \bibinfo{pages}{1}.

\bibitem[{\citenamefont{Chen} \emph{et~al.}(2009)\citenamefont{Chen, Analytis,
  Chu, Liu, Mo, Qi, Zhang, Lu, Dai, Fang} \emph{et~al.}}]{chen2009experimental}
\bibinfo{author}{\bibnamefont{Chen}, \bibfnamefont{Y.}},
  \bibinfo{author}{\bibfnamefont{J.~G.} \bibnamefont{Analytis}},
  \bibinfo{author}{\bibfnamefont{J.-H.} \bibnamefont{Chu}},
  \bibinfo{author}{\bibfnamefont{Z.}~\bibnamefont{Liu}},
  \bibinfo{author}{\bibfnamefont{S.-K.} \bibnamefont{Mo}},
  \bibinfo{author}{\bibfnamefont{X.-L.} \bibnamefont{Qi}},
  \bibinfo{author}{\bibfnamefont{H.}~\bibnamefont{Zhang}},
  \bibinfo{author}{\bibfnamefont{D.}~\bibnamefont{Lu}},
  \bibinfo{author}{\bibfnamefont{X.}~\bibnamefont{Dai}},
  \bibinfo{author}{\bibfnamefont{Z.}~\bibnamefont{Fang}}, \emph{et~al.},
  \bibinfo{year}{2009}, \bibinfo{journal}{Science}
  \textbf{\bibinfo{volume}{325}}(\bibinfo{number}{5937}), \bibinfo{pages}{178}.

\bibitem[{\citenamefont{Chen} \emph{et~al.}(2010)\citenamefont{Chen, Chu,
  Analytis, Liu, Igarashi, Kuo, Qi, Mo, Moore, Lu}
  \emph{et~al.}}]{chen2010massive}
\bibinfo{author}{\bibnamefont{Chen}, \bibfnamefont{Y.}},
  \bibinfo{author}{\bibfnamefont{J.-H.} \bibnamefont{Chu}},
  \bibinfo{author}{\bibfnamefont{J.}~\bibnamefont{Analytis}},
  \bibinfo{author}{\bibfnamefont{Z.}~\bibnamefont{Liu}},
  \bibinfo{author}{\bibfnamefont{K.}~\bibnamefont{Igarashi}},
  \bibinfo{author}{\bibfnamefont{H.-H.} \bibnamefont{Kuo}},
  \bibinfo{author}{\bibfnamefont{X.}~\bibnamefont{Qi}},
  \bibinfo{author}{\bibfnamefont{S.-K.} \bibnamefont{Mo}},
  \bibinfo{author}{\bibfnamefont{R.}~\bibnamefont{Moore}},
  \bibinfo{author}{\bibfnamefont{D.}~\bibnamefont{Lu}}, \emph{et~al.},
  \bibinfo{year}{2010}, \bibinfo{journal}{Science}
  \textbf{\bibinfo{volume}{329}}(\bibinfo{number}{5992}), \bibinfo{pages}{659}.

\bibitem[{\citenamefont{Chen}
  \emph{et~al.}(2020{\natexlab{b}})\citenamefont{Chen, Chuang, Lee, Zhu, Honz,
  Guan, Wang, Wang, Mao, Zhu} \emph{et~al.}}]{chen2020ferromagnetism}
\bibinfo{author}{\bibnamefont{Chen}, \bibfnamefont{Y.}},
  \bibinfo{author}{\bibfnamefont{Y.-W.} \bibnamefont{Chuang}},
  \bibinfo{author}{\bibfnamefont{S.~H.} \bibnamefont{Lee}},
  \bibinfo{author}{\bibfnamefont{Y.}~\bibnamefont{Zhu}},
  \bibinfo{author}{\bibfnamefont{K.}~\bibnamefont{Honz}},
  \bibinfo{author}{\bibfnamefont{Y.}~\bibnamefont{Guan}},
  \bibinfo{author}{\bibfnamefont{Y.}~\bibnamefont{Wang}},
  \bibinfo{author}{\bibfnamefont{K.}~\bibnamefont{Wang}},
  \bibinfo{author}{\bibfnamefont{Z.}~\bibnamefont{Mao}},
  \bibinfo{author}{\bibfnamefont{J.}~\bibnamefont{Zhu}}, \emph{et~al.},
  \bibinfo{year}{2020}{\natexlab{b}}, \bibinfo{journal}{Physical Review
  Materials} \textbf{\bibinfo{volume}{4}}(\bibinfo{number}{6}),
  \bibinfo{pages}{064411}.

\bibitem[{\citenamefont{Chen}
  \emph{et~al.}(2019{\natexlab{c}})\citenamefont{Chen, Xu, Li, Li, Wang, Zhang,
  Li, Wu, Liang, Chen} \emph{et~al.}}]{chen2019topological}
\bibinfo{author}{\bibnamefont{Chen}, \bibfnamefont{Y.}},
  \bibinfo{author}{\bibfnamefont{L.}~\bibnamefont{Xu}},
  \bibinfo{author}{\bibfnamefont{J.}~\bibnamefont{Li}},
  \bibinfo{author}{\bibfnamefont{Y.}~\bibnamefont{Li}},
  \bibinfo{author}{\bibfnamefont{H.}~\bibnamefont{Wang}},
  \bibinfo{author}{\bibfnamefont{C.}~\bibnamefont{Zhang}},
  \bibinfo{author}{\bibfnamefont{H.}~\bibnamefont{Li}},
  \bibinfo{author}{\bibfnamefont{Y.}~\bibnamefont{Wu}},
  \bibinfo{author}{\bibfnamefont{A.}~\bibnamefont{Liang}},
  \bibinfo{author}{\bibfnamefont{C.}~\bibnamefont{Chen}}, \emph{et~al.},
  \bibinfo{year}{2019}{\natexlab{c}}, \bibinfo{journal}{Physical Review X}
  \textbf{\bibinfo{volume}{9}}(\bibinfo{number}{4}), \bibinfo{pages}{041040}.

\bibitem[{\citenamefont{Cheraghchi and Sabze}(2020)}]{cheraghchi2020spin}
\bibinfo{author}{\bibnamefont{Cheraghchi}, \bibfnamefont{H.}}, and
  \bibinfo{author}{\bibfnamefont{T.}~\bibnamefont{Sabze}},
  \bibinfo{year}{2020}, \bibinfo{journal}{Journal of Magnetism and Magnetic
  Materials} \textbf{\bibinfo{volume}{513}}, \bibinfo{pages}{166923}.

\bibitem[{\citenamefont{Chien} \emph{et~al.}(2005)\citenamefont{Chien, Zhou,
  and Uher}}]{chien2005growth}
\bibinfo{author}{\bibnamefont{Chien}, \bibfnamefont{Y.-J.}},
  \bibinfo{author}{\bibfnamefont{Z.}~\bibnamefont{Zhou}}, and
  \bibinfo{author}{\bibfnamefont{C.}~\bibnamefont{Uher}}, \bibinfo{year}{2005},
  \bibinfo{journal}{Journal of Crystal Growth}
  \textbf{\bibinfo{volume}{283}}(\bibinfo{number}{3-4}), \bibinfo{pages}{309}.

\bibitem[{\citenamefont{Choi} \emph{et~al.}(2021)\citenamefont{Choi, Kim, Peng,
  Thomson, Lewandowski, Polski, Zhang, Arora, Watanabe, Taniguchi}
  \emph{et~al.}}]{choi2021correlation}
\bibinfo{author}{\bibnamefont{Choi}, \bibfnamefont{Y.}},
  \bibinfo{author}{\bibfnamefont{H.}~\bibnamefont{Kim}},
  \bibinfo{author}{\bibfnamefont{Y.}~\bibnamefont{Peng}},
  \bibinfo{author}{\bibfnamefont{A.}~\bibnamefont{Thomson}},
  \bibinfo{author}{\bibfnamefont{C.}~\bibnamefont{Lewandowski}},
  \bibinfo{author}{\bibfnamefont{R.}~\bibnamefont{Polski}},
  \bibinfo{author}{\bibfnamefont{Y.}~\bibnamefont{Zhang}},
  \bibinfo{author}{\bibfnamefont{H.~S.} \bibnamefont{Arora}},
  \bibinfo{author}{\bibfnamefont{K.}~\bibnamefont{Watanabe}},
  \bibinfo{author}{\bibfnamefont{T.}~\bibnamefont{Taniguchi}}, \emph{et~al.},
  \bibinfo{year}{2021}, \bibinfo{journal}{Nature}
  \textbf{\bibinfo{volume}{589}}(\bibinfo{number}{7843}), \bibinfo{pages}{536}.

\bibitem[{\citenamefont{Chu} \emph{et~al.}(2011)\citenamefont{Chu, Shi, and
  Shen}}]{chu2011surface}
\bibinfo{author}{\bibnamefont{Chu}, \bibfnamefont{R.-L.}},
  \bibinfo{author}{\bibfnamefont{J.}~\bibnamefont{Shi}}, and
  \bibinfo{author}{\bibfnamefont{S.-Q.} \bibnamefont{Shen}},
  \bibinfo{year}{2011}, \bibinfo{journal}{Physical Review B}
  \textbf{\bibinfo{volume}{84}}(\bibinfo{number}{8}), \bibinfo{pages}{085312}.

\bibitem[{\citenamefont{Connolly} \emph{et~al.}(2012)\citenamefont{Connolly,
  Puddy, Logoteta, Marconcini, Roy, Griffiths, Jones, Maksym, Macucci, and
  Smith}}]{connolly2012unraveling}
\bibinfo{author}{\bibnamefont{Connolly}, \bibfnamefont{M.}},
  \bibinfo{author}{\bibfnamefont{R.}~\bibnamefont{Puddy}},
  \bibinfo{author}{\bibfnamefont{D.}~\bibnamefont{Logoteta}},
  \bibinfo{author}{\bibfnamefont{P.}~\bibnamefont{Marconcini}},
  \bibinfo{author}{\bibfnamefont{M.}~\bibnamefont{Roy}},
  \bibinfo{author}{\bibfnamefont{J.}~\bibnamefont{Griffiths}},
  \bibinfo{author}{\bibfnamefont{G.}~\bibnamefont{Jones}},
  \bibinfo{author}{\bibfnamefont{P.}~\bibnamefont{Maksym}},
  \bibinfo{author}{\bibfnamefont{M.}~\bibnamefont{Macucci}}, and
  \bibinfo{author}{\bibfnamefont{C.}~\bibnamefont{Smith}},
  \bibinfo{year}{2012}, \bibinfo{journal}{Nano Letters}
  \textbf{\bibinfo{volume}{12}}(\bibinfo{number}{11}), \bibinfo{pages}{5448}.

\bibitem[{\citenamefont{Cui} \emph{et~al.}(2019)\citenamefont{Cui, Shi, Wang,
  Yu, Wu, Luo, Ying, and Chen}}]{cui2019transport}
\bibinfo{author}{\bibnamefont{Cui}, \bibfnamefont{J.}},
  \bibinfo{author}{\bibfnamefont{M.}~\bibnamefont{Shi}},
  \bibinfo{author}{\bibfnamefont{H.}~\bibnamefont{Wang}},
  \bibinfo{author}{\bibfnamefont{F.}~\bibnamefont{Yu}},
  \bibinfo{author}{\bibfnamefont{T.}~\bibnamefont{Wu}},
  \bibinfo{author}{\bibfnamefont{X.}~\bibnamefont{Luo}},
  \bibinfo{author}{\bibfnamefont{J.}~\bibnamefont{Ying}}, and
  \bibinfo{author}{\bibfnamefont{X.}~\bibnamefont{Chen}}, \bibinfo{year}{2019},
  \bibinfo{journal}{Physical Review B}
  \textbf{\bibinfo{volume}{99}}(\bibinfo{number}{15}), \bibinfo{pages}{155125}.

\bibitem[{\citenamefont{Das} \emph{et~al.}(2012)\citenamefont{Das, Ronen, Most,
  Oreg, Heiblum, and Shtrikman}}]{das2012zero}
\bibinfo{author}{\bibnamefont{Das}, \bibfnamefont{A.}},
  \bibinfo{author}{\bibfnamefont{Y.}~\bibnamefont{Ronen}},
  \bibinfo{author}{\bibfnamefont{Y.}~\bibnamefont{Most}},
  \bibinfo{author}{\bibfnamefont{Y.}~\bibnamefont{Oreg}},
  \bibinfo{author}{\bibfnamefont{M.}~\bibnamefont{Heiblum}}, and
  \bibinfo{author}{\bibfnamefont{H.}~\bibnamefont{Shtrikman}},
  \bibinfo{year}{2012}, \bibinfo{journal}{Nature Physics}
  \textbf{\bibinfo{volume}{8}}(\bibinfo{number}{12}), \bibinfo{pages}{887}.

\bibitem[{\citenamefont{Deng} \emph{et~al.}(2012)\citenamefont{Deng, Yu, Huang,
  Larsson, Caroff, and Xu}}]{deng2012anomalous}
\bibinfo{author}{\bibnamefont{Deng}, \bibfnamefont{M.}},
  \bibinfo{author}{\bibfnamefont{C.}~\bibnamefont{Yu}},
  \bibinfo{author}{\bibfnamefont{G.}~\bibnamefont{Huang}},
  \bibinfo{author}{\bibfnamefont{M.}~\bibnamefont{Larsson}},
  \bibinfo{author}{\bibfnamefont{P.}~\bibnamefont{Caroff}}, and
  \bibinfo{author}{\bibfnamefont{H.}~\bibnamefont{Xu}}, \bibinfo{year}{2012},
  \bibinfo{journal}{Nano Letters}
  \textbf{\bibinfo{volume}{12}}(\bibinfo{number}{12}), \bibinfo{pages}{6414}.

\bibitem[{\citenamefont{Deng} \emph{et~al.}(2017)\citenamefont{Deng, Qi, Han,
  Zhang, Xu, and Qiao}}]{deng2017realization}
\bibinfo{author}{\bibnamefont{Deng}, \bibfnamefont{X.}},
  \bibinfo{author}{\bibfnamefont{S.}~\bibnamefont{Qi}},
  \bibinfo{author}{\bibfnamefont{Y.}~\bibnamefont{Han}},
  \bibinfo{author}{\bibfnamefont{K.}~\bibnamefont{Zhang}},
  \bibinfo{author}{\bibfnamefont{X.}~\bibnamefont{Xu}}, and
  \bibinfo{author}{\bibfnamefont{Z.}~\bibnamefont{Qiao}}, \bibinfo{year}{2017},
  \bibinfo{journal}{Physical Review B}
  \textbf{\bibinfo{volume}{95}}(\bibinfo{number}{12}), \bibinfo{pages}{121410}.

\bibitem[{\citenamefont{Deng} \emph{et~al.}(2020)\citenamefont{Deng, Yu, Shi,
  Guo, Xu, Wang, Chen, and Zhang}}]{deng2020quantum}
\bibinfo{author}{\bibnamefont{Deng}, \bibfnamefont{Y.}},
  \bibinfo{author}{\bibfnamefont{Y.}~\bibnamefont{Yu}},
  \bibinfo{author}{\bibfnamefont{M.~Z.} \bibnamefont{Shi}},
  \bibinfo{author}{\bibfnamefont{Z.}~\bibnamefont{Guo}},
  \bibinfo{author}{\bibfnamefont{Z.}~\bibnamefont{Xu}},
  \bibinfo{author}{\bibfnamefont{J.}~\bibnamefont{Wang}},
  \bibinfo{author}{\bibfnamefont{X.~H.} \bibnamefont{Chen}}, and
  \bibinfo{author}{\bibfnamefont{Y.}~\bibnamefont{Zhang}},
  \bibinfo{year}{2020}, \bibinfo{journal}{Science}
  \textbf{\bibinfo{volume}{367}}(\bibinfo{number}{6480}), \bibinfo{pages}{895}.

\bibitem[{\citenamefont{Devakul and Fu}(2022)}]{devakul2022quantum}
\bibinfo{author}{\bibnamefont{Devakul}, \bibfnamefont{T.}}, and
  \bibinfo{author}{\bibfnamefont{L.}~\bibnamefont{Fu}}, \bibinfo{year}{2022},
  \bibinfo{journal}{Physical Review X}
  \textbf{\bibinfo{volume}{12}}(\bibinfo{number}{2}), \bibinfo{pages}{021031}.

\bibitem[{\citenamefont{Dietl and Ohno}(2014)}]{dietl2014dilute}
\bibinfo{author}{\bibnamefont{Dietl}, \bibfnamefont{T.}}, and
  \bibinfo{author}{\bibfnamefont{H.}~\bibnamefont{Ohno}}, \bibinfo{year}{2014},
  \bibinfo{journal}{Reviews of Modern Physics}
  \textbf{\bibinfo{volume}{86}}(\bibinfo{number}{1}), \bibinfo{pages}{187}.

\bibitem[{\citenamefont{Ding} \emph{et~al.}(2020)\citenamefont{Ding, Hu, Ye,
  Feng, Ni, and Cao}}]{ding2020crystal}
\bibinfo{author}{\bibnamefont{Ding}, \bibfnamefont{L.}},
  \bibinfo{author}{\bibfnamefont{C.}~\bibnamefont{Hu}},
  \bibinfo{author}{\bibfnamefont{F.}~\bibnamefont{Ye}},
  \bibinfo{author}{\bibfnamefont{E.}~\bibnamefont{Feng}},
  \bibinfo{author}{\bibfnamefont{N.}~\bibnamefont{Ni}}, and
  \bibinfo{author}{\bibfnamefont{H.}~\bibnamefont{Cao}}, \bibinfo{year}{2020},
  \bibinfo{journal}{Physical Review B}
  \textbf{\bibinfo{volume}{101}}(\bibinfo{number}{2}), \bibinfo{pages}{020412}.

\bibitem[{\citenamefont{Dos~Santos}
  \emph{et~al.}(2007)\citenamefont{Dos~Santos, Peres, and
  Neto}}]{dos2007graphene}
\bibinfo{author}{\bibnamefont{Dos~Santos}, \bibfnamefont{J.~L.}},
  \bibinfo{author}{\bibfnamefont{N.}~\bibnamefont{Peres}}, and
  \bibinfo{author}{\bibfnamefont{A.~C.} \bibnamefont{Neto}},
  \bibinfo{year}{2007}, \bibinfo{journal}{Physical Review Letters}
  \textbf{\bibinfo{volume}{99}}(\bibinfo{number}{25}), \bibinfo{pages}{256802}.

\bibitem[{\citenamefont{Dziom} \emph{et~al.}(2017)\citenamefont{Dziom, Shuvaev,
  Pimenov, Astakhov, Ames, Bendias, B{\"o}ttcher, Tkachov, Hankiewicz,
  Br{\"u}ne} \emph{et~al.}}]{dziom2017observation}
\bibinfo{author}{\bibnamefont{Dziom}, \bibfnamefont{V.}},
  \bibinfo{author}{\bibfnamefont{A.}~\bibnamefont{Shuvaev}},
  \bibinfo{author}{\bibfnamefont{A.}~\bibnamefont{Pimenov}},
  \bibinfo{author}{\bibfnamefont{G.}~\bibnamefont{Astakhov}},
  \bibinfo{author}{\bibfnamefont{C.}~\bibnamefont{Ames}},
  \bibinfo{author}{\bibfnamefont{K.}~\bibnamefont{Bendias}},
  \bibinfo{author}{\bibfnamefont{J.}~\bibnamefont{B{\"o}ttcher}},
  \bibinfo{author}{\bibfnamefont{G.}~\bibnamefont{Tkachov}},
  \bibinfo{author}{\bibfnamefont{E.}~\bibnamefont{Hankiewicz}},
  \bibinfo{author}{\bibfnamefont{C.}~\bibnamefont{Br{\"u}ne}}, \emph{et~al.},
  \bibinfo{year}{2017}, \bibinfo{journal}{Nature Communications}
  \textbf{\bibinfo{volume}{8}}(\bibinfo{number}{1}), \bibinfo{pages}{1}.

\bibitem[{\citenamefont{Essin} \emph{et~al.}(2009)\citenamefont{Essin, Moore,
  and Vanderbilt}}]{essin2009magnetoelectric}
\bibinfo{author}{\bibnamefont{Essin}, \bibfnamefont{A.~M.}},
  \bibinfo{author}{\bibfnamefont{J.~E.} \bibnamefont{Moore}}, and
  \bibinfo{author}{\bibfnamefont{D.}~\bibnamefont{Vanderbilt}},
  \bibinfo{year}{2009}, \bibinfo{journal}{Physical Review Letters}
  \textbf{\bibinfo{volume}{102}}(\bibinfo{number}{14}),
  \bibinfo{pages}{146805}.

\bibitem[{\citenamefont{Ezawa}(2012)}]{ezawa2012valley}
\bibinfo{author}{\bibnamefont{Ezawa}, \bibfnamefont{M.}}, \bibinfo{year}{2012},
  \bibinfo{journal}{Physical Review Letters}
  \textbf{\bibinfo{volume}{109}}(\bibinfo{number}{5}), \bibinfo{pages}{055502}.

\bibitem[{\citenamefont{Ezawa}(2013)}]{ezawa2013spin}
\bibinfo{author}{\bibnamefont{Ezawa}, \bibfnamefont{M.}}, \bibinfo{year}{2013},
  \bibinfo{journal}{Physical Review B}
  \textbf{\bibinfo{volume}{87}}(\bibinfo{number}{15}), \bibinfo{pages}{155415}.

\bibitem[{\citenamefont{Fang and Fu}(2019)}]{fang2019new}
\bibinfo{author}{\bibnamefont{Fang}, \bibfnamefont{C.}}, and
  \bibinfo{author}{\bibfnamefont{L.}~\bibnamefont{Fu}}, \bibinfo{year}{2019},
  \bibinfo{journal}{Science advances}
  \textbf{\bibinfo{volume}{5}}(\bibinfo{number}{12}),
  \bibinfo{pages}{eaat2374}.

\bibitem[{\citenamefont{Fang} \emph{et~al.}(2014)\citenamefont{Fang, Gilbert,
  and Bernevig}}]{fang2014large}
\bibinfo{author}{\bibnamefont{Fang}, \bibfnamefont{C.}},
  \bibinfo{author}{\bibfnamefont{M.~J.} \bibnamefont{Gilbert}}, and
  \bibinfo{author}{\bibfnamefont{B.~A.} \bibnamefont{Bernevig}},
  \bibinfo{year}{2014}, \bibinfo{journal}{Physical Review Letters}
  \textbf{\bibinfo{volume}{112}}(\bibinfo{number}{4}), \bibinfo{pages}{046801}.

\bibitem[{\citenamefont{Feng} \emph{et~al.}(2016)\citenamefont{Feng, Feng,
  Wang, Ou, Hao, Liu, Zhang, Zhang, Lin, Liao}
  \emph{et~al.}}]{feng2016thickness}
\bibinfo{author}{\bibnamefont{Feng}, \bibfnamefont{X.}},
  \bibinfo{author}{\bibfnamefont{Y.}~\bibnamefont{Feng}},
  \bibinfo{author}{\bibfnamefont{J.}~\bibnamefont{Wang}},
  \bibinfo{author}{\bibfnamefont{Y.}~\bibnamefont{Ou}},
  \bibinfo{author}{\bibfnamefont{Z.}~\bibnamefont{Hao}},
  \bibinfo{author}{\bibfnamefont{C.}~\bibnamefont{Liu}},
  \bibinfo{author}{\bibfnamefont{Z.}~\bibnamefont{Zhang}},
  \bibinfo{author}{\bibfnamefont{L.}~\bibnamefont{Zhang}},
  \bibinfo{author}{\bibfnamefont{C.}~\bibnamefont{Lin}},
  \bibinfo{author}{\bibfnamefont{J.}~\bibnamefont{Liao}}, \emph{et~al.},
  \bibinfo{year}{2016}, \bibinfo{journal}{Advanced Materials}
  \textbf{\bibinfo{volume}{28}}(\bibinfo{number}{30}), \bibinfo{pages}{6386}.

\bibitem[{\citenamefont{Feng} \emph{et~al.}(2015)\citenamefont{Feng, Feng, Ou,
  Wang, Liu, Zhang, Zhao, Jiang, Zhang, He}
  \emph{et~al.}}]{feng2015observation}
\bibinfo{author}{\bibnamefont{Feng}, \bibfnamefont{Y.}},
  \bibinfo{author}{\bibfnamefont{X.}~\bibnamefont{Feng}},
  \bibinfo{author}{\bibfnamefont{Y.}~\bibnamefont{Ou}},
  \bibinfo{author}{\bibfnamefont{J.}~\bibnamefont{Wang}},
  \bibinfo{author}{\bibfnamefont{C.}~\bibnamefont{Liu}},
  \bibinfo{author}{\bibfnamefont{L.}~\bibnamefont{Zhang}},
  \bibinfo{author}{\bibfnamefont{D.}~\bibnamefont{Zhao}},
  \bibinfo{author}{\bibfnamefont{G.}~\bibnamefont{Jiang}},
  \bibinfo{author}{\bibfnamefont{S.-C.} \bibnamefont{Zhang}},
  \bibinfo{author}{\bibfnamefont{K.}~\bibnamefont{He}}, \emph{et~al.},
  \bibinfo{year}{2015}, \bibinfo{journal}{Physical Review Letters}
  \textbf{\bibinfo{volume}{115}}(\bibinfo{number}{12}),
  \bibinfo{pages}{126801}.

\bibitem[{\citenamefont{Fijalkowski}
  \emph{et~al.}(2021)\citenamefont{Fijalkowski, Liu, Hartl, Winnerlein, Mandal,
  Coschizza, Fothergill, Grauer, Schreyeck, Brunner}
  \emph{et~al.}}]{fijalkowski2021any}
\bibinfo{author}{\bibnamefont{Fijalkowski}, \bibfnamefont{K.}},
  \bibinfo{author}{\bibfnamefont{N.}~\bibnamefont{Liu}},
  \bibinfo{author}{\bibfnamefont{M.}~\bibnamefont{Hartl}},
  \bibinfo{author}{\bibfnamefont{M.}~\bibnamefont{Winnerlein}},
  \bibinfo{author}{\bibfnamefont{P.}~\bibnamefont{Mandal}},
  \bibinfo{author}{\bibfnamefont{A.}~\bibnamefont{Coschizza}},
  \bibinfo{author}{\bibfnamefont{A.}~\bibnamefont{Fothergill}},
  \bibinfo{author}{\bibfnamefont{S.}~\bibnamefont{Grauer}},
  \bibinfo{author}{\bibfnamefont{S.}~\bibnamefont{Schreyeck}},
  \bibinfo{author}{\bibfnamefont{K.}~\bibnamefont{Brunner}}, \emph{et~al.},
  \bibinfo{year}{2021}, \bibinfo{journal}{Physical Review B}
  \textbf{\bibinfo{volume}{103}}(\bibinfo{number}{23}),
  \bibinfo{pages}{235111}.

\bibitem[{\citenamefont{Fox} \emph{et~al.}(2018)\citenamefont{Fox, Rosen, Yang,
  Jones, Elmquist, Kou, Pan, Wang, and Goldhaber-Gordon}}]{fox2018part}
\bibinfo{author}{\bibnamefont{Fox}, \bibfnamefont{E.~J.}},
  \bibinfo{author}{\bibfnamefont{I.~T.} \bibnamefont{Rosen}},
  \bibinfo{author}{\bibfnamefont{Y.}~\bibnamefont{Yang}},
  \bibinfo{author}{\bibfnamefont{G.~R.} \bibnamefont{Jones}},
  \bibinfo{author}{\bibfnamefont{R.~E.} \bibnamefont{Elmquist}},
  \bibinfo{author}{\bibfnamefont{X.}~\bibnamefont{Kou}},
  \bibinfo{author}{\bibfnamefont{L.}~\bibnamefont{Pan}},
  \bibinfo{author}{\bibfnamefont{K.~L.} \bibnamefont{Wang}}, and
  \bibinfo{author}{\bibfnamefont{D.}~\bibnamefont{Goldhaber-Gordon}},
  \bibinfo{year}{2018}, \bibinfo{journal}{Physical Review B}
  \textbf{\bibinfo{volume}{98}}(\bibinfo{number}{7}), \bibinfo{pages}{075145}.

\bibitem[{\citenamefont{Fradkin} \emph{et~al.}(1986)\citenamefont{Fradkin,
  Dagotto, and Boyanovsky}}]{fradkin1986physical}
\bibinfo{author}{\bibnamefont{Fradkin}, \bibfnamefont{E.}},
  \bibinfo{author}{\bibfnamefont{E.}~\bibnamefont{Dagotto}}, and
  \bibinfo{author}{\bibfnamefont{D.}~\bibnamefont{Boyanovsky}},
  \bibinfo{year}{1986}, \bibinfo{journal}{Physical review letters}
  \textbf{\bibinfo{volume}{57}}(\bibinfo{number}{23}), \bibinfo{pages}{2967}.

\bibitem[{\citenamefont{Fu and Kane}(2007)}]{fu2007topological}
\bibinfo{author}{\bibnamefont{Fu}, \bibfnamefont{L.}}, and
  \bibinfo{author}{\bibfnamefont{C.~L.} \bibnamefont{Kane}},
  \bibinfo{year}{2007}, \bibinfo{journal}{Physical Review B}
  \textbf{\bibinfo{volume}{76}}(\bibinfo{number}{4}), \bibinfo{pages}{045302}.

\bibitem[{\citenamefont{Fu} \emph{et~al.}(2007)\citenamefont{Fu, Kane, and
  Mele}}]{fu2007topologicala}
\bibinfo{author}{\bibnamefont{Fu}, \bibfnamefont{L.}},
  \bibinfo{author}{\bibfnamefont{C.~L.} \bibnamefont{Kane}}, and
  \bibinfo{author}{\bibfnamefont{E.~J.} \bibnamefont{Mele}},
  \bibinfo{year}{2007}, \bibinfo{journal}{Physical Review Letters}
  \textbf{\bibinfo{volume}{98}}(\bibinfo{number}{10}), \bibinfo{pages}{106803}.

\bibitem[{\citenamefont{Gao} \emph{et~al.}(2021)\citenamefont{Gao, Liu, Hu,
  Qiu, Tzschaschel, Ghosh, Ho, B{\'e}rub{\'e}, Chen, Sun}
  \emph{et~al.}}]{gao2021layer}
\bibinfo{author}{\bibnamefont{Gao}, \bibfnamefont{A.}},
  \bibinfo{author}{\bibfnamefont{Y.-F.} \bibnamefont{Liu}},
  \bibinfo{author}{\bibfnamefont{C.}~\bibnamefont{Hu}},
  \bibinfo{author}{\bibfnamefont{J.-X.} \bibnamefont{Qiu}},
  \bibinfo{author}{\bibfnamefont{C.}~\bibnamefont{Tzschaschel}},
  \bibinfo{author}{\bibfnamefont{B.}~\bibnamefont{Ghosh}},
  \bibinfo{author}{\bibfnamefont{S.-C.} \bibnamefont{Ho}},
  \bibinfo{author}{\bibfnamefont{D.}~\bibnamefont{B{\'e}rub{\'e}}},
  \bibinfo{author}{\bibfnamefont{R.}~\bibnamefont{Chen}},
  \bibinfo{author}{\bibfnamefont{H.}~\bibnamefont{Sun}}, \emph{et~al.},
  \bibinfo{year}{2021}, \bibinfo{journal}{Nature}
  \textbf{\bibinfo{volume}{595}}(\bibinfo{number}{7868}), \bibinfo{pages}{521}.

\bibitem[{\citenamefont{Garrity and Vanderbilt}(2013)}]{garrity2013chern}
\bibinfo{author}{\bibnamefont{Garrity}, \bibfnamefont{K.~F.}}, and
  \bibinfo{author}{\bibfnamefont{D.}~\bibnamefont{Vanderbilt}},
  \bibinfo{year}{2013}, \bibinfo{journal}{Physical review letters}
  \textbf{\bibinfo{volume}{110}}(\bibinfo{number}{11}),
  \bibinfo{pages}{116802}.

\bibitem[{\citenamefont{Garrity and Vanderbilt}(2014)}]{garrity2014chern}
\bibinfo{author}{\bibnamefont{Garrity}, \bibfnamefont{K.~F.}}, and
  \bibinfo{author}{\bibfnamefont{D.}~\bibnamefont{Vanderbilt}},
  \bibinfo{year}{2014}, \bibinfo{journal}{Physical Review B}
  \textbf{\bibinfo{volume}{90}}(\bibinfo{number}{12}), \bibinfo{pages}{121103}.

\bibitem[{\citenamefont{Ge} \emph{et~al.}(2020)\citenamefont{Ge, Liu, Li, Li,
  Luo, Wu, Xu, and Wang}}]{ge2020high}
\bibinfo{author}{\bibnamefont{Ge}, \bibfnamefont{J.}},
  \bibinfo{author}{\bibfnamefont{Y.}~\bibnamefont{Liu}},
  \bibinfo{author}{\bibfnamefont{J.}~\bibnamefont{Li}},
  \bibinfo{author}{\bibfnamefont{H.}~\bibnamefont{Li}},
  \bibinfo{author}{\bibfnamefont{T.}~\bibnamefont{Luo}},
  \bibinfo{author}{\bibfnamefont{Y.}~\bibnamefont{Wu}},
  \bibinfo{author}{\bibfnamefont{Y.}~\bibnamefont{Xu}}, and
  \bibinfo{author}{\bibfnamefont{J.}~\bibnamefont{Wang}}, \bibinfo{year}{2020},
  \bibinfo{journal}{National Science Review}
  \textbf{\bibinfo{volume}{7}}(\bibinfo{number}{8}), \bibinfo{pages}{1280}.

\bibitem[{\citenamefont{Ge} \emph{et~al.}(2021)\citenamefont{Ge, Sass, Yan,
  Lee, Mao, and Wu}}]{ge2021direct}
\bibinfo{author}{\bibnamefont{Ge}, \bibfnamefont{W.}},
  \bibinfo{author}{\bibfnamefont{P.~M.} \bibnamefont{Sass}},
  \bibinfo{author}{\bibfnamefont{J.}~\bibnamefont{Yan}},
  \bibinfo{author}{\bibfnamefont{S.~H.} \bibnamefont{Lee}},
  \bibinfo{author}{\bibfnamefont{Z.}~\bibnamefont{Mao}}, and
  \bibinfo{author}{\bibfnamefont{W.}~\bibnamefont{Wu}}, \bibinfo{year}{2021},
  \bibinfo{journal}{Physical Review B}
  \textbf{\bibinfo{volume}{103}}(\bibinfo{number}{13}),
  \bibinfo{pages}{134403}.

\bibitem[{\citenamefont{Giovannetti}
  \emph{et~al.}(2007)\citenamefont{Giovannetti, Khomyakov, Brocks, Kelly, and
  Van Den~Brink}}]{giovannetti2007substrate}
\bibinfo{author}{\bibnamefont{Giovannetti}, \bibfnamefont{G.}},
  \bibinfo{author}{\bibfnamefont{P.~A.} \bibnamefont{Khomyakov}},
  \bibinfo{author}{\bibfnamefont{G.}~\bibnamefont{Brocks}},
  \bibinfo{author}{\bibfnamefont{P.~J.} \bibnamefont{Kelly}}, and
  \bibinfo{author}{\bibfnamefont{J.}~\bibnamefont{Van Den~Brink}},
  \bibinfo{year}{2007}, \bibinfo{journal}{Physical Review B}
  \textbf{\bibinfo{volume}{76}}(\bibinfo{number}{7}), \bibinfo{pages}{073103}.

\bibitem[{\citenamefont{Girvin and Prange}(1987)}]{girvin1987quantum}
\bibinfo{author}{\bibnamefont{Girvin}, \bibfnamefont{S.}}, and
  \bibinfo{author}{\bibfnamefont{R.}~\bibnamefont{Prange}},
  \bibinfo{year}{1987}.

\bibitem[{\citenamefont{Gong} \emph{et~al.}(2019)\citenamefont{Gong, Guo, Li,
  Zhu, Liao, Liu, Zhang, Gu, Tang, Feng} \emph{et~al.}}]{gong2019experimental}
\bibinfo{author}{\bibnamefont{Gong}, \bibfnamefont{Y.}},
  \bibinfo{author}{\bibfnamefont{J.}~\bibnamefont{Guo}},
  \bibinfo{author}{\bibfnamefont{J.}~\bibnamefont{Li}},
  \bibinfo{author}{\bibfnamefont{K.}~\bibnamefont{Zhu}},
  \bibinfo{author}{\bibfnamefont{M.}~\bibnamefont{Liao}},
  \bibinfo{author}{\bibfnamefont{X.}~\bibnamefont{Liu}},
  \bibinfo{author}{\bibfnamefont{Q.}~\bibnamefont{Zhang}},
  \bibinfo{author}{\bibfnamefont{L.}~\bibnamefont{Gu}},
  \bibinfo{author}{\bibfnamefont{L.}~\bibnamefont{Tang}},
  \bibinfo{author}{\bibfnamefont{X.}~\bibnamefont{Feng}}, \emph{et~al.},
  \bibinfo{year}{2019}, \bibinfo{journal}{Chinese Physics Letters}
  \textbf{\bibinfo{volume}{36}}(\bibinfo{number}{7}), \bibinfo{pages}{076801}.

\bibitem[{\citenamefont{Goodenough}(1955)}]{goodenough1955theory}
\bibinfo{author}{\bibnamefont{Goodenough}, \bibfnamefont{J.~B.}},
  \bibinfo{year}{1955}, \bibinfo{journal}{Physical Review}
  \textbf{\bibinfo{volume}{100}}(\bibinfo{number}{2}), \bibinfo{pages}{564}.

\bibitem[{\citenamefont{Gordon} \emph{et~al.}(2019)\citenamefont{Gordon, Sun,
  Hu, Linn, Li, Liu, Liu, Mackey, Liu, Ni} \emph{et~al.}}]{gordon2019strongly}
\bibinfo{author}{\bibnamefont{Gordon}, \bibfnamefont{K.~N.}},
  \bibinfo{author}{\bibfnamefont{H.}~\bibnamefont{Sun}},
  \bibinfo{author}{\bibfnamefont{C.}~\bibnamefont{Hu}},
  \bibinfo{author}{\bibfnamefont{A.~G.} \bibnamefont{Linn}},
  \bibinfo{author}{\bibfnamefont{H.}~\bibnamefont{Li}},
  \bibinfo{author}{\bibfnamefont{Y.}~\bibnamefont{Liu}},
  \bibinfo{author}{\bibfnamefont{P.}~\bibnamefont{Liu}},
  \bibinfo{author}{\bibfnamefont{S.}~\bibnamefont{Mackey}},
  \bibinfo{author}{\bibfnamefont{Q.}~\bibnamefont{Liu}},
  \bibinfo{author}{\bibfnamefont{N.}~\bibnamefont{Ni}}, \emph{et~al.},
  \bibinfo{year}{2019}, \bibinfo{journal}{arXiv:1910.13943} .

\bibitem[{\citenamefont{G{\"o}tz} \emph{et~al.}(2018)\citenamefont{G{\"o}tz,
  Fijalkowski, Pesel, Hartl, Schreyeck, Winnerlein, Grauer, Scherer, Brunner,
  Gould} \emph{et~al.}}]{gotz2018precision}
\bibinfo{author}{\bibnamefont{G{\"o}tz}, \bibfnamefont{M.}},
  \bibinfo{author}{\bibfnamefont{K.~M.} \bibnamefont{Fijalkowski}},
  \bibinfo{author}{\bibfnamefont{E.}~\bibnamefont{Pesel}},
  \bibinfo{author}{\bibfnamefont{M.}~\bibnamefont{Hartl}},
  \bibinfo{author}{\bibfnamefont{S.}~\bibnamefont{Schreyeck}},
  \bibinfo{author}{\bibfnamefont{M.}~\bibnamefont{Winnerlein}},
  \bibinfo{author}{\bibfnamefont{S.}~\bibnamefont{Grauer}},
  \bibinfo{author}{\bibfnamefont{H.}~\bibnamefont{Scherer}},
  \bibinfo{author}{\bibfnamefont{K.}~\bibnamefont{Brunner}},
  \bibinfo{author}{\bibfnamefont{C.}~\bibnamefont{Gould}}, \emph{et~al.},
  \bibinfo{year}{2018}, \bibinfo{journal}{Applied Physics Letters}
  \textbf{\bibinfo{volume}{112}}(\bibinfo{number}{7}), \bibinfo{pages}{072102}.

\bibitem[{\citenamefont{Grauer} \emph{et~al.}(2017)\citenamefont{Grauer,
  Fijalkowski, Schreyeck, Winnerlein, Brunner, Thomale, Gould, and
  Molenkamp}}]{grauer2017scaling}
\bibinfo{author}{\bibnamefont{Grauer}, \bibfnamefont{S.}},
  \bibinfo{author}{\bibfnamefont{K.}~\bibnamefont{Fijalkowski}},
  \bibinfo{author}{\bibfnamefont{S.}~\bibnamefont{Schreyeck}},
  \bibinfo{author}{\bibfnamefont{M.}~\bibnamefont{Winnerlein}},
  \bibinfo{author}{\bibfnamefont{K.}~\bibnamefont{Brunner}},
  \bibinfo{author}{\bibfnamefont{R.}~\bibnamefont{Thomale}},
  \bibinfo{author}{\bibfnamefont{C.}~\bibnamefont{Gould}}, and
  \bibinfo{author}{\bibfnamefont{L.}~\bibnamefont{Molenkamp}},
  \bibinfo{year}{2017}, \bibinfo{journal}{Physical Review Letters}
  \textbf{\bibinfo{volume}{118}}(\bibinfo{number}{24}),
  \bibinfo{pages}{246801}.

\bibitem[{\citenamefont{Grauer} \emph{et~al.}(2015)\citenamefont{Grauer,
  Schreyeck, Winnerlein, Brunner, Gould, and
  Molenkamp}}]{grauer2015coincidence}
\bibinfo{author}{\bibnamefont{Grauer}, \bibfnamefont{S.}},
  \bibinfo{author}{\bibfnamefont{S.}~\bibnamefont{Schreyeck}},
  \bibinfo{author}{\bibfnamefont{M.}~\bibnamefont{Winnerlein}},
  \bibinfo{author}{\bibfnamefont{K.}~\bibnamefont{Brunner}},
  \bibinfo{author}{\bibfnamefont{C.}~\bibnamefont{Gould}}, and
  \bibinfo{author}{\bibfnamefont{L.}~\bibnamefont{Molenkamp}},
  \bibinfo{year}{2015}, \bibinfo{journal}{Physical Review B}
  \textbf{\bibinfo{volume}{92}}(\bibinfo{number}{20}), \bibinfo{pages}{201304}.

\bibitem[{\citenamefont{Haim} \emph{et~al.}(2019)\citenamefont{Haim, Ilan, and
  Alicea}}]{haim2019quantum}
\bibinfo{author}{\bibnamefont{Haim}, \bibfnamefont{A.}},
  \bibinfo{author}{\bibfnamefont{R.}~\bibnamefont{Ilan}}, and
  \bibinfo{author}{\bibfnamefont{J.}~\bibnamefont{Alicea}},
  \bibinfo{year}{2019}, \bibinfo{journal}{Physical Review Letters}
  \textbf{\bibinfo{volume}{123}}(\bibinfo{number}{4}), \bibinfo{pages}{046801}.

\bibitem[{\citenamefont{Haldane}(1988)}]{haldane1988model}
\bibinfo{author}{\bibnamefont{Haldane}, \bibfnamefont{F.~D.~M.}},
  \bibinfo{year}{1988}, \bibinfo{journal}{Physical Review Letters}
  \textbf{\bibinfo{volume}{61}}(\bibinfo{number}{18}), \bibinfo{pages}{2015}.

\bibitem[{\citenamefont{Hall}(1880)}]{hall1880new}
\bibinfo{author}{\bibnamefont{Hall}, \bibfnamefont{E.~H.}},
  \bibinfo{year}{1880}, \bibinfo{journal}{American Journal of Science}
  \textbf{\bibinfo{volume}{3}}(\bibinfo{number}{117}), \bibinfo{pages}{161}.

\bibitem[{\citenamefont{Halperin}(1987)}]{halperin1987possible}
\bibinfo{author}{\bibnamefont{Halperin}, \bibfnamefont{B.~I.}},
  \bibinfo{year}{1987}, \bibinfo{journal}{Japanese Journal of Applied Physics}
  \textbf{\bibinfo{volume}{26}}(\bibinfo{number}{S3-3}), \bibinfo{pages}{1913}.

\bibitem[{\citenamefont{Hao} \emph{et~al.}(2019)\citenamefont{Hao, Liu, Feng,
  Ma, Schwier, Arita, Kumar, Hu, Zeng, Wang} \emph{et~al.}}]{hao2019gapless}
\bibinfo{author}{\bibnamefont{Hao}, \bibfnamefont{Y.-J.}},
  \bibinfo{author}{\bibfnamefont{P.}~\bibnamefont{Liu}},
  \bibinfo{author}{\bibfnamefont{Y.}~\bibnamefont{Feng}},
  \bibinfo{author}{\bibfnamefont{X.-M.} \bibnamefont{Ma}},
  \bibinfo{author}{\bibfnamefont{E.~F.} \bibnamefont{Schwier}},
  \bibinfo{author}{\bibfnamefont{M.}~\bibnamefont{Arita}},
  \bibinfo{author}{\bibfnamefont{S.}~\bibnamefont{Kumar}},
  \bibinfo{author}{\bibfnamefont{C.}~\bibnamefont{Hu}},
  \bibinfo{author}{\bibfnamefont{M.}~\bibnamefont{Zeng}},
  \bibinfo{author}{\bibfnamefont{Y.}~\bibnamefont{Wang}}, \emph{et~al.},
  \bibinfo{year}{2019}, \bibinfo{journal}{Physical Review X}
  \textbf{\bibinfo{volume}{9}}(\bibinfo{number}{4}), \bibinfo{pages}{041038}.

\bibitem[{\citenamefont{Hasan and Kane}(2010)}]{hasan2010colloquium}
\bibinfo{author}{\bibnamefont{Hasan}, \bibfnamefont{M.~Z.}}, and
  \bibinfo{author}{\bibfnamefont{C.~L.} \bibnamefont{Kane}},
  \bibinfo{year}{2010}, \bibinfo{journal}{Reviews of Modern Physics}
  \textbf{\bibinfo{volume}{82}}(\bibinfo{number}{4}), \bibinfo{pages}{3045}.

\bibitem[{\citenamefont{He} \emph{et~al.}(2018)\citenamefont{He, Wang, and
  Xue}}]{he2018topological}
\bibinfo{author}{\bibnamefont{He}, \bibfnamefont{K.}},
  \bibinfo{author}{\bibfnamefont{Y.}~\bibnamefont{Wang}}, and
  \bibinfo{author}{\bibfnamefont{Q.-K.} \bibnamefont{Xue}},
  \bibinfo{year}{2018}, \bibinfo{journal}{Annual Review of Condensed Matter
  Physics} \textbf{\bibinfo{volume}{9}}, \bibinfo{pages}{329}.

\bibitem[{\citenamefont{He} \emph{et~al.}(2017)\citenamefont{He, Pan, Stern,
  Burks, Che, Yin, Wang, Lian, Zhou, Choi} \emph{et~al.}}]{he2017chiral}
\bibinfo{author}{\bibnamefont{He}, \bibfnamefont{Q.~L.}},
  \bibinfo{author}{\bibfnamefont{L.}~\bibnamefont{Pan}},
  \bibinfo{author}{\bibfnamefont{A.~L.} \bibnamefont{Stern}},
  \bibinfo{author}{\bibfnamefont{E.~C.} \bibnamefont{Burks}},
  \bibinfo{author}{\bibfnamefont{X.}~\bibnamefont{Che}},
  \bibinfo{author}{\bibfnamefont{G.}~\bibnamefont{Yin}},
  \bibinfo{author}{\bibfnamefont{J.}~\bibnamefont{Wang}},
  \bibinfo{author}{\bibfnamefont{B.}~\bibnamefont{Lian}},
  \bibinfo{author}{\bibfnamefont{Q.}~\bibnamefont{Zhou}},
  \bibinfo{author}{\bibfnamefont{E.~S.} \bibnamefont{Choi}}, \emph{et~al.},
  \bibinfo{year}{2017}, \bibinfo{journal}{Science}
  \textbf{\bibinfo{volume}{357}}(\bibinfo{number}{6348}), \bibinfo{pages}{294}.

\bibitem[{\citenamefont{He} \emph{et~al.}(2020)\citenamefont{He,
  Goldhaber-Gordon, and Law}}]{he2020giant}
\bibinfo{author}{\bibnamefont{He}, \bibfnamefont{W.-Y.}},
  \bibinfo{author}{\bibfnamefont{D.}~\bibnamefont{Goldhaber-Gordon}}, and
  \bibinfo{author}{\bibfnamefont{K.~T.} \bibnamefont{Law}},
  \bibinfo{year}{2020}, \bibinfo{journal}{Nature Communications}
  \textbf{\bibinfo{volume}{11}}(\bibinfo{number}{1}), \bibinfo{pages}{1}.

\bibitem[{\citenamefont{Hofstadter}(1976)}]{hofstadter1976energy}
\bibinfo{author}{\bibnamefont{Hofstadter}, \bibfnamefont{D.~R.}},
  \bibinfo{year}{1976}, \bibinfo{journal}{Physical Review B}
  \textbf{\bibinfo{volume}{14}}(\bibinfo{number}{6}), \bibinfo{pages}{2239}.

\bibitem[{\citenamefont{H{\"o}gl} \emph{et~al.}(2020)\citenamefont{H{\"o}gl,
  Frank, Zollner, Kochan, Gmitra, and Fabian}}]{hogl2020quantum}
\bibinfo{author}{\bibnamefont{H{\"o}gl}, \bibfnamefont{P.}},
  \bibinfo{author}{\bibfnamefont{T.}~\bibnamefont{Frank}},
  \bibinfo{author}{\bibfnamefont{K.}~\bibnamefont{Zollner}},
  \bibinfo{author}{\bibfnamefont{D.}~\bibnamefont{Kochan}},
  \bibinfo{author}{\bibfnamefont{M.}~\bibnamefont{Gmitra}}, and
  \bibinfo{author}{\bibfnamefont{J.}~\bibnamefont{Fabian}},
  \bibinfo{year}{2020}, \bibinfo{journal}{Physical Review Letters}
  \textbf{\bibinfo{volume}{124}}(\bibinfo{number}{13}),
  \bibinfo{pages}{136403}.

\bibitem[{\citenamefont{Hor} \emph{et~al.}(2010)\citenamefont{Hor, Roushan,
  Beidenkopf, Seo, Qu, Checkelsky, Wray, Hsieh, Xia, Xu}
  \emph{et~al.}}]{hor2010development}
\bibinfo{author}{\bibnamefont{Hor}, \bibfnamefont{Y.~S.}},
  \bibinfo{author}{\bibfnamefont{P.}~\bibnamefont{Roushan}},
  \bibinfo{author}{\bibfnamefont{H.}~\bibnamefont{Beidenkopf}},
  \bibinfo{author}{\bibfnamefont{J.}~\bibnamefont{Seo}},
  \bibinfo{author}{\bibfnamefont{D.}~\bibnamefont{Qu}},
  \bibinfo{author}{\bibfnamefont{J.~G.} \bibnamefont{Checkelsky}},
  \bibinfo{author}{\bibfnamefont{L.~A.} \bibnamefont{Wray}},
  \bibinfo{author}{\bibfnamefont{D.}~\bibnamefont{Hsieh}},
  \bibinfo{author}{\bibfnamefont{Y.}~\bibnamefont{Xia}},
  \bibinfo{author}{\bibfnamefont{S.-Y.} \bibnamefont{Xu}}, \emph{et~al.},
  \bibinfo{year}{2010}, \bibinfo{journal}{Physical Review B}
  \textbf{\bibinfo{volume}{81}}(\bibinfo{number}{19}), \bibinfo{pages}{195203}.

\bibitem[{\citenamefont{Hsu} \emph{et~al.}(2017)\citenamefont{Hsu, Fang, Wu,
  Huang, Crisostomo, Gu, Zhu, Lin, Bansil, Chuang}
  \emph{et~al.}}]{hsu2017quantum}
\bibinfo{author}{\bibnamefont{Hsu}, \bibfnamefont{C.-H.}},
  \bibinfo{author}{\bibfnamefont{Y.}~\bibnamefont{Fang}},
  \bibinfo{author}{\bibfnamefont{S.}~\bibnamefont{Wu}},
  \bibinfo{author}{\bibfnamefont{Z.-Q.} \bibnamefont{Huang}},
  \bibinfo{author}{\bibfnamefont{C.~P.} \bibnamefont{Crisostomo}},
  \bibinfo{author}{\bibfnamefont{Y.-M.} \bibnamefont{Gu}},
  \bibinfo{author}{\bibfnamefont{Z.-Z.} \bibnamefont{Zhu}},
  \bibinfo{author}{\bibfnamefont{H.}~\bibnamefont{Lin}},
  \bibinfo{author}{\bibfnamefont{A.}~\bibnamefont{Bansil}},
  \bibinfo{author}{\bibfnamefont{F.-C.} \bibnamefont{Chuang}}, \emph{et~al.},
  \bibinfo{year}{2017}, \bibinfo{journal}{Physical Review B}
  \textbf{\bibinfo{volume}{96}}(\bibinfo{number}{16}), \bibinfo{pages}{165426}.

\bibitem[{\citenamefont{Hu} \emph{et~al.}(2020{\natexlab{a}})\citenamefont{Hu,
  Ding, Gordon, Ghosh, Tien, Li, Linn, Lien, Huang, Mackey}
  \emph{et~al.}}]{hu2020realization}
\bibinfo{author}{\bibnamefont{Hu}, \bibfnamefont{C.}},
  \bibinfo{author}{\bibfnamefont{L.}~\bibnamefont{Ding}},
  \bibinfo{author}{\bibfnamefont{K.~N.} \bibnamefont{Gordon}},
  \bibinfo{author}{\bibfnamefont{B.}~\bibnamefont{Ghosh}},
  \bibinfo{author}{\bibfnamefont{H.-J.} \bibnamefont{Tien}},
  \bibinfo{author}{\bibfnamefont{H.}~\bibnamefont{Li}},
  \bibinfo{author}{\bibfnamefont{A.~G.} \bibnamefont{Linn}},
  \bibinfo{author}{\bibfnamefont{S.-W.} \bibnamefont{Lien}},
  \bibinfo{author}{\bibfnamefont{C.-Y.} \bibnamefont{Huang}},
  \bibinfo{author}{\bibfnamefont{S.}~\bibnamefont{Mackey}}, \emph{et~al.},
  \bibinfo{year}{2020}{\natexlab{a}}, \bibinfo{journal}{Science Advances}
  \textbf{\bibinfo{volume}{6}}(\bibinfo{number}{30}),
  \bibinfo{pages}{eaba4275}.

\bibitem[{\citenamefont{Hu} \emph{et~al.}(2020{\natexlab{b}})\citenamefont{Hu,
  Gordon, Liu, Liu, Zhou, Hao, Narayan, Emmanouilidou, Sun, Liu}
  \emph{et~al.}}]{hu2020van}
\bibinfo{author}{\bibnamefont{Hu}, \bibfnamefont{C.}},
  \bibinfo{author}{\bibfnamefont{K.~N.} \bibnamefont{Gordon}},
  \bibinfo{author}{\bibfnamefont{P.}~\bibnamefont{Liu}},
  \bibinfo{author}{\bibfnamefont{J.}~\bibnamefont{Liu}},
  \bibinfo{author}{\bibfnamefont{X.}~\bibnamefont{Zhou}},
  \bibinfo{author}{\bibfnamefont{P.}~\bibnamefont{Hao}},
  \bibinfo{author}{\bibfnamefont{D.}~\bibnamefont{Narayan}},
  \bibinfo{author}{\bibfnamefont{E.}~\bibnamefont{Emmanouilidou}},
  \bibinfo{author}{\bibfnamefont{H.}~\bibnamefont{Sun}},
  \bibinfo{author}{\bibfnamefont{Y.}~\bibnamefont{Liu}}, \emph{et~al.},
  \bibinfo{year}{2020}{\natexlab{b}}, \bibinfo{journal}{Nature Communications}
  \textbf{\bibinfo{volume}{11}}(\bibinfo{number}{1}), \bibinfo{pages}{1}.

\bibitem[{\citenamefont{Hu} \emph{et~al.}(2020{\natexlab{c}})\citenamefont{Hu,
  Xu, Shi, Luo, Peng, Wang, Ying, Wu, Liu, Zhang}
  \emph{et~al.}}]{hu2020universal}
\bibinfo{author}{\bibnamefont{Hu}, \bibfnamefont{Y.}},
  \bibinfo{author}{\bibfnamefont{L.}~\bibnamefont{Xu}},
  \bibinfo{author}{\bibfnamefont{M.}~\bibnamefont{Shi}},
  \bibinfo{author}{\bibfnamefont{A.}~\bibnamefont{Luo}},
  \bibinfo{author}{\bibfnamefont{S.}~\bibnamefont{Peng}},
  \bibinfo{author}{\bibfnamefont{Z.}~\bibnamefont{Wang}},
  \bibinfo{author}{\bibfnamefont{J.}~\bibnamefont{Ying}},
  \bibinfo{author}{\bibfnamefont{T.}~\bibnamefont{Wu}},
  \bibinfo{author}{\bibfnamefont{Z.}~\bibnamefont{Liu}},
  \bibinfo{author}{\bibfnamefont{C.}~\bibnamefont{Zhang}}, \emph{et~al.},
  \bibinfo{year}{2020}{\natexlab{c}}, \bibinfo{journal}{Physical Review B}
  \textbf{\bibinfo{volume}{101}}(\bibinfo{number}{16}),
  \bibinfo{pages}{161113}.

\bibitem[{\citenamefont{Huan} \emph{et~al.}(2021)\citenamefont{Huan, Zhang,
  Jiang, Su, Wang, Zhang, Yang, Liu, Wang, Yu}
  \emph{et~al.}}]{huan2021multiple}
\bibinfo{author}{\bibnamefont{Huan}, \bibfnamefont{S.}},
  \bibinfo{author}{\bibfnamefont{S.}~\bibnamefont{Zhang}},
  \bibinfo{author}{\bibfnamefont{Z.}~\bibnamefont{Jiang}},
  \bibinfo{author}{\bibfnamefont{H.}~\bibnamefont{Su}},
  \bibinfo{author}{\bibfnamefont{H.}~\bibnamefont{Wang}},
  \bibinfo{author}{\bibfnamefont{X.}~\bibnamefont{Zhang}},
  \bibinfo{author}{\bibfnamefont{Y.}~\bibnamefont{Yang}},
  \bibinfo{author}{\bibfnamefont{Z.}~\bibnamefont{Liu}},
  \bibinfo{author}{\bibfnamefont{X.}~\bibnamefont{Wang}},
  \bibinfo{author}{\bibfnamefont{N.}~\bibnamefont{Yu}}, \emph{et~al.},
  \bibinfo{year}{2021}, \bibinfo{journal}{Physical Review Letters}
  \textbf{\bibinfo{volume}{126}}(\bibinfo{number}{24}),
  \bibinfo{pages}{246601}.

\bibitem[{\citenamefont{Huang} \emph{et~al.}(2017)\citenamefont{Huang, Zhou,
  Wu, Deng, Jena, and Kan}}]{huang2017quantum}
\bibinfo{author}{\bibnamefont{Huang}, \bibfnamefont{C.}},
  \bibinfo{author}{\bibfnamefont{J.}~\bibnamefont{Zhou}},
  \bibinfo{author}{\bibfnamefont{H.}~\bibnamefont{Wu}},
  \bibinfo{author}{\bibfnamefont{K.}~\bibnamefont{Deng}},
  \bibinfo{author}{\bibfnamefont{P.}~\bibnamefont{Jena}}, and
  \bibinfo{author}{\bibfnamefont{E.}~\bibnamefont{Kan}}, \bibinfo{year}{2017},
  \bibinfo{journal}{Physical Review B}
  \textbf{\bibinfo{volume}{95}}(\bibinfo{number}{4}), \bibinfo{pages}{045113}.

\bibitem[{\citenamefont{Huang} \emph{et~al.}(2018)\citenamefont{Huang,
  Setiawan, and Sau}}]{huang2018disorder}
\bibinfo{author}{\bibnamefont{Huang}, \bibfnamefont{Y.}},
  \bibinfo{author}{\bibfnamefont{F.}~\bibnamefont{Setiawan}}, and
  \bibinfo{author}{\bibfnamefont{J.~D.} \bibnamefont{Sau}},
  \bibinfo{year}{2018}, \bibinfo{journal}{Physical Review B}
  \textbf{\bibinfo{volume}{97}}(\bibinfo{number}{10}), \bibinfo{pages}{100501}.

\bibitem[{\citenamefont{Huang} \emph{et~al.}(2020)\citenamefont{Huang, Du, Yan,
  and Wu}}]{huang2020native}
\bibinfo{author}{\bibnamefont{Huang}, \bibfnamefont{Z.}},
  \bibinfo{author}{\bibfnamefont{M.-H.} \bibnamefont{Du}},
  \bibinfo{author}{\bibfnamefont{J.}~\bibnamefont{Yan}}, and
  \bibinfo{author}{\bibfnamefont{W.}~\bibnamefont{Wu}}, \bibinfo{year}{2020},
  \bibinfo{journal}{Physical Review Materials}
  \textbf{\bibinfo{volume}{4}}(\bibinfo{number}{12}), \bibinfo{pages}{121202}.

\bibitem[{\citenamefont{Huckestein}(1995)}]{huckestein1995scaling}
\bibinfo{author}{\bibnamefont{Huckestein}, \bibfnamefont{B.}},
  \bibinfo{year}{1995}, \bibinfo{journal}{Reviews of Modern Physics}
  \textbf{\bibinfo{volume}{67}}(\bibinfo{number}{2}), \bibinfo{pages}{357}.

\bibitem[{\citenamefont{Jeckelmann and
  Jeanneret}(2001)}]{jeckelmann2001quantum}
\bibinfo{author}{\bibnamefont{Jeckelmann}, \bibfnamefont{B.}}, and
  \bibinfo{author}{\bibfnamefont{B.}~\bibnamefont{Jeanneret}},
  \bibinfo{year}{2001}, \bibinfo{journal}{Reports on Progress in Physics}
  \textbf{\bibinfo{volume}{64}}(\bibinfo{number}{12}), \bibinfo{pages}{1603}.

\bibitem[{\citenamefont{Jeon} \emph{et~al.}(2017)\citenamefont{Jeon, Xie, Li,
  Wang, Bernevig, and Yazdani}}]{jeon2017distinguishing}
\bibinfo{author}{\bibnamefont{Jeon}, \bibfnamefont{S.}},
  \bibinfo{author}{\bibfnamefont{Y.}~\bibnamefont{Xie}},
  \bibinfo{author}{\bibfnamefont{J.}~\bibnamefont{Li}},
  \bibinfo{author}{\bibfnamefont{Z.}~\bibnamefont{Wang}},
  \bibinfo{author}{\bibfnamefont{B.~A.} \bibnamefont{Bernevig}}, and
  \bibinfo{author}{\bibfnamefont{A.}~\bibnamefont{Yazdani}},
  \bibinfo{year}{2017}, \bibinfo{journal}{Science}
  \textbf{\bibinfo{volume}{358}}(\bibinfo{number}{6364}), \bibinfo{pages}{772}.

\bibitem[{\citenamefont{Ji and Wen}(2018)}]{ji20181}
\bibinfo{author}{\bibnamefont{Ji}, \bibfnamefont{W.}}, and
  \bibinfo{author}{\bibfnamefont{X.-G.} \bibnamefont{Wen}},
  \bibinfo{year}{2018}, \bibinfo{journal}{Physical Review Letters}
  \textbf{\bibinfo{volume}{120}}(\bibinfo{number}{10}),
  \bibinfo{pages}{107002}.

\bibitem[{\citenamefont{Ji} \emph{et~al.}(2016)\citenamefont{Ji, Zhang, Ding,
  Zhang, Li, Li, Ren, Wang, Zhang, Hu} \emph{et~al.}}]{ji2016giant}
\bibinfo{author}{\bibnamefont{Ji}, \bibfnamefont{W.-x.}},
  \bibinfo{author}{\bibfnamefont{C.-w.} \bibnamefont{Zhang}},
  \bibinfo{author}{\bibfnamefont{M.}~\bibnamefont{Ding}},
  \bibinfo{author}{\bibfnamefont{B.-m.} \bibnamefont{Zhang}},
  \bibinfo{author}{\bibfnamefont{P.}~\bibnamefont{Li}},
  \bibinfo{author}{\bibfnamefont{F.}~\bibnamefont{Li}},
  \bibinfo{author}{\bibfnamefont{M.-j.} \bibnamefont{Ren}},
  \bibinfo{author}{\bibfnamefont{P.-j.} \bibnamefont{Wang}},
  \bibinfo{author}{\bibfnamefont{R.-w.} \bibnamefont{Zhang}},
  \bibinfo{author}{\bibfnamefont{S.-j.} \bibnamefont{Hu}}, \emph{et~al.},
  \bibinfo{year}{2016}, \bibinfo{journal}{New Journal of Physics}
  \textbf{\bibinfo{volume}{18}}(\bibinfo{number}{8}), \bibinfo{pages}{083002}.

\bibitem[{\citenamefont{Jiang} \emph{et~al.}(2018)\citenamefont{Jiang, Feng,
  Wu, Li, Bai, Li, Zhang, Gu, Feng, Zhang} \emph{et~al.}}]{jiang2018quantum}
\bibinfo{author}{\bibnamefont{Jiang}, \bibfnamefont{G.}},
  \bibinfo{author}{\bibfnamefont{Y.}~\bibnamefont{Feng}},
  \bibinfo{author}{\bibfnamefont{W.}~\bibnamefont{Wu}},
  \bibinfo{author}{\bibfnamefont{S.}~\bibnamefont{Li}},
  \bibinfo{author}{\bibfnamefont{Y.}~\bibnamefont{Bai}},
  \bibinfo{author}{\bibfnamefont{Y.}~\bibnamefont{Li}},
  \bibinfo{author}{\bibfnamefont{Q.}~\bibnamefont{Zhang}},
  \bibinfo{author}{\bibfnamefont{L.}~\bibnamefont{Gu}},
  \bibinfo{author}{\bibfnamefont{X.}~\bibnamefont{Feng}},
  \bibinfo{author}{\bibfnamefont{D.}~\bibnamefont{Zhang}}, \emph{et~al.},
  \bibinfo{year}{2018}, \bibinfo{journal}{Chinese Physics Letters}
  \textbf{\bibinfo{volume}{35}}(\bibinfo{number}{7}), \bibinfo{pages}{076802}.

\bibitem[{\citenamefont{Jiang}
  \emph{et~al.}(2012{\natexlab{a}})\citenamefont{Jiang, Qiao, Liu, and
  Niu}}]{jiang2012quantum}
\bibinfo{author}{\bibnamefont{Jiang}, \bibfnamefont{H.}},
  \bibinfo{author}{\bibfnamefont{Z.}~\bibnamefont{Qiao}},
  \bibinfo{author}{\bibfnamefont{H.}~\bibnamefont{Liu}}, and
  \bibinfo{author}{\bibfnamefont{Q.}~\bibnamefont{Niu}},
  \bibinfo{year}{2012}{\natexlab{a}}, \bibinfo{journal}{Physical Review B}
  \textbf{\bibinfo{volume}{85}}(\bibinfo{number}{4}), \bibinfo{pages}{045445}.

\bibitem[{\citenamefont{Jiang} \emph{et~al.}(2020)\citenamefont{Jiang, Xiao,
  Wang, Shin, Andreoli, Zhang, Xiao, Zhao, Kayyalha, Zhang}
  \emph{et~al.}}]{jiang2020concurrence}
\bibinfo{author}{\bibnamefont{Jiang}, \bibfnamefont{J.}},
  \bibinfo{author}{\bibfnamefont{D.}~\bibnamefont{Xiao}},
  \bibinfo{author}{\bibfnamefont{F.}~\bibnamefont{Wang}},
  \bibinfo{author}{\bibfnamefont{J.-H.} \bibnamefont{Shin}},
  \bibinfo{author}{\bibfnamefont{D.}~\bibnamefont{Andreoli}},
  \bibinfo{author}{\bibfnamefont{J.}~\bibnamefont{Zhang}},
  \bibinfo{author}{\bibfnamefont{R.}~\bibnamefont{Xiao}},
  \bibinfo{author}{\bibfnamefont{Y.-F.} \bibnamefont{Zhao}},
  \bibinfo{author}{\bibfnamefont{M.}~\bibnamefont{Kayyalha}},
  \bibinfo{author}{\bibfnamefont{L.}~\bibnamefont{Zhang}}, \emph{et~al.},
  \bibinfo{year}{2020}, \bibinfo{journal}{Nature Materials}
  \textbf{\bibinfo{volume}{19}}(\bibinfo{number}{7}), \bibinfo{pages}{732}.

\bibitem[{\citenamefont{Jiang} \emph{et~al.}(2021)\citenamefont{Jiang, Wang,
  Malinowski, Liu, Shi, Lin, Fei, Song, Graf, Chikara}
  \emph{et~al.}}]{jiang2021quantum}
\bibinfo{author}{\bibnamefont{Jiang}, \bibfnamefont{Q.}},
  \bibinfo{author}{\bibfnamefont{C.}~\bibnamefont{Wang}},
  \bibinfo{author}{\bibfnamefont{P.}~\bibnamefont{Malinowski}},
  \bibinfo{author}{\bibfnamefont{Z.}~\bibnamefont{Liu}},
  \bibinfo{author}{\bibfnamefont{Y.}~\bibnamefont{Shi}},
  \bibinfo{author}{\bibfnamefont{Z.}~\bibnamefont{Lin}},
  \bibinfo{author}{\bibfnamefont{Z.}~\bibnamefont{Fei}},
  \bibinfo{author}{\bibfnamefont{T.}~\bibnamefont{Song}},
  \bibinfo{author}{\bibfnamefont{D.}~\bibnamefont{Graf}},
  \bibinfo{author}{\bibfnamefont{S.}~\bibnamefont{Chikara}}, \emph{et~al.},
  \bibinfo{year}{2021}, \bibinfo{journal}{Physical Review B}
  \textbf{\bibinfo{volume}{103}}(\bibinfo{number}{20}),
  \bibinfo{pages}{205111}.

\bibitem[{\citenamefont{Jiang} \emph{et~al.}(2015)\citenamefont{Jiang, Song,
  Li, Chen, Greene, He, Wang, Chen, Ma, and Xue}}]{jiang2015mass}
\bibinfo{author}{\bibnamefont{Jiang}, \bibfnamefont{Y.}},
  \bibinfo{author}{\bibfnamefont{C.}~\bibnamefont{Song}},
  \bibinfo{author}{\bibfnamefont{Z.}~\bibnamefont{Li}},
  \bibinfo{author}{\bibfnamefont{M.}~\bibnamefont{Chen}},
  \bibinfo{author}{\bibfnamefont{R.~L.} \bibnamefont{Greene}},
  \bibinfo{author}{\bibfnamefont{K.}~\bibnamefont{He}},
  \bibinfo{author}{\bibfnamefont{L.}~\bibnamefont{Wang}},
  \bibinfo{author}{\bibfnamefont{X.}~\bibnamefont{Chen}},
  \bibinfo{author}{\bibfnamefont{X.}~\bibnamefont{Ma}}, and
  \bibinfo{author}{\bibfnamefont{Q.-K.} \bibnamefont{Xue}},
  \bibinfo{year}{2015}, \bibinfo{journal}{Physical Review B}
  \textbf{\bibinfo{volume}{92}}(\bibinfo{number}{19}), \bibinfo{pages}{195418}.

\bibitem[{\citenamefont{Jiang}
  \emph{et~al.}(2012{\natexlab{b}})\citenamefont{Jiang, Wang, Chen, Li, Song,
  He, Wang, Chen, Ma, and Xue}}]{jiang2012landau}
\bibinfo{author}{\bibnamefont{Jiang}, \bibfnamefont{Y.}},
  \bibinfo{author}{\bibfnamefont{Y.}~\bibnamefont{Wang}},
  \bibinfo{author}{\bibfnamefont{M.}~\bibnamefont{Chen}},
  \bibinfo{author}{\bibfnamefont{Z.}~\bibnamefont{Li}},
  \bibinfo{author}{\bibfnamefont{C.}~\bibnamefont{Song}},
  \bibinfo{author}{\bibfnamefont{K.}~\bibnamefont{He}},
  \bibinfo{author}{\bibfnamefont{L.}~\bibnamefont{Wang}},
  \bibinfo{author}{\bibfnamefont{X.}~\bibnamefont{Chen}},
  \bibinfo{author}{\bibfnamefont{X.}~\bibnamefont{Ma}}, and
  \bibinfo{author}{\bibfnamefont{Q.-K.} \bibnamefont{Xue}},
  \bibinfo{year}{2012}{\natexlab{b}}, \bibinfo{journal}{Physical Review
  Letters} \textbf{\bibinfo{volume}{108}}(\bibinfo{number}{1}),
  \bibinfo{pages}{016401}.

\bibitem[{\citenamefont{Jin and Jhi}(2015)}]{jin2015quantum}
\bibinfo{author}{\bibnamefont{Jin}, \bibfnamefont{K.-H.}}, and
  \bibinfo{author}{\bibfnamefont{S.-H.} \bibnamefont{Jhi}},
  \bibinfo{year}{2015}, \bibinfo{journal}{Scientific Reports}
  \textbf{\bibinfo{volume}{5}}(\bibinfo{number}{1}), \bibinfo{pages}{1}.

\bibitem[{\citenamefont{Jo} \emph{et~al.}(2020)\citenamefont{Jo, Wang, Slager,
  Yan, Wu, Lee, Schrunk, Vishwanath, and Kaminski}}]{jo2020intrinsic}
\bibinfo{author}{\bibnamefont{Jo}, \bibfnamefont{N.~H.}},
  \bibinfo{author}{\bibfnamefont{L.-L.} \bibnamefont{Wang}},
  \bibinfo{author}{\bibfnamefont{R.-J.} \bibnamefont{Slager}},
  \bibinfo{author}{\bibfnamefont{J.}~\bibnamefont{Yan}},
  \bibinfo{author}{\bibfnamefont{Y.}~\bibnamefont{Wu}},
  \bibinfo{author}{\bibfnamefont{K.}~\bibnamefont{Lee}},
  \bibinfo{author}{\bibfnamefont{B.}~\bibnamefont{Schrunk}},
  \bibinfo{author}{\bibfnamefont{A.}~\bibnamefont{Vishwanath}}, and
  \bibinfo{author}{\bibfnamefont{A.}~\bibnamefont{Kaminski}},
  \bibinfo{year}{2020}, \bibinfo{journal}{Physical Review B}
  \textbf{\bibinfo{volume}{102}}(\bibinfo{number}{4}), \bibinfo{pages}{045130}.

\bibitem[{\citenamefont{Jungwirth} \emph{et~al.}(2006)\citenamefont{Jungwirth,
  Sinova, Ma{\v{s}}ek, Ku{\v{c}}era, and MacDonald}}]{jungwirth2006theory}
\bibinfo{author}{\bibnamefont{Jungwirth}, \bibfnamefont{T.}},
  \bibinfo{author}{\bibfnamefont{J.}~\bibnamefont{Sinova}},
  \bibinfo{author}{\bibfnamefont{J.}~\bibnamefont{Ma{\v{s}}ek}},
  \bibinfo{author}{\bibfnamefont{J.}~\bibnamefont{Ku{\v{c}}era}}, and
  \bibinfo{author}{\bibfnamefont{A.}~\bibnamefont{MacDonald}},
  \bibinfo{year}{2006}, \bibinfo{journal}{Reviews of Modern Physics}
  \textbf{\bibinfo{volume}{78}}(\bibinfo{number}{3}), \bibinfo{pages}{809}.

\bibitem[{\citenamefont{Kagerer} \emph{et~al.}(2020)\citenamefont{Kagerer,
  Fornari, Buchberger, Morelh{\~a}o, Vidal, Tcakaev, Zabolotnyy, Weschke,
  Hinkov, Kamp} \emph{et~al.}}]{kagerer2020molecular}
\bibinfo{author}{\bibnamefont{Kagerer}, \bibfnamefont{P.}},
  \bibinfo{author}{\bibfnamefont{C.}~\bibnamefont{Fornari}},
  \bibinfo{author}{\bibfnamefont{S.}~\bibnamefont{Buchberger}},
  \bibinfo{author}{\bibfnamefont{S.}~\bibnamefont{Morelh{\~a}o}},
  \bibinfo{author}{\bibfnamefont{R.}~\bibnamefont{Vidal}},
  \bibinfo{author}{\bibfnamefont{A.}~\bibnamefont{Tcakaev}},
  \bibinfo{author}{\bibfnamefont{V.}~\bibnamefont{Zabolotnyy}},
  \bibinfo{author}{\bibfnamefont{E.}~\bibnamefont{Weschke}},
  \bibinfo{author}{\bibfnamefont{V.}~\bibnamefont{Hinkov}},
  \bibinfo{author}{\bibfnamefont{M.}~\bibnamefont{Kamp}}, \emph{et~al.},
  \bibinfo{year}{2020}, \bibinfo{journal}{Journal of Applied Physics}
  \textbf{\bibinfo{volume}{128}}(\bibinfo{number}{13}),
  \bibinfo{pages}{135303}.

\bibitem[{\citenamefont{Kaloni} \emph{et~al.}(2014)\citenamefont{Kaloni, Singh,
  and Schwingenschl{\"o}gl}}]{kaloni2014prediction}
\bibinfo{author}{\bibnamefont{Kaloni}, \bibfnamefont{T.~P.}},
  \bibinfo{author}{\bibfnamefont{N.}~\bibnamefont{Singh}}, and
  \bibinfo{author}{\bibfnamefont{U.}~\bibnamefont{Schwingenschl{\"o}gl}},
  \bibinfo{year}{2014}, \bibinfo{journal}{Physical Review B}
  \textbf{\bibinfo{volume}{89}}(\bibinfo{number}{3}), \bibinfo{pages}{035409}.

\bibitem[{\citenamefont{Kanamori}(1959)}]{kanamori1959superexchange}
\bibinfo{author}{\bibnamefont{Kanamori}, \bibfnamefont{J.}},
  \bibinfo{year}{1959}, \bibinfo{journal}{Journal of Physics and Chemistry of
  Solids} \textbf{\bibinfo{volume}{10}}(\bibinfo{number}{2-3}),
  \bibinfo{pages}{87}.

\bibitem[{\citenamefont{Kandala} \emph{et~al.}(2015)\citenamefont{Kandala,
  Richardella, Kempinger, Liu, and Samarth}}]{kandala2015giant}
\bibinfo{author}{\bibnamefont{Kandala}, \bibfnamefont{A.}},
  \bibinfo{author}{\bibfnamefont{A.}~\bibnamefont{Richardella}},
  \bibinfo{author}{\bibfnamefont{S.}~\bibnamefont{Kempinger}},
  \bibinfo{author}{\bibfnamefont{C.-X.} \bibnamefont{Liu}}, and
  \bibinfo{author}{\bibfnamefont{N.}~\bibnamefont{Samarth}},
  \bibinfo{year}{2015}, \bibinfo{journal}{Nature Communications}
  \textbf{\bibinfo{volume}{6}}(\bibinfo{number}{1}), \bibinfo{pages}{1}.

\bibitem[{\citenamefont{Kane and Mele}(2005)}]{kane2005quantum}
\bibinfo{author}{\bibnamefont{Kane}, \bibfnamefont{C.~L.}}, and
  \bibinfo{author}{\bibfnamefont{E.~J.} \bibnamefont{Mele}},
  \bibinfo{year}{2005}, \bibinfo{journal}{Physical Review Letters}
  \textbf{\bibinfo{volume}{95}}(\bibinfo{number}{22}), \bibinfo{pages}{226801}.

\bibitem[{\citenamefont{Kang and Vafek}(2019)}]{kang2019strong}
\bibinfo{author}{\bibnamefont{Kang}, \bibfnamefont{J.}}, and
  \bibinfo{author}{\bibfnamefont{O.}~\bibnamefont{Vafek}},
  \bibinfo{year}{2019}, \bibinfo{journal}{Physical Review Letters}
  \textbf{\bibinfo{volume}{122}}(\bibinfo{number}{24}),
  \bibinfo{pages}{246401}.

\bibitem[{\citenamefont{Kasuya}(1956)}]{kasuya1956theory}
\bibinfo{author}{\bibnamefont{Kasuya}, \bibfnamefont{T.}},
  \bibinfo{year}{1956}, \bibinfo{journal}{Progress of Theoretical Physics}
  \textbf{\bibinfo{volume}{16}}(\bibinfo{number}{1}), \bibinfo{pages}{45}.

\bibitem[{\citenamefont{Kawaji}(1996)}]{kawaji1996breakdown}
\bibinfo{author}{\bibnamefont{Kawaji}, \bibfnamefont{S.}},
  \bibinfo{year}{1996}, \bibinfo{journal}{Semiconductor Science and Technology}
  \textbf{\bibinfo{volume}{11}}(\bibinfo{number}{11S}), \bibinfo{pages}{1546}.

\bibitem[{\citenamefont{Kawamura} \emph{et~al.}(2018)\citenamefont{Kawamura,
  Mogi, Yoshimi, Tsukazaki, Kozuka, Takahashi, Kawasaki, and
  Tokura}}]{kawamura2018topological}
\bibinfo{author}{\bibnamefont{Kawamura}, \bibfnamefont{M.}},
  \bibinfo{author}{\bibfnamefont{M.}~\bibnamefont{Mogi}},
  \bibinfo{author}{\bibfnamefont{R.}~\bibnamefont{Yoshimi}},
  \bibinfo{author}{\bibfnamefont{A.}~\bibnamefont{Tsukazaki}},
  \bibinfo{author}{\bibfnamefont{Y.}~\bibnamefont{Kozuka}},
  \bibinfo{author}{\bibfnamefont{K.~S.} \bibnamefont{Takahashi}},
  \bibinfo{author}{\bibfnamefont{M.}~\bibnamefont{Kawasaki}}, and
  \bibinfo{author}{\bibfnamefont{Y.}~\bibnamefont{Tokura}},
  \bibinfo{year}{2018}, \bibinfo{journal}{Physical Review B}
  \textbf{\bibinfo{volume}{98}}(\bibinfo{number}{14}), \bibinfo{pages}{140404}.

\bibitem[{\citenamefont{Kawamura} \emph{et~al.}(2020)\citenamefont{Kawamura,
  Mogi, Yoshimi, Tsukazaki, Kozuka, Takahashi, Kawasaki, and
  Tokura}}]{kawamura2020current}
\bibinfo{author}{\bibnamefont{Kawamura}, \bibfnamefont{M.}},
  \bibinfo{author}{\bibfnamefont{M.}~\bibnamefont{Mogi}},
  \bibinfo{author}{\bibfnamefont{R.}~\bibnamefont{Yoshimi}},
  \bibinfo{author}{\bibfnamefont{A.}~\bibnamefont{Tsukazaki}},
  \bibinfo{author}{\bibfnamefont{Y.}~\bibnamefont{Kozuka}},
  \bibinfo{author}{\bibfnamefont{K.~S.} \bibnamefont{Takahashi}},
  \bibinfo{author}{\bibfnamefont{M.}~\bibnamefont{Kawasaki}}, and
  \bibinfo{author}{\bibfnamefont{Y.}~\bibnamefont{Tokura}},
  \bibinfo{year}{2020}, \bibinfo{journal}{Physical Review B}
  \textbf{\bibinfo{volume}{102}}(\bibinfo{number}{4}), \bibinfo{pages}{041301}.

\bibitem[{\citenamefont{Kawamura} \emph{et~al.}(2017)\citenamefont{Kawamura,
  Yoshimi, Tsukazaki, Takahashi, Kawasaki, and Tokura}}]{kawamura2017current}
\bibinfo{author}{\bibnamefont{Kawamura}, \bibfnamefont{M.}},
  \bibinfo{author}{\bibfnamefont{R.}~\bibnamefont{Yoshimi}},
  \bibinfo{author}{\bibfnamefont{A.}~\bibnamefont{Tsukazaki}},
  \bibinfo{author}{\bibfnamefont{K.~S.} \bibnamefont{Takahashi}},
  \bibinfo{author}{\bibfnamefont{M.}~\bibnamefont{Kawasaki}}, and
  \bibinfo{author}{\bibfnamefont{Y.}~\bibnamefont{Tokura}},
  \bibinfo{year}{2017}, \bibinfo{journal}{Physical Review Letters}
  \textbf{\bibinfo{volume}{119}}(\bibinfo{number}{1}), \bibinfo{pages}{016803}.

\bibitem[{\citenamefont{Kayyalha} \emph{et~al.}(2020)\citenamefont{Kayyalha,
  Xiao, Zhang, Shin, Jiang, Wang, Zhao, Xiao, Zhang, Fijalkowski}
  \emph{et~al.}}]{kayyalha2020absence}
\bibinfo{author}{\bibnamefont{Kayyalha}, \bibfnamefont{M.}},
  \bibinfo{author}{\bibfnamefont{D.}~\bibnamefont{Xiao}},
  \bibinfo{author}{\bibfnamefont{R.}~\bibnamefont{Zhang}},
  \bibinfo{author}{\bibfnamefont{J.}~\bibnamefont{Shin}},
  \bibinfo{author}{\bibfnamefont{J.}~\bibnamefont{Jiang}},
  \bibinfo{author}{\bibfnamefont{F.}~\bibnamefont{Wang}},
  \bibinfo{author}{\bibfnamefont{Y.-F.} \bibnamefont{Zhao}},
  \bibinfo{author}{\bibfnamefont{R.}~\bibnamefont{Xiao}},
  \bibinfo{author}{\bibfnamefont{L.}~\bibnamefont{Zhang}},
  \bibinfo{author}{\bibfnamefont{K.~M.} \bibnamefont{Fijalkowski}},
  \emph{et~al.}, \bibinfo{year}{2020}, \bibinfo{journal}{Science}
  \textbf{\bibinfo{volume}{367}}(\bibinfo{number}{6473}), \bibinfo{pages}{64}.

\bibitem[{\citenamefont{Kim}
  \emph{et~al.}(2017{\natexlab{a}})\citenamefont{Kim, Jhi, MacDonald, and
  Wu}}]{kim2017ordering}
\bibinfo{author}{\bibnamefont{Kim}, \bibfnamefont{J.}},
  \bibinfo{author}{\bibfnamefont{S.-H.} \bibnamefont{Jhi}},
  \bibinfo{author}{\bibfnamefont{A.}~\bibnamefont{MacDonald}}, and
  \bibinfo{author}{\bibfnamefont{R.}~\bibnamefont{Wu}},
  \bibinfo{year}{2017}{\natexlab{a}}, \bibinfo{journal}{Physical Review B}
  \textbf{\bibinfo{volume}{96}}(\bibinfo{number}{14}), \bibinfo{pages}{140410}.

\bibitem[{\citenamefont{Kim}
  \emph{et~al.}(2017{\natexlab{b}})\citenamefont{Kim, DaSilva, Huang,
  Fallahazad, Larentis, Taniguchi, Watanabe, LeRoy, MacDonald, and
  Tutuc}}]{kim2017tunable}
\bibinfo{author}{\bibnamefont{Kim}, \bibfnamefont{K.}},
  \bibinfo{author}{\bibfnamefont{A.}~\bibnamefont{DaSilva}},
  \bibinfo{author}{\bibfnamefont{S.}~\bibnamefont{Huang}},
  \bibinfo{author}{\bibfnamefont{B.}~\bibnamefont{Fallahazad}},
  \bibinfo{author}{\bibfnamefont{S.}~\bibnamefont{Larentis}},
  \bibinfo{author}{\bibfnamefont{T.}~\bibnamefont{Taniguchi}},
  \bibinfo{author}{\bibfnamefont{K.}~\bibnamefont{Watanabe}},
  \bibinfo{author}{\bibfnamefont{B.~J.} \bibnamefont{LeRoy}},
  \bibinfo{author}{\bibfnamefont{A.~H.} \bibnamefont{MacDonald}}, and
  \bibinfo{author}{\bibfnamefont{E.}~\bibnamefont{Tutuc}},
  \bibinfo{year}{2017}{\natexlab{b}}, \bibinfo{journal}{Proceedings of the
  National Academy of Sciences}
  \textbf{\bibinfo{volume}{114}}(\bibinfo{number}{13}), \bibinfo{pages}{3364}.

\bibitem[{\citenamefont{Kim} \emph{et~al.}(2019)\citenamefont{Kim, Kurebayashi,
  and Nomura}}]{kim2019electrically}
\bibinfo{author}{\bibnamefont{Kim}, \bibfnamefont{S.}},
  \bibinfo{author}{\bibfnamefont{D.}~\bibnamefont{Kurebayashi}}, and
  \bibinfo{author}{\bibfnamefont{K.}~\bibnamefont{Nomura}},
  \bibinfo{year}{2019}, \bibinfo{journal}{Journal of the Physical Society of
  Japan} \textbf{\bibinfo{volume}{88}}(\bibinfo{number}{8}),
  \bibinfo{pages}{083704}.

\bibitem[{\citenamefont{Klimovskikh}
  \emph{et~al.}(2020)\citenamefont{Klimovskikh, Otrokov, Estyunin, Eremeev,
  Filnov, Koroleva, Shevchenko, Voroshnin, Rybkin, Rusinov}
  \emph{et~al.}}]{klimovskikh2020tunable}
\bibinfo{author}{\bibnamefont{Klimovskikh}, \bibfnamefont{I.~I.}},
  \bibinfo{author}{\bibfnamefont{M.~M.} \bibnamefont{Otrokov}},
  \bibinfo{author}{\bibfnamefont{D.}~\bibnamefont{Estyunin}},
  \bibinfo{author}{\bibfnamefont{S.~V.} \bibnamefont{Eremeev}},
  \bibinfo{author}{\bibfnamefont{S.~O.} \bibnamefont{Filnov}},
  \bibinfo{author}{\bibfnamefont{A.}~\bibnamefont{Koroleva}},
  \bibinfo{author}{\bibfnamefont{E.}~\bibnamefont{Shevchenko}},
  \bibinfo{author}{\bibfnamefont{V.}~\bibnamefont{Voroshnin}},
  \bibinfo{author}{\bibfnamefont{A.~G.} \bibnamefont{Rybkin}},
  \bibinfo{author}{\bibfnamefont{I.~P.} \bibnamefont{Rusinov}}, \emph{et~al.},
  \bibinfo{year}{2020}, \bibinfo{journal}{npj Quantum Materials}
  \textbf{\bibinfo{volume}{5}}(\bibinfo{number}{1}), \bibinfo{pages}{1}.

\bibitem[{\citenamefont{Klinovaja} \emph{et~al.}(2015)\citenamefont{Klinovaja,
  Tserkovnyak, and Loss}}]{klinovaja2015integer}
\bibinfo{author}{\bibnamefont{Klinovaja}, \bibfnamefont{J.}},
  \bibinfo{author}{\bibfnamefont{Y.}~\bibnamefont{Tserkovnyak}}, and
  \bibinfo{author}{\bibfnamefont{D.}~\bibnamefont{Loss}}, \bibinfo{year}{2015},
  \bibinfo{journal}{Physical Review B}
  \textbf{\bibinfo{volume}{91}}(\bibinfo{number}{8}), \bibinfo{pages}{085426}.

\bibitem[{\citenamefont{Klitzing} \emph{et~al.}(1980)\citenamefont{Klitzing,
  Dorda, and Pepper}}]{klitzing1980new}
\bibinfo{author}{\bibnamefont{Klitzing}, \bibfnamefont{K.~v.}},
  \bibinfo{author}{\bibfnamefont{G.}~\bibnamefont{Dorda}}, and
  \bibinfo{author}{\bibfnamefont{M.}~\bibnamefont{Pepper}},
  \bibinfo{year}{1980}, \bibinfo{journal}{Physical Review Letters}
  \textbf{\bibinfo{volume}{45}}(\bibinfo{number}{6}), \bibinfo{pages}{494}.

\bibitem[{\citenamefont{K{\"o}nig} \emph{et~al.}(2007)\citenamefont{K{\"o}nig,
  Wiedmann, Br{\"u}ne, Roth, Buhmann, Molenkamp, Qi, and
  Zhang}}]{konig2007quantum}
\bibinfo{author}{\bibnamefont{K{\"o}nig}, \bibfnamefont{M.}},
  \bibinfo{author}{\bibfnamefont{S.}~\bibnamefont{Wiedmann}},
  \bibinfo{author}{\bibfnamefont{C.}~\bibnamefont{Br{\"u}ne}},
  \bibinfo{author}{\bibfnamefont{A.}~\bibnamefont{Roth}},
  \bibinfo{author}{\bibfnamefont{H.}~\bibnamefont{Buhmann}},
  \bibinfo{author}{\bibfnamefont{L.~W.} \bibnamefont{Molenkamp}},
  \bibinfo{author}{\bibfnamefont{X.-L.} \bibnamefont{Qi}}, and
  \bibinfo{author}{\bibfnamefont{S.-C.} \bibnamefont{Zhang}},
  \bibinfo{year}{2007}, \bibinfo{journal}{Science}
  \textbf{\bibinfo{volume}{318}}(\bibinfo{number}{5851}), \bibinfo{pages}{766}.

\bibitem[{\citenamefont{Koroteev} \emph{et~al.}(2004)\citenamefont{Koroteev,
  Bihlmayer, Gayone, Chulkov, Bl{\"u}gel, Echenique, and
  Hofmann}}]{koroteev2004strong}
\bibinfo{author}{\bibnamefont{Koroteev}, \bibfnamefont{Y.~M.}},
  \bibinfo{author}{\bibfnamefont{G.}~\bibnamefont{Bihlmayer}},
  \bibinfo{author}{\bibfnamefont{J.}~\bibnamefont{Gayone}},
  \bibinfo{author}{\bibfnamefont{E.~V.} \bibnamefont{Chulkov}},
  \bibinfo{author}{\bibfnamefont{S.}~\bibnamefont{Bl{\"u}gel}},
  \bibinfo{author}{\bibfnamefont{P.~M.} \bibnamefont{Echenique}}, and
  \bibinfo{author}{\bibfnamefont{P.}~\bibnamefont{Hofmann}},
  \bibinfo{year}{2004}, \bibinfo{journal}{Physical Review Letters}
  \textbf{\bibinfo{volume}{93}}(\bibinfo{number}{4}), \bibinfo{pages}{046403}.

\bibitem[{\citenamefont{Kou} \emph{et~al.}(2014)\citenamefont{Kou, Guo, Fan,
  Pan, Lang, Jiang, Shao, Nie, Murata, Tang} \emph{et~al.}}]{kou2014scale}
\bibinfo{author}{\bibnamefont{Kou}, \bibfnamefont{X.}},
  \bibinfo{author}{\bibfnamefont{S.-T.} \bibnamefont{Guo}},
  \bibinfo{author}{\bibfnamefont{Y.}~\bibnamefont{Fan}},
  \bibinfo{author}{\bibfnamefont{L.}~\bibnamefont{Pan}},
  \bibinfo{author}{\bibfnamefont{M.}~\bibnamefont{Lang}},
  \bibinfo{author}{\bibfnamefont{Y.}~\bibnamefont{Jiang}},
  \bibinfo{author}{\bibfnamefont{Q.}~\bibnamefont{Shao}},
  \bibinfo{author}{\bibfnamefont{T.}~\bibnamefont{Nie}},
  \bibinfo{author}{\bibfnamefont{K.}~\bibnamefont{Murata}},
  \bibinfo{author}{\bibfnamefont{J.}~\bibnamefont{Tang}}, \emph{et~al.},
  \bibinfo{year}{2014}, \bibinfo{journal}{Physical Review Letters}
  \textbf{\bibinfo{volume}{113}}(\bibinfo{number}{13}),
  \bibinfo{pages}{137201}.

\bibitem[{\citenamefont{Kou} \emph{et~al.}(2015)\citenamefont{Kou, Pan, Wang,
  Fan, Choi, Lee, Nie, Murata, Shao, Zhang} \emph{et~al.}}]{kou2015metal}
\bibinfo{author}{\bibnamefont{Kou}, \bibfnamefont{X.}},
  \bibinfo{author}{\bibfnamefont{L.}~\bibnamefont{Pan}},
  \bibinfo{author}{\bibfnamefont{J.}~\bibnamefont{Wang}},
  \bibinfo{author}{\bibfnamefont{Y.}~\bibnamefont{Fan}},
  \bibinfo{author}{\bibfnamefont{E.~S.} \bibnamefont{Choi}},
  \bibinfo{author}{\bibfnamefont{W.-L.} \bibnamefont{Lee}},
  \bibinfo{author}{\bibfnamefont{T.}~\bibnamefont{Nie}},
  \bibinfo{author}{\bibfnamefont{K.}~\bibnamefont{Murata}},
  \bibinfo{author}{\bibfnamefont{Q.}~\bibnamefont{Shao}},
  \bibinfo{author}{\bibfnamefont{S.-C.} \bibnamefont{Zhang}}, \emph{et~al.},
  \bibinfo{year}{2015}, \bibinfo{journal}{Nature Communications}
  \textbf{\bibinfo{volume}{6}}(\bibinfo{number}{1}), \bibinfo{pages}{1}.

\bibitem[{\citenamefont{Krupin} \emph{et~al.}(2005)\citenamefont{Krupin,
  Bihlmayer, Starke, Gorovikov, Prieto, D{\"o}brich, Bl{\"u}gel, and
  Kaindl}}]{krupin2005rashba}
\bibinfo{author}{\bibnamefont{Krupin}, \bibfnamefont{O.}},
  \bibinfo{author}{\bibfnamefont{G.}~\bibnamefont{Bihlmayer}},
  \bibinfo{author}{\bibfnamefont{K.}~\bibnamefont{Starke}},
  \bibinfo{author}{\bibfnamefont{S.}~\bibnamefont{Gorovikov}},
  \bibinfo{author}{\bibfnamefont{J.}~\bibnamefont{Prieto}},
  \bibinfo{author}{\bibfnamefont{K.}~\bibnamefont{D{\"o}brich}},
  \bibinfo{author}{\bibfnamefont{S.}~\bibnamefont{Bl{\"u}gel}}, and
  \bibinfo{author}{\bibfnamefont{G.}~\bibnamefont{Kaindl}},
  \bibinfo{year}{2005}, \bibinfo{journal}{Physical Review B}
  \textbf{\bibinfo{volume}{71}}(\bibinfo{number}{20}), \bibinfo{pages}{201403}.

\bibitem[{\citenamefont{Kuroda} \emph{et~al.}(2017)\citenamefont{Kuroda,
  Tomita, Suzuki, Bareille, Nugroho, Goswami, Ochi, Ikhlas, Nakayama, Akebi}
  \emph{et~al.}}]{kuroda2017evidence}
\bibinfo{author}{\bibnamefont{Kuroda}, \bibfnamefont{K.}},
  \bibinfo{author}{\bibfnamefont{T.}~\bibnamefont{Tomita}},
  \bibinfo{author}{\bibfnamefont{M.-T.} \bibnamefont{Suzuki}},
  \bibinfo{author}{\bibfnamefont{C.}~\bibnamefont{Bareille}},
  \bibinfo{author}{\bibfnamefont{A.}~\bibnamefont{Nugroho}},
  \bibinfo{author}{\bibfnamefont{P.}~\bibnamefont{Goswami}},
  \bibinfo{author}{\bibfnamefont{M.}~\bibnamefont{Ochi}},
  \bibinfo{author}{\bibfnamefont{M.}~\bibnamefont{Ikhlas}},
  \bibinfo{author}{\bibfnamefont{M.}~\bibnamefont{Nakayama}},
  \bibinfo{author}{\bibfnamefont{S.}~\bibnamefont{Akebi}}, \emph{et~al.},
  \bibinfo{year}{2017}, \bibinfo{journal}{Nature Materials}
  \textbf{\bibinfo{volume}{16}}(\bibinfo{number}{11}), \bibinfo{pages}{1090}.

\bibitem[{\citenamefont{Kwan} \emph{et~al.}(2021)\citenamefont{Kwan, Hu, Simon,
  and Parameswaran}}]{kwan2021exciton}
\bibinfo{author}{\bibnamefont{Kwan}, \bibfnamefont{Y.~H.}},
  \bibinfo{author}{\bibfnamefont{Y.}~\bibnamefont{Hu}},
  \bibinfo{author}{\bibfnamefont{S.~H.} \bibnamefont{Simon}}, and
  \bibinfo{author}{\bibfnamefont{S.}~\bibnamefont{Parameswaran}},
  \bibinfo{year}{2021}, \bibinfo{journal}{Physical Review Letters}
  \textbf{\bibinfo{volume}{126}}(\bibinfo{number}{13}),
  \bibinfo{pages}{137601}.

\bibitem[{\citenamefont{Lachman} \emph{et~al.}(2017)\citenamefont{Lachman,
  Mogi, Sarkar, Uri, Bagani, Anahory, Myasoedov, Huber, Tsukazaki, Kawasaki}
  \emph{et~al.}}]{lachman2017observation}
\bibinfo{author}{\bibnamefont{Lachman}, \bibfnamefont{E.~O.}},
  \bibinfo{author}{\bibfnamefont{M.}~\bibnamefont{Mogi}},
  \bibinfo{author}{\bibfnamefont{J.}~\bibnamefont{Sarkar}},
  \bibinfo{author}{\bibfnamefont{A.}~\bibnamefont{Uri}},
  \bibinfo{author}{\bibfnamefont{K.}~\bibnamefont{Bagani}},
  \bibinfo{author}{\bibfnamefont{Y.}~\bibnamefont{Anahory}},
  \bibinfo{author}{\bibfnamefont{Y.}~\bibnamefont{Myasoedov}},
  \bibinfo{author}{\bibfnamefont{M.~E.} \bibnamefont{Huber}},
  \bibinfo{author}{\bibfnamefont{A.}~\bibnamefont{Tsukazaki}},
  \bibinfo{author}{\bibfnamefont{M.}~\bibnamefont{Kawasaki}}, \emph{et~al.},
  \bibinfo{year}{2017}, \bibinfo{journal}{npj Quantum Materials}
  \textbf{\bibinfo{volume}{2}}(\bibinfo{number}{1}), \bibinfo{pages}{1}.

\bibitem[{\citenamefont{Lachman} \emph{et~al.}(2015)\citenamefont{Lachman,
  Young, Richardella, Cuppens, Naren, Anahory, Meltzer, Kandala, Kempinger,
  Myasoedov} \emph{et~al.}}]{lachman2015visualization}
\bibinfo{author}{\bibnamefont{Lachman}, \bibfnamefont{E.~O.}},
  \bibinfo{author}{\bibfnamefont{A.~F.} \bibnamefont{Young}},
  \bibinfo{author}{\bibfnamefont{A.}~\bibnamefont{Richardella}},
  \bibinfo{author}{\bibfnamefont{J.}~\bibnamefont{Cuppens}},
  \bibinfo{author}{\bibfnamefont{H.}~\bibnamefont{Naren}},
  \bibinfo{author}{\bibfnamefont{Y.}~\bibnamefont{Anahory}},
  \bibinfo{author}{\bibfnamefont{A.~Y.} \bibnamefont{Meltzer}},
  \bibinfo{author}{\bibfnamefont{A.}~\bibnamefont{Kandala}},
  \bibinfo{author}{\bibfnamefont{S.}~\bibnamefont{Kempinger}},
  \bibinfo{author}{\bibfnamefont{Y.}~\bibnamefont{Myasoedov}}, \emph{et~al.},
  \bibinfo{year}{2015}, \bibinfo{journal}{Science Advances}
  \textbf{\bibinfo{volume}{1}}(\bibinfo{number}{10}),
  \bibinfo{pages}{e1500740}.

\bibitem[{\citenamefont{Landau}(1930)}]{landau1930diamagnetismus}
\bibinfo{author}{\bibnamefont{Landau}, \bibfnamefont{L.}},
  \bibinfo{year}{1930}, \bibinfo{journal}{Zeitschrift f{\"u}r Physik}
  \textbf{\bibinfo{volume}{64}}(\bibinfo{number}{9}), \bibinfo{pages}{629}.

\bibitem[{\citenamefont{Landau and Lifshitz}(2013)}]{landau2013quantum}
\bibinfo{author}{\bibnamefont{Landau}, \bibfnamefont{L.~D.}}, and
  \bibinfo{author}{\bibfnamefont{E.~M.} \bibnamefont{Lifshitz}},
  \bibinfo{year}{2013}, \emph{\bibinfo{title}{Quantum mechanics:
  non-relativistic theory}}, volume~\bibinfo{volume}{3}
  (\bibinfo{publisher}{Elsevier}).

\bibitem[{\citenamefont{Lapano} \emph{et~al.}(2020)\citenamefont{Lapano,
  Nuckols, Mazza, Pai, Zhang, Lawrie, Moore, Eres, Lee, Du}
  \emph{et~al.}}]{lapano2020adsorption}
\bibinfo{author}{\bibnamefont{Lapano}, \bibfnamefont{J.}},
  \bibinfo{author}{\bibfnamefont{L.}~\bibnamefont{Nuckols}},
  \bibinfo{author}{\bibfnamefont{A.~R.} \bibnamefont{Mazza}},
  \bibinfo{author}{\bibfnamefont{Y.-Y.} \bibnamefont{Pai}},
  \bibinfo{author}{\bibfnamefont{J.}~\bibnamefont{Zhang}},
  \bibinfo{author}{\bibfnamefont{B.}~\bibnamefont{Lawrie}},
  \bibinfo{author}{\bibfnamefont{R.~G.} \bibnamefont{Moore}},
  \bibinfo{author}{\bibfnamefont{G.}~\bibnamefont{Eres}},
  \bibinfo{author}{\bibfnamefont{H.~N.} \bibnamefont{Lee}},
  \bibinfo{author}{\bibfnamefont{M.-H.} \bibnamefont{Du}}, \emph{et~al.},
  \bibinfo{year}{2020}, \bibinfo{journal}{Physical Review Materials}
  \textbf{\bibinfo{volume}{4}}(\bibinfo{number}{11}), \bibinfo{pages}{111201}.

\bibitem[{\citenamefont{Ledwith} \emph{et~al.}(2020)\citenamefont{Ledwith,
  Tarnopolsky, Khalaf, and Vishwanath}}]{ledwith2020fractional}
\bibinfo{author}{\bibnamefont{Ledwith}, \bibfnamefont{P.~J.}},
  \bibinfo{author}{\bibfnamefont{G.}~\bibnamefont{Tarnopolsky}},
  \bibinfo{author}{\bibfnamefont{E.}~\bibnamefont{Khalaf}}, and
  \bibinfo{author}{\bibfnamefont{A.}~\bibnamefont{Vishwanath}},
  \bibinfo{year}{2020}, \bibinfo{journal}{Physical Review Research}
  \textbf{\bibinfo{volume}{2}}(\bibinfo{number}{2}), \bibinfo{pages}{023237}.

\bibitem[{\citenamefont{Lee and Chalker}(1994)}]{lee1994unified}
\bibinfo{author}{\bibnamefont{Lee}, \bibfnamefont{D.~K.}}, and
  \bibinfo{author}{\bibfnamefont{J.}~\bibnamefont{Chalker}},
  \bibinfo{year}{1994}, \bibinfo{journal}{Physical Review Letters}
  \textbf{\bibinfo{volume}{72}}(\bibinfo{number}{10}), \bibinfo{pages}{1510}.

\bibitem[{\citenamefont{Lee} \emph{et~al.}(2015)\citenamefont{Lee, Kim, Lee,
  Billinge, Zhong, Schneeloch, Liu, Valla, Tranquada, Gu}
  \emph{et~al.}}]{lee2015imaging}
\bibinfo{author}{\bibnamefont{Lee}, \bibfnamefont{I.}},
  \bibinfo{author}{\bibfnamefont{C.~K.} \bibnamefont{Kim}},
  \bibinfo{author}{\bibfnamefont{J.}~\bibnamefont{Lee}},
  \bibinfo{author}{\bibfnamefont{S.~J.} \bibnamefont{Billinge}},
  \bibinfo{author}{\bibfnamefont{R.}~\bibnamefont{Zhong}},
  \bibinfo{author}{\bibfnamefont{J.~A.} \bibnamefont{Schneeloch}},
  \bibinfo{author}{\bibfnamefont{T.}~\bibnamefont{Liu}},
  \bibinfo{author}{\bibfnamefont{T.}~\bibnamefont{Valla}},
  \bibinfo{author}{\bibfnamefont{J.~M.} \bibnamefont{Tranquada}},
  \bibinfo{author}{\bibfnamefont{G.}~\bibnamefont{Gu}}, \emph{et~al.},
  \bibinfo{year}{2015}, \bibinfo{journal}{Proceedings of the National Academy
  of Sciences} \textbf{\bibinfo{volume}{112}}(\bibinfo{number}{5}),
  \bibinfo{pages}{1316}.

\bibitem[{\citenamefont{Lee} \emph{et~al.}(2021)\citenamefont{Lee, Graf, Min,
  Zhu, Yi, Ciocys, Wang, Choi, Basnet, Fereidouni}
  \emph{et~al.}}]{lee2021evidence}
\bibinfo{author}{\bibnamefont{Lee}, \bibfnamefont{S.~H.}},
  \bibinfo{author}{\bibfnamefont{D.}~\bibnamefont{Graf}},
  \bibinfo{author}{\bibfnamefont{L.}~\bibnamefont{Min}},
  \bibinfo{author}{\bibfnamefont{Y.}~\bibnamefont{Zhu}},
  \bibinfo{author}{\bibfnamefont{H.}~\bibnamefont{Yi}},
  \bibinfo{author}{\bibfnamefont{S.}~\bibnamefont{Ciocys}},
  \bibinfo{author}{\bibfnamefont{Y.}~\bibnamefont{Wang}},
  \bibinfo{author}{\bibfnamefont{E.~S.} \bibnamefont{Choi}},
  \bibinfo{author}{\bibfnamefont{R.}~\bibnamefont{Basnet}},
  \bibinfo{author}{\bibfnamefont{A.}~\bibnamefont{Fereidouni}}, \emph{et~al.},
  \bibinfo{year}{2021}, \bibinfo{journal}{Physical Review X}
  \textbf{\bibinfo{volume}{11}}(\bibinfo{number}{3}), \bibinfo{pages}{031032}.

\bibitem[{\citenamefont{Lee} \emph{et~al.}(2019)\citenamefont{Lee, Zhu, Wang,
  Miao, Pillsbury, Yi, Kempinger, Hu, Heikes, Quarterman}
  \emph{et~al.}}]{lee2019spin}
\bibinfo{author}{\bibnamefont{Lee}, \bibfnamefont{S.~H.}},
  \bibinfo{author}{\bibfnamefont{Y.}~\bibnamefont{Zhu}},
  \bibinfo{author}{\bibfnamefont{Y.}~\bibnamefont{Wang}},
  \bibinfo{author}{\bibfnamefont{L.}~\bibnamefont{Miao}},
  \bibinfo{author}{\bibfnamefont{T.}~\bibnamefont{Pillsbury}},
  \bibinfo{author}{\bibfnamefont{H.}~\bibnamefont{Yi}},
  \bibinfo{author}{\bibfnamefont{S.}~\bibnamefont{Kempinger}},
  \bibinfo{author}{\bibfnamefont{J.}~\bibnamefont{Hu}},
  \bibinfo{author}{\bibfnamefont{C.~A.} \bibnamefont{Heikes}},
  \bibinfo{author}{\bibfnamefont{P.}~\bibnamefont{Quarterman}}, \emph{et~al.},
  \bibinfo{year}{2019}, \bibinfo{journal}{Physical Review Research}
  \textbf{\bibinfo{volume}{1}}(\bibinfo{number}{1}), \bibinfo{pages}{012011}.

\bibitem[{\citenamefont{Lei} \emph{et~al.}(2020)\citenamefont{Lei, Chen, and
  MacDonald}}]{lei2020magnetized}
\bibinfo{author}{\bibnamefont{Lei}, \bibfnamefont{C.}},
  \bibinfo{author}{\bibfnamefont{S.}~\bibnamefont{Chen}}, and
  \bibinfo{author}{\bibfnamefont{A.~H.} \bibnamefont{MacDonald}},
  \bibinfo{year}{2020}, \bibinfo{journal}{Proceedings of the National Academy
  of Sciences} \textbf{\bibinfo{volume}{117}}(\bibinfo{number}{44}),
  \bibinfo{pages}{27224}.

\bibitem[{\citenamefont{Lei and MacDonald}(2021)}]{lei2021gate}
\bibinfo{author}{\bibnamefont{Lei}, \bibfnamefont{C.}}, and
  \bibinfo{author}{\bibfnamefont{A.~H.} \bibnamefont{MacDonald}},
  \bibinfo{year}{2021}, \bibinfo{journal}{Physical Review Materials}
  \textbf{\bibinfo{volume}{5}}(\bibinfo{number}{5}), \bibinfo{pages}{L051201}.

\bibitem[{\citenamefont{Li} \emph{et~al.}(2010{\natexlab{a}})\citenamefont{Li,
  Luican, Dos~Santos, Neto, Reina, Kong, and Andrei}}]{li2010observation}
\bibinfo{author}{\bibnamefont{Li}, \bibfnamefont{G.}},
  \bibinfo{author}{\bibfnamefont{A.}~\bibnamefont{Luican}},
  \bibinfo{author}{\bibfnamefont{J.~L.} \bibnamefont{Dos~Santos}},
  \bibinfo{author}{\bibfnamefont{A.~C.} \bibnamefont{Neto}},
  \bibinfo{author}{\bibfnamefont{A.}~\bibnamefont{Reina}},
  \bibinfo{author}{\bibfnamefont{J.}~\bibnamefont{Kong}}, and
  \bibinfo{author}{\bibfnamefont{E.}~\bibnamefont{Andrei}},
  \bibinfo{year}{2010}{\natexlab{a}}, \bibinfo{journal}{Nature Physics}
  \textbf{\bibinfo{volume}{6}}(\bibinfo{number}{2}), \bibinfo{pages}{109}.

\bibitem[{\citenamefont{Li} \emph{et~al.}(2021{\natexlab{a}})\citenamefont{Li,
  Chen, Jiang, and Xie}}]{li2021coexistence}
\bibinfo{author}{\bibnamefont{Li}, \bibfnamefont{H.}},
  \bibinfo{author}{\bibfnamefont{C.-Z.} \bibnamefont{Chen}},
  \bibinfo{author}{\bibfnamefont{H.}~\bibnamefont{Jiang}}, and
  \bibinfo{author}{\bibfnamefont{X.}~\bibnamefont{Xie}},
  \bibinfo{year}{2021}{\natexlab{a}}, \bibinfo{journal}{Physical Review
  Letters} \textbf{\bibinfo{volume}{127}}(\bibinfo{number}{23}),
  \bibinfo{pages}{236402}.

\bibitem[{\citenamefont{Li} \emph{et~al.}(2019{\natexlab{a}})\citenamefont{Li,
  Gao, Duan, Xu, Zhu, Tian, Gao, Fan, Rao, Huang} \emph{et~al.}}]{li2019dirac}
\bibinfo{author}{\bibnamefont{Li}, \bibfnamefont{H.}},
  \bibinfo{author}{\bibfnamefont{S.-Y.} \bibnamefont{Gao}},
  \bibinfo{author}{\bibfnamefont{S.-F.} \bibnamefont{Duan}},
  \bibinfo{author}{\bibfnamefont{Y.-F.} \bibnamefont{Xu}},
  \bibinfo{author}{\bibfnamefont{K.-J.} \bibnamefont{Zhu}},
  \bibinfo{author}{\bibfnamefont{S.-J.} \bibnamefont{Tian}},
  \bibinfo{author}{\bibfnamefont{J.-C.} \bibnamefont{Gao}},
  \bibinfo{author}{\bibfnamefont{W.-H.} \bibnamefont{Fan}},
  \bibinfo{author}{\bibfnamefont{Z.-C.} \bibnamefont{Rao}},
  \bibinfo{author}{\bibfnamefont{J.-R.} \bibnamefont{Huang}}, \emph{et~al.},
  \bibinfo{year}{2019}{\natexlab{a}}, \bibinfo{journal}{Physical Review X}
  \textbf{\bibinfo{volume}{9}}(\bibinfo{number}{4}), \bibinfo{pages}{041039}.

\bibitem[{\citenamefont{Li} \emph{et~al.}(2021{\natexlab{b}})\citenamefont{Li,
  Jiang, Chen, and Xie}}]{li2021critical}
\bibinfo{author}{\bibnamefont{Li}, \bibfnamefont{H.}},
  \bibinfo{author}{\bibfnamefont{H.}~\bibnamefont{Jiang}},
  \bibinfo{author}{\bibfnamefont{C.-Z.} \bibnamefont{Chen}}, and
  \bibinfo{author}{\bibfnamefont{X.}~\bibnamefont{Xie}},
  \bibinfo{year}{2021}{\natexlab{b}}, \bibinfo{journal}{Physical Review
  Letters} \textbf{\bibinfo{volume}{126}}(\bibinfo{number}{15}),
  \bibinfo{pages}{156601}.

\bibitem[{\citenamefont{Li} \emph{et~al.}(2021{\natexlab{c}})\citenamefont{Li,
  Li, Regan, Wang, Zhao, Kahn, Yumigeta, Blei, Taniguchi, Watanabe}
  \emph{et~al.}}]{li2021imaging}
\bibinfo{author}{\bibnamefont{Li}, \bibfnamefont{H.}},
  \bibinfo{author}{\bibfnamefont{S.}~\bibnamefont{Li}},
  \bibinfo{author}{\bibfnamefont{E.~C.} \bibnamefont{Regan}},
  \bibinfo{author}{\bibfnamefont{D.}~\bibnamefont{Wang}},
  \bibinfo{author}{\bibfnamefont{W.}~\bibnamefont{Zhao}},
  \bibinfo{author}{\bibfnamefont{S.}~\bibnamefont{Kahn}},
  \bibinfo{author}{\bibfnamefont{K.}~\bibnamefont{Yumigeta}},
  \bibinfo{author}{\bibfnamefont{M.}~\bibnamefont{Blei}},
  \bibinfo{author}{\bibfnamefont{T.}~\bibnamefont{Taniguchi}},
  \bibinfo{author}{\bibfnamefont{K.}~\bibnamefont{Watanabe}}, \emph{et~al.},
  \bibinfo{year}{2021}{\natexlab{c}}, \bibinfo{journal}{arXiv:2106.10599} .

\bibitem[{\citenamefont{Li} \emph{et~al.}(2020)\citenamefont{Li, Liu, Liu,
  Zhang, Xu, Yu, Wu, Zhang, and Fan}}]{li2020antiferromagnetic}
\bibinfo{author}{\bibnamefont{Li}, \bibfnamefont{H.}},
  \bibinfo{author}{\bibfnamefont{S.}~\bibnamefont{Liu}},
  \bibinfo{author}{\bibfnamefont{C.}~\bibnamefont{Liu}},
  \bibinfo{author}{\bibfnamefont{J.}~\bibnamefont{Zhang}},
  \bibinfo{author}{\bibfnamefont{Y.}~\bibnamefont{Xu}},
  \bibinfo{author}{\bibfnamefont{R.}~\bibnamefont{Yu}},
  \bibinfo{author}{\bibfnamefont{Y.}~\bibnamefont{Wu}},
  \bibinfo{author}{\bibfnamefont{Y.}~\bibnamefont{Zhang}}, and
  \bibinfo{author}{\bibfnamefont{S.}~\bibnamefont{Fan}}, \bibinfo{year}{2020},
  \bibinfo{journal}{Physical Chemistry Chemical Physics}
  \textbf{\bibinfo{volume}{22}}(\bibinfo{number}{2}), \bibinfo{pages}{556}.

\bibitem[{\citenamefont{Li} \emph{et~al.}(2019{\natexlab{b}})\citenamefont{Li,
  Li, Du, Wang, Gu, Zhang, He, Duan, and Xu}}]{li2019intrinsic}
\bibinfo{author}{\bibnamefont{Li}, \bibfnamefont{J.}},
  \bibinfo{author}{\bibfnamefont{Y.}~\bibnamefont{Li}},
  \bibinfo{author}{\bibfnamefont{S.}~\bibnamefont{Du}},
  \bibinfo{author}{\bibfnamefont{Z.}~\bibnamefont{Wang}},
  \bibinfo{author}{\bibfnamefont{B.-L.} \bibnamefont{Gu}},
  \bibinfo{author}{\bibfnamefont{S.-C.} \bibnamefont{Zhang}},
  \bibinfo{author}{\bibfnamefont{K.}~\bibnamefont{He}},
  \bibinfo{author}{\bibfnamefont{W.}~\bibnamefont{Duan}}, and
  \bibinfo{author}{\bibfnamefont{Y.}~\bibnamefont{Xu}},
  \bibinfo{year}{2019}{\natexlab{b}}, \bibinfo{journal}{Science Advances}
  \textbf{\bibinfo{volume}{5}}(\bibinfo{number}{6}), \bibinfo{pages}{eaaw5685}.

\bibitem[{\citenamefont{Li} \emph{et~al.}(2016{\natexlab{a}})\citenamefont{Li,
  Wang, McFaul, Zern, Ren, Watanabe, Taniguchi, Qiao, and Zhu}}]{li2016gate}
\bibinfo{author}{\bibnamefont{Li}, \bibfnamefont{J.}},
  \bibinfo{author}{\bibfnamefont{K.}~\bibnamefont{Wang}},
  \bibinfo{author}{\bibfnamefont{K.~J.} \bibnamefont{McFaul}},
  \bibinfo{author}{\bibfnamefont{Z.}~\bibnamefont{Zern}},
  \bibinfo{author}{\bibfnamefont{Y.}~\bibnamefont{Ren}},
  \bibinfo{author}{\bibfnamefont{K.}~\bibnamefont{Watanabe}},
  \bibinfo{author}{\bibfnamefont{T.}~\bibnamefont{Taniguchi}},
  \bibinfo{author}{\bibfnamefont{Z.}~\bibnamefont{Qiao}}, and
  \bibinfo{author}{\bibfnamefont{J.}~\bibnamefont{Zhu}},
  \bibinfo{year}{2016}{\natexlab{a}}, \bibinfo{journal}{Nature Nanotechnology}
  \textbf{\bibinfo{volume}{11}}(\bibinfo{number}{12}), \bibinfo{pages}{1060}.

\bibitem[{\citenamefont{Li} \emph{et~al.}(2018)\citenamefont{Li, Zhang, Yin,
  Zhang, Watanabe, Taniguchi, Liu, and Zhu}}]{li2018valley}
\bibinfo{author}{\bibnamefont{Li}, \bibfnamefont{J.}},
  \bibinfo{author}{\bibfnamefont{R.-X.} \bibnamefont{Zhang}},
  \bibinfo{author}{\bibfnamefont{Z.}~\bibnamefont{Yin}},
  \bibinfo{author}{\bibfnamefont{J.}~\bibnamefont{Zhang}},
  \bibinfo{author}{\bibfnamefont{K.}~\bibnamefont{Watanabe}},
  \bibinfo{author}{\bibfnamefont{T.}~\bibnamefont{Taniguchi}},
  \bibinfo{author}{\bibfnamefont{C.}~\bibnamefont{Liu}}, and
  \bibinfo{author}{\bibfnamefont{J.}~\bibnamefont{Zhu}}, \bibinfo{year}{2018},
  \bibinfo{journal}{Science}
  \textbf{\bibinfo{volume}{362}}(\bibinfo{number}{6419}),
  \bibinfo{pages}{1149}.

\bibitem[{\citenamefont{Li} \emph{et~al.}(2015{\natexlab{a}})\citenamefont{Li,
  Chang, Wu, Tao, Zhao, Chan, Moodera, Li, and Zhu}}]{li2015experimental}
\bibinfo{author}{\bibnamefont{Li}, \bibfnamefont{M.}},
  \bibinfo{author}{\bibfnamefont{C.-Z.} \bibnamefont{Chang}},
  \bibinfo{author}{\bibfnamefont{L.}~\bibnamefont{Wu}},
  \bibinfo{author}{\bibfnamefont{J.}~\bibnamefont{Tao}},
  \bibinfo{author}{\bibfnamefont{W.}~\bibnamefont{Zhao}},
  \bibinfo{author}{\bibfnamefont{M.~H.} \bibnamefont{Chan}},
  \bibinfo{author}{\bibfnamefont{J.~S.} \bibnamefont{Moodera}},
  \bibinfo{author}{\bibfnamefont{J.}~\bibnamefont{Li}}, and
  \bibinfo{author}{\bibfnamefont{Y.}~\bibnamefont{Zhu}},
  \bibinfo{year}{2015}{\natexlab{a}}, \bibinfo{journal}{Physical Review
  Letters} \textbf{\bibinfo{volume}{114}}(\bibinfo{number}{14}),
  \bibinfo{pages}{146802}.

\bibitem[{\citenamefont{Li}(2019)}]{li2019stanene}
\bibinfo{author}{\bibnamefont{Li}, \bibfnamefont{P.}}, \bibinfo{year}{2019},
  \bibinfo{journal}{Physical Chemistry Chemical Physics}
  \textbf{\bibinfo{volume}{21}}(\bibinfo{number}{21}), \bibinfo{pages}{11150}.

\bibitem[{\citenamefont{Li} \emph{et~al.}(2021{\natexlab{d}})\citenamefont{Li,
  Jiang, Li, Zhang, Kang, Zhu, Watanabe, Taniguchi, Chowdhury, Fu}
  \emph{et~al.}}]{li2021continuous}
\bibinfo{author}{\bibnamefont{Li}, \bibfnamefont{T.}},
  \bibinfo{author}{\bibfnamefont{S.}~\bibnamefont{Jiang}},
  \bibinfo{author}{\bibfnamefont{L.}~\bibnamefont{Li}},
  \bibinfo{author}{\bibfnamefont{Y.}~\bibnamefont{Zhang}},
  \bibinfo{author}{\bibfnamefont{K.}~\bibnamefont{Kang}},
  \bibinfo{author}{\bibfnamefont{J.}~\bibnamefont{Zhu}},
  \bibinfo{author}{\bibfnamefont{K.}~\bibnamefont{Watanabe}},
  \bibinfo{author}{\bibfnamefont{T.}~\bibnamefont{Taniguchi}},
  \bibinfo{author}{\bibfnamefont{D.}~\bibnamefont{Chowdhury}},
  \bibinfo{author}{\bibfnamefont{L.}~\bibnamefont{Fu}}, \emph{et~al.},
  \bibinfo{year}{2021}{\natexlab{d}}, \bibinfo{journal}{Nature}
  \textbf{\bibinfo{volume}{597}}(\bibinfo{number}{7876}), \bibinfo{pages}{350}.

\bibitem[{\citenamefont{Li} \emph{et~al.}(2021{\natexlab{e}})\citenamefont{Li,
  Jiang, Shen, Zhang, Li, Tao, Devakul, Watanabe, Taniguchi, Fu}
  \emph{et~al.}}]{li2021quantum}
\bibinfo{author}{\bibnamefont{Li}, \bibfnamefont{T.}},
  \bibinfo{author}{\bibfnamefont{S.}~\bibnamefont{Jiang}},
  \bibinfo{author}{\bibfnamefont{B.}~\bibnamefont{Shen}},
  \bibinfo{author}{\bibfnamefont{Y.}~\bibnamefont{Zhang}},
  \bibinfo{author}{\bibfnamefont{L.}~\bibnamefont{Li}},
  \bibinfo{author}{\bibfnamefont{Z.}~\bibnamefont{Tao}},
  \bibinfo{author}{\bibfnamefont{T.}~\bibnamefont{Devakul}},
  \bibinfo{author}{\bibfnamefont{K.}~\bibnamefont{Watanabe}},
  \bibinfo{author}{\bibfnamefont{T.}~\bibnamefont{Taniguchi}},
  \bibinfo{author}{\bibfnamefont{L.}~\bibnamefont{Fu}}, \emph{et~al.},
  \bibinfo{year}{2021}{\natexlab{e}}, \bibinfo{journal}{Nature}
  \textbf{\bibinfo{volume}{600}}(\bibinfo{number}{7890}), \bibinfo{pages}{641}.

\bibitem[{\citenamefont{Li} \emph{et~al.}(2016{\natexlab{b}})\citenamefont{Li,
  Claassen, Chang, Moritz, Jia, Zhang, Rebec, Lee, Hashimoto, Lu}
  \emph{et~al.}}]{li2016origin}
\bibinfo{author}{\bibnamefont{Li}, \bibfnamefont{W.}},
  \bibinfo{author}{\bibfnamefont{M.}~\bibnamefont{Claassen}},
  \bibinfo{author}{\bibfnamefont{C.-Z.} \bibnamefont{Chang}},
  \bibinfo{author}{\bibfnamefont{B.}~\bibnamefont{Moritz}},
  \bibinfo{author}{\bibfnamefont{T.}~\bibnamefont{Jia}},
  \bibinfo{author}{\bibfnamefont{C.}~\bibnamefont{Zhang}},
  \bibinfo{author}{\bibfnamefont{S.}~\bibnamefont{Rebec}},
  \bibinfo{author}{\bibfnamefont{J.}~\bibnamefont{Lee}},
  \bibinfo{author}{\bibfnamefont{M.}~\bibnamefont{Hashimoto}},
  \bibinfo{author}{\bibfnamefont{D.-H.} \bibnamefont{Lu}}, \emph{et~al.},
  \bibinfo{year}{2016}{\natexlab{b}}, \bibinfo{journal}{Scientific Reports}
  \textbf{\bibinfo{volume}{6}}(\bibinfo{number}{1}), \bibinfo{pages}{1}.

\bibitem[{\citenamefont{Li} \emph{et~al.}(2015{\natexlab{b}})\citenamefont{Li,
  West, Huang, Li, Zhang, and Duan}}]{li2015theory}
\bibinfo{author}{\bibnamefont{Li}, \bibfnamefont{Y.}},
  \bibinfo{author}{\bibfnamefont{D.}~\bibnamefont{West}},
  \bibinfo{author}{\bibfnamefont{H.}~\bibnamefont{Huang}},
  \bibinfo{author}{\bibfnamefont{J.}~\bibnamefont{Li}},
  \bibinfo{author}{\bibfnamefont{S.}~\bibnamefont{Zhang}}, and
  \bibinfo{author}{\bibfnamefont{W.}~\bibnamefont{Duan}},
  \bibinfo{year}{2015}{\natexlab{b}}, \bibinfo{journal}{Physical Review B}
  \textbf{\bibinfo{volume}{92}}(\bibinfo{number}{20}), \bibinfo{pages}{201403}.

\bibitem[{\citenamefont{Li} \emph{et~al.}(2010{\natexlab{b}})\citenamefont{Li,
  Wang, Zhu, Liu, Ye, Chen, Wang, He, Wang, Ma}
  \emph{et~al.}}]{li2010intrinsic}
\bibinfo{author}{\bibnamefont{Li}, \bibfnamefont{Y.-Y.}},
  \bibinfo{author}{\bibfnamefont{G.}~\bibnamefont{Wang}},
  \bibinfo{author}{\bibfnamefont{X.-G.} \bibnamefont{Zhu}},
  \bibinfo{author}{\bibfnamefont{M.-H.} \bibnamefont{Liu}},
  \bibinfo{author}{\bibfnamefont{C.}~\bibnamefont{Ye}},
  \bibinfo{author}{\bibfnamefont{X.}~\bibnamefont{Chen}},
  \bibinfo{author}{\bibfnamefont{Y.-Y.} \bibnamefont{Wang}},
  \bibinfo{author}{\bibfnamefont{K.}~\bibnamefont{He}},
  \bibinfo{author}{\bibfnamefont{L.-L.} \bibnamefont{Wang}},
  \bibinfo{author}{\bibfnamefont{X.-C.} \bibnamefont{Ma}}, \emph{et~al.},
  \bibinfo{year}{2010}{\natexlab{b}}, \bibinfo{journal}{Advanced Materials}
  \textbf{\bibinfo{volume}{22}}(\bibinfo{number}{36}), \bibinfo{pages}{4002}.

\bibitem[{\citenamefont{Lian} \emph{et~al.}(2021)\citenamefont{Lian, Song,
  Regnault, Efetov, Yazdani, and Bernevig}}]{lian2021twisted}
\bibinfo{author}{\bibnamefont{Lian}, \bibfnamefont{B.}},
  \bibinfo{author}{\bibfnamefont{Z.-D.} \bibnamefont{Song}},
  \bibinfo{author}{\bibfnamefont{N.}~\bibnamefont{Regnault}},
  \bibinfo{author}{\bibfnamefont{D.~K.} \bibnamefont{Efetov}},
  \bibinfo{author}{\bibfnamefont{A.}~\bibnamefont{Yazdani}}, and
  \bibinfo{author}{\bibfnamefont{B.~A.} \bibnamefont{Bernevig}},
  \bibinfo{year}{2021}, \bibinfo{journal}{Physical Review B}
  \textbf{\bibinfo{volume}{103}}(\bibinfo{number}{20}),
  \bibinfo{pages}{205414}.

\bibitem[{\citenamefont{Lian} \emph{et~al.}(2018)\citenamefont{Lian, Wang, Sun,
  Vaezi, and Zhang}}]{lian2018quantum}
\bibinfo{author}{\bibnamefont{Lian}, \bibfnamefont{B.}},
  \bibinfo{author}{\bibfnamefont{J.}~\bibnamefont{Wang}},
  \bibinfo{author}{\bibfnamefont{X.-Q.} \bibnamefont{Sun}},
  \bibinfo{author}{\bibfnamefont{A.}~\bibnamefont{Vaezi}}, and
  \bibinfo{author}{\bibfnamefont{S.-C.} \bibnamefont{Zhang}},
  \bibinfo{year}{2018}, \bibinfo{journal}{Physical Review B}
  \textbf{\bibinfo{volume}{97}}(\bibinfo{number}{12}), \bibinfo{pages}{125408}.

\bibitem[{\citenamefont{Liang} \emph{et~al.}(2020)\citenamefont{Liang, Luo,
  Shi, Zhang, Nie, Ying, He, Wu, Wang, Xu} \emph{et~al.}}]{liang2020mapping}
\bibinfo{author}{\bibnamefont{Liang}, \bibfnamefont{Z.}},
  \bibinfo{author}{\bibfnamefont{A.}~\bibnamefont{Luo}},
  \bibinfo{author}{\bibfnamefont{M.}~\bibnamefont{Shi}},
  \bibinfo{author}{\bibfnamefont{Q.}~\bibnamefont{Zhang}},
  \bibinfo{author}{\bibfnamefont{S.}~\bibnamefont{Nie}},
  \bibinfo{author}{\bibfnamefont{J.}~\bibnamefont{Ying}},
  \bibinfo{author}{\bibfnamefont{J.-F.} \bibnamefont{He}},
  \bibinfo{author}{\bibfnamefont{T.}~\bibnamefont{Wu}},
  \bibinfo{author}{\bibfnamefont{Z.}~\bibnamefont{Wang}},
  \bibinfo{author}{\bibfnamefont{G.}~\bibnamefont{Xu}}, \emph{et~al.},
  \bibinfo{year}{2020}, \bibinfo{journal}{Physical Review B}
  \textbf{\bibinfo{volume}{102}}(\bibinfo{number}{16}),
  \bibinfo{pages}{161115}.

\bibitem[{\citenamefont{de~Lima} \emph{et~al.}(2018)\citenamefont{de~Lima,
  Ferreira, and Miwa}}]{de2018quantum}
\bibinfo{author}{\bibnamefont{de~Lima}, \bibfnamefont{F.~C.}},
  \bibinfo{author}{\bibfnamefont{G.~J.} \bibnamefont{Ferreira}}, and
  \bibinfo{author}{\bibfnamefont{R.}~\bibnamefont{Miwa}}, \bibinfo{year}{2018},
  \bibinfo{journal}{Physical Chemistry Chemical Physics}
  \textbf{\bibinfo{volume}{20}}(\bibinfo{number}{35}), \bibinfo{pages}{22652}.

\bibitem[{\citenamefont{Lin}
  \emph{et~al.}(2022{\natexlab{a}})\citenamefont{Lin, Zhang, Morissette, Wang,
  Liu, Rhodes, Watanabe, Taniguchi, Hone, and Li}}]{lin2022spin}
\bibinfo{author}{\bibnamefont{Lin}, \bibfnamefont{J.-X.}},
  \bibinfo{author}{\bibfnamefont{Y.-H.} \bibnamefont{Zhang}},
  \bibinfo{author}{\bibfnamefont{E.}~\bibnamefont{Morissette}},
  \bibinfo{author}{\bibfnamefont{Z.}~\bibnamefont{Wang}},
  \bibinfo{author}{\bibfnamefont{S.}~\bibnamefont{Liu}},
  \bibinfo{author}{\bibfnamefont{D.}~\bibnamefont{Rhodes}},
  \bibinfo{author}{\bibfnamefont{K.}~\bibnamefont{Watanabe}},
  \bibinfo{author}{\bibfnamefont{T.}~\bibnamefont{Taniguchi}},
  \bibinfo{author}{\bibfnamefont{J.}~\bibnamefont{Hone}}, and
  \bibinfo{author}{\bibfnamefont{J.}~\bibnamefont{Li}},
  \bibinfo{year}{2022}{\natexlab{a}}, \bibinfo{journal}{Science}
  \textbf{\bibinfo{volume}{375}}(\bibinfo{number}{6579}), \bibinfo{pages}{437}.

\bibitem[{\citenamefont{Lin}
  \emph{et~al.}(2022{\natexlab{b}})\citenamefont{Lin, Feng, Wang, Lian, Li, Wu,
  Liu, Wang, Zhang, Wang} \emph{et~al.}}]{lin2022influence}
\bibinfo{author}{\bibnamefont{Lin}, \bibfnamefont{W.}},
  \bibinfo{author}{\bibfnamefont{Y.}~\bibnamefont{Feng}},
  \bibinfo{author}{\bibfnamefont{Y.}~\bibnamefont{Wang}},
  \bibinfo{author}{\bibfnamefont{Z.}~\bibnamefont{Lian}},
  \bibinfo{author}{\bibfnamefont{H.}~\bibnamefont{Li}},
  \bibinfo{author}{\bibfnamefont{Y.}~\bibnamefont{Wu}},
  \bibinfo{author}{\bibfnamefont{C.}~\bibnamefont{Liu}},
  \bibinfo{author}{\bibfnamefont{Y.}~\bibnamefont{Wang}},
  \bibinfo{author}{\bibfnamefont{J.}~\bibnamefont{Zhang}},
  \bibinfo{author}{\bibfnamefont{Y.}~\bibnamefont{Wang}}, \emph{et~al.},
  \bibinfo{year}{2022}{\natexlab{b}}, \bibinfo{journal}{Physical Review B}
  \textbf{\bibinfo{volume}{105}}(\bibinfo{number}{16}),
  \bibinfo{pages}{165411}.

\bibitem[{\citenamefont{Lippertz} \emph{et~al.}(2022)\citenamefont{Lippertz,
  Bliesener, Uday, Pereira, Taskin, and Ando}}]{lippertz2022current}
\bibinfo{author}{\bibnamefont{Lippertz}, \bibfnamefont{G.}},
  \bibinfo{author}{\bibfnamefont{A.}~\bibnamefont{Bliesener}},
  \bibinfo{author}{\bibfnamefont{A.}~\bibnamefont{Uday}},
  \bibinfo{author}{\bibfnamefont{L.~M.} \bibnamefont{Pereira}},
  \bibinfo{author}{\bibfnamefont{A.}~\bibnamefont{Taskin}}, and
  \bibinfo{author}{\bibfnamefont{Y.}~\bibnamefont{Ando}}, \bibinfo{year}{2022},
  \bibinfo{journal}{Physical Review B}
  \textbf{\bibinfo{volume}{106}}(\bibinfo{number}{4}), \bibinfo{pages}{045419}.

\bibitem[{\citenamefont{Liu}
  \emph{et~al.}(2008{\natexlab{a}})\citenamefont{Liu, Hughes, Qi, Wang, and
  Zhang}}]{liu2008quantumSpin}
\bibinfo{author}{\bibnamefont{Liu}, \bibfnamefont{C.}},
  \bibinfo{author}{\bibfnamefont{T.~L.} \bibnamefont{Hughes}},
  \bibinfo{author}{\bibfnamefont{X.-L.} \bibnamefont{Qi}},
  \bibinfo{author}{\bibfnamefont{K.}~\bibnamefont{Wang}}, and
  \bibinfo{author}{\bibfnamefont{S.-C.} \bibnamefont{Zhang}},
  \bibinfo{year}{2008}{\natexlab{a}}, \bibinfo{journal}{Physical Review
  Letters} \textbf{\bibinfo{volume}{100}}(\bibinfo{number}{23}),
  \bibinfo{pages}{236601}.

\bibitem[{\citenamefont{Liu}
  \emph{et~al.}(2020{\natexlab{a}})\citenamefont{Liu, Ou, Feng, Jiang, Wu, Li,
  Cheng, He, Ma, Xue} \emph{et~al.}}]{liu2020distinct}
\bibinfo{author}{\bibnamefont{Liu}, \bibfnamefont{C.}},
  \bibinfo{author}{\bibfnamefont{Y.}~\bibnamefont{Ou}},
  \bibinfo{author}{\bibfnamefont{Y.}~\bibnamefont{Feng}},
  \bibinfo{author}{\bibfnamefont{G.}~\bibnamefont{Jiang}},
  \bibinfo{author}{\bibfnamefont{W.}~\bibnamefont{Wu}},
  \bibinfo{author}{\bibfnamefont{S.}~\bibnamefont{Li}},
  \bibinfo{author}{\bibfnamefont{Z.}~\bibnamefont{Cheng}},
  \bibinfo{author}{\bibfnamefont{K.}~\bibnamefont{He}},
  \bibinfo{author}{\bibfnamefont{X.}~\bibnamefont{Ma}},
  \bibinfo{author}{\bibfnamefont{Q.}~\bibnamefont{Xue}}, \emph{et~al.},
  \bibinfo{year}{2020}{\natexlab{a}}, \bibinfo{journal}{Physical Review X}
  \textbf{\bibinfo{volume}{10}}(\bibinfo{number}{4}), \bibinfo{pages}{041063}.

\bibitem[{\citenamefont{Liu}
  \emph{et~al.}(2020{\natexlab{b}})\citenamefont{Liu, Wang, Li, Wu, Li, Li, He,
  Xu, Zhang, and Wang}}]{liu2020robust}
\bibinfo{author}{\bibnamefont{Liu}, \bibfnamefont{C.}},
  \bibinfo{author}{\bibfnamefont{Y.}~\bibnamefont{Wang}},
  \bibinfo{author}{\bibfnamefont{H.}~\bibnamefont{Li}},
  \bibinfo{author}{\bibfnamefont{Y.}~\bibnamefont{Wu}},
  \bibinfo{author}{\bibfnamefont{Y.}~\bibnamefont{Li}},
  \bibinfo{author}{\bibfnamefont{J.}~\bibnamefont{Li}},
  \bibinfo{author}{\bibfnamefont{K.}~\bibnamefont{He}},
  \bibinfo{author}{\bibfnamefont{Y.}~\bibnamefont{Xu}},
  \bibinfo{author}{\bibfnamefont{J.}~\bibnamefont{Zhang}}, and
  \bibinfo{author}{\bibfnamefont{Y.}~\bibnamefont{Wang}},
  \bibinfo{year}{2020}{\natexlab{b}}, \bibinfo{journal}{Nature Materials}
  \textbf{\bibinfo{volume}{19}}(\bibinfo{number}{5}), \bibinfo{pages}{522}.

\bibitem[{\citenamefont{Liu}
  \emph{et~al.}(2021{\natexlab{a}})\citenamefont{Liu, Wang, Yang, Mao, Li, Li,
  Li, Zhu, Wang, Li} \emph{et~al.}}]{liu2021magnetic}
\bibinfo{author}{\bibnamefont{Liu}, \bibfnamefont{C.}},
  \bibinfo{author}{\bibfnamefont{Y.}~\bibnamefont{Wang}},
  \bibinfo{author}{\bibfnamefont{M.}~\bibnamefont{Yang}},
  \bibinfo{author}{\bibfnamefont{J.}~\bibnamefont{Mao}},
  \bibinfo{author}{\bibfnamefont{H.}~\bibnamefont{Li}},
  \bibinfo{author}{\bibfnamefont{Y.}~\bibnamefont{Li}},
  \bibinfo{author}{\bibfnamefont{J.}~\bibnamefont{Li}},
  \bibinfo{author}{\bibfnamefont{H.}~\bibnamefont{Zhu}},
  \bibinfo{author}{\bibfnamefont{J.}~\bibnamefont{Wang}},
  \bibinfo{author}{\bibfnamefont{L.}~\bibnamefont{Li}}, \emph{et~al.},
  \bibinfo{year}{2021}{\natexlab{a}}, \bibinfo{journal}{Nature Communications}
  \textbf{\bibinfo{volume}{12}}(\bibinfo{number}{1}), \bibinfo{pages}{1}.

\bibitem[{\citenamefont{Liu} \emph{et~al.}(2015)\citenamefont{Liu, Zhou, and
  Yao}}]{liu2015valley}
\bibinfo{author}{\bibnamefont{Liu}, \bibfnamefont{C.-C.}},
  \bibinfo{author}{\bibfnamefont{J.-J.} \bibnamefont{Zhou}}, and
  \bibinfo{author}{\bibfnamefont{Y.}~\bibnamefont{Yao}}, \bibinfo{year}{2015},
  \bibinfo{journal}{Physical Review B}
  \textbf{\bibinfo{volume}{91}}(\bibinfo{number}{16}), \bibinfo{pages}{165430}.

\bibitem[{\citenamefont{Liu}
  \emph{et~al.}(2008{\natexlab{b}})\citenamefont{Liu, Qi, Dai, Fang, and
  Zhang}}]{liu2008quantum}
\bibinfo{author}{\bibnamefont{Liu}, \bibfnamefont{C.-X.}},
  \bibinfo{author}{\bibfnamefont{X.-L.} \bibnamefont{Qi}},
  \bibinfo{author}{\bibfnamefont{X.}~\bibnamefont{Dai}},
  \bibinfo{author}{\bibfnamefont{Z.}~\bibnamefont{Fang}}, and
  \bibinfo{author}{\bibfnamefont{S.-C.} \bibnamefont{Zhang}},
  \bibinfo{year}{2008}{\natexlab{b}}, \bibinfo{journal}{Physical Review
  Letters} \textbf{\bibinfo{volume}{101}}(\bibinfo{number}{14}),
  \bibinfo{pages}{146802}.

\bibitem[{\citenamefont{Liu}
  \emph{et~al.}(2016{\natexlab{a}})\citenamefont{Liu, Zhang, and
  Qi}}]{liu2016quantum}
\bibinfo{author}{\bibnamefont{Liu}, \bibfnamefont{C.-X.}},
  \bibinfo{author}{\bibfnamefont{S.-C.} \bibnamefont{Zhang}}, and
  \bibinfo{author}{\bibfnamefont{X.-L.} \bibnamefont{Qi}},
  \bibinfo{year}{2016}{\natexlab{a}}, \bibinfo{journal}{Annual Review of
  Condensed Matter Physics} \textbf{\bibinfo{volume}{7}}, \bibinfo{pages}{301}.

\bibitem[{\citenamefont{Liu}
  \emph{et~al.}(2019{\natexlab{a}})\citenamefont{Liu, Liang, Liu, Xu, Li, Chen,
  Pei, Shi, Mo, Dudin} \emph{et~al.}}]{liu2019magnetic}
\bibinfo{author}{\bibnamefont{Liu}, \bibfnamefont{D.}},
  \bibinfo{author}{\bibfnamefont{A.}~\bibnamefont{Liang}},
  \bibinfo{author}{\bibfnamefont{E.}~\bibnamefont{Liu}},
  \bibinfo{author}{\bibfnamefont{Q.}~\bibnamefont{Xu}},
  \bibinfo{author}{\bibfnamefont{Y.}~\bibnamefont{Li}},
  \bibinfo{author}{\bibfnamefont{C.}~\bibnamefont{Chen}},
  \bibinfo{author}{\bibfnamefont{D.}~\bibnamefont{Pei}},
  \bibinfo{author}{\bibfnamefont{W.}~\bibnamefont{Shi}},
  \bibinfo{author}{\bibfnamefont{S.}~\bibnamefont{Mo}},
  \bibinfo{author}{\bibfnamefont{P.}~\bibnamefont{Dudin}}, \emph{et~al.},
  \bibinfo{year}{2019}{\natexlab{a}}, \bibinfo{journal}{Science}
  \textbf{\bibinfo{volume}{365}}(\bibinfo{number}{6459}),
  \bibinfo{pages}{1282}.

\bibitem[{\citenamefont{Liu and Dai}(2021{\natexlab{a}})}]{liu2021orbital}
\bibinfo{author}{\bibnamefont{Liu}, \bibfnamefont{J.}}, and
  \bibinfo{author}{\bibfnamefont{X.}~\bibnamefont{Dai}},
  \bibinfo{year}{2021}{\natexlab{a}}, \bibinfo{journal}{Nature Reviews Physics}
  \textbf{\bibinfo{volume}{3}}(\bibinfo{number}{5}), \bibinfo{pages}{367}.

\bibitem[{\citenamefont{Liu and Dai}(2021{\natexlab{b}})}]{liu2021theories}
\bibinfo{author}{\bibnamefont{Liu}, \bibfnamefont{J.}}, and
  \bibinfo{author}{\bibfnamefont{X.}~\bibnamefont{Dai}},
  \bibinfo{year}{2021}{\natexlab{b}}, \bibinfo{journal}{Physical Review B}
  \textbf{\bibinfo{volume}{103}}(\bibinfo{number}{3}), \bibinfo{pages}{035427}.

\bibitem[{\citenamefont{Liu}
  \emph{et~al.}(2019{\natexlab{b}})\citenamefont{Liu, Liu, and
  Dai}}]{liu2019pseudo}
\bibinfo{author}{\bibnamefont{Liu}, \bibfnamefont{J.}},
  \bibinfo{author}{\bibfnamefont{J.}~\bibnamefont{Liu}}, and
  \bibinfo{author}{\bibfnamefont{X.}~\bibnamefont{Dai}},
  \bibinfo{year}{2019}{\natexlab{b}}, \bibinfo{journal}{Physical Review B}
  \textbf{\bibinfo{volume}{99}}(\bibinfo{number}{15}), \bibinfo{pages}{155415}.

\bibitem[{\citenamefont{Liu}
  \emph{et~al.}(2019{\natexlab{c}})\citenamefont{Liu, Ma, Gao, and
  Dai}}]{liu2019quantum}
\bibinfo{author}{\bibnamefont{Liu}, \bibfnamefont{J.}},
  \bibinfo{author}{\bibfnamefont{Z.}~\bibnamefont{Ma}},
  \bibinfo{author}{\bibfnamefont{J.}~\bibnamefont{Gao}}, and
  \bibinfo{author}{\bibfnamefont{X.}~\bibnamefont{Dai}},
  \bibinfo{year}{2019}{\natexlab{c}}, \bibinfo{journal}{Physical Review X}
  \textbf{\bibinfo{volume}{9}}(\bibinfo{number}{3}), \bibinfo{pages}{031021}.

\bibitem[{\citenamefont{Liu}
  \emph{et~al.}(2016{\natexlab{b}})\citenamefont{Liu, Wang, Richardella,
  Kandala, Li, Yazdani, Samarth, and Ong}}]{liu2016large}
\bibinfo{author}{\bibnamefont{Liu}, \bibfnamefont{M.}},
  \bibinfo{author}{\bibfnamefont{W.}~\bibnamefont{Wang}},
  \bibinfo{author}{\bibfnamefont{A.~R.} \bibnamefont{Richardella}},
  \bibinfo{author}{\bibfnamefont{A.}~\bibnamefont{Kandala}},
  \bibinfo{author}{\bibfnamefont{J.}~\bibnamefont{Li}},
  \bibinfo{author}{\bibfnamefont{A.}~\bibnamefont{Yazdani}},
  \bibinfo{author}{\bibfnamefont{N.}~\bibnamefont{Samarth}}, and
  \bibinfo{author}{\bibfnamefont{N.~P.} \bibnamefont{Ong}},
  \bibinfo{year}{2016}{\natexlab{b}}, \bibinfo{journal}{Science Advances}
  \textbf{\bibinfo{volume}{2}}(\bibinfo{number}{7}), \bibinfo{pages}{e1600167}.

\bibitem[{\citenamefont{Liu} \emph{et~al.}(2009)\citenamefont{Liu, Liu, Xu, Qi,
  and Zhang}}]{liu2009magnetic}
\bibinfo{author}{\bibnamefont{Liu}, \bibfnamefont{Q.}},
  \bibinfo{author}{\bibfnamefont{C.-X.} \bibnamefont{Liu}},
  \bibinfo{author}{\bibfnamefont{C.}~\bibnamefont{Xu}},
  \bibinfo{author}{\bibfnamefont{X.-L.} \bibnamefont{Qi}}, and
  \bibinfo{author}{\bibfnamefont{S.-C.} \bibnamefont{Zhang}},
  \bibinfo{year}{2009}, \bibinfo{journal}{Physical Review Letters}
  \textbf{\bibinfo{volume}{102}}(\bibinfo{number}{15}),
  \bibinfo{pages}{156603}.

\bibitem[{\citenamefont{Liu}
  \emph{et~al.}(2021{\natexlab{b}})\citenamefont{Liu, Wang, Watanabe,
  Taniguchi, Vafek, and Li}}]{liu2021tuning}
\bibinfo{author}{\bibnamefont{Liu}, \bibfnamefont{X.}},
  \bibinfo{author}{\bibfnamefont{Z.}~\bibnamefont{Wang}},
  \bibinfo{author}{\bibfnamefont{K.}~\bibnamefont{Watanabe}},
  \bibinfo{author}{\bibfnamefont{T.}~\bibnamefont{Taniguchi}},
  \bibinfo{author}{\bibfnamefont{O.}~\bibnamefont{Vafek}}, and
  \bibinfo{author}{\bibfnamefont{J.}~\bibnamefont{Li}},
  \bibinfo{year}{2021}{\natexlab{b}}, \bibinfo{journal}{Science}
  \textbf{\bibinfo{volume}{371}}(\bibinfo{number}{6535}),
  \bibinfo{pages}{1261}.

\bibitem[{\citenamefont{Liu}
  \emph{et~al.}(2021{\natexlab{c}})\citenamefont{Liu, Wang, Zheng, Huang, Wang,
  Chi, Wu, Chakoumakos, McGuire, Sales} \emph{et~al.}}]{liu2021site}
\bibinfo{author}{\bibnamefont{Liu}, \bibfnamefont{Y.}},
  \bibinfo{author}{\bibfnamefont{L.-L.} \bibnamefont{Wang}},
  \bibinfo{author}{\bibfnamefont{Q.}~\bibnamefont{Zheng}},
  \bibinfo{author}{\bibfnamefont{Z.}~\bibnamefont{Huang}},
  \bibinfo{author}{\bibfnamefont{X.}~\bibnamefont{Wang}},
  \bibinfo{author}{\bibfnamefont{M.}~\bibnamefont{Chi}},
  \bibinfo{author}{\bibfnamefont{Y.}~\bibnamefont{Wu}},
  \bibinfo{author}{\bibfnamefont{B.~C.} \bibnamefont{Chakoumakos}},
  \bibinfo{author}{\bibfnamefont{M.~A.} \bibnamefont{McGuire}},
  \bibinfo{author}{\bibfnamefont{B.~C.} \bibnamefont{Sales}}, \emph{et~al.},
  \bibinfo{year}{2021}{\natexlab{c}}, \bibinfo{journal}{Physical Review X}
  \textbf{\bibinfo{volume}{11}}(\bibinfo{number}{2}), \bibinfo{pages}{021033}.

\bibitem[{\citenamefont{Lu}(2019)}]{lu20193d}
\bibinfo{author}{\bibnamefont{Lu}, \bibfnamefont{H.-Z.}}, \bibinfo{year}{2019},
  \bibinfo{journal}{National Science Review}
  \textbf{\bibinfo{volume}{6}}(\bibinfo{number}{2}), \bibinfo{pages}{208}.

\bibitem[{\citenamefont{Lu} \emph{et~al.}(2019)\citenamefont{Lu, Stepanov,
  Yang, Xie, Aamir, Das, Urgell, Watanabe, Taniguchi, Zhang}
  \emph{et~al.}}]{lu2019superconductors}
\bibinfo{author}{\bibnamefont{Lu}, \bibfnamefont{X.}},
  \bibinfo{author}{\bibfnamefont{P.}~\bibnamefont{Stepanov}},
  \bibinfo{author}{\bibfnamefont{W.}~\bibnamefont{Yang}},
  \bibinfo{author}{\bibfnamefont{M.}~\bibnamefont{Xie}},
  \bibinfo{author}{\bibfnamefont{M.~A.} \bibnamefont{Aamir}},
  \bibinfo{author}{\bibfnamefont{I.}~\bibnamefont{Das}},
  \bibinfo{author}{\bibfnamefont{C.}~\bibnamefont{Urgell}},
  \bibinfo{author}{\bibfnamefont{K.}~\bibnamefont{Watanabe}},
  \bibinfo{author}{\bibfnamefont{T.}~\bibnamefont{Taniguchi}},
  \bibinfo{author}{\bibfnamefont{G.}~\bibnamefont{Zhang}}, \emph{et~al.},
  \bibinfo{year}{2019}, \bibinfo{journal}{Nature}
  \textbf{\bibinfo{volume}{574}}(\bibinfo{number}{7780}), \bibinfo{pages}{653}.

\bibitem[{\citenamefont{Luo} \emph{et~al.}(2021)\citenamefont{Luo, Hao, Jia,
  Yao, Meng, Zhai, Wu, Dou, and Zhou}}]{luo2021functionalization}
\bibinfo{author}{\bibnamefont{Luo}, \bibfnamefont{F.}},
  \bibinfo{author}{\bibfnamefont{X.}~\bibnamefont{Hao}},
  \bibinfo{author}{\bibfnamefont{Y.}~\bibnamefont{Jia}},
  \bibinfo{author}{\bibfnamefont{J.}~\bibnamefont{Yao}},
  \bibinfo{author}{\bibfnamefont{Q.}~\bibnamefont{Meng}},
  \bibinfo{author}{\bibfnamefont{S.}~\bibnamefont{Zhai}},
  \bibinfo{author}{\bibfnamefont{J.}~\bibnamefont{Wu}},
  \bibinfo{author}{\bibfnamefont{W.}~\bibnamefont{Dou}}, and
  \bibinfo{author}{\bibfnamefont{M.}~\bibnamefont{Zhou}}, \bibinfo{year}{2021},
  \bibinfo{journal}{Nanoscale}
  \textbf{\bibinfo{volume}{13}}(\bibinfo{number}{4}), \bibinfo{pages}{2527}.

\bibitem[{\citenamefont{Ma} \emph{et~al.}(2020)\citenamefont{Ma, Chen, Schwier,
  Zhang, Hao, Kumar, Lu, Shao, Jin, Zeng} \emph{et~al.}}]{ma2020hybridization}
\bibinfo{author}{\bibnamefont{Ma}, \bibfnamefont{X.-M.}},
  \bibinfo{author}{\bibfnamefont{Z.}~\bibnamefont{Chen}},
  \bibinfo{author}{\bibfnamefont{E.~F.} \bibnamefont{Schwier}},
  \bibinfo{author}{\bibfnamefont{Y.}~\bibnamefont{Zhang}},
  \bibinfo{author}{\bibfnamefont{Y.-J.} \bibnamefont{Hao}},
  \bibinfo{author}{\bibfnamefont{S.}~\bibnamefont{Kumar}},
  \bibinfo{author}{\bibfnamefont{R.}~\bibnamefont{Lu}},
  \bibinfo{author}{\bibfnamefont{J.}~\bibnamefont{Shao}},
  \bibinfo{author}{\bibfnamefont{Y.}~\bibnamefont{Jin}},
  \bibinfo{author}{\bibfnamefont{M.}~\bibnamefont{Zeng}}, \emph{et~al.},
  \bibinfo{year}{2020}, \bibinfo{journal}{Physical Review B}
  \textbf{\bibinfo{volume}{102}}(\bibinfo{number}{24}),
  \bibinfo{pages}{245136}.

\bibitem[{\citenamefont{MacDonald}(1983)}]{macdonald1983landau}
\bibinfo{author}{\bibnamefont{MacDonald}, \bibfnamefont{A.}},
  \bibinfo{year}{1983}, \bibinfo{journal}{Physical Review B}
  \textbf{\bibinfo{volume}{28}}(\bibinfo{number}{12}), \bibinfo{pages}{6713}.

\bibitem[{\citenamefont{MacDonald}(1995)}]{macdonald1995proceedings}
\bibinfo{author}{\bibnamefont{MacDonald}, \bibfnamefont{A.}},
  \bibinfo{year}{1995}, \bibinfo{journal}{Proceedings of the Les Houches Summer
  School on Mesoscopic Physics} .

\bibitem[{\citenamefont{MacDonald} \emph{et~al.}(2012)\citenamefont{MacDonald,
  Jung, and Zhang}}]{macdonald2012pseudospin}
\bibinfo{author}{\bibnamefont{MacDonald}, \bibfnamefont{A.~H.}},
  \bibinfo{author}{\bibfnamefont{J.}~\bibnamefont{Jung}}, and
  \bibinfo{author}{\bibfnamefont{F.}~\bibnamefont{Zhang}},
  \bibinfo{year}{2012}, \bibinfo{journal}{Physica Scripta}
  \textbf{\bibinfo{volume}{2012}}(\bibinfo{number}{T146}),
  \bibinfo{pages}{014012}.

\bibitem[{\citenamefont{Martin and Batista}(2008)}]{martin2008itinerant}
\bibinfo{author}{\bibnamefont{Martin}, \bibfnamefont{I.}}, and
  \bibinfo{author}{\bibfnamefont{C.}~\bibnamefont{Batista}},
  \bibinfo{year}{2008}, \bibinfo{journal}{Physical Review Letters}
  \textbf{\bibinfo{volume}{101}}(\bibinfo{number}{15}),
  \bibinfo{pages}{156402}.

\bibitem[{\citenamefont{Martin} \emph{et~al.}(2008)\citenamefont{Martin,
  Blanter, and Morpurgo}}]{martin2008topological}
\bibinfo{author}{\bibnamefont{Martin}, \bibfnamefont{I.}},
  \bibinfo{author}{\bibfnamefont{Y.~M.} \bibnamefont{Blanter}}, and
  \bibinfo{author}{\bibfnamefont{A.}~\bibnamefont{Morpurgo}},
  \bibinfo{year}{2008}, \bibinfo{journal}{Physical Review Letters}
  \textbf{\bibinfo{volume}{100}}(\bibinfo{number}{3}), \bibinfo{pages}{036804}.

\bibitem[{\citenamefont{Meng} \emph{et~al.}(2021)\citenamefont{Meng, Zhang, Yu,
  Liu, Tian, Dai, and Liu}}]{meng2021multiple}
\bibinfo{author}{\bibnamefont{Meng}, \bibfnamefont{W.}},
  \bibinfo{author}{\bibfnamefont{X.}~\bibnamefont{Zhang}},
  \bibinfo{author}{\bibfnamefont{W.}~\bibnamefont{Yu}},
  \bibinfo{author}{\bibfnamefont{Y.}~\bibnamefont{Liu}},
  \bibinfo{author}{\bibfnamefont{L.}~\bibnamefont{Tian}},
  \bibinfo{author}{\bibfnamefont{X.}~\bibnamefont{Dai}}, and
  \bibinfo{author}{\bibfnamefont{G.}~\bibnamefont{Liu}}, \bibinfo{year}{2021},
  \bibinfo{journal}{Applied Surface Science} \textbf{\bibinfo{volume}{551}},
  \bibinfo{pages}{149390}.

\bibitem[{\citenamefont{Mensz and Tsui}(1989)}]{mensz1989high}
\bibinfo{author}{\bibnamefont{Mensz}, \bibfnamefont{P.}}, and
  \bibinfo{author}{\bibfnamefont{D.}~\bibnamefont{Tsui}}, \bibinfo{year}{1989},
  \bibinfo{journal}{Physical Review B}
  \textbf{\bibinfo{volume}{40}}(\bibinfo{number}{6}), \bibinfo{pages}{3919}.

\bibitem[{\citenamefont{Min} \emph{et~al.}(2008)\citenamefont{Min, Borghi,
  Polini, and MacDonald}}]{min2008pseudospin}
\bibinfo{author}{\bibnamefont{Min}, \bibfnamefont{H.}},
  \bibinfo{author}{\bibfnamefont{G.}~\bibnamefont{Borghi}},
  \bibinfo{author}{\bibfnamefont{M.}~\bibnamefont{Polini}}, and
  \bibinfo{author}{\bibfnamefont{A.~H.} \bibnamefont{MacDonald}},
  \bibinfo{year}{2008}, \bibinfo{journal}{Physical Review B}
  \textbf{\bibinfo{volume}{77}}(\bibinfo{number}{4}), \bibinfo{pages}{041407}.

\bibitem[{\citenamefont{Mogi}
  \emph{et~al.}(2017{\natexlab{a}})\citenamefont{Mogi, Kawamura, Tsukazaki,
  Yoshimi, Takahashi, Kawasaki, and Tokura}}]{mogi2017tailoring}
\bibinfo{author}{\bibnamefont{Mogi}, \bibfnamefont{M.}},
  \bibinfo{author}{\bibfnamefont{M.}~\bibnamefont{Kawamura}},
  \bibinfo{author}{\bibfnamefont{A.}~\bibnamefont{Tsukazaki}},
  \bibinfo{author}{\bibfnamefont{R.}~\bibnamefont{Yoshimi}},
  \bibinfo{author}{\bibfnamefont{K.~S.} \bibnamefont{Takahashi}},
  \bibinfo{author}{\bibfnamefont{M.}~\bibnamefont{Kawasaki}}, and
  \bibinfo{author}{\bibfnamefont{Y.}~\bibnamefont{Tokura}},
  \bibinfo{year}{2017}{\natexlab{a}}, \bibinfo{journal}{Science Advances}
  \textbf{\bibinfo{volume}{3}}(\bibinfo{number}{10}),
  \bibinfo{pages}{eaao1669}.

\bibitem[{\citenamefont{Mogi}
  \emph{et~al.}(2017{\natexlab{b}})\citenamefont{Mogi, Kawamura, Yoshimi,
  Tsukazaki, Kozuka, Shirakawa, Takahashi, Kawasaki, and
  Tokura}}]{mogi2017magnetic}
\bibinfo{author}{\bibnamefont{Mogi}, \bibfnamefont{M.}},
  \bibinfo{author}{\bibfnamefont{M.}~\bibnamefont{Kawamura}},
  \bibinfo{author}{\bibfnamefont{R.}~\bibnamefont{Yoshimi}},
  \bibinfo{author}{\bibfnamefont{A.}~\bibnamefont{Tsukazaki}},
  \bibinfo{author}{\bibfnamefont{Y.}~\bibnamefont{Kozuka}},
  \bibinfo{author}{\bibfnamefont{N.}~\bibnamefont{Shirakawa}},
  \bibinfo{author}{\bibfnamefont{K.}~\bibnamefont{Takahashi}},
  \bibinfo{author}{\bibfnamefont{M.}~\bibnamefont{Kawasaki}}, and
  \bibinfo{author}{\bibfnamefont{Y.}~\bibnamefont{Tokura}},
  \bibinfo{year}{2017}{\natexlab{b}}, \bibinfo{journal}{Nature Materials}
  \textbf{\bibinfo{volume}{16}}(\bibinfo{number}{5}), \bibinfo{pages}{516}.

\bibitem[{\citenamefont{Mogi} \emph{et~al.}(2019)\citenamefont{Mogi, Nakajima,
  Ukleev, Tsukazaki, Yoshimi, Kawamura, Takahashi, Hanashima, Kakurai, Arima}
  \emph{et~al.}}]{mogi2019large}
\bibinfo{author}{\bibnamefont{Mogi}, \bibfnamefont{M.}},
  \bibinfo{author}{\bibfnamefont{T.}~\bibnamefont{Nakajima}},
  \bibinfo{author}{\bibfnamefont{V.}~\bibnamefont{Ukleev}},
  \bibinfo{author}{\bibfnamefont{A.}~\bibnamefont{Tsukazaki}},
  \bibinfo{author}{\bibfnamefont{R.}~\bibnamefont{Yoshimi}},
  \bibinfo{author}{\bibfnamefont{M.}~\bibnamefont{Kawamura}},
  \bibinfo{author}{\bibfnamefont{K.~S.} \bibnamefont{Takahashi}},
  \bibinfo{author}{\bibfnamefont{T.}~\bibnamefont{Hanashima}},
  \bibinfo{author}{\bibfnamefont{K.}~\bibnamefont{Kakurai}},
  \bibinfo{author}{\bibfnamefont{T.-h.} \bibnamefont{Arima}}, \emph{et~al.},
  \bibinfo{year}{2019}, \bibinfo{journal}{Physical Review Letters}
  \textbf{\bibinfo{volume}{123}}(\bibinfo{number}{1}), \bibinfo{pages}{016804}.

\bibitem[{\citenamefont{Mogi} \emph{et~al.}(2015)\citenamefont{Mogi, Yoshimi,
  Tsukazaki, Yasuda, Kozuka, Takahashi, Kawasaki, and
  Tokura}}]{mogi2015magnetic}
\bibinfo{author}{\bibnamefont{Mogi}, \bibfnamefont{M.}},
  \bibinfo{author}{\bibfnamefont{R.}~\bibnamefont{Yoshimi}},
  \bibinfo{author}{\bibfnamefont{A.}~\bibnamefont{Tsukazaki}},
  \bibinfo{author}{\bibfnamefont{K.}~\bibnamefont{Yasuda}},
  \bibinfo{author}{\bibfnamefont{Y.}~\bibnamefont{Kozuka}},
  \bibinfo{author}{\bibfnamefont{K.}~\bibnamefont{Takahashi}},
  \bibinfo{author}{\bibfnamefont{M.}~\bibnamefont{Kawasaki}}, and
  \bibinfo{author}{\bibfnamefont{Y.}~\bibnamefont{Tokura}},
  \bibinfo{year}{2015}, \bibinfo{journal}{Applied Physics Letters}
  \textbf{\bibinfo{volume}{107}}(\bibinfo{number}{18}),
  \bibinfo{pages}{182401}.

\bibitem[{\citenamefont{Moore and Balents}(2007)}]{moore2007topological}
\bibinfo{author}{\bibnamefont{Moore}, \bibfnamefont{J.~E.}}, and
  \bibinfo{author}{\bibfnamefont{L.}~\bibnamefont{Balents}},
  \bibinfo{year}{2007}, \bibinfo{journal}{Physical Review B}
  \textbf{\bibinfo{volume}{75}}(\bibinfo{number}{12}), \bibinfo{pages}{121306}.

\bibitem[{\citenamefont{Morali} \emph{et~al.}(2019)\citenamefont{Morali,
  Batabyal, Nag, Liu, Xu, Sun, Yan, Felser, Avraham, and
  Beidenkopf}}]{morali2019fermi}
\bibinfo{author}{\bibnamefont{Morali}, \bibfnamefont{N.}},
  \bibinfo{author}{\bibfnamefont{R.}~\bibnamefont{Batabyal}},
  \bibinfo{author}{\bibfnamefont{P.~K.} \bibnamefont{Nag}},
  \bibinfo{author}{\bibfnamefont{E.}~\bibnamefont{Liu}},
  \bibinfo{author}{\bibfnamefont{Q.}~\bibnamefont{Xu}},
  \bibinfo{author}{\bibfnamefont{Y.}~\bibnamefont{Sun}},
  \bibinfo{author}{\bibfnamefont{B.}~\bibnamefont{Yan}},
  \bibinfo{author}{\bibfnamefont{C.}~\bibnamefont{Felser}},
  \bibinfo{author}{\bibfnamefont{N.}~\bibnamefont{Avraham}}, and
  \bibinfo{author}{\bibfnamefont{H.}~\bibnamefont{Beidenkopf}},
  \bibinfo{year}{2019}, \bibinfo{journal}{Science}
  \textbf{\bibinfo{volume}{365}}(\bibinfo{number}{6459}),
  \bibinfo{pages}{1286}.

\bibitem[{\citenamefont{Morimoto} \emph{et~al.}(2015)\citenamefont{Morimoto,
  Furusaki, and Nagaosa}}]{morimoto2015topological}
\bibinfo{author}{\bibnamefont{Morimoto}, \bibfnamefont{T.}},
  \bibinfo{author}{\bibfnamefont{A.}~\bibnamefont{Furusaki}}, and
  \bibinfo{author}{\bibfnamefont{N.}~\bibnamefont{Nagaosa}},
  \bibinfo{year}{2015}, \bibinfo{journal}{Physical Review B}
  \textbf{\bibinfo{volume}{92}}(\bibinfo{number}{8}), \bibinfo{pages}{085113}.

\bibitem[{\citenamefont{Mourik} \emph{et~al.}(2012)\citenamefont{Mourik, Zuo,
  Frolov, Plissard, Bakkers, and Kouwenhoven}}]{mourik2012signatures}
\bibinfo{author}{\bibnamefont{Mourik}, \bibfnamefont{V.}},
  \bibinfo{author}{\bibfnamefont{K.}~\bibnamefont{Zuo}},
  \bibinfo{author}{\bibfnamefont{S.~M.} \bibnamefont{Frolov}},
  \bibinfo{author}{\bibfnamefont{S.}~\bibnamefont{Plissard}},
  \bibinfo{author}{\bibfnamefont{E.~P.} \bibnamefont{Bakkers}}, and
  \bibinfo{author}{\bibfnamefont{L.~P.} \bibnamefont{Kouwenhoven}},
  \bibinfo{year}{2012}, \bibinfo{journal}{Science}
  \textbf{\bibinfo{volume}{336}}(\bibinfo{number}{6084}),
  \bibinfo{pages}{1003}.

\bibitem[{\citenamefont{Muechler} \emph{et~al.}(2020)\citenamefont{Muechler,
  Liu, Gayles, Xu, Felser, and Sun}}]{muechler2020emerging}
\bibinfo{author}{\bibnamefont{Muechler}, \bibfnamefont{L.}},
  \bibinfo{author}{\bibfnamefont{E.}~\bibnamefont{Liu}},
  \bibinfo{author}{\bibfnamefont{J.}~\bibnamefont{Gayles}},
  \bibinfo{author}{\bibfnamefont{Q.}~\bibnamefont{Xu}},
  \bibinfo{author}{\bibfnamefont{C.}~\bibnamefont{Felser}}, and
  \bibinfo{author}{\bibfnamefont{Y.}~\bibnamefont{Sun}}, \bibinfo{year}{2020},
  \bibinfo{journal}{Physical Review B}
  \textbf{\bibinfo{volume}{101}}(\bibinfo{number}{11}),
  \bibinfo{pages}{115106}.

\bibitem[{\citenamefont{Murakami} \emph{et~al.}(2019)\citenamefont{Murakami,
  Nambu, Koretsune, Xiangyu, Yamamoto, Brown, and
  Kageyama}}]{murakami2019realization}
\bibinfo{author}{\bibnamefont{Murakami}, \bibfnamefont{T.}},
  \bibinfo{author}{\bibfnamefont{Y.}~\bibnamefont{Nambu}},
  \bibinfo{author}{\bibfnamefont{T.}~\bibnamefont{Koretsune}},
  \bibinfo{author}{\bibfnamefont{G.}~\bibnamefont{Xiangyu}},
  \bibinfo{author}{\bibfnamefont{T.}~\bibnamefont{Yamamoto}},
  \bibinfo{author}{\bibfnamefont{C.~M.} \bibnamefont{Brown}}, and
  \bibinfo{author}{\bibfnamefont{H.}~\bibnamefont{Kageyama}},
  \bibinfo{year}{2019}, \bibinfo{journal}{Physical Review B}
  \textbf{\bibinfo{volume}{100}}(\bibinfo{number}{19}),
  \bibinfo{pages}{195103}.

\bibitem[{\citenamefont{Nadj-Perge}
  \emph{et~al.}(2014)\citenamefont{Nadj-Perge, Drozdov, Li, Chen, Jeon, Seo,
  MacDonald, Bernevig, and Yazdani}}]{nadj2014observation}
\bibinfo{author}{\bibnamefont{Nadj-Perge}, \bibfnamefont{S.}},
  \bibinfo{author}{\bibfnamefont{I.~K.} \bibnamefont{Drozdov}},
  \bibinfo{author}{\bibfnamefont{J.}~\bibnamefont{Li}},
  \bibinfo{author}{\bibfnamefont{H.}~\bibnamefont{Chen}},
  \bibinfo{author}{\bibfnamefont{S.}~\bibnamefont{Jeon}},
  \bibinfo{author}{\bibfnamefont{J.}~\bibnamefont{Seo}},
  \bibinfo{author}{\bibfnamefont{A.~H.} \bibnamefont{MacDonald}},
  \bibinfo{author}{\bibfnamefont{B.~A.} \bibnamefont{Bernevig}}, and
  \bibinfo{author}{\bibfnamefont{A.}~\bibnamefont{Yazdani}},
  \bibinfo{year}{2014}, \bibinfo{journal}{Science}
  \textbf{\bibinfo{volume}{346}}(\bibinfo{number}{6209}), \bibinfo{pages}{602}.

\bibitem[{\citenamefont{Nagaosa} \emph{et~al.}(2010)\citenamefont{Nagaosa,
  Sinova, Onoda, MacDonald, and Ong}}]{nagaosa2010anomalous}
\bibinfo{author}{\bibnamefont{Nagaosa}, \bibfnamefont{N.}},
  \bibinfo{author}{\bibfnamefont{J.}~\bibnamefont{Sinova}},
  \bibinfo{author}{\bibfnamefont{S.}~\bibnamefont{Onoda}},
  \bibinfo{author}{\bibfnamefont{A.~H.} \bibnamefont{MacDonald}}, and
  \bibinfo{author}{\bibfnamefont{N.~P.} \bibnamefont{Ong}},
  \bibinfo{year}{2010}, \bibinfo{journal}{Reviews of Modern Physics}
  \textbf{\bibinfo{volume}{82}}(\bibinfo{number}{2}), \bibinfo{pages}{1539}.

\bibitem[{\citenamefont{Nandkishore and
  Levitov}(2010)}]{nandkishore2010quantum}
\bibinfo{author}{\bibnamefont{Nandkishore}, \bibfnamefont{R.}}, and
  \bibinfo{author}{\bibfnamefont{L.}~\bibnamefont{Levitov}},
  \bibinfo{year}{2010}, \bibinfo{journal}{Physical Review B}
  \textbf{\bibinfo{volume}{82}}(\bibinfo{number}{11}), \bibinfo{pages}{115124}.

\bibitem[{\citenamefont{Nayak} \emph{et~al.}(2008)\citenamefont{Nayak, Simon,
  Stern, Freedman, and Sarma}}]{nayak2008non}
\bibinfo{author}{\bibnamefont{Nayak}, \bibfnamefont{C.}},
  \bibinfo{author}{\bibfnamefont{S.~H.} \bibnamefont{Simon}},
  \bibinfo{author}{\bibfnamefont{A.}~\bibnamefont{Stern}},
  \bibinfo{author}{\bibfnamefont{M.}~\bibnamefont{Freedman}}, and
  \bibinfo{author}{\bibfnamefont{S.~D.} \bibnamefont{Sarma}},
  \bibinfo{year}{2008}, \bibinfo{journal}{Reviews of Modern Physics}
  \textbf{\bibinfo{volume}{80}}(\bibinfo{number}{3}), \bibinfo{pages}{1083}.

\bibitem[{\citenamefont{Neupert} \emph{et~al.}(2011)\citenamefont{Neupert,
  Santos, Chamon, and Mudry}}]{neupert2011fractional}
\bibinfo{author}{\bibnamefont{Neupert}, \bibfnamefont{T.}},
  \bibinfo{author}{\bibfnamefont{L.}~\bibnamefont{Santos}},
  \bibinfo{author}{\bibfnamefont{C.}~\bibnamefont{Chamon}}, and
  \bibinfo{author}{\bibfnamefont{C.}~\bibnamefont{Mudry}},
  \bibinfo{year}{2011}, \bibinfo{journal}{Physical Review Letters}
  \textbf{\bibinfo{volume}{106}}(\bibinfo{number}{23}),
  \bibinfo{pages}{236804}.

\bibitem[{\citenamefont{Nevola} \emph{et~al.}(2020)\citenamefont{Nevola, Li,
  Yan, Moore, Lee, Miao, and Johnson}}]{nevola2020coexistence}
\bibinfo{author}{\bibnamefont{Nevola}, \bibfnamefont{D.}},
  \bibinfo{author}{\bibfnamefont{H.~X.} \bibnamefont{Li}},
  \bibinfo{author}{\bibfnamefont{J.-Q.} \bibnamefont{Yan}},
  \bibinfo{author}{\bibfnamefont{R.}~\bibnamefont{Moore}},
  \bibinfo{author}{\bibfnamefont{H.-N.} \bibnamefont{Lee}},
  \bibinfo{author}{\bibfnamefont{H.}~\bibnamefont{Miao}}, and
  \bibinfo{author}{\bibfnamefont{P.~D.} \bibnamefont{Johnson}},
  \bibinfo{year}{2020}, \bibinfo{journal}{Physical Review Letters}
  \textbf{\bibinfo{volume}{125}}(\bibinfo{number}{11}),
  \bibinfo{pages}{117205}.

\bibitem[{\citenamefont{Nielsen and
  Ninomiya}(1981{\natexlab{a}})}]{nielsen1981absence1}
\bibinfo{author}{\bibnamefont{Nielsen}, \bibfnamefont{H.~B.}}, and
  \bibinfo{author}{\bibfnamefont{M.}~\bibnamefont{Ninomiya}},
  \bibinfo{year}{1981}{\natexlab{a}}, \bibinfo{journal}{Nuclear Physics B}
  \textbf{\bibinfo{volume}{185}}(\bibinfo{number}{1}), \bibinfo{pages}{20}.

\bibitem[{\citenamefont{Nielsen and
  Ninomiya}(1981{\natexlab{b}})}]{nielsen1981absence2}
\bibinfo{author}{\bibnamefont{Nielsen}, \bibfnamefont{H.~B.}}, and
  \bibinfo{author}{\bibfnamefont{M.}~\bibnamefont{Ninomiya}},
  \bibinfo{year}{1981}{\natexlab{b}}, \bibinfo{journal}{Nuclear Physics B}
  \textbf{\bibinfo{volume}{193}}(\bibinfo{number}{1}), \bibinfo{pages}{173}.

\bibitem[{\citenamefont{Ning and Mao}(2020)}]{ning2020recent}
\bibinfo{author}{\bibnamefont{Ning}, \bibfnamefont{W.}}, and
  \bibinfo{author}{\bibfnamefont{Z.}~\bibnamefont{Mao}}, \bibinfo{year}{2020},
  \bibinfo{journal}{APL Materials}
  \textbf{\bibinfo{volume}{8}}(\bibinfo{number}{9}), \bibinfo{pages}{090701}.

\bibitem[{\citenamefont{Niu} \emph{et~al.}(2015)\citenamefont{Niu, Bihlmayer,
  Zhang, Wortmann, Bl{\"u}gel, and Mokrousov}}]{niu2015functionalized}
\bibinfo{author}{\bibnamefont{Niu}, \bibfnamefont{C.}},
  \bibinfo{author}{\bibfnamefont{G.}~\bibnamefont{Bihlmayer}},
  \bibinfo{author}{\bibfnamefont{H.}~\bibnamefont{Zhang}},
  \bibinfo{author}{\bibfnamefont{D.}~\bibnamefont{Wortmann}},
  \bibinfo{author}{\bibfnamefont{S.}~\bibnamefont{Bl{\"u}gel}}, and
  \bibinfo{author}{\bibfnamefont{Y.}~\bibnamefont{Mokrousov}},
  \bibinfo{year}{2015}, \bibinfo{journal}{Physical Review B}
  \textbf{\bibinfo{volume}{91}}(\bibinfo{number}{4}), \bibinfo{pages}{041303}.

\bibitem[{\citenamefont{Nomura and Nagaosa}(2011)}]{nomura2011surface}
\bibinfo{author}{\bibnamefont{Nomura}, \bibfnamefont{K.}}, and
  \bibinfo{author}{\bibfnamefont{N.}~\bibnamefont{Nagaosa}},
  \bibinfo{year}{2011}, \bibinfo{journal}{Physical Review Letters}
  \textbf{\bibinfo{volume}{106}}(\bibinfo{number}{16}),
  \bibinfo{pages}{166802}.

\bibitem[{\citenamefont{Nuckolls} \emph{et~al.}(2020)\citenamefont{Nuckolls,
  Oh, Wong, Lian, Watanabe, Taniguchi, Bernevig, and
  Yazdani}}]{nuckolls2020strongly}
\bibinfo{author}{\bibnamefont{Nuckolls}, \bibfnamefont{K.~P.}},
  \bibinfo{author}{\bibfnamefont{M.}~\bibnamefont{Oh}},
  \bibinfo{author}{\bibfnamefont{D.}~\bibnamefont{Wong}},
  \bibinfo{author}{\bibfnamefont{B.}~\bibnamefont{Lian}},
  \bibinfo{author}{\bibfnamefont{K.}~\bibnamefont{Watanabe}},
  \bibinfo{author}{\bibfnamefont{T.}~\bibnamefont{Taniguchi}},
  \bibinfo{author}{\bibfnamefont{B.~A.} \bibnamefont{Bernevig}}, and
  \bibinfo{author}{\bibfnamefont{A.}~\bibnamefont{Yazdani}},
  \bibinfo{year}{2020}, \bibinfo{journal}{Nature}
  \textbf{\bibinfo{volume}{588}}(\bibinfo{number}{7839}), \bibinfo{pages}{610}.

\bibitem[{\citenamefont{Ohno}(1999)}]{ohno1999properties}
\bibinfo{author}{\bibnamefont{Ohno}, \bibfnamefont{H.}}, \bibinfo{year}{1999},
  \bibinfo{journal}{Journal of Magnetism and Magnetic Materials}
  \textbf{\bibinfo{volume}{200}}(\bibinfo{number}{1-3}), \bibinfo{pages}{110}.

\bibitem[{\citenamefont{Ohno} \emph{et~al.}(2000)\citenamefont{Ohno, Chiba,
  Matsukura, Omiya, Abe, Dietl, Ohno, and Ohtani}}]{ohno2000electric}
\bibinfo{author}{\bibnamefont{Ohno}, \bibfnamefont{H.}},
  \bibinfo{author}{\bibfnamefont{a.~D.} \bibnamefont{Chiba}},
  \bibinfo{author}{\bibfnamefont{a.~F.} \bibnamefont{Matsukura}},
  \bibinfo{author}{\bibfnamefont{T.}~\bibnamefont{Omiya}},
  \bibinfo{author}{\bibfnamefont{E.}~\bibnamefont{Abe}},
  \bibinfo{author}{\bibfnamefont{T.}~\bibnamefont{Dietl}},
  \bibinfo{author}{\bibfnamefont{Y.}~\bibnamefont{Ohno}}, and
  \bibinfo{author}{\bibfnamefont{K.}~\bibnamefont{Ohtani}},
  \bibinfo{year}{2000}, \bibinfo{journal}{Nature}
  \textbf{\bibinfo{volume}{408}}(\bibinfo{number}{6815}), \bibinfo{pages}{944}.

\bibitem[{\citenamefont{Okada} \emph{et~al.}(2016)\citenamefont{Okada,
  Takahashi, Mogi, Yoshimi, Tsukazaki, Takahashi, Ogawa, Kawasaki, and
  Tokura}}]{okada2016terahertz}
\bibinfo{author}{\bibnamefont{Okada}, \bibfnamefont{K.~N.}},
  \bibinfo{author}{\bibfnamefont{Y.}~\bibnamefont{Takahashi}},
  \bibinfo{author}{\bibfnamefont{M.}~\bibnamefont{Mogi}},
  \bibinfo{author}{\bibfnamefont{R.}~\bibnamefont{Yoshimi}},
  \bibinfo{author}{\bibfnamefont{A.}~\bibnamefont{Tsukazaki}},
  \bibinfo{author}{\bibfnamefont{K.~S.} \bibnamefont{Takahashi}},
  \bibinfo{author}{\bibfnamefont{N.}~\bibnamefont{Ogawa}},
  \bibinfo{author}{\bibfnamefont{M.}~\bibnamefont{Kawasaki}}, and
  \bibinfo{author}{\bibfnamefont{Y.}~\bibnamefont{Tokura}},
  \bibinfo{year}{2016}, \bibinfo{journal}{Nature Communications}
  \textbf{\bibinfo{volume}{7}}(\bibinfo{number}{1}), \bibinfo{pages}{1}.

\bibitem[{\citenamefont{Okazaki} \emph{et~al.}(2020)\citenamefont{Okazaki, Oe,
  Kawamura, Yoshimi, Nakamura, Takada, Mogi, Takahashi, Tsukazaki, Kawasaki}
  \emph{et~al.}}]{okazaki2020precise}
\bibinfo{author}{\bibnamefont{Okazaki}, \bibfnamefont{Y.}},
  \bibinfo{author}{\bibfnamefont{T.}~\bibnamefont{Oe}},
  \bibinfo{author}{\bibfnamefont{M.}~\bibnamefont{Kawamura}},
  \bibinfo{author}{\bibfnamefont{R.}~\bibnamefont{Yoshimi}},
  \bibinfo{author}{\bibfnamefont{S.}~\bibnamefont{Nakamura}},
  \bibinfo{author}{\bibfnamefont{S.}~\bibnamefont{Takada}},
  \bibinfo{author}{\bibfnamefont{M.}~\bibnamefont{Mogi}},
  \bibinfo{author}{\bibfnamefont{K.~S.} \bibnamefont{Takahashi}},
  \bibinfo{author}{\bibfnamefont{A.}~\bibnamefont{Tsukazaki}},
  \bibinfo{author}{\bibfnamefont{M.}~\bibnamefont{Kawasaki}}, \emph{et~al.},
  \bibinfo{year}{2020}, \bibinfo{journal}{Applied Physics Letters}
  \textbf{\bibinfo{volume}{116}}(\bibinfo{number}{14}),
  \bibinfo{pages}{143101}.

\bibitem[{\citenamefont{Okazaki} \emph{et~al.}(2021)\citenamefont{Okazaki, Oe,
  Kawamura, Yoshimi, Nakamura, Takada, Mogi, Takahashi, Tsukazaki, Kawasaki}
  \emph{et~al.}}]{okazaki2021quantum}
\bibinfo{author}{\bibnamefont{Okazaki}, \bibfnamefont{Y.}},
  \bibinfo{author}{\bibfnamefont{T.}~\bibnamefont{Oe}},
  \bibinfo{author}{\bibfnamefont{M.}~\bibnamefont{Kawamura}},
  \bibinfo{author}{\bibfnamefont{R.}~\bibnamefont{Yoshimi}},
  \bibinfo{author}{\bibfnamefont{S.}~\bibnamefont{Nakamura}},
  \bibinfo{author}{\bibfnamefont{S.}~\bibnamefont{Takada}},
  \bibinfo{author}{\bibfnamefont{M.}~\bibnamefont{Mogi}},
  \bibinfo{author}{\bibfnamefont{K.~S.} \bibnamefont{Takahashi}},
  \bibinfo{author}{\bibfnamefont{A.}~\bibnamefont{Tsukazaki}},
  \bibinfo{author}{\bibfnamefont{M.}~\bibnamefont{Kawasaki}}, \emph{et~al.},
  \bibinfo{year}{2021}, \bibinfo{journal}{Nature Physics} , \bibinfo{pages}{1}.

\bibitem[{\citenamefont{Onoda and Nagaosa}(2003)}]{onoda2003quantized}
\bibinfo{author}{\bibnamefont{Onoda}, \bibfnamefont{M.}}, and
  \bibinfo{author}{\bibfnamefont{N.}~\bibnamefont{Nagaosa}},
  \bibinfo{year}{2003}, \bibinfo{journal}{Physical Review Letters}
  \textbf{\bibinfo{volume}{90}}(\bibinfo{number}{20}), \bibinfo{pages}{206601}.

\bibitem[{\citenamefont{Otrokov}
  \emph{et~al.}(2019{\natexlab{a}})\citenamefont{Otrokov, Klimovskikh,
  Bentmann, Estyunin, Zeugner, Aliev, Ga{\ss}, Wolter, Koroleva, Shikin}
  \emph{et~al.}}]{otrokov2019prediction}
\bibinfo{author}{\bibnamefont{Otrokov}, \bibfnamefont{M.~M.}},
  \bibinfo{author}{\bibfnamefont{I.~I.} \bibnamefont{Klimovskikh}},
  \bibinfo{author}{\bibfnamefont{H.}~\bibnamefont{Bentmann}},
  \bibinfo{author}{\bibfnamefont{D.}~\bibnamefont{Estyunin}},
  \bibinfo{author}{\bibfnamefont{A.}~\bibnamefont{Zeugner}},
  \bibinfo{author}{\bibfnamefont{Z.~S.} \bibnamefont{Aliev}},
  \bibinfo{author}{\bibfnamefont{S.}~\bibnamefont{Ga{\ss}}},
  \bibinfo{author}{\bibfnamefont{A.}~\bibnamefont{Wolter}},
  \bibinfo{author}{\bibfnamefont{A.}~\bibnamefont{Koroleva}},
  \bibinfo{author}{\bibfnamefont{A.~M.} \bibnamefont{Shikin}}, \emph{et~al.},
  \bibinfo{year}{2019}{\natexlab{a}}, \bibinfo{journal}{Nature}
  \textbf{\bibinfo{volume}{576}}(\bibinfo{number}{7787}), \bibinfo{pages}{416}.

\bibitem[{\citenamefont{Otrokov} \emph{et~al.}(2017)\citenamefont{Otrokov,
  Menshchikova, Vergniory, Rusinov, Vyazovskaya, Koroteev, Bihlmayer, Ernst,
  Echenique, Arnau} \emph{et~al.}}]{otrokov2017highly}
\bibinfo{author}{\bibnamefont{Otrokov}, \bibfnamefont{M.~M.}},
  \bibinfo{author}{\bibfnamefont{T.~V.} \bibnamefont{Menshchikova}},
  \bibinfo{author}{\bibfnamefont{M.~G.} \bibnamefont{Vergniory}},
  \bibinfo{author}{\bibfnamefont{I.~P.} \bibnamefont{Rusinov}},
  \bibinfo{author}{\bibfnamefont{A.~Y.} \bibnamefont{Vyazovskaya}},
  \bibinfo{author}{\bibfnamefont{Y.~M.} \bibnamefont{Koroteev}},
  \bibinfo{author}{\bibfnamefont{G.}~\bibnamefont{Bihlmayer}},
  \bibinfo{author}{\bibfnamefont{A.}~\bibnamefont{Ernst}},
  \bibinfo{author}{\bibfnamefont{P.~M.} \bibnamefont{Echenique}},
  \bibinfo{author}{\bibfnamefont{A.}~\bibnamefont{Arnau}}, \emph{et~al.},
  \bibinfo{year}{2017}, \bibinfo{journal}{2D Materials}
  \textbf{\bibinfo{volume}{4}}(\bibinfo{number}{2}), \bibinfo{pages}{025082}.

\bibitem[{\citenamefont{Otrokov}
  \emph{et~al.}(2019{\natexlab{b}})\citenamefont{Otrokov, Rusinov, Blanco-Rey,
  Hoffmann, Vyazovskaya, Eremeev, Ernst, Echenique, Arnau, and
  Chulkov}}]{otrokov2019unique}
\bibinfo{author}{\bibnamefont{Otrokov}, \bibfnamefont{M.~M.}},
  \bibinfo{author}{\bibfnamefont{I.~P.} \bibnamefont{Rusinov}},
  \bibinfo{author}{\bibfnamefont{M.}~\bibnamefont{Blanco-Rey}},
  \bibinfo{author}{\bibfnamefont{M.}~\bibnamefont{Hoffmann}},
  \bibinfo{author}{\bibfnamefont{A.~Y.} \bibnamefont{Vyazovskaya}},
  \bibinfo{author}{\bibfnamefont{S.~V.} \bibnamefont{Eremeev}},
  \bibinfo{author}{\bibfnamefont{A.}~\bibnamefont{Ernst}},
  \bibinfo{author}{\bibfnamefont{P.~M.} \bibnamefont{Echenique}},
  \bibinfo{author}{\bibfnamefont{A.}~\bibnamefont{Arnau}}, and
  \bibinfo{author}{\bibfnamefont{E.~V.} \bibnamefont{Chulkov}},
  \bibinfo{year}{2019}{\natexlab{b}}, \bibinfo{journal}{Physical Review
  Letters} \textbf{\bibinfo{volume}{122}}(\bibinfo{number}{10}),
  \bibinfo{pages}{107202}.

\bibitem[{\citenamefont{Ou} \emph{et~al.}(2018)\citenamefont{Ou, Liu, Jiang,
  Feng, Zhao, Wu, Wang, Li, Song, Wang} \emph{et~al.}}]{ou2018enhancing}
\bibinfo{author}{\bibnamefont{Ou}, \bibfnamefont{Y.}},
  \bibinfo{author}{\bibfnamefont{C.}~\bibnamefont{Liu}},
  \bibinfo{author}{\bibfnamefont{G.}~\bibnamefont{Jiang}},
  \bibinfo{author}{\bibfnamefont{Y.}~\bibnamefont{Feng}},
  \bibinfo{author}{\bibfnamefont{D.}~\bibnamefont{Zhao}},
  \bibinfo{author}{\bibfnamefont{W.}~\bibnamefont{Wu}},
  \bibinfo{author}{\bibfnamefont{X.-X.} \bibnamefont{Wang}},
  \bibinfo{author}{\bibfnamefont{W.}~\bibnamefont{Li}},
  \bibinfo{author}{\bibfnamefont{C.}~\bibnamefont{Song}},
  \bibinfo{author}{\bibfnamefont{L.-L.} \bibnamefont{Wang}}, \emph{et~al.},
  \bibinfo{year}{2018}, \bibinfo{journal}{Advanced Materials}
  \textbf{\bibinfo{volume}{30}}(\bibinfo{number}{1}), \bibinfo{pages}{1703062}.

\bibitem[{\citenamefont{Ovchinnikov}
  \emph{et~al.}(2021)\citenamefont{Ovchinnikov, Huang, Lin, Fei, Cai, Song, He,
  Jiang, Wang, Li} \emph{et~al.}}]{ovchinnikov2021intertwined}
\bibinfo{author}{\bibnamefont{Ovchinnikov}, \bibfnamefont{D.}},
  \bibinfo{author}{\bibfnamefont{X.}~\bibnamefont{Huang}},
  \bibinfo{author}{\bibfnamefont{Z.}~\bibnamefont{Lin}},
  \bibinfo{author}{\bibfnamefont{Z.}~\bibnamefont{Fei}},
  \bibinfo{author}{\bibfnamefont{J.}~\bibnamefont{Cai}},
  \bibinfo{author}{\bibfnamefont{T.}~\bibnamefont{Song}},
  \bibinfo{author}{\bibfnamefont{M.}~\bibnamefont{He}},
  \bibinfo{author}{\bibfnamefont{Q.}~\bibnamefont{Jiang}},
  \bibinfo{author}{\bibfnamefont{C.}~\bibnamefont{Wang}},
  \bibinfo{author}{\bibfnamefont{H.}~\bibnamefont{Li}}, \emph{et~al.},
  \bibinfo{year}{2021}, \bibinfo{journal}{Nano Letters}
  \textbf{\bibinfo{volume}{21}}(\bibinfo{number}{6}), \bibinfo{pages}{2544}.

\bibitem[{\citenamefont{Pan} \emph{et~al.}(2014)\citenamefont{Pan, Li, Liu,
  Zhu, Qiao, and Yao}}]{pan2014valley}
\bibinfo{author}{\bibnamefont{Pan}, \bibfnamefont{H.}},
  \bibinfo{author}{\bibfnamefont{Z.}~\bibnamefont{Li}},
  \bibinfo{author}{\bibfnamefont{C.-C.} \bibnamefont{Liu}},
  \bibinfo{author}{\bibfnamefont{G.}~\bibnamefont{Zhu}},
  \bibinfo{author}{\bibfnamefont{Z.}~\bibnamefont{Qiao}}, and
  \bibinfo{author}{\bibfnamefont{Y.}~\bibnamefont{Yao}}, \bibinfo{year}{2014},
  \bibinfo{journal}{Physical Review Letters}
  \textbf{\bibinfo{volume}{112}}(\bibinfo{number}{10}),
  \bibinfo{pages}{106802}.

\bibitem[{\citenamefont{Pan} \emph{et~al.}(2022)\citenamefont{Pan, Xie, Wu, and
  Sarma}}]{pan2022topological}
\bibinfo{author}{\bibnamefont{Pan}, \bibfnamefont{H.}},
  \bibinfo{author}{\bibfnamefont{M.}~\bibnamefont{Xie}},
  \bibinfo{author}{\bibfnamefont{F.}~\bibnamefont{Wu}}, and
  \bibinfo{author}{\bibfnamefont{S.~D.} \bibnamefont{Sarma}},
  \bibinfo{year}{2022}, \bibinfo{journal}{Physical Review Letters}
  \textbf{\bibinfo{volume}{129}}(\bibinfo{number}{5}), \bibinfo{pages}{056804}.

\bibitem[{\citenamefont{Pan} \emph{et~al.}(2020)\citenamefont{Pan, Yu, Zhang,
  Du, Janotti, Liu, and Yan}}]{pan2020quantum}
\bibinfo{author}{\bibnamefont{Pan}, \bibfnamefont{J.}},
  \bibinfo{author}{\bibfnamefont{J.}~\bibnamefont{Yu}},
  \bibinfo{author}{\bibfnamefont{Y.-F.} \bibnamefont{Zhang}},
  \bibinfo{author}{\bibfnamefont{S.}~\bibnamefont{Du}},
  \bibinfo{author}{\bibfnamefont{A.}~\bibnamefont{Janotti}},
  \bibinfo{author}{\bibfnamefont{C.-X.} \bibnamefont{Liu}}, and
  \bibinfo{author}{\bibfnamefont{Q.}~\bibnamefont{Yan}}, \bibinfo{year}{2020},
  \bibinfo{journal}{npj Computational Materials}
  \textbf{\bibinfo{volume}{6}}(\bibinfo{number}{1}), \bibinfo{pages}{1}.

\bibitem[{\citenamefont{Peccei and Quinn}(1977)}]{peccei1977cp}
\bibinfo{author}{\bibnamefont{Peccei}, \bibfnamefont{R.~D.}}, and
  \bibinfo{author}{\bibfnamefont{H.~R.} \bibnamefont{Quinn}},
  \bibinfo{year}{1977}, \bibinfo{journal}{Physical Review Letters}
  \textbf{\bibinfo{volume}{38}}(\bibinfo{number}{25}), \bibinfo{pages}{1440}.

\bibitem[{\citenamefont{Peixoto} \emph{et~al.}(2016)\citenamefont{Peixoto,
  Bentmann, Schreyeck, Winnerlein, Seibel, Maa{\ss}, Al-Baidhani, Treiber,
  Schatz, Grauer} \emph{et~al.}}]{peixoto2016impurity}
\bibinfo{author}{\bibnamefont{Peixoto}, \bibfnamefont{T.~R.}},
  \bibinfo{author}{\bibfnamefont{H.}~\bibnamefont{Bentmann}},
  \bibinfo{author}{\bibfnamefont{S.}~\bibnamefont{Schreyeck}},
  \bibinfo{author}{\bibfnamefont{M.}~\bibnamefont{Winnerlein}},
  \bibinfo{author}{\bibfnamefont{C.}~\bibnamefont{Seibel}},
  \bibinfo{author}{\bibfnamefont{H.}~\bibnamefont{Maa{\ss}}},
  \bibinfo{author}{\bibfnamefont{M.}~\bibnamefont{Al-Baidhani}},
  \bibinfo{author}{\bibfnamefont{K.}~\bibnamefont{Treiber}},
  \bibinfo{author}{\bibfnamefont{S.}~\bibnamefont{Schatz}},
  \bibinfo{author}{\bibfnamefont{S.}~\bibnamefont{Grauer}}, \emph{et~al.},
  \bibinfo{year}{2016}, \bibinfo{journal}{Physical Review B}
  \textbf{\bibinfo{volume}{94}}(\bibinfo{number}{19}), \bibinfo{pages}{195140}.

\bibitem[{\citenamefont{Pham and Ganesh}(2020)}]{pham2020designing}
\bibinfo{author}{\bibnamefont{Pham}, \bibfnamefont{A.}}, and
  \bibinfo{author}{\bibfnamefont{P.}~\bibnamefont{Ganesh}},
  \bibinfo{year}{2020}, \bibinfo{journal}{arXiv:2003.05840} .

\bibitem[{\citenamefont{Pierce} \emph{et~al.}(2021)\citenamefont{Pierce, Xie,
  Park, Khalaf, Lee, Cao, Parker, Forrester, Chen, Watanabe}
  \emph{et~al.}}]{pierce2021unconventional}
\bibinfo{author}{\bibnamefont{Pierce}, \bibfnamefont{A.~T.}},
  \bibinfo{author}{\bibfnamefont{Y.}~\bibnamefont{Xie}},
  \bibinfo{author}{\bibfnamefont{J.~M.} \bibnamefont{Park}},
  \bibinfo{author}{\bibfnamefont{E.}~\bibnamefont{Khalaf}},
  \bibinfo{author}{\bibfnamefont{S.~H.} \bibnamefont{Lee}},
  \bibinfo{author}{\bibfnamefont{Y.}~\bibnamefont{Cao}},
  \bibinfo{author}{\bibfnamefont{D.~E.} \bibnamefont{Parker}},
  \bibinfo{author}{\bibfnamefont{P.~R.} \bibnamefont{Forrester}},
  \bibinfo{author}{\bibfnamefont{S.}~\bibnamefont{Chen}},
  \bibinfo{author}{\bibfnamefont{K.}~\bibnamefont{Watanabe}}, \emph{et~al.},
  \bibinfo{year}{2021}, \bibinfo{journal}{Nature Physics}
  \textbf{\bibinfo{volume}{17}}(\bibinfo{number}{11}), \bibinfo{pages}{1210}.

\bibitem[{\citenamefont{Pikulin and Hyart}(2014)}]{pikulin2014interplay}
\bibinfo{author}{\bibnamefont{Pikulin}, \bibfnamefont{D.}}, and
  \bibinfo{author}{\bibfnamefont{T.}~\bibnamefont{Hyart}},
  \bibinfo{year}{2014}, \bibinfo{journal}{Physical Review Letters}
  \textbf{\bibinfo{volume}{112}}(\bibinfo{number}{17}),
  \bibinfo{pages}{176403}.

\bibitem[{\citenamefont{Po} \emph{et~al.}(2019)\citenamefont{Po, Zou, Senthil,
  and Vishwanath}}]{po2019faithful}
\bibinfo{author}{\bibnamefont{Po}, \bibfnamefont{H.~C.}},
  \bibinfo{author}{\bibfnamefont{L.}~\bibnamefont{Zou}},
  \bibinfo{author}{\bibfnamefont{T.}~\bibnamefont{Senthil}}, and
  \bibinfo{author}{\bibfnamefont{A.}~\bibnamefont{Vishwanath}},
  \bibinfo{year}{2019}, \bibinfo{journal}{Physical Review B}
  \textbf{\bibinfo{volume}{99}}(\bibinfo{number}{19}), \bibinfo{pages}{195455}.

\bibitem[{\citenamefont{Po} \emph{et~al.}(2018)\citenamefont{Po, Zou,
  Vishwanath, and Senthil}}]{po2018origin}
\bibinfo{author}{\bibnamefont{Po}, \bibfnamefont{H.~C.}},
  \bibinfo{author}{\bibfnamefont{L.}~\bibnamefont{Zou}},
  \bibinfo{author}{\bibfnamefont{A.}~\bibnamefont{Vishwanath}}, and
  \bibinfo{author}{\bibfnamefont{T.}~\bibnamefont{Senthil}},
  \bibinfo{year}{2018}, \bibinfo{journal}{Physical Review X}
  \textbf{\bibinfo{volume}{8}}(\bibinfo{number}{3}), \bibinfo{pages}{031089}.

\bibitem[{\citenamefont{Poirier and Schopfer}(2009)}]{poirier2009resistance}
\bibinfo{author}{\bibnamefont{Poirier}, \bibfnamefont{W.}}, and
  \bibinfo{author}{\bibfnamefont{F.}~\bibnamefont{Schopfer}},
  \bibinfo{year}{2009}, \bibinfo{journal}{The European Physical Journal Special
  Topics} \textbf{\bibinfo{volume}{172}}(\bibinfo{number}{1}),
  \bibinfo{pages}{207}.

\bibitem[{\citenamefont{Polshyn} \emph{et~al.}(2020)\citenamefont{Polshyn, Zhu,
  Kumar, Zhang, Yang, Tschirhart, Serlin, Watanabe, Taniguchi, MacDonald}
  \emph{et~al.}}]{polshyn2020electrical}
\bibinfo{author}{\bibnamefont{Polshyn}, \bibfnamefont{H.}},
  \bibinfo{author}{\bibfnamefont{J.}~\bibnamefont{Zhu}},
  \bibinfo{author}{\bibfnamefont{M.~A.} \bibnamefont{Kumar}},
  \bibinfo{author}{\bibfnamefont{Y.}~\bibnamefont{Zhang}},
  \bibinfo{author}{\bibfnamefont{F.}~\bibnamefont{Yang}},
  \bibinfo{author}{\bibfnamefont{C.~L.} \bibnamefont{Tschirhart}},
  \bibinfo{author}{\bibfnamefont{M.}~\bibnamefont{Serlin}},
  \bibinfo{author}{\bibfnamefont{K.}~\bibnamefont{Watanabe}},
  \bibinfo{author}{\bibfnamefont{T.}~\bibnamefont{Taniguchi}},
  \bibinfo{author}{\bibfnamefont{A.~H.} \bibnamefont{MacDonald}},
  \emph{et~al.}, \bibinfo{year}{2020}, \bibinfo{journal}{Nature}
  \textbf{\bibinfo{volume}{588}}(\bibinfo{number}{7836}), \bibinfo{pages}{66}.

\bibitem[{\citenamefont{Pournaghavi}
  \emph{et~al.}(2021)\citenamefont{Pournaghavi, Pertsova, MacDonald, and
  Canali}}]{pournaghavi2021nonlocal}
\bibinfo{author}{\bibnamefont{Pournaghavi}, \bibfnamefont{N.}},
  \bibinfo{author}{\bibfnamefont{A.}~\bibnamefont{Pertsova}},
  \bibinfo{author}{\bibfnamefont{A.}~\bibnamefont{MacDonald}}, and
  \bibinfo{author}{\bibfnamefont{C.~M.} \bibnamefont{Canali}},
  \bibinfo{year}{2021}, \bibinfo{journal}{Physical Review B}
  \textbf{\bibinfo{volume}{104}}(\bibinfo{number}{20}),
  \bibinfo{pages}{L201102}.

\bibitem[{\citenamefont{Qi} \emph{et~al.}(2008)\citenamefont{Qi, Hughes, and
  Zhang}}]{qi2008topological}
\bibinfo{author}{\bibnamefont{Qi}, \bibfnamefont{X.-L.}},
  \bibinfo{author}{\bibfnamefont{T.~L.} \bibnamefont{Hughes}}, and
  \bibinfo{author}{\bibfnamefont{S.-C.} \bibnamefont{Zhang}},
  \bibinfo{year}{2008}, \bibinfo{journal}{Physical Review B}
  \textbf{\bibinfo{volume}{78}}(\bibinfo{number}{19}), \bibinfo{pages}{195424}.

\bibitem[{\citenamefont{Qi} \emph{et~al.}(2010)\citenamefont{Qi, Hughes, and
  Zhang}}]{qi2010chiral}
\bibinfo{author}{\bibnamefont{Qi}, \bibfnamefont{X.-L.}},
  \bibinfo{author}{\bibfnamefont{T.~L.} \bibnamefont{Hughes}}, and
  \bibinfo{author}{\bibfnamefont{S.-C.} \bibnamefont{Zhang}},
  \bibinfo{year}{2010}, \bibinfo{journal}{Physical Review B}
  \textbf{\bibinfo{volume}{82}}(\bibinfo{number}{18}), \bibinfo{pages}{184516}.

\bibitem[{\citenamefont{Qi} \emph{et~al.}(2006)\citenamefont{Qi, Wu, and
  Zhang}}]{qi2006topological}
\bibinfo{author}{\bibnamefont{Qi}, \bibfnamefont{X.-L.}},
  \bibinfo{author}{\bibfnamefont{Y.-S.} \bibnamefont{Wu}}, and
  \bibinfo{author}{\bibfnamefont{S.-C.} \bibnamefont{Zhang}},
  \bibinfo{year}{2006}, \bibinfo{journal}{Physical Review B}
  \textbf{\bibinfo{volume}{74}}(\bibinfo{number}{8}), \bibinfo{pages}{085308}.

\bibitem[{\citenamefont{Qi and Zhang}(2011)}]{qi2011topological}
\bibinfo{author}{\bibnamefont{Qi}, \bibfnamefont{X.-L.}}, and
  \bibinfo{author}{\bibfnamefont{S.-C.} \bibnamefont{Zhang}},
  \bibinfo{year}{2011}, \bibinfo{journal}{Reviews of Modern Physics}
  \textbf{\bibinfo{volume}{83}}(\bibinfo{number}{4}), \bibinfo{pages}{1057}.

\bibitem[{\citenamefont{Qian} \emph{et~al.}(2019)\citenamefont{Qian, Zhang, Wu,
  and Lin}}]{qian2019robust}
\bibinfo{author}{\bibnamefont{Qian}, \bibfnamefont{J.}},
  \bibinfo{author}{\bibfnamefont{J.}~\bibnamefont{Zhang}},
  \bibinfo{author}{\bibfnamefont{Q.}~\bibnamefont{Wu}}, and
  \bibinfo{author}{\bibfnamefont{Z.}~\bibnamefont{Lin}}, \bibinfo{year}{2019},
  \bibinfo{journal}{Applied Physics Letters}
  \textbf{\bibinfo{volume}{114}}(\bibinfo{number}{5}), \bibinfo{pages}{053105}.

\bibitem[{\citenamefont{Qiao} \emph{et~al.}(2016)\citenamefont{Qiao, Han,
  Zhang, Wang, Deng, Jiang, Yang, Wang, and Niu}}]{qiao2016anderson}
\bibinfo{author}{\bibnamefont{Qiao}, \bibfnamefont{Z.}},
  \bibinfo{author}{\bibfnamefont{Y.}~\bibnamefont{Han}},
  \bibinfo{author}{\bibfnamefont{L.}~\bibnamefont{Zhang}},
  \bibinfo{author}{\bibfnamefont{K.}~\bibnamefont{Wang}},
  \bibinfo{author}{\bibfnamefont{X.}~\bibnamefont{Deng}},
  \bibinfo{author}{\bibfnamefont{H.}~\bibnamefont{Jiang}},
  \bibinfo{author}{\bibfnamefont{S.~A.} \bibnamefont{Yang}},
  \bibinfo{author}{\bibfnamefont{J.}~\bibnamefont{Wang}}, and
  \bibinfo{author}{\bibfnamefont{Q.}~\bibnamefont{Niu}}, \bibinfo{year}{2016},
  \bibinfo{journal}{Physical Review Letters}
  \textbf{\bibinfo{volume}{117}}(\bibinfo{number}{5}), \bibinfo{pages}{056802}.

\bibitem[{\citenamefont{Qiao} \emph{et~al.}(2012)\citenamefont{Qiao, Jiang, Li,
  Yao, and Niu}}]{qiao2012microscopic}
\bibinfo{author}{\bibnamefont{Qiao}, \bibfnamefont{Z.}},
  \bibinfo{author}{\bibfnamefont{H.}~\bibnamefont{Jiang}},
  \bibinfo{author}{\bibfnamefont{X.}~\bibnamefont{Li}},
  \bibinfo{author}{\bibfnamefont{Y.}~\bibnamefont{Yao}}, and
  \bibinfo{author}{\bibfnamefont{Q.}~\bibnamefont{Niu}}, \bibinfo{year}{2012},
  \bibinfo{journal}{Physical Review B}
  \textbf{\bibinfo{volume}{85}}(\bibinfo{number}{11}), \bibinfo{pages}{115439}.

\bibitem[{\citenamefont{Qiao} \emph{et~al.}(2010)\citenamefont{Qiao, Yang,
  Feng, Tse, Ding, Yao, Wang, and Niu}}]{qiao2010quantum}
\bibinfo{author}{\bibnamefont{Qiao}, \bibfnamefont{Z.}},
  \bibinfo{author}{\bibfnamefont{S.~A.} \bibnamefont{Yang}},
  \bibinfo{author}{\bibfnamefont{W.}~\bibnamefont{Feng}},
  \bibinfo{author}{\bibfnamefont{W.-K.} \bibnamefont{Tse}},
  \bibinfo{author}{\bibfnamefont{J.}~\bibnamefont{Ding}},
  \bibinfo{author}{\bibfnamefont{Y.}~\bibnamefont{Yao}},
  \bibinfo{author}{\bibfnamefont{J.}~\bibnamefont{Wang}}, and
  \bibinfo{author}{\bibfnamefont{Q.}~\bibnamefont{Niu}}, \bibinfo{year}{2010},
  \bibinfo{journal}{Physical Review B}
  \textbf{\bibinfo{volume}{82}}(\bibinfo{number}{16}), \bibinfo{pages}{161414}.

\bibitem[{\citenamefont{Quhe} \emph{et~al.}(2012)\citenamefont{Quhe, Zheng,
  Luo, Liu, Qin, Zhou, Yu, Nagase, Mei, Gao} \emph{et~al.}}]{quhe2012tunable}
\bibinfo{author}{\bibnamefont{Quhe}, \bibfnamefont{R.}},
  \bibinfo{author}{\bibfnamefont{J.}~\bibnamefont{Zheng}},
  \bibinfo{author}{\bibfnamefont{G.}~\bibnamefont{Luo}},
  \bibinfo{author}{\bibfnamefont{Q.}~\bibnamefont{Liu}},
  \bibinfo{author}{\bibfnamefont{R.}~\bibnamefont{Qin}},
  \bibinfo{author}{\bibfnamefont{J.}~\bibnamefont{Zhou}},
  \bibinfo{author}{\bibfnamefont{D.}~\bibnamefont{Yu}},
  \bibinfo{author}{\bibfnamefont{S.}~\bibnamefont{Nagase}},
  \bibinfo{author}{\bibfnamefont{W.-N.} \bibnamefont{Mei}},
  \bibinfo{author}{\bibfnamefont{Z.}~\bibnamefont{Gao}}, \emph{et~al.},
  \bibinfo{year}{2012}, \bibinfo{journal}{NPG Asia Materials}
  \textbf{\bibinfo{volume}{4}}(\bibinfo{number}{2}), \bibinfo{pages}{e6}.

\bibitem[{\citenamefont{Raghu} \emph{et~al.}(2008)\citenamefont{Raghu, Qi,
  Honerkamp, and Zhang}}]{raghu2008topological}
\bibinfo{author}{\bibnamefont{Raghu}, \bibfnamefont{S.}},
  \bibinfo{author}{\bibfnamefont{X.-L.} \bibnamefont{Qi}},
  \bibinfo{author}{\bibfnamefont{C.}~\bibnamefont{Honerkamp}}, and
  \bibinfo{author}{\bibfnamefont{S.-C.} \bibnamefont{Zhang}},
  \bibinfo{year}{2008}, \bibinfo{journal}{Physical Review Letters}
  \textbf{\bibinfo{volume}{100}}(\bibinfo{number}{15}),
  \bibinfo{pages}{156401}.

\bibitem[{\citenamefont{Repellin and Senthil}(2020)}]{repellin2020chern}
\bibinfo{author}{\bibnamefont{Repellin}, \bibfnamefont{C.}}, and
  \bibinfo{author}{\bibfnamefont{T.}~\bibnamefont{Senthil}},
  \bibinfo{year}{2020}, \bibinfo{journal}{Physical Review Research}
  \textbf{\bibinfo{volume}{2}}(\bibinfo{number}{2}), \bibinfo{pages}{023238}.

\bibitem[{\citenamefont{Ribeiro-Palau}
  \emph{et~al.}(2018)\citenamefont{Ribeiro-Palau, Zhang, Watanabe, Taniguchi,
  Hone, and Dean}}]{ribeiro2018twistable}
\bibinfo{author}{\bibnamefont{Ribeiro-Palau}, \bibfnamefont{R.}},
  \bibinfo{author}{\bibfnamefont{C.}~\bibnamefont{Zhang}},
  \bibinfo{author}{\bibfnamefont{K.}~\bibnamefont{Watanabe}},
  \bibinfo{author}{\bibfnamefont{T.}~\bibnamefont{Taniguchi}},
  \bibinfo{author}{\bibfnamefont{J.}~\bibnamefont{Hone}}, and
  \bibinfo{author}{\bibfnamefont{C.~R.} \bibnamefont{Dean}},
  \bibinfo{year}{2018}, \bibinfo{journal}{Science}
  \textbf{\bibinfo{volume}{361}}(\bibinfo{number}{6403}), \bibinfo{pages}{690}.

\bibitem[{\citenamefont{Rienks} \emph{et~al.}(2019)\citenamefont{Rienks,
  Wimmer, S{\'a}nchez-Barriga, Caha, Mandal, Ruzicka, Ney, Steiner, Volobuev,
  Groi{\ss}} \emph{et~al.}}]{rienks2019large}
\bibinfo{author}{\bibnamefont{Rienks}, \bibfnamefont{E.~D.}},
  \bibinfo{author}{\bibfnamefont{S.}~\bibnamefont{Wimmer}},
  \bibinfo{author}{\bibfnamefont{J.}~\bibnamefont{S{\'a}nchez-Barriga}},
  \bibinfo{author}{\bibfnamefont{O.}~\bibnamefont{Caha}},
  \bibinfo{author}{\bibfnamefont{P.~S.} \bibnamefont{Mandal}},
  \bibinfo{author}{\bibfnamefont{J.}~\bibnamefont{Ruzicka}},
  \bibinfo{author}{\bibfnamefont{A.}~\bibnamefont{Ney}},
  \bibinfo{author}{\bibfnamefont{H.}~\bibnamefont{Steiner}},
  \bibinfo{author}{\bibfnamefont{V.~V.} \bibnamefont{Volobuev}},
  \bibinfo{author}{\bibfnamefont{H.}~\bibnamefont{Groi{\ss}}}, \emph{et~al.},
  \bibinfo{year}{2019}, \bibinfo{journal}{Nature}
  \textbf{\bibinfo{volume}{576}}(\bibinfo{number}{7787}), \bibinfo{pages}{423}.

\bibitem[{\citenamefont{Rokhinson} \emph{et~al.}(2012)\citenamefont{Rokhinson,
  Liu, and Furdyna}}]{rokhinson2012fractional}
\bibinfo{author}{\bibnamefont{Rokhinson}, \bibfnamefont{L.~P.}},
  \bibinfo{author}{\bibfnamefont{X.}~\bibnamefont{Liu}}, and
  \bibinfo{author}{\bibfnamefont{J.~K.} \bibnamefont{Furdyna}},
  \bibinfo{year}{2012}, \bibinfo{journal}{Nature Physics}
  \textbf{\bibinfo{volume}{8}}(\bibinfo{number}{11}), \bibinfo{pages}{795}.

\bibitem[{\citenamefont{Rosen} \emph{et~al.}(2017)\citenamefont{Rosen, Fox,
  Kou, Pan, Wang, and Goldhaber-Gordon}}]{rosen2017chiral}
\bibinfo{author}{\bibnamefont{Rosen}, \bibfnamefont{I.~T.}},
  \bibinfo{author}{\bibfnamefont{E.~J.} \bibnamefont{Fox}},
  \bibinfo{author}{\bibfnamefont{X.}~\bibnamefont{Kou}},
  \bibinfo{author}{\bibfnamefont{L.}~\bibnamefont{Pan}},
  \bibinfo{author}{\bibfnamefont{K.~L.} \bibnamefont{Wang}}, and
  \bibinfo{author}{\bibfnamefont{D.}~\bibnamefont{Goldhaber-Gordon}},
  \bibinfo{year}{2017}, \bibinfo{journal}{npj Quantum Materials}
  \textbf{\bibinfo{volume}{2}}(\bibinfo{number}{1}), \bibinfo{pages}{1}.

\bibitem[{\citenamefont{Roy}(2009)}]{roy2009topological}
\bibinfo{author}{\bibnamefont{Roy}, \bibfnamefont{R.}}, \bibinfo{year}{2009},
  \bibinfo{journal}{Physical Review B}
  \textbf{\bibinfo{volume}{79}}(\bibinfo{number}{19}), \bibinfo{pages}{195322}.

\bibitem[{\citenamefont{Ruderman and Kittel}(1954)}]{ruderman1954indirect}
\bibinfo{author}{\bibnamefont{Ruderman}, \bibfnamefont{M.~A.}}, and
  \bibinfo{author}{\bibfnamefont{C.}~\bibnamefont{Kittel}},
  \bibinfo{year}{1954}, \bibinfo{journal}{Physical Review}
  \textbf{\bibinfo{volume}{96}}(\bibinfo{number}{1}), \bibinfo{pages}{99}.

\bibitem[{\citenamefont{Saito} \emph{et~al.}(2021)\citenamefont{Saito, Ge,
  Rademaker, Watanabe, Taniguchi, Abanin, and Young}}]{saito2021hofstadter}
\bibinfo{author}{\bibnamefont{Saito}, \bibfnamefont{Y.}},
  \bibinfo{author}{\bibfnamefont{J.}~\bibnamefont{Ge}},
  \bibinfo{author}{\bibfnamefont{L.}~\bibnamefont{Rademaker}},
  \bibinfo{author}{\bibfnamefont{K.}~\bibnamefont{Watanabe}},
  \bibinfo{author}{\bibfnamefont{T.}~\bibnamefont{Taniguchi}},
  \bibinfo{author}{\bibfnamefont{D.~A.} \bibnamefont{Abanin}}, and
  \bibinfo{author}{\bibfnamefont{A.~F.} \bibnamefont{Young}},
  \bibinfo{year}{2021}, \bibinfo{journal}{Nature Physics}
  \textbf{\bibinfo{volume}{17}}(\bibinfo{number}{4}), \bibinfo{pages}{478}.

\bibitem[{\citenamefont{San-Jose} \emph{et~al.}(2014)\citenamefont{San-Jose,
  Guti{\'e}rrez-Rubio, Sturla, and Guinea}}]{san2014spontaneous}
\bibinfo{author}{\bibnamefont{San-Jose}, \bibfnamefont{P.}},
  \bibinfo{author}{\bibfnamefont{A.}~\bibnamefont{Guti{\'e}rrez-Rubio}},
  \bibinfo{author}{\bibfnamefont{M.}~\bibnamefont{Sturla}}, and
  \bibinfo{author}{\bibfnamefont{F.}~\bibnamefont{Guinea}},
  \bibinfo{year}{2014}, \bibinfo{journal}{Physical Review B}
  \textbf{\bibinfo{volume}{90}}(\bibinfo{number}{7}), \bibinfo{pages}{075428}.

\bibitem[{\citenamefont{Sass}
  \emph{et~al.}(2020{\natexlab{a}})\citenamefont{Sass, Ge, Yan, Obeysekera,
  Yang, and Wu}}]{sass2020magnetic}
\bibinfo{author}{\bibnamefont{Sass}, \bibfnamefont{P.~M.}},
  \bibinfo{author}{\bibfnamefont{W.}~\bibnamefont{Ge}},
  \bibinfo{author}{\bibfnamefont{J.}~\bibnamefont{Yan}},
  \bibinfo{author}{\bibfnamefont{D.}~\bibnamefont{Obeysekera}},
  \bibinfo{author}{\bibfnamefont{J.}~\bibnamefont{Yang}}, and
  \bibinfo{author}{\bibfnamefont{W.}~\bibnamefont{Wu}},
  \bibinfo{year}{2020}{\natexlab{a}}, \bibinfo{journal}{Nano Letters}
  \textbf{\bibinfo{volume}{20}}(\bibinfo{number}{4}), \bibinfo{pages}{2609}.

\bibitem[{\citenamefont{Sass}
  \emph{et~al.}(2020{\natexlab{b}})\citenamefont{Sass, Kim, Vanderbilt, Yan,
  and Wu}}]{sass2020robust}
\bibinfo{author}{\bibnamefont{Sass}, \bibfnamefont{P.~M.}},
  \bibinfo{author}{\bibfnamefont{J.}~\bibnamefont{Kim}},
  \bibinfo{author}{\bibfnamefont{D.}~\bibnamefont{Vanderbilt}},
  \bibinfo{author}{\bibfnamefont{J.}~\bibnamefont{Yan}}, and
  \bibinfo{author}{\bibfnamefont{W.}~\bibnamefont{Wu}},
  \bibinfo{year}{2020}{\natexlab{b}}, \bibinfo{journal}{Physical Review
  Letters} \textbf{\bibinfo{volume}{125}}(\bibinfo{number}{3}),
  \bibinfo{pages}{037201}.

\bibitem[{\citenamefont{Sato} \emph{et~al.}(2010)\citenamefont{Sato, Bergqvist,
  Kudrnovsk{\`y}, Dederichs, Eriksson, Turek, Sanyal, Bouzerar,
  Katayama-Yoshida, Dinh} \emph{et~al.}}]{sato2010first}
\bibinfo{author}{\bibnamefont{Sato}, \bibfnamefont{K.}},
  \bibinfo{author}{\bibfnamefont{L.}~\bibnamefont{Bergqvist}},
  \bibinfo{author}{\bibfnamefont{J.}~\bibnamefont{Kudrnovsk{\`y}}},
  \bibinfo{author}{\bibfnamefont{P.~H.} \bibnamefont{Dederichs}},
  \bibinfo{author}{\bibfnamefont{O.}~\bibnamefont{Eriksson}},
  \bibinfo{author}{\bibfnamefont{I.}~\bibnamefont{Turek}},
  \bibinfo{author}{\bibfnamefont{B.}~\bibnamefont{Sanyal}},
  \bibinfo{author}{\bibfnamefont{G.}~\bibnamefont{Bouzerar}},
  \bibinfo{author}{\bibfnamefont{H.}~\bibnamefont{Katayama-Yoshida}},
  \bibinfo{author}{\bibfnamefont{V.}~\bibnamefont{Dinh}}, \emph{et~al.},
  \bibinfo{year}{2010}, \bibinfo{journal}{Reviews of Modern Physics}
  \textbf{\bibinfo{volume}{82}}(\bibinfo{number}{2}), \bibinfo{pages}{1633}.

\bibitem[{\citenamefont{Schopfer and Poirier}(2007)}]{schopfer2007testing}
\bibinfo{author}{\bibnamefont{Schopfer}, \bibfnamefont{F.}}, and
  \bibinfo{author}{\bibfnamefont{W.}~\bibnamefont{Poirier}},
  \bibinfo{year}{2007}, \bibinfo{journal}{Journal of Applied Physics}
  \textbf{\bibinfo{volume}{102}}(\bibinfo{number}{5}), \bibinfo{pages}{054903}.

\bibitem[{\citenamefont{Sekine and Nomura}(2021)}]{sekine2021axion}
\bibinfo{author}{\bibnamefont{Sekine}, \bibfnamefont{A.}}, and
  \bibinfo{author}{\bibfnamefont{K.}~\bibnamefont{Nomura}},
  \bibinfo{year}{2021}, \bibinfo{journal}{Journal of Applied Physics}
  \textbf{\bibinfo{volume}{129}}(\bibinfo{number}{14}),
  \bibinfo{pages}{141101}.

\bibitem[{\citenamefont{Semenoff}(1984)}]{semenoff1984condensed}
\bibinfo{author}{\bibnamefont{Semenoff}, \bibfnamefont{G.~W.}},
  \bibinfo{year}{1984}, \bibinfo{journal}{Physical Review Letters}
  \textbf{\bibinfo{volume}{53}}(\bibinfo{number}{26}), \bibinfo{pages}{2449}.

\bibitem[{\citenamefont{Seo} \emph{et~al.}(2019)\citenamefont{Seo, Kotov, and
  Uchoa}}]{seo2019ferromagnetic}
\bibinfo{author}{\bibnamefont{Seo}, \bibfnamefont{K.}},
  \bibinfo{author}{\bibfnamefont{V.~N.} \bibnamefont{Kotov}}, and
  \bibinfo{author}{\bibfnamefont{B.}~\bibnamefont{Uchoa}},
  \bibinfo{year}{2019}, \bibinfo{journal}{Physical Review Letters}
  \textbf{\bibinfo{volume}{122}}(\bibinfo{number}{24}),
  \bibinfo{pages}{246402}.

\bibitem[{\citenamefont{Serlin} \emph{et~al.}(2020)\citenamefont{Serlin,
  Tschirhart, Polshyn, Zhang, Zhu, Watanabe, Taniguchi, Balents, and
  Young}}]{serlin2020intrinsic}
\bibinfo{author}{\bibnamefont{Serlin}, \bibfnamefont{M.}},
  \bibinfo{author}{\bibfnamefont{C.}~\bibnamefont{Tschirhart}},
  \bibinfo{author}{\bibfnamefont{H.}~\bibnamefont{Polshyn}},
  \bibinfo{author}{\bibfnamefont{Y.}~\bibnamefont{Zhang}},
  \bibinfo{author}{\bibfnamefont{J.}~\bibnamefont{Zhu}},
  \bibinfo{author}{\bibfnamefont{K.}~\bibnamefont{Watanabe}},
  \bibinfo{author}{\bibfnamefont{T.}~\bibnamefont{Taniguchi}},
  \bibinfo{author}{\bibfnamefont{L.}~\bibnamefont{Balents}}, and
  \bibinfo{author}{\bibfnamefont{A.}~\bibnamefont{Young}},
  \bibinfo{year}{2020}, \bibinfo{journal}{Science}
  \textbf{\bibinfo{volume}{367}}(\bibinfo{number}{6480}), \bibinfo{pages}{900}.

\bibitem[{\citenamefont{Sharpe} \emph{et~al.}(2019)\citenamefont{Sharpe, Fox,
  Barnard, Finney, Watanabe, Taniguchi, Kastner, and
  Goldhaber-Gordon}}]{sharpe2019emergent}
\bibinfo{author}{\bibnamefont{Sharpe}, \bibfnamefont{A.~L.}},
  \bibinfo{author}{\bibfnamefont{E.~J.} \bibnamefont{Fox}},
  \bibinfo{author}{\bibfnamefont{A.~W.} \bibnamefont{Barnard}},
  \bibinfo{author}{\bibfnamefont{J.}~\bibnamefont{Finney}},
  \bibinfo{author}{\bibfnamefont{K.}~\bibnamefont{Watanabe}},
  \bibinfo{author}{\bibfnamefont{T.}~\bibnamefont{Taniguchi}},
  \bibinfo{author}{\bibfnamefont{M.}~\bibnamefont{Kastner}}, and
  \bibinfo{author}{\bibfnamefont{D.}~\bibnamefont{Goldhaber-Gordon}},
  \bibinfo{year}{2019}, \bibinfo{journal}{Science}
  \textbf{\bibinfo{volume}{365}}(\bibinfo{number}{6453}), \bibinfo{pages}{605}.

\bibitem[{\citenamefont{Sharpe} \emph{et~al.}(2021)\citenamefont{Sharpe, Fox,
  Barnard, Finney, Watanabe, Taniguchi, Kastner, and
  Goldhaber-Gordon}}]{sharpe2021evidence}
\bibinfo{author}{\bibnamefont{Sharpe}, \bibfnamefont{A.~L.}},
  \bibinfo{author}{\bibfnamefont{E.~J.} \bibnamefont{Fox}},
  \bibinfo{author}{\bibfnamefont{A.~W.} \bibnamefont{Barnard}},
  \bibinfo{author}{\bibfnamefont{J.}~\bibnamefont{Finney}},
  \bibinfo{author}{\bibfnamefont{K.}~\bibnamefont{Watanabe}},
  \bibinfo{author}{\bibfnamefont{T.}~\bibnamefont{Taniguchi}},
  \bibinfo{author}{\bibfnamefont{M.~A.} \bibnamefont{Kastner}}, and
  \bibinfo{author}{\bibfnamefont{D.}~\bibnamefont{Goldhaber-Gordon}},
  \bibinfo{year}{2021}, \bibinfo{journal}{Nano Letters} .

\bibitem[{\citenamefont{Shi} \emph{et~al.}(2007)\citenamefont{Shi, Vignale,
  Xiao, and Niu}}]{shi2007quantum}
\bibinfo{author}{\bibnamefont{Shi}, \bibfnamefont{J.}},
  \bibinfo{author}{\bibfnamefont{G.}~\bibnamefont{Vignale}},
  \bibinfo{author}{\bibfnamefont{D.}~\bibnamefont{Xiao}}, and
  \bibinfo{author}{\bibfnamefont{Q.}~\bibnamefont{Niu}}, \bibinfo{year}{2007},
  \bibinfo{journal}{Physical Review Letters}
  \textbf{\bibinfo{volume}{99}}(\bibinfo{number}{19}), \bibinfo{pages}{197202}.

\bibitem[{\citenamefont{Shi} \emph{et~al.}(2021)\citenamefont{Shi, Zhu, and
  MacDonald}}]{shi2021moire}
\bibinfo{author}{\bibnamefont{Shi}, \bibfnamefont{J.}},
  \bibinfo{author}{\bibfnamefont{J.}~\bibnamefont{Zhu}}, and
  \bibinfo{author}{\bibfnamefont{A.}~\bibnamefont{MacDonald}},
  \bibinfo{year}{2021}, \bibinfo{journal}{Physical Review B}
  \textbf{\bibinfo{volume}{103}}(\bibinfo{number}{7}), \bibinfo{pages}{075122}.

\bibitem[{\citenamefont{Shi} \emph{et~al.}(2019)\citenamefont{Shi, Lei, Zhu,
  Ma, Cui, Sun, Ying, and Chen}}]{shi2019magnetic}
\bibinfo{author}{\bibnamefont{Shi}, \bibfnamefont{M.}},
  \bibinfo{author}{\bibfnamefont{B.}~\bibnamefont{Lei}},
  \bibinfo{author}{\bibfnamefont{C.}~\bibnamefont{Zhu}},
  \bibinfo{author}{\bibfnamefont{D.}~\bibnamefont{Ma}},
  \bibinfo{author}{\bibfnamefont{J.}~\bibnamefont{Cui}},
  \bibinfo{author}{\bibfnamefont{Z.}~\bibnamefont{Sun}},
  \bibinfo{author}{\bibfnamefont{J.}~\bibnamefont{Ying}}, and
  \bibinfo{author}{\bibfnamefont{X.}~\bibnamefont{Chen}}, \bibinfo{year}{2019},
  \bibinfo{journal}{Physical Review B}
  \textbf{\bibinfo{volume}{100}}(\bibinfo{number}{15}),
  \bibinfo{pages}{155144}.

\bibitem[{\citenamefont{Singh and Deshmukh}(2009)}]{singh2009nonequilibrium}
\bibinfo{author}{\bibnamefont{Singh}, \bibfnamefont{V.}}, and
  \bibinfo{author}{\bibfnamefont{M.~M.} \bibnamefont{Deshmukh}},
  \bibinfo{year}{2009}, \bibinfo{journal}{Physical Review B}
  \textbf{\bibinfo{volume}{80}}(\bibinfo{number}{8}), \bibinfo{pages}{081404}.

\bibitem[{\citenamefont{{\'S}liwa} \emph{et~al.}(2021)\citenamefont{{\'S}liwa,
  Autieri, Majewski, and Dietl}}]{sliwa2021superexchange}
\bibinfo{author}{\bibnamefont{{\'S}liwa}, \bibfnamefont{C.}},
  \bibinfo{author}{\bibfnamefont{C.}~\bibnamefont{Autieri}},
  \bibinfo{author}{\bibfnamefont{J.~A.} \bibnamefont{Majewski}}, and
  \bibinfo{author}{\bibfnamefont{T.}~\bibnamefont{Dietl}},
  \bibinfo{year}{2021}, \bibinfo{journal}{Physical Review B}
  \textbf{\bibinfo{volume}{104}}(\bibinfo{number}{22}),
  \bibinfo{pages}{L220404}.

\bibitem[{\citenamefont{Song} \emph{et~al.}(2019)\citenamefont{Song, Wang, Shi,
  Li, Fang, and Bernevig}}]{song2019all}
\bibinfo{author}{\bibnamefont{Song}, \bibfnamefont{Z.}},
  \bibinfo{author}{\bibfnamefont{Z.}~\bibnamefont{Wang}},
  \bibinfo{author}{\bibfnamefont{W.}~\bibnamefont{Shi}},
  \bibinfo{author}{\bibfnamefont{G.}~\bibnamefont{Li}},
  \bibinfo{author}{\bibfnamefont{C.}~\bibnamefont{Fang}}, and
  \bibinfo{author}{\bibfnamefont{B.~A.} \bibnamefont{Bernevig}},
  \bibinfo{year}{2019}, \bibinfo{journal}{Physical Review Letters}
  \textbf{\bibinfo{volume}{123}}(\bibinfo{number}{3}), \bibinfo{pages}{036401}.

\bibitem[{\citenamefont{Song}
  \emph{et~al.}(2021{\natexlab{a}})\citenamefont{Song, Lian, Queiroz, Ilan,
  Bernevig, and Stern}}]{song2021delocalization}
\bibinfo{author}{\bibnamefont{Song}, \bibfnamefont{Z.-D.}},
  \bibinfo{author}{\bibfnamefont{B.}~\bibnamefont{Lian}},
  \bibinfo{author}{\bibfnamefont{R.}~\bibnamefont{Queiroz}},
  \bibinfo{author}{\bibfnamefont{R.}~\bibnamefont{Ilan}},
  \bibinfo{author}{\bibfnamefont{B.~A.} \bibnamefont{Bernevig}}, and
  \bibinfo{author}{\bibfnamefont{A.}~\bibnamefont{Stern}},
  \bibinfo{year}{2021}{\natexlab{a}}, \bibinfo{journal}{Physical Review
  Letters} \textbf{\bibinfo{volume}{127}}(\bibinfo{number}{1}),
  \bibinfo{pages}{016602}.

\bibitem[{\citenamefont{Song}
  \emph{et~al.}(2021{\natexlab{b}})\citenamefont{Song, Lian, Regnault, and
  Bernevig}}]{song2021twisted}
\bibinfo{author}{\bibnamefont{Song}, \bibfnamefont{Z.-D.}},
  \bibinfo{author}{\bibfnamefont{B.}~\bibnamefont{Lian}},
  \bibinfo{author}{\bibfnamefont{N.}~\bibnamefont{Regnault}}, and
  \bibinfo{author}{\bibfnamefont{B.~A.} \bibnamefont{Bernevig}},
  \bibinfo{year}{2021}{\natexlab{b}}, \bibinfo{journal}{Physical Review B}
  \textbf{\bibinfo{volume}{103}}(\bibinfo{number}{20}),
  \bibinfo{pages}{205412}.

\bibitem[{\citenamefont{Stormer} \emph{et~al.}(1999)\citenamefont{Stormer,
  Tsui, and Gossard}}]{stormer1999fractional}
\bibinfo{author}{\bibnamefont{Stormer}, \bibfnamefont{H.~L.}},
  \bibinfo{author}{\bibfnamefont{D.~C.} \bibnamefont{Tsui}}, and
  \bibinfo{author}{\bibfnamefont{A.~C.} \bibnamefont{Gossard}},
  \bibinfo{year}{1999}, \bibinfo{journal}{Reviews of Modern Physics}
  \textbf{\bibinfo{volume}{71}}(\bibinfo{number}{2}), \bibinfo{pages}{S298}.

\bibitem[{\citenamefont{Streda}(1982)}]{streda1982theory}
\bibinfo{author}{\bibnamefont{Streda}, \bibfnamefont{P.}},
  \bibinfo{year}{1982}, \bibinfo{journal}{Journal of Physics C: Solid State
  Physics} \textbf{\bibinfo{volume}{15}}(\bibinfo{number}{22}),
  \bibinfo{pages}{L717}.

\bibitem[{\citenamefont{Sun} \emph{et~al.}(2011)\citenamefont{Sun, Gu, Katsura,
  and Sarma}}]{sun2011nearly}
\bibinfo{author}{\bibnamefont{Sun}, \bibfnamefont{K.}},
  \bibinfo{author}{\bibfnamefont{Z.}~\bibnamefont{Gu}},
  \bibinfo{author}{\bibfnamefont{H.}~\bibnamefont{Katsura}}, and
  \bibinfo{author}{\bibfnamefont{S.~D.} \bibnamefont{Sarma}},
  \bibinfo{year}{2011}, \bibinfo{journal}{Physical Review Letters}
  \textbf{\bibinfo{volume}{106}}(\bibinfo{number}{23}),
  \bibinfo{pages}{236803}.

\bibitem[{\citenamefont{Swatek} \emph{et~al.}(2020)\citenamefont{Swatek, Wu,
  Wang, Lee, Schrunk, Yan, and Kaminski}}]{swatek2020gapless}
\bibinfo{author}{\bibnamefont{Swatek}, \bibfnamefont{P.}},
  \bibinfo{author}{\bibfnamefont{Y.}~\bibnamefont{Wu}},
  \bibinfo{author}{\bibfnamefont{L.-L.} \bibnamefont{Wang}},
  \bibinfo{author}{\bibfnamefont{K.}~\bibnamefont{Lee}},
  \bibinfo{author}{\bibfnamefont{B.}~\bibnamefont{Schrunk}},
  \bibinfo{author}{\bibfnamefont{J.}~\bibnamefont{Yan}}, and
  \bibinfo{author}{\bibfnamefont{A.}~\bibnamefont{Kaminski}},
  \bibinfo{year}{2020}, \bibinfo{journal}{Physical Review B}
  \textbf{\bibinfo{volume}{101}}(\bibinfo{number}{16}),
  \bibinfo{pages}{161109}.

\bibitem[{\citenamefont{Tai} \emph{et~al.}(2021)\citenamefont{Tai, Chong,
  Zhang, Zhang, Deng, Eckberg, Qiu, Dai, He, Wu}
  \emph{et~al.}}]{tai2021polarity}
\bibinfo{author}{\bibnamefont{Tai}, \bibfnamefont{L.}},
  \bibinfo{author}{\bibfnamefont{S.~K.} \bibnamefont{Chong}},
  \bibinfo{author}{\bibfnamefont{H.}~\bibnamefont{Zhang}},
  \bibinfo{author}{\bibfnamefont{P.}~\bibnamefont{Zhang}},
  \bibinfo{author}{\bibfnamefont{P.}~\bibnamefont{Deng}},
  \bibinfo{author}{\bibfnamefont{C.}~\bibnamefont{Eckberg}},
  \bibinfo{author}{\bibfnamefont{G.}~\bibnamefont{Qiu}},
  \bibinfo{author}{\bibfnamefont{B.}~\bibnamefont{Dai}},
  \bibinfo{author}{\bibfnamefont{H.}~\bibnamefont{He}},
  \bibinfo{author}{\bibfnamefont{D.}~\bibnamefont{Wu}}, \emph{et~al.},
  \bibinfo{year}{2021}, \bibinfo{journal}{arXiv:2103.09878} .

\bibitem[{\citenamefont{Tang} \emph{et~al.}(2017)\citenamefont{Tang, Chang,
  Zhao, Liu, Jiang, Liu, McCartney, Smith, Chen, Moodera}
  \emph{et~al.}}]{tang2017above}
\bibinfo{author}{\bibnamefont{Tang}, \bibfnamefont{C.}},
  \bibinfo{author}{\bibfnamefont{C.-Z.} \bibnamefont{Chang}},
  \bibinfo{author}{\bibfnamefont{G.}~\bibnamefont{Zhao}},
  \bibinfo{author}{\bibfnamefont{Y.}~\bibnamefont{Liu}},
  \bibinfo{author}{\bibfnamefont{Z.}~\bibnamefont{Jiang}},
  \bibinfo{author}{\bibfnamefont{C.-X.} \bibnamefont{Liu}},
  \bibinfo{author}{\bibfnamefont{M.~R.} \bibnamefont{McCartney}},
  \bibinfo{author}{\bibfnamefont{D.~J.} \bibnamefont{Smith}},
  \bibinfo{author}{\bibfnamefont{T.}~\bibnamefont{Chen}},
  \bibinfo{author}{\bibfnamefont{J.~S.} \bibnamefont{Moodera}}, \emph{et~al.},
  \bibinfo{year}{2017}, \bibinfo{journal}{Science Advances}
  \textbf{\bibinfo{volume}{3}}(\bibinfo{number}{6}), \bibinfo{pages}{e1700307}.

\bibitem[{\citenamefont{Tang} \emph{et~al.}(2011)\citenamefont{Tang, Mei, and
  Wen}}]{tang2011high}
\bibinfo{author}{\bibnamefont{Tang}, \bibfnamefont{E.}},
  \bibinfo{author}{\bibfnamefont{J.-W.} \bibnamefont{Mei}}, and
  \bibinfo{author}{\bibfnamefont{X.-G.} \bibnamefont{Wen}},
  \bibinfo{year}{2011}, \bibinfo{journal}{Physical Review Letters}
  \textbf{\bibinfo{volume}{106}}(\bibinfo{number}{23}),
  \bibinfo{pages}{236802}.

\bibitem[{\citenamefont{Tao} \emph{et~al.}(2022)\citenamefont{Tao, Shen, Jiang,
  Li, Li, Ma, Zhao, Hu, Pistunova, Watanabe} \emph{et~al.}}]{tao2022valley}
\bibinfo{author}{\bibnamefont{Tao}, \bibfnamefont{Z.}},
  \bibinfo{author}{\bibfnamefont{B.}~\bibnamefont{Shen}},
  \bibinfo{author}{\bibfnamefont{S.}~\bibnamefont{Jiang}},
  \bibinfo{author}{\bibfnamefont{T.}~\bibnamefont{Li}},
  \bibinfo{author}{\bibfnamefont{L.}~\bibnamefont{Li}},
  \bibinfo{author}{\bibfnamefont{L.}~\bibnamefont{Ma}},
  \bibinfo{author}{\bibfnamefont{W.}~\bibnamefont{Zhao}},
  \bibinfo{author}{\bibfnamefont{J.}~\bibnamefont{Hu}},
  \bibinfo{author}{\bibfnamefont{K.}~\bibnamefont{Pistunova}},
  \bibinfo{author}{\bibfnamefont{K.}~\bibnamefont{Watanabe}}, \emph{et~al.},
  \bibinfo{year}{2022}, \bibinfo{journal}{arXiv preprint arXiv:2208.07452} .

\bibitem[{\citenamefont{Tcakaev} \emph{et~al.}(2020)\citenamefont{Tcakaev,
  Zabolotnyy, Green, Peixoto, Stier, Dettbarn, Schreyeck, Winnerlein, Vidal,
  Schatz} \emph{et~al.}}]{tcakaev2020comparing}
\bibinfo{author}{\bibnamefont{Tcakaev}, \bibfnamefont{A.}},
  \bibinfo{author}{\bibfnamefont{V.}~\bibnamefont{Zabolotnyy}},
  \bibinfo{author}{\bibfnamefont{R.}~\bibnamefont{Green}},
  \bibinfo{author}{\bibfnamefont{T.}~\bibnamefont{Peixoto}},
  \bibinfo{author}{\bibfnamefont{F.}~\bibnamefont{Stier}},
  \bibinfo{author}{\bibfnamefont{M.}~\bibnamefont{Dettbarn}},
  \bibinfo{author}{\bibfnamefont{S.}~\bibnamefont{Schreyeck}},
  \bibinfo{author}{\bibfnamefont{M.}~\bibnamefont{Winnerlein}},
  \bibinfo{author}{\bibfnamefont{R.~C.} \bibnamefont{Vidal}},
  \bibinfo{author}{\bibfnamefont{S.}~\bibnamefont{Schatz}}, \emph{et~al.},
  \bibinfo{year}{2020}, \bibinfo{journal}{Physical Review B}
  \textbf{\bibinfo{volume}{101}}(\bibinfo{number}{4}), \bibinfo{pages}{045127}.

\bibitem[{\citenamefont{Thonhauser}(2011)}]{thonhauser2011theory}
\bibinfo{author}{\bibnamefont{Thonhauser}, \bibfnamefont{T.}},
  \bibinfo{year}{2011}, \bibinfo{journal}{International Journal of Modern
  Physics B} \textbf{\bibinfo{volume}{25}}(\bibinfo{number}{11}),
  \bibinfo{pages}{1429}.

\bibitem[{\citenamefont{Thonhauser}
  \emph{et~al.}(2005)\citenamefont{Thonhauser, Ceresoli, Vanderbilt, and
  Resta}}]{thonhauser2005orbital}
\bibinfo{author}{\bibnamefont{Thonhauser}, \bibfnamefont{T.}},
  \bibinfo{author}{\bibfnamefont{D.}~\bibnamefont{Ceresoli}},
  \bibinfo{author}{\bibfnamefont{D.}~\bibnamefont{Vanderbilt}}, and
  \bibinfo{author}{\bibfnamefont{R.}~\bibnamefont{Resta}},
  \bibinfo{year}{2005}, \bibinfo{journal}{Physical Review Letters}
  \textbf{\bibinfo{volume}{95}}(\bibinfo{number}{13}), \bibinfo{pages}{137205}.

\bibitem[{\citenamefont{Thouless} \emph{et~al.}(1982)\citenamefont{Thouless,
  Kohmoto, Nightingale, and den Nijs}}]{thouless1982quantized}
\bibinfo{author}{\bibnamefont{Thouless}, \bibfnamefont{D.~J.}},
  \bibinfo{author}{\bibfnamefont{M.}~\bibnamefont{Kohmoto}},
  \bibinfo{author}{\bibfnamefont{M.~P.} \bibnamefont{Nightingale}}, and
  \bibinfo{author}{\bibfnamefont{M.}~\bibnamefont{den Nijs}},
  \bibinfo{year}{1982}, \bibinfo{journal}{Physical Review Letters}
  \textbf{\bibinfo{volume}{49}}(\bibinfo{number}{6}), \bibinfo{pages}{405}.

\bibitem[{\citenamefont{Tian} \emph{et~al.}(2020)\citenamefont{Tian, Gao, Nie,
  Qian, Gong, Fu, Li, Fan, Zhang, Kondo} \emph{et~al.}}]{tian2020magnetic}
\bibinfo{author}{\bibnamefont{Tian}, \bibfnamefont{S.}},
  \bibinfo{author}{\bibfnamefont{S.}~\bibnamefont{Gao}},
  \bibinfo{author}{\bibfnamefont{S.}~\bibnamefont{Nie}},
  \bibinfo{author}{\bibfnamefont{Y.}~\bibnamefont{Qian}},
  \bibinfo{author}{\bibfnamefont{C.}~\bibnamefont{Gong}},
  \bibinfo{author}{\bibfnamefont{Y.}~\bibnamefont{Fu}},
  \bibinfo{author}{\bibfnamefont{H.}~\bibnamefont{Li}},
  \bibinfo{author}{\bibfnamefont{W.}~\bibnamefont{Fan}},
  \bibinfo{author}{\bibfnamefont{P.}~\bibnamefont{Zhang}},
  \bibinfo{author}{\bibfnamefont{T.}~\bibnamefont{Kondo}}, \emph{et~al.},
  \bibinfo{year}{2020}, \bibinfo{journal}{Physical Review B}
  \textbf{\bibinfo{volume}{102}}(\bibinfo{number}{3}), \bibinfo{pages}{035144}.

\bibitem[{\citenamefont{Tokura} \emph{et~al.}(2019)\citenamefont{Tokura,
  Yasuda, and Tsukazaki}}]{tokura2019magnetic}
\bibinfo{author}{\bibnamefont{Tokura}, \bibfnamefont{Y.}},
  \bibinfo{author}{\bibfnamefont{K.}~\bibnamefont{Yasuda}}, and
  \bibinfo{author}{\bibfnamefont{A.}~\bibnamefont{Tsukazaki}},
  \bibinfo{year}{2019}, \bibinfo{journal}{Nature Reviews Physics}
  \textbf{\bibinfo{volume}{1}}(\bibinfo{number}{2}), \bibinfo{pages}{126}.

\bibitem[{\citenamefont{Trang} \emph{et~al.}(2021)\citenamefont{Trang, Li, Yin,
  Hwang, Akhgar, Di~Bernardo, Grubisic-Cabo, Tadich, Fuhrer, Mo}
  \emph{et~al.}}]{trang2021crossover}
\bibinfo{author}{\bibnamefont{Trang}, \bibfnamefont{C.~X.}},
  \bibinfo{author}{\bibfnamefont{Q.}~\bibnamefont{Li}},
  \bibinfo{author}{\bibfnamefont{Y.}~\bibnamefont{Yin}},
  \bibinfo{author}{\bibfnamefont{J.}~\bibnamefont{Hwang}},
  \bibinfo{author}{\bibfnamefont{G.}~\bibnamefont{Akhgar}},
  \bibinfo{author}{\bibfnamefont{I.}~\bibnamefont{Di~Bernardo}},
  \bibinfo{author}{\bibfnamefont{A.}~\bibnamefont{Grubisic-Cabo}},
  \bibinfo{author}{\bibfnamefont{A.}~\bibnamefont{Tadich}},
  \bibinfo{author}{\bibfnamefont{M.~S.} \bibnamefont{Fuhrer}},
  \bibinfo{author}{\bibfnamefont{S.-K.} \bibnamefont{Mo}}, \emph{et~al.},
  \bibinfo{year}{2021}, \bibinfo{journal}{ACS Nano}
  \textbf{\bibinfo{volume}{15}}(\bibinfo{number}{8}), \bibinfo{pages}{13444}.

\bibitem[{\citenamefont{Tschirhart}
  \emph{et~al.}(2022)\citenamefont{Tschirhart, Redekop, Li, Li, Jiang, Arp,
  Sheekey, Taniguchi, Watanabe, Mak} \emph{et~al.}}]{tschirhart2022intrinsic}
\bibinfo{author}{\bibnamefont{Tschirhart}, \bibfnamefont{C.}},
  \bibinfo{author}{\bibfnamefont{E.}~\bibnamefont{Redekop}},
  \bibinfo{author}{\bibfnamefont{L.}~\bibnamefont{Li}},
  \bibinfo{author}{\bibfnamefont{T.}~\bibnamefont{Li}},
  \bibinfo{author}{\bibfnamefont{S.}~\bibnamefont{Jiang}},
  \bibinfo{author}{\bibfnamefont{T.}~\bibnamefont{Arp}},
  \bibinfo{author}{\bibfnamefont{O.}~\bibnamefont{Sheekey}},
  \bibinfo{author}{\bibfnamefont{T.}~\bibnamefont{Taniguchi}},
  \bibinfo{author}{\bibfnamefont{K.}~\bibnamefont{Watanabe}},
  \bibinfo{author}{\bibfnamefont{K.~F.} \bibnamefont{Mak}}, \emph{et~al.},
  \bibinfo{year}{2022}, \bibinfo{journal}{arXiv:2205.02823} .

\bibitem[{\citenamefont{Tschirhart}
  \emph{et~al.}(2021)\citenamefont{Tschirhart, Serlin, Polshyn, Shragai, Xia,
  Zhu, Zhang, Watanabe, Taniguchi, Huber}
  \emph{et~al.}}]{tschirhart2021imaging}
\bibinfo{author}{\bibnamefont{Tschirhart}, \bibfnamefont{C.}},
  \bibinfo{author}{\bibfnamefont{M.}~\bibnamefont{Serlin}},
  \bibinfo{author}{\bibfnamefont{H.}~\bibnamefont{Polshyn}},
  \bibinfo{author}{\bibfnamefont{A.}~\bibnamefont{Shragai}},
  \bibinfo{author}{\bibfnamefont{Z.}~\bibnamefont{Xia}},
  \bibinfo{author}{\bibfnamefont{J.}~\bibnamefont{Zhu}},
  \bibinfo{author}{\bibfnamefont{Y.}~\bibnamefont{Zhang}},
  \bibinfo{author}{\bibfnamefont{K.}~\bibnamefont{Watanabe}},
  \bibinfo{author}{\bibfnamefont{T.}~\bibnamefont{Taniguchi}},
  \bibinfo{author}{\bibfnamefont{M.}~\bibnamefont{Huber}}, \emph{et~al.},
  \bibinfo{year}{2021}, \bibinfo{journal}{Science}
  \textbf{\bibinfo{volume}{372}}(\bibinfo{number}{6548}),
  \bibinfo{pages}{1323}.

\bibitem[{\citenamefont{Tse} \emph{et~al.}(2011)\citenamefont{Tse, Qiao, Yao,
  MacDonald, and Niu}}]{tse2011quantum}
\bibinfo{author}{\bibnamefont{Tse}, \bibfnamefont{W.-K.}},
  \bibinfo{author}{\bibfnamefont{Z.}~\bibnamefont{Qiao}},
  \bibinfo{author}{\bibfnamefont{Y.}~\bibnamefont{Yao}},
  \bibinfo{author}{\bibfnamefont{A.}~\bibnamefont{MacDonald}}, and
  \bibinfo{author}{\bibfnamefont{Q.}~\bibnamefont{Niu}}, \bibinfo{year}{2011},
  \bibinfo{journal}{Physical Review B}
  \textbf{\bibinfo{volume}{83}}(\bibinfo{number}{15}), \bibinfo{pages}{155447}.

\bibitem[{\citenamefont{Turner} \emph{et~al.}(2012)\citenamefont{Turner, Zhang,
  Mong, and Vishwanath}}]{turner2012quantized}
\bibinfo{author}{\bibnamefont{Turner}, \bibfnamefont{A.~M.}},
  \bibinfo{author}{\bibfnamefont{Y.}~\bibnamefont{Zhang}},
  \bibinfo{author}{\bibfnamefont{R.~S.} \bibnamefont{Mong}}, and
  \bibinfo{author}{\bibfnamefont{A.}~\bibnamefont{Vishwanath}},
  \bibinfo{year}{2012}, \bibinfo{journal}{Physical Review B}
  \textbf{\bibinfo{volume}{85}}(\bibinfo{number}{16}), \bibinfo{pages}{165120}.

\bibitem[{\citenamefont{Upadhyaya and Tserkovnyak}(2016)}]{upadhyaya2016domain}
\bibinfo{author}{\bibnamefont{Upadhyaya}, \bibfnamefont{P.}}, and
  \bibinfo{author}{\bibfnamefont{Y.}~\bibnamefont{Tserkovnyak}},
  \bibinfo{year}{2016}, \bibinfo{journal}{Physical Review B}
  \textbf{\bibinfo{volume}{94}}(\bibinfo{number}{2}), \bibinfo{pages}{020411}.

\bibitem[{\citenamefont{Van~Vleck}(1932)}]{van1932theory}
\bibinfo{author}{\bibnamefont{Van~Vleck}, \bibfnamefont{J.~H.}},
  \bibinfo{year}{1932}, \emph{\bibinfo{title}{The theory of electric and
  magnetic susceptibilities}} (\bibinfo{publisher}{Clarendon Press}).

\bibitem[{\citenamefont{Van~Vleck}(1953)}]{van1953models}
\bibinfo{author}{\bibnamefont{Van~Vleck}, \bibfnamefont{J.~H.}},
  \bibinfo{year}{1953}, \bibinfo{journal}{Reviews of Modern Physics}
  \textbf{\bibinfo{volume}{25}}(\bibinfo{number}{1}), \bibinfo{pages}{220}.

\bibitem[{\citenamefont{Varnava and Vanderbilt}(2018)}]{varnava2018surfaces}
\bibinfo{author}{\bibnamefont{Varnava}, \bibfnamefont{N.}}, and
  \bibinfo{author}{\bibfnamefont{D.}~\bibnamefont{Vanderbilt}},
  \bibinfo{year}{2018}, \bibinfo{journal}{Physical Review B}
  \textbf{\bibinfo{volume}{98}}(\bibinfo{number}{24}), \bibinfo{pages}{245117}.

\bibitem[{\citenamefont{Varnava} \emph{et~al.}(2021)\citenamefont{Varnava,
  Wilson, Pixley, and Vanderbilt}}]{varnava2021controllable}
\bibinfo{author}{\bibnamefont{Varnava}, \bibfnamefont{N.}},
  \bibinfo{author}{\bibfnamefont{J.~H.} \bibnamefont{Wilson}},
  \bibinfo{author}{\bibfnamefont{J.}~\bibnamefont{Pixley}}, and
  \bibinfo{author}{\bibfnamefont{D.}~\bibnamefont{Vanderbilt}},
  \bibinfo{year}{2021}, \bibinfo{journal}{Nature communications}
  \textbf{\bibinfo{volume}{12}}(\bibinfo{number}{1}), \bibinfo{pages}{1}.

\bibitem[{\citenamefont{Vidal}
  \emph{et~al.}(2019{\natexlab{a}})\citenamefont{Vidal, Bentmann, Peixoto,
  Zeugner, Moser, Min, Schatz, Ki{\ss}ner, {\"U}nzelmann, Fornari}
  \emph{et~al.}}]{vidal2019surface}
\bibinfo{author}{\bibnamefont{Vidal}, \bibfnamefont{R.}},
  \bibinfo{author}{\bibfnamefont{H.}~\bibnamefont{Bentmann}},
  \bibinfo{author}{\bibfnamefont{T.}~\bibnamefont{Peixoto}},
  \bibinfo{author}{\bibfnamefont{A.}~\bibnamefont{Zeugner}},
  \bibinfo{author}{\bibfnamefont{S.}~\bibnamefont{Moser}},
  \bibinfo{author}{\bibfnamefont{C.-H.} \bibnamefont{Min}},
  \bibinfo{author}{\bibfnamefont{S.}~\bibnamefont{Schatz}},
  \bibinfo{author}{\bibfnamefont{K.}~\bibnamefont{Ki{\ss}ner}},
  \bibinfo{author}{\bibfnamefont{M.}~\bibnamefont{{\"U}nzelmann}},
  \bibinfo{author}{\bibfnamefont{C.}~\bibnamefont{Fornari}}, \emph{et~al.},
  \bibinfo{year}{2019}{\natexlab{a}}, \bibinfo{journal}{Physical Review B}
  \textbf{\bibinfo{volume}{100}}(\bibinfo{number}{12}),
  \bibinfo{pages}{121104}.

\bibitem[{\citenamefont{Vidal}
  \emph{et~al.}(2019{\natexlab{b}})\citenamefont{Vidal, Zeugner, Facio, Ray,
  Haghighi, Wolter, Bohorquez, Caglieris, Moser, Figgemeier}
  \emph{et~al.}}]{vidal2019topological}
\bibinfo{author}{\bibnamefont{Vidal}, \bibfnamefont{R.~C.}},
  \bibinfo{author}{\bibfnamefont{A.}~\bibnamefont{Zeugner}},
  \bibinfo{author}{\bibfnamefont{J.~I.} \bibnamefont{Facio}},
  \bibinfo{author}{\bibfnamefont{R.}~\bibnamefont{Ray}},
  \bibinfo{author}{\bibfnamefont{M.~H.} \bibnamefont{Haghighi}},
  \bibinfo{author}{\bibfnamefont{A.~U.} \bibnamefont{Wolter}},
  \bibinfo{author}{\bibfnamefont{L.~T.~C.} \bibnamefont{Bohorquez}},
  \bibinfo{author}{\bibfnamefont{F.}~\bibnamefont{Caglieris}},
  \bibinfo{author}{\bibfnamefont{S.}~\bibnamefont{Moser}},
  \bibinfo{author}{\bibfnamefont{T.}~\bibnamefont{Figgemeier}}, \emph{et~al.},
  \bibinfo{year}{2019}{\natexlab{b}}, \bibinfo{journal}{Physical Review X}
  \textbf{\bibinfo{volume}{9}}(\bibinfo{number}{4}), \bibinfo{pages}{041065}.

\bibitem[{\citenamefont{Vila} \emph{et~al.}(2021)\citenamefont{Vila, Garcia,
  and Roche}}]{vila2021valley}
\bibinfo{author}{\bibnamefont{Vila}, \bibfnamefont{M.}},
  \bibinfo{author}{\bibfnamefont{J.~H.} \bibnamefont{Garcia}}, and
  \bibinfo{author}{\bibfnamefont{S.}~\bibnamefont{Roche}},
  \bibinfo{year}{2021}, \bibinfo{journal}{Physical Review B}
  \textbf{\bibinfo{volume}{104}}(\bibinfo{number}{16}),
  \bibinfo{pages}{L161113}.

\bibitem[{\citenamefont{Wang} \emph{et~al.}(2017)\citenamefont{Wang, Sun, Lu,
  and Xie}}]{wang20173d}
\bibinfo{author}{\bibnamefont{Wang}, \bibfnamefont{C.}},
  \bibinfo{author}{\bibfnamefont{H.-P.} \bibnamefont{Sun}},
  \bibinfo{author}{\bibfnamefont{H.-Z.} \bibnamefont{Lu}}, and
  \bibinfo{author}{\bibfnamefont{X.}~\bibnamefont{Xie}}, \bibinfo{year}{2017},
  \bibinfo{journal}{Physical review letters}
  \textbf{\bibinfo{volume}{119}}(\bibinfo{number}{13}),
  \bibinfo{pages}{136806}.

\bibitem[{\citenamefont{Wang}
  \emph{et~al.}(2018{\natexlab{a}})\citenamefont{Wang, Kong, Fan, Chen, Zhu,
  Liu, Cao, Sun, Du, Schneeloch} \emph{et~al.}}]{wang2018evidence}
\bibinfo{author}{\bibnamefont{Wang}, \bibfnamefont{D.}},
  \bibinfo{author}{\bibfnamefont{L.}~\bibnamefont{Kong}},
  \bibinfo{author}{\bibfnamefont{P.}~\bibnamefont{Fan}},
  \bibinfo{author}{\bibfnamefont{H.}~\bibnamefont{Chen}},
  \bibinfo{author}{\bibfnamefont{S.}~\bibnamefont{Zhu}},
  \bibinfo{author}{\bibfnamefont{W.}~\bibnamefont{Liu}},
  \bibinfo{author}{\bibfnamefont{L.}~\bibnamefont{Cao}},
  \bibinfo{author}{\bibfnamefont{Y.}~\bibnamefont{Sun}},
  \bibinfo{author}{\bibfnamefont{S.}~\bibnamefont{Du}},
  \bibinfo{author}{\bibfnamefont{J.}~\bibnamefont{Schneeloch}}, \emph{et~al.},
  \bibinfo{year}{2018}{\natexlab{a}}, \bibinfo{journal}{Science}
  \textbf{\bibinfo{volume}{362}}(\bibinfo{number}{6412}), \bibinfo{pages}{333}.

\bibitem[{\citenamefont{Wang} \emph{et~al.}(2019)\citenamefont{Wang, Xiao,
  Yuan, Jiang, Zhao, Zhang, Yao, Liu, Zhang, Liu}
  \emph{et~al.}}]{wang2019observation}
\bibinfo{author}{\bibnamefont{Wang}, \bibfnamefont{F.}},
  \bibinfo{author}{\bibfnamefont{D.}~\bibnamefont{Xiao}},
  \bibinfo{author}{\bibfnamefont{W.}~\bibnamefont{Yuan}},
  \bibinfo{author}{\bibfnamefont{J.}~\bibnamefont{Jiang}},
  \bibinfo{author}{\bibfnamefont{Y.-F.} \bibnamefont{Zhao}},
  \bibinfo{author}{\bibfnamefont{L.}~\bibnamefont{Zhang}},
  \bibinfo{author}{\bibfnamefont{Y.}~\bibnamefont{Yao}},
  \bibinfo{author}{\bibfnamefont{W.}~\bibnamefont{Liu}},
  \bibinfo{author}{\bibfnamefont{Z.}~\bibnamefont{Zhang}},
  \bibinfo{author}{\bibfnamefont{C.}~\bibnamefont{Liu}}, \emph{et~al.},
  \bibinfo{year}{2019}, \bibinfo{journal}{Nano Letters}
  \textbf{\bibinfo{volume}{19}}(\bibinfo{number}{5}), \bibinfo{pages}{2945}.

\bibitem[{\citenamefont{Wang}
  \emph{et~al.}(2018{\natexlab{b}})\citenamefont{Wang, Zhang, Jiang, Zhao, Yu,
  Liu, Li, Chan, Sun, Zhang} \emph{et~al.}}]{wang2018chromium}
\bibinfo{author}{\bibnamefont{Wang}, \bibfnamefont{F.}},
  \bibinfo{author}{\bibfnamefont{H.}~\bibnamefont{Zhang}},
  \bibinfo{author}{\bibfnamefont{J.}~\bibnamefont{Jiang}},
  \bibinfo{author}{\bibfnamefont{Y.-F.} \bibnamefont{Zhao}},
  \bibinfo{author}{\bibfnamefont{J.}~\bibnamefont{Yu}},
  \bibinfo{author}{\bibfnamefont{W.}~\bibnamefont{Liu}},
  \bibinfo{author}{\bibfnamefont{D.}~\bibnamefont{Li}},
  \bibinfo{author}{\bibfnamefont{M.~H.} \bibnamefont{Chan}},
  \bibinfo{author}{\bibfnamefont{J.}~\bibnamefont{Sun}},
  \bibinfo{author}{\bibfnamefont{Z.}~\bibnamefont{Zhang}}, \emph{et~al.},
  \bibinfo{year}{2018}{\natexlab{b}}, \bibinfo{journal}{Physical Review B}
  \textbf{\bibinfo{volume}{97}}(\bibinfo{number}{11}), \bibinfo{pages}{115414}.

\bibitem[{\citenamefont{Wang} \emph{et~al.}(2010)\citenamefont{Wang, Zhu, Wen,
  Chen, He, Wang, Ma, Liu, Dai, Fang} \emph{et~al.}}]{wang2010atomically}
\bibinfo{author}{\bibnamefont{Wang}, \bibfnamefont{G.}},
  \bibinfo{author}{\bibfnamefont{X.}~\bibnamefont{Zhu}},
  \bibinfo{author}{\bibfnamefont{J.}~\bibnamefont{Wen}},
  \bibinfo{author}{\bibfnamefont{X.}~\bibnamefont{Chen}},
  \bibinfo{author}{\bibfnamefont{K.}~\bibnamefont{He}},
  \bibinfo{author}{\bibfnamefont{L.}~\bibnamefont{Wang}},
  \bibinfo{author}{\bibfnamefont{X.}~\bibnamefont{Ma}},
  \bibinfo{author}{\bibfnamefont{Y.}~\bibnamefont{Liu}},
  \bibinfo{author}{\bibfnamefont{X.}~\bibnamefont{Dai}},
  \bibinfo{author}{\bibfnamefont{Z.}~\bibnamefont{Fang}}, \emph{et~al.},
  \bibinfo{year}{2010}, \bibinfo{journal}{Nano Research}
  \textbf{\bibinfo{volume}{3}}(\bibinfo{number}{12}), \bibinfo{pages}{874}.

\bibitem[{\citenamefont{Wang}
  \emph{et~al.}(2015{\natexlab{a}})\citenamefont{Wang, Lian, Qi, and
  Zhang}}]{wang2015quantized}
\bibinfo{author}{\bibnamefont{Wang}, \bibfnamefont{J.}},
  \bibinfo{author}{\bibfnamefont{B.}~\bibnamefont{Lian}},
  \bibinfo{author}{\bibfnamefont{X.-L.} \bibnamefont{Qi}}, and
  \bibinfo{author}{\bibfnamefont{S.-C.} \bibnamefont{Zhang}},
  \bibinfo{year}{2015}{\natexlab{a}}, \bibinfo{journal}{Physical Review B}
  \textbf{\bibinfo{volume}{92}}(\bibinfo{number}{8}), \bibinfo{pages}{081107}.

\bibitem[{\citenamefont{Wang}
  \emph{et~al.}(2013{\natexlab{a}})\citenamefont{Wang, Lian, Zhang, and
  Zhang}}]{wang2013anomalous}
\bibinfo{author}{\bibnamefont{Wang}, \bibfnamefont{J.}},
  \bibinfo{author}{\bibfnamefont{B.}~\bibnamefont{Lian}},
  \bibinfo{author}{\bibfnamefont{H.}~\bibnamefont{Zhang}}, and
  \bibinfo{author}{\bibfnamefont{S.-C.} \bibnamefont{Zhang}},
  \bibinfo{year}{2013}{\natexlab{a}}, \bibinfo{journal}{Physical Review
  Letters} \textbf{\bibinfo{volume}{111}}(\bibinfo{number}{8}),
  \bibinfo{pages}{086803}.

\bibitem[{\citenamefont{Wang}
  \emph{et~al.}(2014{\natexlab{a}})\citenamefont{Wang, Lian, and
  Zhang}}]{wang2014universal}
\bibinfo{author}{\bibnamefont{Wang}, \bibfnamefont{J.}},
  \bibinfo{author}{\bibfnamefont{B.}~\bibnamefont{Lian}}, and
  \bibinfo{author}{\bibfnamefont{S.-C.} \bibnamefont{Zhang}},
  \bibinfo{year}{2014}{\natexlab{a}}, \bibinfo{journal}{Physical Review B}
  \textbf{\bibinfo{volume}{89}}(\bibinfo{number}{8}), \bibinfo{pages}{085106}.

\bibitem[{\citenamefont{Wang}
  \emph{et~al.}(2015{\natexlab{b}})\citenamefont{Wang, Zhou, Lian, and
  Zhang}}]{wang2015chiral}
\bibinfo{author}{\bibnamefont{Wang}, \bibfnamefont{J.}},
  \bibinfo{author}{\bibfnamefont{Q.}~\bibnamefont{Zhou}},
  \bibinfo{author}{\bibfnamefont{B.}~\bibnamefont{Lian}}, and
  \bibinfo{author}{\bibfnamefont{S.-C.} \bibnamefont{Zhang}},
  \bibinfo{year}{2015}{\natexlab{b}}, \bibinfo{journal}{Physical Review B}
  \textbf{\bibinfo{volume}{92}}(\bibinfo{number}{6}), \bibinfo{pages}{064520}.

\bibitem[{\citenamefont{Wang} \emph{et~al.}(2021)\citenamefont{Wang, Ge, Li,
  Liu, Xu, and Wang}}]{wang2021intrinsic}
\bibinfo{author}{\bibnamefont{Wang}, \bibfnamefont{P.}},
  \bibinfo{author}{\bibfnamefont{J.}~\bibnamefont{Ge}},
  \bibinfo{author}{\bibfnamefont{J.}~\bibnamefont{Li}},
  \bibinfo{author}{\bibfnamefont{Y.}~\bibnamefont{Liu}},
  \bibinfo{author}{\bibfnamefont{Y.}~\bibnamefont{Xu}}, and
  \bibinfo{author}{\bibfnamefont{J.}~\bibnamefont{Wang}}, \bibinfo{year}{2021},
  \bibinfo{journal}{The Innovation} , \bibinfo{pages}{100098}.

\bibitem[{\citenamefont{Wang}
  \emph{et~al.}(2014{\natexlab{b}})\citenamefont{Wang, Liu, Zhang, Samarth,
  Zhang, and Liu}}]{wang2014quantum}
\bibinfo{author}{\bibnamefont{Wang}, \bibfnamefont{Q.-Z.}},
  \bibinfo{author}{\bibfnamefont{X.}~\bibnamefont{Liu}},
  \bibinfo{author}{\bibfnamefont{H.-J.} \bibnamefont{Zhang}},
  \bibinfo{author}{\bibfnamefont{N.}~\bibnamefont{Samarth}},
  \bibinfo{author}{\bibfnamefont{S.-C.} \bibnamefont{Zhang}}, and
  \bibinfo{author}{\bibfnamefont{C.-X.} \bibnamefont{Liu}},
  \bibinfo{year}{2014}{\natexlab{b}}, \bibinfo{journal}{Physical Review
  Letters} \textbf{\bibinfo{volume}{113}}(\bibinfo{number}{14}),
  \bibinfo{pages}{147201}.

\bibitem[{\citenamefont{Wang} \emph{et~al.}(2020)\citenamefont{Wang, Xiao, Dou,
  Cao, Zhao, Samarth, Chang, Connolly, and Smith}}]{wang2020demonstration}
\bibinfo{author}{\bibnamefont{Wang}, \bibfnamefont{S.-W.}},
  \bibinfo{author}{\bibfnamefont{D.}~\bibnamefont{Xiao}},
  \bibinfo{author}{\bibfnamefont{Z.}~\bibnamefont{Dou}},
  \bibinfo{author}{\bibfnamefont{M.}~\bibnamefont{Cao}},
  \bibinfo{author}{\bibfnamefont{Y.-F.} \bibnamefont{Zhao}},
  \bibinfo{author}{\bibfnamefont{N.}~\bibnamefont{Samarth}},
  \bibinfo{author}{\bibfnamefont{C.-Z.} \bibnamefont{Chang}},
  \bibinfo{author}{\bibfnamefont{M.~R.} \bibnamefont{Connolly}}, and
  \bibinfo{author}{\bibfnamefont{C.~G.} \bibnamefont{Smith}},
  \bibinfo{year}{2020}, \bibinfo{journal}{Physical Review Letters}
  \textbf{\bibinfo{volume}{125}}(\bibinfo{number}{12}),
  \bibinfo{pages}{126801}.

\bibitem[{\citenamefont{Wang}
  \emph{et~al.}(2018{\natexlab{c}})\citenamefont{Wang, Ou, Liu, Wang, He, Xue,
  and Wu}}]{wang2018direct}
\bibinfo{author}{\bibnamefont{Wang}, \bibfnamefont{W.}},
  \bibinfo{author}{\bibfnamefont{Y.}~\bibnamefont{Ou}},
  \bibinfo{author}{\bibfnamefont{C.}~\bibnamefont{Liu}},
  \bibinfo{author}{\bibfnamefont{Y.}~\bibnamefont{Wang}},
  \bibinfo{author}{\bibfnamefont{K.}~\bibnamefont{He}},
  \bibinfo{author}{\bibfnamefont{Q.-K.} \bibnamefont{Xue}}, and
  \bibinfo{author}{\bibfnamefont{W.}~\bibnamefont{Wu}},
  \bibinfo{year}{2018}{\natexlab{c}}, \bibinfo{journal}{Nature Physics}
  \textbf{\bibinfo{volume}{14}}(\bibinfo{number}{8}), \bibinfo{pages}{791}.

\bibitem[{\citenamefont{Wang}
  \emph{et~al.}(2013{\natexlab{b}})\citenamefont{Wang, Liu, and
  Liu}}]{wang2013quantum}
\bibinfo{author}{\bibnamefont{Wang}, \bibfnamefont{Z.}},
  \bibinfo{author}{\bibfnamefont{Z.}~\bibnamefont{Liu}}, and
  \bibinfo{author}{\bibfnamefont{F.}~\bibnamefont{Liu}},
  \bibinfo{year}{2013}{\natexlab{b}}, \bibinfo{journal}{Physical Review
  Letters} \textbf{\bibinfo{volume}{110}}(\bibinfo{number}{19}),
  \bibinfo{pages}{196801}.

\bibitem[{\citenamefont{Watanabe} \emph{et~al.}(2019)\citenamefont{Watanabe,
  Yoshimi, Kawamura, Mogi, Tsukazaki, Yu, Nakajima, Takahashi, Kawasaki, and
  Tokura}}]{watanabe2019quantum}
\bibinfo{author}{\bibnamefont{Watanabe}, \bibfnamefont{R.}},
  \bibinfo{author}{\bibfnamefont{R.}~\bibnamefont{Yoshimi}},
  \bibinfo{author}{\bibfnamefont{M.}~\bibnamefont{Kawamura}},
  \bibinfo{author}{\bibfnamefont{M.}~\bibnamefont{Mogi}},
  \bibinfo{author}{\bibfnamefont{A.}~\bibnamefont{Tsukazaki}},
  \bibinfo{author}{\bibfnamefont{X.}~\bibnamefont{Yu}},
  \bibinfo{author}{\bibfnamefont{K.}~\bibnamefont{Nakajima}},
  \bibinfo{author}{\bibfnamefont{K.~S.} \bibnamefont{Takahashi}},
  \bibinfo{author}{\bibfnamefont{M.}~\bibnamefont{Kawasaki}}, and
  \bibinfo{author}{\bibfnamefont{Y.}~\bibnamefont{Tokura}},
  \bibinfo{year}{2019}, \bibinfo{journal}{Applied Physics Letters}
  \textbf{\bibinfo{volume}{115}}(\bibinfo{number}{10}),
  \bibinfo{pages}{102403}.

\bibitem[{\citenamefont{Wei} \emph{et~al.}(1988)\citenamefont{Wei, Tsui,
  Paalanen, and Pruisken}}]{wei1988experiments}
\bibinfo{author}{\bibnamefont{Wei}, \bibfnamefont{H.}},
  \bibinfo{author}{\bibfnamefont{D.}~\bibnamefont{Tsui}},
  \bibinfo{author}{\bibfnamefont{M.}~\bibnamefont{Paalanen}}, and
  \bibinfo{author}{\bibfnamefont{A.}~\bibnamefont{Pruisken}},
  \bibinfo{year}{1988}, \bibinfo{journal}{Physical Review Letters}
  \textbf{\bibinfo{volume}{61}}(\bibinfo{number}{11}), \bibinfo{pages}{1294}.

\bibitem[{\citenamefont{Wei} \emph{et~al.}(1986)\citenamefont{Wei, Tsui, and
  Pruisken}}]{wei1986localization}
\bibinfo{author}{\bibnamefont{Wei}, \bibfnamefont{H.}},
  \bibinfo{author}{\bibfnamefont{D.}~\bibnamefont{Tsui}}, and
  \bibinfo{author}{\bibfnamefont{A.}~\bibnamefont{Pruisken}},
  \bibinfo{year}{1986}, \bibinfo{journal}{Physical Review B}
  \textbf{\bibinfo{volume}{33}}(\bibinfo{number}{2}), \bibinfo{pages}{1488}.

\bibitem[{\citenamefont{Weng} \emph{et~al.}(2015)\citenamefont{Weng, Yu, Hu,
  Dai, and Fang}}]{weng2015quantum}
\bibinfo{author}{\bibnamefont{Weng}, \bibfnamefont{H.}},
  \bibinfo{author}{\bibfnamefont{R.}~\bibnamefont{Yu}},
  \bibinfo{author}{\bibfnamefont{X.}~\bibnamefont{Hu}},
  \bibinfo{author}{\bibfnamefont{X.}~\bibnamefont{Dai}}, and
  \bibinfo{author}{\bibfnamefont{Z.}~\bibnamefont{Fang}}, \bibinfo{year}{2015},
  \bibinfo{journal}{Advances in Physics}
  \textbf{\bibinfo{volume}{64}}(\bibinfo{number}{3}), \bibinfo{pages}{227}.

\bibitem[{\citenamefont{Wieder and Bernevig}(2018)}]{wieder2018axion}
\bibinfo{author}{\bibnamefont{Wieder}, \bibfnamefont{B.~J.}}, and
  \bibinfo{author}{\bibfnamefont{B.~A.} \bibnamefont{Bernevig}},
  \bibinfo{year}{2018}, \bibinfo{journal}{arXiv:1810.02373} .

\bibitem[{\citenamefont{Wilczek}(1987)}]{wilczek1987two}
\bibinfo{author}{\bibnamefont{Wilczek}, \bibfnamefont{F.}},
  \bibinfo{year}{1987}, \bibinfo{journal}{Physical Review Letters}
  \textbf{\bibinfo{volume}{58}}(\bibinfo{number}{18}), \bibinfo{pages}{1799}.

\bibitem[{\citenamefont{Wimmer} \emph{et~al.}(2021)\citenamefont{Wimmer,
  S{\'a}nchez-Barriga, K{\"u}ppers, Ney, Schierle, Freyse, Caha,
  Michali{\v{c}}ka, Liebmann, Primetzhofer} \emph{et~al.}}]{wimmer2021mn}
\bibinfo{author}{\bibnamefont{Wimmer}, \bibfnamefont{S.}},
  \bibinfo{author}{\bibfnamefont{J.}~\bibnamefont{S{\'a}nchez-Barriga}},
  \bibinfo{author}{\bibfnamefont{P.}~\bibnamefont{K{\"u}ppers}},
  \bibinfo{author}{\bibfnamefont{A.}~\bibnamefont{Ney}},
  \bibinfo{author}{\bibfnamefont{E.}~\bibnamefont{Schierle}},
  \bibinfo{author}{\bibfnamefont{F.}~\bibnamefont{Freyse}},
  \bibinfo{author}{\bibfnamefont{O.}~\bibnamefont{Caha}},
  \bibinfo{author}{\bibfnamefont{J.}~\bibnamefont{Michali{\v{c}}ka}},
  \bibinfo{author}{\bibfnamefont{M.}~\bibnamefont{Liebmann}},
  \bibinfo{author}{\bibfnamefont{D.}~\bibnamefont{Primetzhofer}},
  \emph{et~al.}, \bibinfo{year}{2021}, \bibinfo{journal}{Advanced Materials}
  \textbf{\bibinfo{volume}{33}}(\bibinfo{number}{42}),
  \bibinfo{pages}{2102935}.

\bibitem[{\citenamefont{Wu} \emph{et~al.}(2018)\citenamefont{Wu, Lovorn, Tutuc,
  and MacDonald}}]{wu2018hubbard}
\bibinfo{author}{\bibnamefont{Wu}, \bibfnamefont{F.}},
  \bibinfo{author}{\bibfnamefont{T.}~\bibnamefont{Lovorn}},
  \bibinfo{author}{\bibfnamefont{E.}~\bibnamefont{Tutuc}}, and
  \bibinfo{author}{\bibfnamefont{A.~H.} \bibnamefont{MacDonald}},
  \bibinfo{year}{2018}, \bibinfo{journal}{Physical Review Letters}
  \textbf{\bibinfo{volume}{121}}(\bibinfo{number}{2}), \bibinfo{pages}{026402}.

\bibitem[{\citenamefont{Wu} \emph{et~al.}(2019{\natexlab{a}})\citenamefont{Wu,
  Lovorn, Tutuc, Martin, and MacDonald}}]{wu2019topological}
\bibinfo{author}{\bibnamefont{Wu}, \bibfnamefont{F.}},
  \bibinfo{author}{\bibfnamefont{T.}~\bibnamefont{Lovorn}},
  \bibinfo{author}{\bibfnamefont{E.}~\bibnamefont{Tutuc}},
  \bibinfo{author}{\bibfnamefont{I.}~\bibnamefont{Martin}}, and
  \bibinfo{author}{\bibfnamefont{A.}~\bibnamefont{MacDonald}},
  \bibinfo{year}{2019}{\natexlab{a}}, \bibinfo{journal}{Physical Review
  Letters} \textbf{\bibinfo{volume}{122}}(\bibinfo{number}{8}),
  \bibinfo{pages}{086402}.

\bibitem[{\citenamefont{Wu and Sarma}(2020)}]{wu2020collective}
\bibinfo{author}{\bibnamefont{Wu}, \bibfnamefont{F.}}, and
  \bibinfo{author}{\bibfnamefont{S.~D.} \bibnamefont{Sarma}},
  \bibinfo{year}{2020}, \bibinfo{journal}{Physical Review Letters}
  \textbf{\bibinfo{volume}{124}}(\bibinfo{number}{4}), \bibinfo{pages}{046403}.

\bibitem[{\citenamefont{Wu} \emph{et~al.}(2019{\natexlab{b}})\citenamefont{Wu,
  Liu, Sasase, Ienaga, Obata, Yukawa, Horiba, Kumigashira, Okuma, Inoshita}
  \emph{et~al.}}]{wu2019natural}
\bibinfo{author}{\bibnamefont{Wu}, \bibfnamefont{J.}},
  \bibinfo{author}{\bibfnamefont{F.}~\bibnamefont{Liu}},
  \bibinfo{author}{\bibfnamefont{M.}~\bibnamefont{Sasase}},
  \bibinfo{author}{\bibfnamefont{K.}~\bibnamefont{Ienaga}},
  \bibinfo{author}{\bibfnamefont{Y.}~\bibnamefont{Obata}},
  \bibinfo{author}{\bibfnamefont{R.}~\bibnamefont{Yukawa}},
  \bibinfo{author}{\bibfnamefont{K.}~\bibnamefont{Horiba}},
  \bibinfo{author}{\bibfnamefont{H.}~\bibnamefont{Kumigashira}},
  \bibinfo{author}{\bibfnamefont{S.}~\bibnamefont{Okuma}},
  \bibinfo{author}{\bibfnamefont{T.}~\bibnamefont{Inoshita}}, \emph{et~al.},
  \bibinfo{year}{2019}{\natexlab{b}}, \bibinfo{journal}{Science Advances}
  \textbf{\bibinfo{volume}{5}}(\bibinfo{number}{11}),
  \bibinfo{pages}{eaax9989}.

\bibitem[{\citenamefont{Wu} \emph{et~al.}(2014{\natexlab{a}})\citenamefont{Wu,
  Liu, and Liu}}]{wu2014topological}
\bibinfo{author}{\bibnamefont{Wu}, \bibfnamefont{J.}},
  \bibinfo{author}{\bibfnamefont{J.}~\bibnamefont{Liu}}, and
  \bibinfo{author}{\bibfnamefont{X.-J.} \bibnamefont{Liu}},
  \bibinfo{year}{2014}{\natexlab{a}}, \bibinfo{journal}{Physical Review
  Letters} \textbf{\bibinfo{volume}{113}}(\bibinfo{number}{13}),
  \bibinfo{pages}{136403}.

\bibitem[{\citenamefont{Wu} \emph{et~al.}(2016)\citenamefont{Wu, Salehi,
  Koirala, Moon, Oh, and Armitage}}]{wu2016quantized}
\bibinfo{author}{\bibnamefont{Wu}, \bibfnamefont{L.}},
  \bibinfo{author}{\bibfnamefont{M.}~\bibnamefont{Salehi}},
  \bibinfo{author}{\bibfnamefont{N.}~\bibnamefont{Koirala}},
  \bibinfo{author}{\bibfnamefont{J.}~\bibnamefont{Moon}},
  \bibinfo{author}{\bibfnamefont{S.}~\bibnamefont{Oh}}, and
  \bibinfo{author}{\bibfnamefont{N.}~\bibnamefont{Armitage}},
  \bibinfo{year}{2016}, \bibinfo{journal}{Science}
  \textbf{\bibinfo{volume}{354}}(\bibinfo{number}{6316}),
  \bibinfo{pages}{1124}.

\bibitem[{\citenamefont{Wu}(2017)}]{wu2017high}
\bibinfo{author}{\bibnamefont{Wu}, \bibfnamefont{M.}}, \bibinfo{year}{2017},
  \bibinfo{journal}{2D Materials}
  \textbf{\bibinfo{volume}{4}}(\bibinfo{number}{2}), \bibinfo{pages}{021014}.

\bibitem[{\citenamefont{Wu} \emph{et~al.}(2021)\citenamefont{Wu, Zhang,
  Watanabe, Taniguchi, and Andrei}}]{wu2021chern}
\bibinfo{author}{\bibnamefont{Wu}, \bibfnamefont{S.}},
  \bibinfo{author}{\bibfnamefont{Z.}~\bibnamefont{Zhang}},
  \bibinfo{author}{\bibfnamefont{K.}~\bibnamefont{Watanabe}},
  \bibinfo{author}{\bibfnamefont{T.}~\bibnamefont{Taniguchi}}, and
  \bibinfo{author}{\bibfnamefont{E.~Y.} \bibnamefont{Andrei}},
  \bibinfo{year}{2021}, \bibinfo{journal}{Nature Materials}
  \textbf{\bibinfo{volume}{20}}(\bibinfo{number}{4}), \bibinfo{pages}{488}.

\bibitem[{\citenamefont{Wu} \emph{et~al.}(2014{\natexlab{b}})\citenamefont{Wu,
  Shan, and Yan}}]{wu2014prediction}
\bibinfo{author}{\bibnamefont{Wu}, \bibfnamefont{S.-C.}},
  \bibinfo{author}{\bibfnamefont{G.}~\bibnamefont{Shan}}, and
  \bibinfo{author}{\bibfnamefont{B.}~\bibnamefont{Yan}},
  \bibinfo{year}{2014}{\natexlab{b}}, \bibinfo{journal}{Physical Review
  Letters} \textbf{\bibinfo{volume}{113}}(\bibinfo{number}{25}),
  \bibinfo{pages}{256401}.

\bibitem[{\citenamefont{Wu} \emph{et~al.}(2020{\natexlab{a}})\citenamefont{Wu,
  Li, Ma, Zhang, Liu, Zhou, Shao, Wang, Hao, Feng}
  \emph{et~al.}}]{wu2020distinct}
\bibinfo{author}{\bibnamefont{Wu}, \bibfnamefont{X.}},
  \bibinfo{author}{\bibfnamefont{J.}~\bibnamefont{Li}},
  \bibinfo{author}{\bibfnamefont{X.-M.} \bibnamefont{Ma}},
  \bibinfo{author}{\bibfnamefont{Y.}~\bibnamefont{Zhang}},
  \bibinfo{author}{\bibfnamefont{Y.}~\bibnamefont{Liu}},
  \bibinfo{author}{\bibfnamefont{C.-S.} \bibnamefont{Zhou}},
  \bibinfo{author}{\bibfnamefont{J.}~\bibnamefont{Shao}},
  \bibinfo{author}{\bibfnamefont{Q.}~\bibnamefont{Wang}},
  \bibinfo{author}{\bibfnamefont{Y.-J.} \bibnamefont{Hao}},
  \bibinfo{author}{\bibfnamefont{Y.}~\bibnamefont{Feng}}, \emph{et~al.},
  \bibinfo{year}{2020}{\natexlab{a}}, \bibinfo{journal}{Physical Review X}
  \textbf{\bibinfo{volume}{10}}(\bibinfo{number}{3}), \bibinfo{pages}{031013}.

\bibitem[{\citenamefont{Wu} \emph{et~al.}(2020{\natexlab{b}})\citenamefont{Wu,
  Xiao, Chen, Sun, Zhang, Chan, Samarth, Xie, Lin, and Chang}}]{wu2020scaling}
\bibinfo{author}{\bibnamefont{Wu}, \bibfnamefont{X.}},
  \bibinfo{author}{\bibfnamefont{D.}~\bibnamefont{Xiao}},
  \bibinfo{author}{\bibfnamefont{C.-Z.} \bibnamefont{Chen}},
  \bibinfo{author}{\bibfnamefont{J.}~\bibnamefont{Sun}},
  \bibinfo{author}{\bibfnamefont{L.}~\bibnamefont{Zhang}},
  \bibinfo{author}{\bibfnamefont{M.~H.} \bibnamefont{Chan}},
  \bibinfo{author}{\bibfnamefont{N.}~\bibnamefont{Samarth}},
  \bibinfo{author}{\bibfnamefont{X.}~\bibnamefont{Xie}},
  \bibinfo{author}{\bibfnamefont{X.}~\bibnamefont{Lin}}, and
  \bibinfo{author}{\bibfnamefont{C.-Z.} \bibnamefont{Chang}},
  \bibinfo{year}{2020}{\natexlab{b}}, \bibinfo{journal}{Nature Communications}
  \textbf{\bibinfo{volume}{11}}(\bibinfo{number}{1}), \bibinfo{pages}{1}.

\bibitem[{\citenamefont{Xia} \emph{et~al.}(2009)\citenamefont{Xia, Qian, Hsieh,
  Wray, Pal, Lin, Bansil, Grauer, Hor, Cava}
  \emph{et~al.}}]{xia2009observation}
\bibinfo{author}{\bibnamefont{Xia}, \bibfnamefont{Y.}},
  \bibinfo{author}{\bibfnamefont{D.}~\bibnamefont{Qian}},
  \bibinfo{author}{\bibfnamefont{D.}~\bibnamefont{Hsieh}},
  \bibinfo{author}{\bibfnamefont{L.}~\bibnamefont{Wray}},
  \bibinfo{author}{\bibfnamefont{A.}~\bibnamefont{Pal}},
  \bibinfo{author}{\bibfnamefont{H.}~\bibnamefont{Lin}},
  \bibinfo{author}{\bibfnamefont{A.}~\bibnamefont{Bansil}},
  \bibinfo{author}{\bibfnamefont{D.}~\bibnamefont{Grauer}},
  \bibinfo{author}{\bibfnamefont{Y.~S.} \bibnamefont{Hor}},
  \bibinfo{author}{\bibfnamefont{R.~J.} \bibnamefont{Cava}}, \emph{et~al.},
  \bibinfo{year}{2009}, \bibinfo{journal}{Nature Physics}
  \textbf{\bibinfo{volume}{5}}(\bibinfo{number}{6}), \bibinfo{pages}{398}.

\bibitem[{\citenamefont{Xiao} \emph{et~al.}(2010)\citenamefont{Xiao, Chang, and
  Niu}}]{xiao2010berry}
\bibinfo{author}{\bibnamefont{Xiao}, \bibfnamefont{D.}},
  \bibinfo{author}{\bibfnamefont{M.-C.} \bibnamefont{Chang}}, and
  \bibinfo{author}{\bibfnamefont{Q.}~\bibnamefont{Niu}}, \bibinfo{year}{2010},
  \bibinfo{journal}{Reviews of Modern Physics}
  \textbf{\bibinfo{volume}{82}}(\bibinfo{number}{3}), \bibinfo{pages}{1959}.

\bibitem[{\citenamefont{Xiao} \emph{et~al.}(2018)\citenamefont{Xiao, Jiang,
  Shin, Wang, Wang, Zhao, Liu, Wu, Chan, Samarth}
  \emph{et~al.}}]{xiao2018realization}
\bibinfo{author}{\bibnamefont{Xiao}, \bibfnamefont{D.}},
  \bibinfo{author}{\bibfnamefont{J.}~\bibnamefont{Jiang}},
  \bibinfo{author}{\bibfnamefont{J.-H.} \bibnamefont{Shin}},
  \bibinfo{author}{\bibfnamefont{W.}~\bibnamefont{Wang}},
  \bibinfo{author}{\bibfnamefont{F.}~\bibnamefont{Wang}},
  \bibinfo{author}{\bibfnamefont{Y.-F.} \bibnamefont{Zhao}},
  \bibinfo{author}{\bibfnamefont{C.}~\bibnamefont{Liu}},
  \bibinfo{author}{\bibfnamefont{W.}~\bibnamefont{Wu}},
  \bibinfo{author}{\bibfnamefont{M.~H.} \bibnamefont{Chan}},
  \bibinfo{author}{\bibfnamefont{N.}~\bibnamefont{Samarth}}, \emph{et~al.},
  \bibinfo{year}{2018}, \bibinfo{journal}{Physical Review Letters}
  \textbf{\bibinfo{volume}{120}}(\bibinfo{number}{5}), \bibinfo{pages}{056801}.

\bibitem[{\citenamefont{Xiao} \emph{et~al.}(2012)\citenamefont{Xiao, Liu, Feng,
  Xu, and Yao}}]{xiao2012coupled}
\bibinfo{author}{\bibnamefont{Xiao}, \bibfnamefont{D.}},
  \bibinfo{author}{\bibfnamefont{G.-B.} \bibnamefont{Liu}},
  \bibinfo{author}{\bibfnamefont{W.}~\bibnamefont{Feng}},
  \bibinfo{author}{\bibfnamefont{X.}~\bibnamefont{Xu}}, and
  \bibinfo{author}{\bibfnamefont{W.}~\bibnamefont{Yao}}, \bibinfo{year}{2012},
  \bibinfo{journal}{Physical Review Letters}
  \textbf{\bibinfo{volume}{108}}(\bibinfo{number}{19}),
  \bibinfo{pages}{196802}.

\bibitem[{\citenamefont{Xiao} \emph{et~al.}(2005)\citenamefont{Xiao, Shi, and
  Niu}}]{xiao2005berry}
\bibinfo{author}{\bibnamefont{Xiao}, \bibfnamefont{D.}},
  \bibinfo{author}{\bibfnamefont{J.}~\bibnamefont{Shi}}, and
  \bibinfo{author}{\bibfnamefont{Q.}~\bibnamefont{Niu}}, \bibinfo{year}{2005},
  \bibinfo{journal}{Physical Review Letters}
  \textbf{\bibinfo{volume}{95}}(\bibinfo{number}{13}), \bibinfo{pages}{137204}.

\bibitem[{\citenamefont{Xiao} \emph{et~al.}(2007)\citenamefont{Xiao, Yao, and
  Niu}}]{xiao2007valley}
\bibinfo{author}{\bibnamefont{Xiao}, \bibfnamefont{D.}},
  \bibinfo{author}{\bibfnamefont{W.}~\bibnamefont{Yao}}, and
  \bibinfo{author}{\bibfnamefont{Q.}~\bibnamefont{Niu}}, \bibinfo{year}{2007},
  \bibinfo{journal}{Physical Review Letters}
  \textbf{\bibinfo{volume}{99}}(\bibinfo{number}{23}), \bibinfo{pages}{236809}.

\bibitem[{\citenamefont{Xiao} \emph{et~al.}(2011)\citenamefont{Xiao, Zhu, Ran,
  Nagaosa, and Okamoto}}]{xiao2011interface}
\bibinfo{author}{\bibnamefont{Xiao}, \bibfnamefont{D.}},
  \bibinfo{author}{\bibfnamefont{W.}~\bibnamefont{Zhu}},
  \bibinfo{author}{\bibfnamefont{Y.}~\bibnamefont{Ran}},
  \bibinfo{author}{\bibfnamefont{N.}~\bibnamefont{Nagaosa}}, and
  \bibinfo{author}{\bibfnamefont{S.}~\bibnamefont{Okamoto}},
  \bibinfo{year}{2011}, \bibinfo{journal}{Nature Communications}
  \textbf{\bibinfo{volume}{2}}(\bibinfo{number}{1}), \bibinfo{pages}{1}.

\bibitem[{\citenamefont{Xie and MacDonald}(2020)}]{xie2020nature}
\bibinfo{author}{\bibnamefont{Xie}, \bibfnamefont{M.}}, and
  \bibinfo{author}{\bibfnamefont{A.~H.} \bibnamefont{MacDonald}},
  \bibinfo{year}{2020}, \bibinfo{journal}{Physical Review Letters}
  \textbf{\bibinfo{volume}{124}}(\bibinfo{number}{9}), \bibinfo{pages}{097601}.

\bibitem[{\citenamefont{Xie} \emph{et~al.}(2021)\citenamefont{Xie, Pierce,
  Park, Parker, Khalaf, Ledwith, Cao, Lee, Chen, Forrester}
  \emph{et~al.}}]{xie2021fractional}
\bibinfo{author}{\bibnamefont{Xie}, \bibfnamefont{Y.}},
  \bibinfo{author}{\bibfnamefont{A.~T.} \bibnamefont{Pierce}},
  \bibinfo{author}{\bibfnamefont{J.~M.} \bibnamefont{Park}},
  \bibinfo{author}{\bibfnamefont{D.~E.} \bibnamefont{Parker}},
  \bibinfo{author}{\bibfnamefont{E.}~\bibnamefont{Khalaf}},
  \bibinfo{author}{\bibfnamefont{P.}~\bibnamefont{Ledwith}},
  \bibinfo{author}{\bibfnamefont{Y.}~\bibnamefont{Cao}},
  \bibinfo{author}{\bibfnamefont{S.~H.} \bibnamefont{Lee}},
  \bibinfo{author}{\bibfnamefont{S.}~\bibnamefont{Chen}},
  \bibinfo{author}{\bibfnamefont{P.~R.} \bibnamefont{Forrester}},
  \emph{et~al.}, \bibinfo{year}{2021}, \bibinfo{journal}{Nature}
  \textbf{\bibinfo{volume}{600}}(\bibinfo{number}{7889}), \bibinfo{pages}{439}.

\bibitem[{\citenamefont{Xie} \emph{et~al.}(2020)\citenamefont{Xie, Gao, Ng, and
  Law}}]{xie2020creating}
\bibinfo{author}{\bibnamefont{Xie}, \bibfnamefont{Y.-M.}},
  \bibinfo{author}{\bibfnamefont{X.-J.} \bibnamefont{Gao}},
  \bibinfo{author}{\bibfnamefont{T.-K.} \bibnamefont{Ng}}, and
  \bibinfo{author}{\bibfnamefont{K.}~\bibnamefont{Law}}, \bibinfo{year}{2020},
  \bibinfo{journal}{arXiv:2012.15523} .

\bibitem[{\citenamefont{Xie}
  \emph{et~al.}(2022{\natexlab{a}})\citenamefont{Xie, Zhang, Hu, Mak, and
  Law}}]{xie2022valley}
\bibinfo{author}{\bibnamefont{Xie}, \bibfnamefont{Y.-M.}},
  \bibinfo{author}{\bibfnamefont{C.-P.} \bibnamefont{Zhang}},
  \bibinfo{author}{\bibfnamefont{J.-X.} \bibnamefont{Hu}},
  \bibinfo{author}{\bibfnamefont{K.~F.} \bibnamefont{Mak}}, and
  \bibinfo{author}{\bibfnamefont{K.}~\bibnamefont{Law}},
  \bibinfo{year}{2022}{\natexlab{a}}, \bibinfo{journal}{Physical Review
  Letters} \textbf{\bibinfo{volume}{128}}(\bibinfo{number}{2}),
  \bibinfo{pages}{026402}.

\bibitem[{\citenamefont{Xie}
  \emph{et~al.}(2022{\natexlab{b}})\citenamefont{Xie, Zhang, and
  Law}}]{xie2022topological}
\bibinfo{author}{\bibnamefont{Xie}, \bibfnamefont{Y.-M.}},
  \bibinfo{author}{\bibfnamefont{C.-P.} \bibnamefont{Zhang}}, and
  \bibinfo{author}{\bibfnamefont{K.}~\bibnamefont{Law}},
  \bibinfo{year}{2022}{\natexlab{b}}, \bibinfo{journal}{arXiv:2206.11666} .

\bibitem[{\citenamefont{Xiong} \emph{et~al.}(2001)\citenamefont{Xiong, Wang,
  Niu, Tian, and Wang}}]{xiong2001metallic}
\bibinfo{author}{\bibnamefont{Xiong}, \bibfnamefont{G.}},
  \bibinfo{author}{\bibfnamefont{S.-D.} \bibnamefont{Wang}},
  \bibinfo{author}{\bibfnamefont{Q.}~\bibnamefont{Niu}},
  \bibinfo{author}{\bibfnamefont{D.-C.} \bibnamefont{Tian}}, and
  \bibinfo{author}{\bibfnamefont{X.}~\bibnamefont{Wang}}, \bibinfo{year}{2001},
  \bibinfo{journal}{Physical Review Letters}
  \textbf{\bibinfo{volume}{87}}(\bibinfo{number}{21}), \bibinfo{pages}{216802}.

\bibitem[{\citenamefont{Xu} \emph{et~al.}(2011)\citenamefont{Xu, Weng, Wang,
  Dai, and Fang}}]{xu2011chern}
\bibinfo{author}{\bibnamefont{Xu}, \bibfnamefont{G.}},
  \bibinfo{author}{\bibfnamefont{H.}~\bibnamefont{Weng}},
  \bibinfo{author}{\bibfnamefont{Z.}~\bibnamefont{Wang}},
  \bibinfo{author}{\bibfnamefont{X.}~\bibnamefont{Dai}}, and
  \bibinfo{author}{\bibfnamefont{Z.}~\bibnamefont{Fang}}, \bibinfo{year}{2011},
  \bibinfo{journal}{Physical Review Letters}
  \textbf{\bibinfo{volume}{107}}(\bibinfo{number}{18}),
  \bibinfo{pages}{186806}.

\bibitem[{\citenamefont{Xu} \emph{et~al.}(2020{\natexlab{a}})\citenamefont{Xu,
  Mao, Wang, Li, Chen, Xia, Li, Pei, Zhang, Zheng}
  \emph{et~al.}}]{xu2020persistent}
\bibinfo{author}{\bibnamefont{Xu}, \bibfnamefont{L.}},
  \bibinfo{author}{\bibfnamefont{Y.}~\bibnamefont{Mao}},
  \bibinfo{author}{\bibfnamefont{H.}~\bibnamefont{Wang}},
  \bibinfo{author}{\bibfnamefont{J.}~\bibnamefont{Li}},
  \bibinfo{author}{\bibfnamefont{Y.}~\bibnamefont{Chen}},
  \bibinfo{author}{\bibfnamefont{Y.}~\bibnamefont{Xia}},
  \bibinfo{author}{\bibfnamefont{Y.}~\bibnamefont{Li}},
  \bibinfo{author}{\bibfnamefont{D.}~\bibnamefont{Pei}},
  \bibinfo{author}{\bibfnamefont{J.}~\bibnamefont{Zhang}},
  \bibinfo{author}{\bibfnamefont{H.}~\bibnamefont{Zheng}}, \emph{et~al.},
  \bibinfo{year}{2020}{\natexlab{a}}, \bibinfo{journal}{Science Bulletin}
  \textbf{\bibinfo{volume}{65}}(\bibinfo{number}{24}), \bibinfo{pages}{2086}.

\bibitem[{\citenamefont{Xu} \emph{et~al.}(2020{\natexlab{b}})\citenamefont{Xu,
  Liu, Rhodes, Watanabe, Taniguchi, Hone, Elser, Mak, and
  Shan}}]{xu2020correlated}
\bibinfo{author}{\bibnamefont{Xu}, \bibfnamefont{Y.}},
  \bibinfo{author}{\bibfnamefont{S.}~\bibnamefont{Liu}},
  \bibinfo{author}{\bibfnamefont{D.~A.} \bibnamefont{Rhodes}},
  \bibinfo{author}{\bibfnamefont{K.}~\bibnamefont{Watanabe}},
  \bibinfo{author}{\bibfnamefont{T.}~\bibnamefont{Taniguchi}},
  \bibinfo{author}{\bibfnamefont{J.}~\bibnamefont{Hone}},
  \bibinfo{author}{\bibfnamefont{V.}~\bibnamefont{Elser}},
  \bibinfo{author}{\bibfnamefont{K.~F.} \bibnamefont{Mak}}, and
  \bibinfo{author}{\bibfnamefont{J.}~\bibnamefont{Shan}},
  \bibinfo{year}{2020}{\natexlab{b}}, \bibinfo{journal}{Nature}
  \textbf{\bibinfo{volume}{587}}(\bibinfo{number}{7833}), \bibinfo{pages}{214}.

\bibitem[{\citenamefont{Xu} \emph{et~al.}(2019)\citenamefont{Xu, Song, Wang,
  Weng, and Dai}}]{xu2019higher}
\bibinfo{author}{\bibnamefont{Xu}, \bibfnamefont{Y.}},
  \bibinfo{author}{\bibfnamefont{Z.}~\bibnamefont{Song}},
  \bibinfo{author}{\bibfnamefont{Z.}~\bibnamefont{Wang}},
  \bibinfo{author}{\bibfnamefont{H.}~\bibnamefont{Weng}}, and
  \bibinfo{author}{\bibfnamefont{X.}~\bibnamefont{Dai}}, \bibinfo{year}{2019},
  \bibinfo{journal}{Physical Review Letters}
  \textbf{\bibinfo{volume}{122}}(\bibinfo{number}{25}),
  \bibinfo{pages}{256402}.

\bibitem[{\citenamefont{Xu} \emph{et~al.}(2013)\citenamefont{Xu, Yan, Zhang,
  Wang, Xu, Tang, Duan, and Zhang}}]{xu2013large}
\bibinfo{author}{\bibnamefont{Xu}, \bibfnamefont{Y.}},
  \bibinfo{author}{\bibfnamefont{B.}~\bibnamefont{Yan}},
  \bibinfo{author}{\bibfnamefont{H.-J.} \bibnamefont{Zhang}},
  \bibinfo{author}{\bibfnamefont{J.}~\bibnamefont{Wang}},
  \bibinfo{author}{\bibfnamefont{G.}~\bibnamefont{Xu}},
  \bibinfo{author}{\bibfnamefont{P.}~\bibnamefont{Tang}},
  \bibinfo{author}{\bibfnamefont{W.}~\bibnamefont{Duan}}, and
  \bibinfo{author}{\bibfnamefont{S.-C.} \bibnamefont{Zhang}},
  \bibinfo{year}{2013}, \bibinfo{journal}{Physical Review Letters}
  \textbf{\bibinfo{volume}{111}}(\bibinfo{number}{13}),
  \bibinfo{pages}{136804}.

\bibitem[{\citenamefont{Xue and MacDonald}(2018)}]{xue2018time}
\bibinfo{author}{\bibnamefont{Xue}, \bibfnamefont{F.}}, and
  \bibinfo{author}{\bibfnamefont{A.~H.} \bibnamefont{MacDonald}},
  \bibinfo{year}{2018}, \bibinfo{journal}{Physical Review Letters}
  \textbf{\bibinfo{volume}{120}}(\bibinfo{number}{18}),
  \bibinfo{pages}{186802}.

\bibitem[{\citenamefont{Yan}
  \emph{et~al.}(2021{\natexlab{a}})\citenamefont{Yan, Fernandez-Mulligan, Mei,
  Lee, Protic, Fukumori, Yan, Liu, Mao, and Yang}}]{yan2021origins}
\bibinfo{author}{\bibnamefont{Yan}, \bibfnamefont{C.}},
  \bibinfo{author}{\bibfnamefont{S.}~\bibnamefont{Fernandez-Mulligan}},
  \bibinfo{author}{\bibfnamefont{R.}~\bibnamefont{Mei}},
  \bibinfo{author}{\bibfnamefont{S.~H.} \bibnamefont{Lee}},
  \bibinfo{author}{\bibfnamefont{N.}~\bibnamefont{Protic}},
  \bibinfo{author}{\bibfnamefont{R.}~\bibnamefont{Fukumori}},
  \bibinfo{author}{\bibfnamefont{B.}~\bibnamefont{Yan}},
  \bibinfo{author}{\bibfnamefont{C.}~\bibnamefont{Liu}},
  \bibinfo{author}{\bibfnamefont{Z.}~\bibnamefont{Mao}}, and
  \bibinfo{author}{\bibfnamefont{S.}~\bibnamefont{Yang}},
  \bibinfo{year}{2021}{\natexlab{a}}, \bibinfo{journal}{Physical Review B}
  \textbf{\bibinfo{volume}{104}}(\bibinfo{number}{4}),
  \bibinfo{pages}{L041102}.

\bibitem[{\citenamefont{Yan}
  \emph{et~al.}(2021{\natexlab{b}})\citenamefont{Yan, Zhu, Fernandez-Mulligan,
  Green, Mei, Yan, Liu, Mao, and Yang}}]{yan2021delicate}
\bibinfo{author}{\bibnamefont{Yan}, \bibfnamefont{C.}},
  \bibinfo{author}{\bibfnamefont{Y.}~\bibnamefont{Zhu}},
  \bibinfo{author}{\bibfnamefont{S.}~\bibnamefont{Fernandez-Mulligan}},
  \bibinfo{author}{\bibfnamefont{E.}~\bibnamefont{Green}},
  \bibinfo{author}{\bibfnamefont{R.}~\bibnamefont{Mei}},
  \bibinfo{author}{\bibfnamefont{B.}~\bibnamefont{Yan}},
  \bibinfo{author}{\bibfnamefont{C.}~\bibnamefont{Liu}},
  \bibinfo{author}{\bibfnamefont{Z.}~\bibnamefont{Mao}}, and
  \bibinfo{author}{\bibfnamefont{S.}~\bibnamefont{Yang}},
  \bibinfo{year}{2021}{\natexlab{b}}, \bibinfo{journal}{arXiv:2107.08137} .

\bibitem[{\citenamefont{Yan}
  \emph{et~al.}(2021{\natexlab{c}})\citenamefont{Yan, Yang, Song, Song, Wang,
  Yi, and Shi}}]{yan2021site}
\bibinfo{author}{\bibnamefont{Yan}, \bibfnamefont{D.}},
  \bibinfo{author}{\bibfnamefont{M.}~\bibnamefont{Yang}},
  \bibinfo{author}{\bibfnamefont{P.}~\bibnamefont{Song}},
  \bibinfo{author}{\bibfnamefont{Y.}~\bibnamefont{Song}},
  \bibinfo{author}{\bibfnamefont{C.}~\bibnamefont{Wang}},
  \bibinfo{author}{\bibfnamefont{C.}~\bibnamefont{Yi}}, and
  \bibinfo{author}{\bibfnamefont{Y.}~\bibnamefont{Shi}},
  \bibinfo{year}{2021}{\natexlab{c}}, \bibinfo{journal}{Physical Review B}
  \textbf{\bibinfo{volume}{103}}(\bibinfo{number}{22}),
  \bibinfo{pages}{224412}.

\bibitem[{\citenamefont{Yan} \emph{et~al.}(2020)\citenamefont{Yan, Liu, Parker,
  Wu, Aczel, Matsuda, McGuire, and Sales}}]{yan2020type}
\bibinfo{author}{\bibnamefont{Yan}, \bibfnamefont{J.-Q.}},
  \bibinfo{author}{\bibfnamefont{Y.}~\bibnamefont{Liu}},
  \bibinfo{author}{\bibfnamefont{D.~S.} \bibnamefont{Parker}},
  \bibinfo{author}{\bibfnamefont{Y.}~\bibnamefont{Wu}},
  \bibinfo{author}{\bibfnamefont{A.}~\bibnamefont{Aczel}},
  \bibinfo{author}{\bibfnamefont{M.}~\bibnamefont{Matsuda}},
  \bibinfo{author}{\bibfnamefont{M.~A.} \bibnamefont{McGuire}}, and
  \bibinfo{author}{\bibfnamefont{B.~C.} \bibnamefont{Sales}},
  \bibinfo{year}{2020}, \bibinfo{journal}{Physical Review Materials}
  \textbf{\bibinfo{volume}{4}}(\bibinfo{number}{5}), \bibinfo{pages}{054202}.

\bibitem[{\citenamefont{Yan}
  \emph{et~al.}(2019{\natexlab{a}})\citenamefont{Yan, Okamoto, McGuire, May,
  McQueeney, and Sales}}]{yan2019evolution}
\bibinfo{author}{\bibnamefont{Yan}, \bibfnamefont{J.-Q.}},
  \bibinfo{author}{\bibfnamefont{S.}~\bibnamefont{Okamoto}},
  \bibinfo{author}{\bibfnamefont{M.~A.} \bibnamefont{McGuire}},
  \bibinfo{author}{\bibfnamefont{A.~F.} \bibnamefont{May}},
  \bibinfo{author}{\bibfnamefont{R.~J.} \bibnamefont{McQueeney}}, and
  \bibinfo{author}{\bibfnamefont{B.~C.} \bibnamefont{Sales}},
  \bibinfo{year}{2019}{\natexlab{a}}, \bibinfo{journal}{Physical Review B}
  \textbf{\bibinfo{volume}{100}}(\bibinfo{number}{10}),
  \bibinfo{pages}{104409}.

\bibitem[{\citenamefont{Yan}
  \emph{et~al.}(2019{\natexlab{b}})\citenamefont{Yan, Zhang, Heitmann, Huang,
  Chen, Cheng, Wu, Vaknin, Sales, and McQueeney}}]{yan2019crystal}
\bibinfo{author}{\bibnamefont{Yan}, \bibfnamefont{J.-Q.}},
  \bibinfo{author}{\bibfnamefont{Q.}~\bibnamefont{Zhang}},
  \bibinfo{author}{\bibfnamefont{T.}~\bibnamefont{Heitmann}},
  \bibinfo{author}{\bibfnamefont{Z.}~\bibnamefont{Huang}},
  \bibinfo{author}{\bibfnamefont{K.}~\bibnamefont{Chen}},
  \bibinfo{author}{\bibfnamefont{J.-G.} \bibnamefont{Cheng}},
  \bibinfo{author}{\bibfnamefont{W.}~\bibnamefont{Wu}},
  \bibinfo{author}{\bibfnamefont{D.}~\bibnamefont{Vaknin}},
  \bibinfo{author}{\bibfnamefont{B.~C.} \bibnamefont{Sales}}, and
  \bibinfo{author}{\bibfnamefont{R.~J.} \bibnamefont{McQueeney}},
  \bibinfo{year}{2019}{\natexlab{b}}, \bibinfo{journal}{Physical Review
  Materials} \textbf{\bibinfo{volume}{3}}(\bibinfo{number}{6}),
  \bibinfo{pages}{064202}.

\bibitem[{\citenamefont{Yang} \emph{et~al.}(2021)\citenamefont{Yang, Xu, Zhu,
  Niu, Xu, Peng, Cheng, Jia, Huang, Xu} \emph{et~al.}}]{yang2021odd}
\bibinfo{author}{\bibnamefont{Yang}, \bibfnamefont{S.}},
  \bibinfo{author}{\bibfnamefont{X.}~\bibnamefont{Xu}},
  \bibinfo{author}{\bibfnamefont{Y.}~\bibnamefont{Zhu}},
  \bibinfo{author}{\bibfnamefont{R.}~\bibnamefont{Niu}},
  \bibinfo{author}{\bibfnamefont{C.}~\bibnamefont{Xu}},
  \bibinfo{author}{\bibfnamefont{Y.}~\bibnamefont{Peng}},
  \bibinfo{author}{\bibfnamefont{X.}~\bibnamefont{Cheng}},
  \bibinfo{author}{\bibfnamefont{X.}~\bibnamefont{Jia}},
  \bibinfo{author}{\bibfnamefont{Y.}~\bibnamefont{Huang}},
  \bibinfo{author}{\bibfnamefont{X.}~\bibnamefont{Xu}}, \emph{et~al.},
  \bibinfo{year}{2021}, \bibinfo{journal}{Physical Review X}
  \textbf{\bibinfo{volume}{11}}(\bibinfo{number}{1}), \bibinfo{pages}{011003}.

\bibitem[{\citenamefont{Yankowitz} \emph{et~al.}(2019)\citenamefont{Yankowitz,
  Chen, Polshyn, Zhang, Watanabe, Taniguchi, Graf, Young, and
  Dean}}]{yankowitz2019tuning}
\bibinfo{author}{\bibnamefont{Yankowitz}, \bibfnamefont{M.}},
  \bibinfo{author}{\bibfnamefont{S.}~\bibnamefont{Chen}},
  \bibinfo{author}{\bibfnamefont{H.}~\bibnamefont{Polshyn}},
  \bibinfo{author}{\bibfnamefont{Y.}~\bibnamefont{Zhang}},
  \bibinfo{author}{\bibfnamefont{K.}~\bibnamefont{Watanabe}},
  \bibinfo{author}{\bibfnamefont{T.}~\bibnamefont{Taniguchi}},
  \bibinfo{author}{\bibfnamefont{D.}~\bibnamefont{Graf}},
  \bibinfo{author}{\bibfnamefont{A.~F.} \bibnamefont{Young}}, and
  \bibinfo{author}{\bibfnamefont{C.~R.} \bibnamefont{Dean}},
  \bibinfo{year}{2019}, \bibinfo{journal}{Science}
  \textbf{\bibinfo{volume}{363}}(\bibinfo{number}{6431}),
  \bibinfo{pages}{1059}.

\bibitem[{\citenamefont{Yankowitz} \emph{et~al.}(2012)\citenamefont{Yankowitz,
  Xue, Cormode, Sanchez-Yamagishi, Watanabe, Taniguchi, Jarillo-Herrero,
  Jacquod, and LeRoy}}]{yankowitz2012emergence}
\bibinfo{author}{\bibnamefont{Yankowitz}, \bibfnamefont{M.}},
  \bibinfo{author}{\bibfnamefont{J.}~\bibnamefont{Xue}},
  \bibinfo{author}{\bibfnamefont{D.}~\bibnamefont{Cormode}},
  \bibinfo{author}{\bibfnamefont{J.~D.} \bibnamefont{Sanchez-Yamagishi}},
  \bibinfo{author}{\bibfnamefont{K.}~\bibnamefont{Watanabe}},
  \bibinfo{author}{\bibfnamefont{T.}~\bibnamefont{Taniguchi}},
  \bibinfo{author}{\bibfnamefont{P.}~\bibnamefont{Jarillo-Herrero}},
  \bibinfo{author}{\bibfnamefont{P.}~\bibnamefont{Jacquod}}, and
  \bibinfo{author}{\bibfnamefont{B.~J.} \bibnamefont{LeRoy}},
  \bibinfo{year}{2012}, \bibinfo{journal}{Nature Physics}
  \textbf{\bibinfo{volume}{8}}(\bibinfo{number}{5}), \bibinfo{pages}{382}.

\bibitem[{\citenamefont{Yasuda} \emph{et~al.}(2017)\citenamefont{Yasuda, Mogi,
  Yoshimi, Tsukazaki, Takahashi, Kawasaki, Kagawa, and
  Tokura}}]{yasuda2017quantized}
\bibinfo{author}{\bibnamefont{Yasuda}, \bibfnamefont{K.}},
  \bibinfo{author}{\bibfnamefont{M.}~\bibnamefont{Mogi}},
  \bibinfo{author}{\bibfnamefont{R.}~\bibnamefont{Yoshimi}},
  \bibinfo{author}{\bibfnamefont{A.}~\bibnamefont{Tsukazaki}},
  \bibinfo{author}{\bibfnamefont{K.}~\bibnamefont{Takahashi}},
  \bibinfo{author}{\bibfnamefont{M.}~\bibnamefont{Kawasaki}},
  \bibinfo{author}{\bibfnamefont{F.}~\bibnamefont{Kagawa}}, and
  \bibinfo{author}{\bibfnamefont{Y.}~\bibnamefont{Tokura}},
  \bibinfo{year}{2017}, \bibinfo{journal}{Science}
  \textbf{\bibinfo{volume}{358}}(\bibinfo{number}{6368}),
  \bibinfo{pages}{1311}.

\bibitem[{\citenamefont{Ying} \emph{et~al.}(2022)\citenamefont{Ying, Zhang,
  Chen, Jia, Fei, Zhang, Zhang, Wang, and Song}}]{ying2022experimental}
\bibinfo{author}{\bibnamefont{Ying}, \bibfnamefont{Z.}},
  \bibinfo{author}{\bibfnamefont{S.}~\bibnamefont{Zhang}},
  \bibinfo{author}{\bibfnamefont{B.}~\bibnamefont{Chen}},
  \bibinfo{author}{\bibfnamefont{B.}~\bibnamefont{Jia}},
  \bibinfo{author}{\bibfnamefont{F.}~\bibnamefont{Fei}},
  \bibinfo{author}{\bibfnamefont{M.}~\bibnamefont{Zhang}},
  \bibinfo{author}{\bibfnamefont{H.}~\bibnamefont{Zhang}},
  \bibinfo{author}{\bibfnamefont{X.}~\bibnamefont{Wang}}, and
  \bibinfo{author}{\bibfnamefont{F.}~\bibnamefont{Song}}, \bibinfo{year}{2022},
  \bibinfo{journal}{Physical Review B}
  \textbf{\bibinfo{volume}{105}}(\bibinfo{number}{8}), \bibinfo{pages}{085412}.

\bibitem[{\citenamefont{Yosida}(1957)}]{yosida1957magnetic}
\bibinfo{author}{\bibnamefont{Yosida}, \bibfnamefont{K.}},
  \bibinfo{year}{1957}, \bibinfo{journal}{Physical Review}
  \textbf{\bibinfo{volume}{106}}(\bibinfo{number}{5}), \bibinfo{pages}{893}.

\bibitem[{\citenamefont{You} \emph{et~al.}(2019)\citenamefont{You, Chen, Zhang,
  Sheng, Yang, and Su}}]{you2019two}
\bibinfo{author}{\bibnamefont{You}, \bibfnamefont{J.-Y.}},
  \bibinfo{author}{\bibfnamefont{C.}~\bibnamefont{Chen}},
  \bibinfo{author}{\bibfnamefont{Z.}~\bibnamefont{Zhang}},
  \bibinfo{author}{\bibfnamefont{X.-L.} \bibnamefont{Sheng}},
  \bibinfo{author}{\bibfnamefont{S.~A.} \bibnamefont{Yang}}, and
  \bibinfo{author}{\bibfnamefont{G.}~\bibnamefont{Su}}, \bibinfo{year}{2019},
  \bibinfo{journal}{Physical Review B}
  \textbf{\bibinfo{volume}{100}}(\bibinfo{number}{6}), \bibinfo{pages}{064408}.

\bibitem[{\citenamefont{Yu} \emph{et~al.}(2010)\citenamefont{Yu, Zhang, Zhang,
  Zhang, Dai, and Fang}}]{yu2010quantized}
\bibinfo{author}{\bibnamefont{Yu}, \bibfnamefont{R.}},
  \bibinfo{author}{\bibfnamefont{W.}~\bibnamefont{Zhang}},
  \bibinfo{author}{\bibfnamefont{H.-J.} \bibnamefont{Zhang}},
  \bibinfo{author}{\bibfnamefont{S.-C.} \bibnamefont{Zhang}},
  \bibinfo{author}{\bibfnamefont{X.}~\bibnamefont{Dai}}, and
  \bibinfo{author}{\bibfnamefont{Z.}~\bibnamefont{Fang}}, \bibinfo{year}{2010},
  \bibinfo{journal}{Science}
  \textbf{\bibinfo{volume}{329}}(\bibinfo{number}{5987}), \bibinfo{pages}{61}.

\bibitem[{\citenamefont{Yuan} \emph{et~al.}(2022)\citenamefont{Yuan, Zhou,
  Yang, Zhao, Zhang, Yan, Zhuo, Mei, Chan, Kayyalha}
  \emph{et~al.}}]{yuan2022electrical}
\bibinfo{author}{\bibnamefont{Yuan}, \bibfnamefont{W.}},
  \bibinfo{author}{\bibfnamefont{L.-J.} \bibnamefont{Zhou}},
  \bibinfo{author}{\bibfnamefont{K.}~\bibnamefont{Yang}},
  \bibinfo{author}{\bibfnamefont{Y.-F.} \bibnamefont{Zhao}},
  \bibinfo{author}{\bibfnamefont{R.}~\bibnamefont{Zhang}},
  \bibinfo{author}{\bibfnamefont{Z.}~\bibnamefont{Yan}},
  \bibinfo{author}{\bibfnamefont{D.}~\bibnamefont{Zhuo}},
  \bibinfo{author}{\bibfnamefont{R.}~\bibnamefont{Mei}},
  \bibinfo{author}{\bibfnamefont{M.~H.} \bibnamefont{Chan}},
  \bibinfo{author}{\bibfnamefont{M.}~\bibnamefont{Kayyalha}}, \emph{et~al.},
  \bibinfo{year}{2022}, \bibinfo{journal}{arXiv:2205.01581} .

\bibitem[{\citenamefont{Yuan} \emph{et~al.}(2020)\citenamefont{Yuan, Wang, Li,
  Li, Ji, Hao, Wu, He, Wang, Xu} \emph{et~al.}}]{yuan2020electronic}
\bibinfo{author}{\bibnamefont{Yuan}, \bibfnamefont{Y.}},
  \bibinfo{author}{\bibfnamefont{X.}~\bibnamefont{Wang}},
  \bibinfo{author}{\bibfnamefont{H.}~\bibnamefont{Li}},
  \bibinfo{author}{\bibfnamefont{J.}~\bibnamefont{Li}},
  \bibinfo{author}{\bibfnamefont{Y.}~\bibnamefont{Ji}},
  \bibinfo{author}{\bibfnamefont{Z.}~\bibnamefont{Hao}},
  \bibinfo{author}{\bibfnamefont{Y.}~\bibnamefont{Wu}},
  \bibinfo{author}{\bibfnamefont{K.}~\bibnamefont{He}},
  \bibinfo{author}{\bibfnamefont{Y.}~\bibnamefont{Wang}},
  \bibinfo{author}{\bibfnamefont{Y.}~\bibnamefont{Xu}}, \emph{et~al.},
  \bibinfo{year}{2020}, \bibinfo{journal}{Nano Letters}
  \textbf{\bibinfo{volume}{20}}(\bibinfo{number}{5}), \bibinfo{pages}{3271}.

\bibitem[{\citenamefont{Yue} \emph{et~al.}(2019)\citenamefont{Yue, Xu, Song,
  Weng, Lu, Fang, and Dai}}]{yue2019symmetry}
\bibinfo{author}{\bibnamefont{Yue}, \bibfnamefont{C.}},
  \bibinfo{author}{\bibfnamefont{Y.}~\bibnamefont{Xu}},
  \bibinfo{author}{\bibfnamefont{Z.}~\bibnamefont{Song}},
  \bibinfo{author}{\bibfnamefont{H.}~\bibnamefont{Weng}},
  \bibinfo{author}{\bibfnamefont{Y.-M.} \bibnamefont{Lu}},
  \bibinfo{author}{\bibfnamefont{C.}~\bibnamefont{Fang}}, and
  \bibinfo{author}{\bibfnamefont{X.}~\bibnamefont{Dai}}, \bibinfo{year}{2019},
  \bibinfo{journal}{Nature Physics}
  \textbf{\bibinfo{volume}{15}}(\bibinfo{number}{6}), \bibinfo{pages}{577}.

\bibitem[{\citenamefont{Zener}(1951)}]{zener1951interaction}
\bibinfo{author}{\bibnamefont{Zener}, \bibfnamefont{C.}}, \bibinfo{year}{1951},
  \bibinfo{journal}{Physical Review}
  \textbf{\bibinfo{volume}{82}}(\bibinfo{number}{3}), \bibinfo{pages}{403}.

\bibitem[{\citenamefont{Zener and Heikes}(1953)}]{zener1953exchange}
\bibinfo{author}{\bibnamefont{Zener}, \bibfnamefont{C.}}, and
  \bibinfo{author}{\bibfnamefont{R.}~\bibnamefont{Heikes}},
  \bibinfo{year}{1953}, \bibinfo{journal}{Reviews of Modern Physics}
  \textbf{\bibinfo{volume}{25}}(\bibinfo{number}{1}), \bibinfo{pages}{191}.

\bibitem[{\citenamefont{Zeng} \emph{et~al.}(2018)\citenamefont{Zeng, Lei,
  Chaudhary, and MacDonald}}]{zeng2018quantum}
\bibinfo{author}{\bibnamefont{Zeng}, \bibfnamefont{Y.}},
  \bibinfo{author}{\bibfnamefont{C.}~\bibnamefont{Lei}},
  \bibinfo{author}{\bibfnamefont{G.}~\bibnamefont{Chaudhary}}, and
  \bibinfo{author}{\bibfnamefont{A.~H.} \bibnamefont{MacDonald}},
  \bibinfo{year}{2018}, \bibinfo{journal}{Physical Review B}
  \textbf{\bibinfo{volume}{97}}(\bibinfo{number}{8}), \bibinfo{pages}{081102}.

\bibitem[{\citenamefont{Zeng} \emph{et~al.}(2022)\citenamefont{Zeng, Xue, and
  MacDonald}}]{zeng2022plane}
\bibinfo{author}{\bibnamefont{Zeng}, \bibfnamefont{Y.}},
  \bibinfo{author}{\bibfnamefont{F.}~\bibnamefont{Xue}}, and
  \bibinfo{author}{\bibfnamefont{A.~H.} \bibnamefont{MacDonald}},
  \bibinfo{year}{2022}, \bibinfo{journal}{Physical Review B}
  \textbf{\bibinfo{volume}{105}}(\bibinfo{number}{12}),
  \bibinfo{pages}{125102}.

\bibitem[{\citenamefont{Zhang}
  \emph{et~al.}(2019{\natexlab{a}})\citenamefont{Zhang, Shi, Zhu, Xing, Zhang,
  and Wang}}]{zhang2019topological}
\bibinfo{author}{\bibnamefont{Zhang}, \bibfnamefont{D.}},
  \bibinfo{author}{\bibfnamefont{M.}~\bibnamefont{Shi}},
  \bibinfo{author}{\bibfnamefont{T.}~\bibnamefont{Zhu}},
  \bibinfo{author}{\bibfnamefont{D.}~\bibnamefont{Xing}},
  \bibinfo{author}{\bibfnamefont{H.}~\bibnamefont{Zhang}}, and
  \bibinfo{author}{\bibfnamefont{J.}~\bibnamefont{Wang}},
  \bibinfo{year}{2019}{\natexlab{a}}, \bibinfo{journal}{Physical Review
  Letters} \textbf{\bibinfo{volume}{122}}(\bibinfo{number}{20}),
  \bibinfo{pages}{206401}.

\bibitem[{\citenamefont{Zhang} \emph{et~al.}(2011)\citenamefont{Zhang, Jung,
  Fiete, Niu, and MacDonald}}]{zhang2011spontaneous}
\bibinfo{author}{\bibnamefont{Zhang}, \bibfnamefont{F.}},
  \bibinfo{author}{\bibfnamefont{J.}~\bibnamefont{Jung}},
  \bibinfo{author}{\bibfnamefont{G.~A.} \bibnamefont{Fiete}},
  \bibinfo{author}{\bibfnamefont{Q.}~\bibnamefont{Niu}}, and
  \bibinfo{author}{\bibfnamefont{A.~H.} \bibnamefont{MacDonald}},
  \bibinfo{year}{2011}, \bibinfo{journal}{Physical Review Letters}
  \textbf{\bibinfo{volume}{106}}(\bibinfo{number}{15}),
  \bibinfo{pages}{156801}.

\bibitem[{\citenamefont{Zhang} \emph{et~al.}(2012)\citenamefont{Zhang, Lazo,
  Bl{\"u}gel, Heinze, and Mokrousov}}]{zhang2012electrically}
\bibinfo{author}{\bibnamefont{Zhang}, \bibfnamefont{H.}},
  \bibinfo{author}{\bibfnamefont{C.}~\bibnamefont{Lazo}},
  \bibinfo{author}{\bibfnamefont{S.}~\bibnamefont{Bl{\"u}gel}},
  \bibinfo{author}{\bibfnamefont{S.}~\bibnamefont{Heinze}}, and
  \bibinfo{author}{\bibfnamefont{Y.}~\bibnamefont{Mokrousov}},
  \bibinfo{year}{2012}, \bibinfo{journal}{Physical Review Letters}
  \textbf{\bibinfo{volume}{108}}(\bibinfo{number}{5}), \bibinfo{pages}{056802}.

\bibitem[{\citenamefont{Zhang} \emph{et~al.}(2009)\citenamefont{Zhang, Liu, Qi,
  Dai, Fang, and Zhang}}]{zhang2009topological}
\bibinfo{author}{\bibnamefont{Zhang}, \bibfnamefont{H.}},
  \bibinfo{author}{\bibfnamefont{C.-X.} \bibnamefont{Liu}},
  \bibinfo{author}{\bibfnamefont{X.-L.} \bibnamefont{Qi}},
  \bibinfo{author}{\bibfnamefont{X.}~\bibnamefont{Dai}},
  \bibinfo{author}{\bibfnamefont{Z.}~\bibnamefont{Fang}}, and
  \bibinfo{author}{\bibfnamefont{S.-C.} \bibnamefont{Zhang}},
  \bibinfo{year}{2009}, \bibinfo{journal}{Nature Physics}
  \textbf{\bibinfo{volume}{5}}(\bibinfo{number}{6}), \bibinfo{pages}{438}.

\bibitem[{\citenamefont{Zhang}
  \emph{et~al.}(2019{\natexlab{b}})\citenamefont{Zhang, Ning, Yang, Zhang,
  Zhang, and Xu}}]{zhang2019possible}
\bibinfo{author}{\bibnamefont{Zhang}, \bibfnamefont{H.}},
  \bibinfo{author}{\bibfnamefont{Y.}~\bibnamefont{Ning}},
  \bibinfo{author}{\bibfnamefont{W.}~\bibnamefont{Yang}},
  \bibinfo{author}{\bibfnamefont{J.}~\bibnamefont{Zhang}},
  \bibinfo{author}{\bibfnamefont{R.}~\bibnamefont{Zhang}}, and
  \bibinfo{author}{\bibfnamefont{X.}~\bibnamefont{Xu}},
  \bibinfo{year}{2019}{\natexlab{b}}, \bibinfo{journal}{Physical Chemistry
  Chemical Physics} \textbf{\bibinfo{volume}{21}}(\bibinfo{number}{31}),
  \bibinfo{pages}{17087}.

\bibitem[{\citenamefont{Zhang}
  \emph{et~al.}(2019{\natexlab{c}})\citenamefont{Zhang, Qin, Chen, Cui, Zhang,
  and Xu}}]{zhang2019converting}
\bibinfo{author}{\bibnamefont{Zhang}, \bibfnamefont{H.}},
  \bibinfo{author}{\bibfnamefont{W.}~\bibnamefont{Qin}},
  \bibinfo{author}{\bibfnamefont{M.}~\bibnamefont{Chen}},
  \bibinfo{author}{\bibfnamefont{P.}~\bibnamefont{Cui}},
  \bibinfo{author}{\bibfnamefont{Z.}~\bibnamefont{Zhang}}, and
  \bibinfo{author}{\bibfnamefont{X.}~\bibnamefont{Xu}},
  \bibinfo{year}{2019}{\natexlab{c}}, \bibinfo{journal}{Physical Review B}
  \textbf{\bibinfo{volume}{99}}(\bibinfo{number}{16}), \bibinfo{pages}{165410}.

\bibitem[{\citenamefont{Zhang} \emph{et~al.}(2014)\citenamefont{Zhang, Xu,
  Wang, Chang, and Zhang}}]{zhang2014quantumSpin}
\bibinfo{author}{\bibnamefont{Zhang}, \bibfnamefont{H.}},
  \bibinfo{author}{\bibfnamefont{Y.}~\bibnamefont{Xu}},
  \bibinfo{author}{\bibfnamefont{J.}~\bibnamefont{Wang}},
  \bibinfo{author}{\bibfnamefont{K.}~\bibnamefont{Chang}}, and
  \bibinfo{author}{\bibfnamefont{S.-C.} \bibnamefont{Zhang}},
  \bibinfo{year}{2014}, \bibinfo{journal}{Physical Review Letters}
  \textbf{\bibinfo{volume}{112}}(\bibinfo{number}{21}),
  \bibinfo{pages}{216803}.

\bibitem[{\citenamefont{Zhang}
  \emph{et~al.}(2016{\natexlab{a}})\citenamefont{Zhang, Zhang, Zhao, Zhou, and
  Yang}}]{zhang2016quantumanomalous}
\bibinfo{author}{\bibnamefont{Zhang}, \bibfnamefont{H.}},
  \bibinfo{author}{\bibfnamefont{J.}~\bibnamefont{Zhang}},
  \bibinfo{author}{\bibfnamefont{B.}~\bibnamefont{Zhao}},
  \bibinfo{author}{\bibfnamefont{T.}~\bibnamefont{Zhou}}, and
  \bibinfo{author}{\bibfnamefont{Z.}~\bibnamefont{Yang}},
  \bibinfo{year}{2016}{\natexlab{a}}, \bibinfo{journal}{Applied Physics
  Letters} \textbf{\bibinfo{volume}{108}}(\bibinfo{number}{8}),
  \bibinfo{pages}{082104}.

\bibitem[{\citenamefont{Zhang}
  \emph{et~al.}(2016{\natexlab{b}})\citenamefont{Zhang, Zhou, Zhang, Zhao, Yao,
  and Yang}}]{zhang2016quantum}
\bibinfo{author}{\bibnamefont{Zhang}, \bibfnamefont{H.}},
  \bibinfo{author}{\bibfnamefont{T.}~\bibnamefont{Zhou}},
  \bibinfo{author}{\bibfnamefont{J.}~\bibnamefont{Zhang}},
  \bibinfo{author}{\bibfnamefont{B.}~\bibnamefont{Zhao}},
  \bibinfo{author}{\bibfnamefont{Y.}~\bibnamefont{Yao}}, and
  \bibinfo{author}{\bibfnamefont{Z.}~\bibnamefont{Yang}},
  \bibinfo{year}{2016}{\natexlab{b}}, \bibinfo{journal}{Physical Review B}
  \textbf{\bibinfo{volume}{94}}(\bibinfo{number}{23}), \bibinfo{pages}{235409}.

\bibitem[{\citenamefont{Zhang}
  \emph{et~al.}(2013{\natexlab{a}})\citenamefont{Zhang, Chang, Tang, Zhang,
  Feng, Li, Wang, Chen, Liu, Duan} \emph{et~al.}}]{zhang2013topology}
\bibinfo{author}{\bibnamefont{Zhang}, \bibfnamefont{J.}},
  \bibinfo{author}{\bibfnamefont{C.-Z.} \bibnamefont{Chang}},
  \bibinfo{author}{\bibfnamefont{P.}~\bibnamefont{Tang}},
  \bibinfo{author}{\bibfnamefont{Z.}~\bibnamefont{Zhang}},
  \bibinfo{author}{\bibfnamefont{X.}~\bibnamefont{Feng}},
  \bibinfo{author}{\bibfnamefont{K.}~\bibnamefont{Li}},
  \bibinfo{author}{\bibfnamefont{L.-l.} \bibnamefont{Wang}},
  \bibinfo{author}{\bibfnamefont{X.}~\bibnamefont{Chen}},
  \bibinfo{author}{\bibfnamefont{C.}~\bibnamefont{Liu}},
  \bibinfo{author}{\bibfnamefont{W.}~\bibnamefont{Duan}}, \emph{et~al.},
  \bibinfo{year}{2013}{\natexlab{a}}, \bibinfo{journal}{Science}
  \textbf{\bibinfo{volume}{339}}(\bibinfo{number}{6127}),
  \bibinfo{pages}{1582}.

\bibitem[{\citenamefont{Zhang}
  \emph{et~al.}(2013{\natexlab{b}})\citenamefont{Zhang, Zhao, and
  Yang}}]{zhang2013abundant}
\bibinfo{author}{\bibnamefont{Zhang}, \bibfnamefont{J.}},
  \bibinfo{author}{\bibfnamefont{B.}~\bibnamefont{Zhao}}, and
  \bibinfo{author}{\bibfnamefont{Z.}~\bibnamefont{Yang}},
  \bibinfo{year}{2013}{\natexlab{b}}, \bibinfo{journal}{Physical Review B}
  \textbf{\bibinfo{volume}{88}}(\bibinfo{number}{16}), \bibinfo{pages}{165422}.

\bibitem[{\citenamefont{Zhang} \emph{et~al.}(2015)\citenamefont{Zhang, Zhao,
  Yao, and Yang}}]{zhang2015robust}
\bibinfo{author}{\bibnamefont{Zhang}, \bibfnamefont{J.}},
  \bibinfo{author}{\bibfnamefont{B.}~\bibnamefont{Zhao}},
  \bibinfo{author}{\bibfnamefont{Y.}~\bibnamefont{Yao}}, and
  \bibinfo{author}{\bibfnamefont{Z.}~\bibnamefont{Yang}}, \bibinfo{year}{2015},
  \bibinfo{journal}{Physical Review B}
  \textbf{\bibinfo{volume}{92}}(\bibinfo{number}{16}), \bibinfo{pages}{165418}.

\bibitem[{\citenamefont{Zhang}
  \emph{et~al.}(2018{\natexlab{a}})\citenamefont{Zhang, Li, Liu, and
  Zhu}}]{zhang2018realizing}
\bibinfo{author}{\bibnamefont{Zhang}, \bibfnamefont{K.-C.}},
  \bibinfo{author}{\bibfnamefont{Y.-F.} \bibnamefont{Li}},
  \bibinfo{author}{\bibfnamefont{Y.}~\bibnamefont{Liu}}, and
  \bibinfo{author}{\bibfnamefont{Y.}~\bibnamefont{Zhu}},
  \bibinfo{year}{2018}{\natexlab{a}}, \bibinfo{journal}{Journal of Physics:
  Condensed Matter} \textbf{\bibinfo{volume}{31}}(\bibinfo{number}{4}),
  \bibinfo{pages}{045802}.

\bibitem[{\citenamefont{Zhang}
  \emph{et~al.}(2016{\natexlab{c}})\citenamefont{Zhang, Hsu, and
  Liu}}]{zhang2016electrically}
\bibinfo{author}{\bibnamefont{Zhang}, \bibfnamefont{R.-X.}},
  \bibinfo{author}{\bibfnamefont{H.-C.} \bibnamefont{Hsu}}, and
  \bibinfo{author}{\bibfnamefont{C.-X.} \bibnamefont{Liu}},
  \bibinfo{year}{2016}{\natexlab{c}}, \bibinfo{journal}{Physical Review B}
  \textbf{\bibinfo{volume}{93}}(\bibinfo{number}{23}), \bibinfo{pages}{235315}.

\bibitem[{\citenamefont{Zhang} \emph{et~al.}(2020)\citenamefont{Zhang, Wu, and
  Sarma}}]{zhang2020mobius}
\bibinfo{author}{\bibnamefont{Zhang}, \bibfnamefont{R.-X.}},
  \bibinfo{author}{\bibfnamefont{F.}~\bibnamefont{Wu}}, and
  \bibinfo{author}{\bibfnamefont{S.~D.} \bibnamefont{Sarma}},
  \bibinfo{year}{2020}, \bibinfo{journal}{Physical Review Letters}
  \textbf{\bibinfo{volume}{124}}(\bibinfo{number}{13}),
  \bibinfo{pages}{136407}.

\bibitem[{\citenamefont{Zhang}
  \emph{et~al.}(2019{\natexlab{d}})\citenamefont{Zhang, Wang, Wang, Wei, Chen,
  Wang, Shi, Wang, Jia, Ouyang} \emph{et~al.}}]{zhang2019experimental}
\bibinfo{author}{\bibnamefont{Zhang}, \bibfnamefont{S.}},
  \bibinfo{author}{\bibfnamefont{R.}~\bibnamefont{Wang}},
  \bibinfo{author}{\bibfnamefont{X.}~\bibnamefont{Wang}},
  \bibinfo{author}{\bibfnamefont{B.}~\bibnamefont{Wei}},
  \bibinfo{author}{\bibfnamefont{B.}~\bibnamefont{Chen}},
  \bibinfo{author}{\bibfnamefont{H.}~\bibnamefont{Wang}},
  \bibinfo{author}{\bibfnamefont{G.}~\bibnamefont{Shi}},
  \bibinfo{author}{\bibfnamefont{F.}~\bibnamefont{Wang}},
  \bibinfo{author}{\bibfnamefont{B.}~\bibnamefont{Jia}},
  \bibinfo{author}{\bibfnamefont{Y.}~\bibnamefont{Ouyang}}, \emph{et~al.},
  \bibinfo{year}{2019}{\natexlab{d}}, \bibinfo{journal}{Nano Letters}
  \textbf{\bibinfo{volume}{20}}(\bibinfo{number}{1}), \bibinfo{pages}{709}.

\bibitem[{\citenamefont{Zhang}
  \emph{et~al.}(2018{\natexlab{b}})\citenamefont{Zhang, West, Lee, Qiu, Chang,
  Moodera, San~Hor, Zhang, and Wu}}]{zhang2018electronic}
\bibinfo{author}{\bibnamefont{Zhang}, \bibfnamefont{W.}},
  \bibinfo{author}{\bibfnamefont{D.}~\bibnamefont{West}},
  \bibinfo{author}{\bibfnamefont{S.~H.} \bibnamefont{Lee}},
  \bibinfo{author}{\bibfnamefont{Y.}~\bibnamefont{Qiu}},
  \bibinfo{author}{\bibfnamefont{C.-Z.} \bibnamefont{Chang}},
  \bibinfo{author}{\bibfnamefont{J.~S.} \bibnamefont{Moodera}},
  \bibinfo{author}{\bibfnamefont{Y.}~\bibnamefont{San~Hor}},
  \bibinfo{author}{\bibfnamefont{S.}~\bibnamefont{Zhang}}, and
  \bibinfo{author}{\bibfnamefont{W.}~\bibnamefont{Wu}},
  \bibinfo{year}{2018}{\natexlab{b}}, \bibinfo{journal}{Physical Review B}
  \textbf{\bibinfo{volume}{98}}(\bibinfo{number}{11}), \bibinfo{pages}{115165}.

\bibitem[{\citenamefont{Zhang and Zhang}(2012)}]{zhang2012chiral}
\bibinfo{author}{\bibnamefont{Zhang}, \bibfnamefont{X.}}, and
  \bibinfo{author}{\bibfnamefont{S.-C.} \bibnamefont{Zhang}},
  \bibinfo{year}{2012}, in \emph{\bibinfo{booktitle}{Micro-and Nanotechnology
  Sensors, Systems, and Applications IV}} (\bibinfo{organization}{International
  Society for Optics and Photonics}), volume \bibinfo{volume}{8373}, p.
  \bibinfo{pages}{837309}.

\bibitem[{\citenamefont{Zhang}
  \emph{et~al.}(2013{\natexlab{c}})\citenamefont{Zhang, Liu, and
  Liu}}]{zhang2013quantum}
\bibinfo{author}{\bibnamefont{Zhang}, \bibfnamefont{X.-L.}},
  \bibinfo{author}{\bibfnamefont{L.-F.} \bibnamefont{Liu}}, and
  \bibinfo{author}{\bibfnamefont{W.-M.} \bibnamefont{Liu}},
  \bibinfo{year}{2013}{\natexlab{c}}, \bibinfo{journal}{Scientific Reports}
  \textbf{\bibinfo{volume}{3}}(\bibinfo{number}{1}), \bibinfo{pages}{1}.

\bibitem[{\citenamefont{Zhang} \emph{et~al.}(2021)\citenamefont{Zhang, Devakul,
  and Fu}}]{zhang2021spin}
\bibinfo{author}{\bibnamefont{Zhang}, \bibfnamefont{Y.}},
  \bibinfo{author}{\bibfnamefont{T.}~\bibnamefont{Devakul}}, and
  \bibinfo{author}{\bibfnamefont{L.}~\bibnamefont{Fu}}, \bibinfo{year}{2021},
  \bibinfo{journal}{Proceedings of the National Academy of Sciences}
  \textbf{\bibinfo{volume}{118}}(\bibinfo{number}{36}).

\bibitem[{\citenamefont{Zhang} \emph{et~al.}(2010)\citenamefont{Zhang, He,
  Chang, Song, Wang, Chen, Jia, Fang, Dai, Shan}
  \emph{et~al.}}]{zhang2010crossover}
\bibinfo{author}{\bibnamefont{Zhang}, \bibfnamefont{Y.}},
  \bibinfo{author}{\bibfnamefont{K.}~\bibnamefont{He}},
  \bibinfo{author}{\bibfnamefont{C.-Z.} \bibnamefont{Chang}},
  \bibinfo{author}{\bibfnamefont{C.-L.} \bibnamefont{Song}},
  \bibinfo{author}{\bibfnamefont{L.-L.} \bibnamefont{Wang}},
  \bibinfo{author}{\bibfnamefont{X.}~\bibnamefont{Chen}},
  \bibinfo{author}{\bibfnamefont{J.-F.} \bibnamefont{Jia}},
  \bibinfo{author}{\bibfnamefont{Z.}~\bibnamefont{Fang}},
  \bibinfo{author}{\bibfnamefont{X.}~\bibnamefont{Dai}},
  \bibinfo{author}{\bibfnamefont{W.-Y.} \bibnamefont{Shan}}, \emph{et~al.},
  \bibinfo{year}{2010}, \bibinfo{journal}{Nature Physics}
  \textbf{\bibinfo{volume}{6}}(\bibinfo{number}{8}), \bibinfo{pages}{584}.

\bibitem[{\citenamefont{Zhang}
  \emph{et~al.}(2019{\natexlab{e}})\citenamefont{Zhang, Mao, Cao,
  Jarillo-Herrero, and Senthil}}]{zhang2019nearly}
\bibinfo{author}{\bibnamefont{Zhang}, \bibfnamefont{Y.-H.}},
  \bibinfo{author}{\bibfnamefont{D.}~\bibnamefont{Mao}},
  \bibinfo{author}{\bibfnamefont{Y.}~\bibnamefont{Cao}},
  \bibinfo{author}{\bibfnamefont{P.}~\bibnamefont{Jarillo-Herrero}}, and
  \bibinfo{author}{\bibfnamefont{T.}~\bibnamefont{Senthil}},
  \bibinfo{year}{2019}{\natexlab{e}}, \bibinfo{journal}{Physical Review B}
  \textbf{\bibinfo{volume}{99}}(\bibinfo{number}{7}), \bibinfo{pages}{075127}.

\bibitem[{\citenamefont{Zhao and Liu}(2021)}]{zhao2021routes}
\bibinfo{author}{\bibnamefont{Zhao}, \bibfnamefont{Y.}}, and
  \bibinfo{author}{\bibfnamefont{Q.}~\bibnamefont{Liu}}, \bibinfo{year}{2021},
  \bibinfo{journal}{Applied Physics Letters}
  \textbf{\bibinfo{volume}{119}}(\bibinfo{number}{6}), \bibinfo{pages}{060502}.

\bibitem[{\citenamefont{Zhao} \emph{et~al.}(2020)\citenamefont{Zhao, Zhang,
  Mei, Zhou, Yi, Zhang, Yu, Xiao, Wang, Samarth}
  \emph{et~al.}}]{zhao2020tuning}
\bibinfo{author}{\bibnamefont{Zhao}, \bibfnamefont{Y.-F.}},
  \bibinfo{author}{\bibfnamefont{R.}~\bibnamefont{Zhang}},
  \bibinfo{author}{\bibfnamefont{R.}~\bibnamefont{Mei}},
  \bibinfo{author}{\bibfnamefont{L.-J.} \bibnamefont{Zhou}},
  \bibinfo{author}{\bibfnamefont{H.}~\bibnamefont{Yi}},
  \bibinfo{author}{\bibfnamefont{Y.-Q.} \bibnamefont{Zhang}},
  \bibinfo{author}{\bibfnamefont{J.}~\bibnamefont{Yu}},
  \bibinfo{author}{\bibfnamefont{R.}~\bibnamefont{Xiao}},
  \bibinfo{author}{\bibfnamefont{K.}~\bibnamefont{Wang}},
  \bibinfo{author}{\bibfnamefont{N.}~\bibnamefont{Samarth}}, \emph{et~al.},
  \bibinfo{year}{2020}, \bibinfo{journal}{Nature}
  \textbf{\bibinfo{volume}{588}}(\bibinfo{number}{7838}), \bibinfo{pages}{419}.

\bibitem[{\citenamefont{Zhao} \emph{et~al.}(2022)\citenamefont{Zhao, Zhang,
  Zhou, Mei, Yan, Chan, Liu, and Chang}}]{zhao2022zero}
\bibinfo{author}{\bibnamefont{Zhao}, \bibfnamefont{Y.-F.}},
  \bibinfo{author}{\bibfnamefont{R.}~\bibnamefont{Zhang}},
  \bibinfo{author}{\bibfnamefont{L.-J.} \bibnamefont{Zhou}},
  \bibinfo{author}{\bibfnamefont{R.}~\bibnamefont{Mei}},
  \bibinfo{author}{\bibfnamefont{Z.-J.} \bibnamefont{Yan}},
  \bibinfo{author}{\bibfnamefont{M.~H.} \bibnamefont{Chan}},
  \bibinfo{author}{\bibfnamefont{C.-X.} \bibnamefont{Liu}}, and
  \bibinfo{author}{\bibfnamefont{C.-Z.} \bibnamefont{Chang}},
  \bibinfo{year}{2022}, \bibinfo{journal}{Physical Review Letters}
  \textbf{\bibinfo{volume}{128}}(\bibinfo{number}{21}),
  \bibinfo{pages}{216801}.

\bibitem[{\citenamefont{Zhao} \emph{et~al.}(2021)\citenamefont{Zhao, Zhou,
  Wang, Wang, Song, Ovchinnikov, Yi, Mei, Wang, Chan}
  \emph{et~al.}}]{zhao2021even}
\bibinfo{author}{\bibnamefont{Zhao}, \bibfnamefont{Y.-F.}},
  \bibinfo{author}{\bibfnamefont{L.-J.} \bibnamefont{Zhou}},
  \bibinfo{author}{\bibfnamefont{F.}~\bibnamefont{Wang}},
  \bibinfo{author}{\bibfnamefont{G.}~\bibnamefont{Wang}},
  \bibinfo{author}{\bibfnamefont{T.}~\bibnamefont{Song}},
  \bibinfo{author}{\bibfnamefont{D.}~\bibnamefont{Ovchinnikov}},
  \bibinfo{author}{\bibfnamefont{H.}~\bibnamefont{Yi}},
  \bibinfo{author}{\bibfnamefont{R.}~\bibnamefont{Mei}},
  \bibinfo{author}{\bibfnamefont{K.}~\bibnamefont{Wang}},
  \bibinfo{author}{\bibfnamefont{M.~H.} \bibnamefont{Chan}}, \emph{et~al.},
  \bibinfo{year}{2021}, \bibinfo{journal}{Nano Letters}
  \textbf{\bibinfo{volume}{21}}(\bibinfo{number}{18}), \bibinfo{pages}{7691}.

\bibitem[{\citenamefont{Zhou} \emph{et~al.}(2017)\citenamefont{Zhou, Sun, and
  Jena}}]{zhou2017valley}
\bibinfo{author}{\bibnamefont{Zhou}, \bibfnamefont{J.}},
  \bibinfo{author}{\bibfnamefont{Q.}~\bibnamefont{Sun}}, and
  \bibinfo{author}{\bibfnamefont{P.}~\bibnamefont{Jena}}, \bibinfo{year}{2017},
  \bibinfo{journal}{Physical Review Letters}
  \textbf{\bibinfo{volume}{119}}(\bibinfo{number}{4}), \bibinfo{pages}{046403}.

\bibitem[{\citenamefont{Zhou} \emph{et~al.}(2005)\citenamefont{Zhou, Chien, and
  Uher}}]{zhou2005thin}
\bibinfo{author}{\bibnamefont{Zhou}, \bibfnamefont{Z.}},
  \bibinfo{author}{\bibfnamefont{Y.-J.} \bibnamefont{Chien}}, and
  \bibinfo{author}{\bibfnamefont{C.}~\bibnamefont{Uher}}, \bibinfo{year}{2005},
  \bibinfo{journal}{Applied Physics Letters}
  \textbf{\bibinfo{volume}{87}}(\bibinfo{number}{11}), \bibinfo{pages}{112503}.

\bibitem[{\citenamefont{Zhou} \emph{et~al.}(2006)\citenamefont{Zhou, Chien, and
  Uher}}]{zhou2006thin}
\bibinfo{author}{\bibnamefont{Zhou}, \bibfnamefont{Z.}},
  \bibinfo{author}{\bibfnamefont{Y.-J.} \bibnamefont{Chien}}, and
  \bibinfo{author}{\bibfnamefont{C.}~\bibnamefont{Uher}}, \bibinfo{year}{2006},
  \bibinfo{journal}{Physical Review B}
  \textbf{\bibinfo{volume}{74}}(\bibinfo{number}{22}), \bibinfo{pages}{224418}.

\bibitem[{\citenamefont{Zhu}
  \emph{et~al.}(2020{\natexlab{a}})\citenamefont{Zhu, Su, and
  MacDonald}}]{zhu2020voltage}
\bibinfo{author}{\bibnamefont{Zhu}, \bibfnamefont{J.}},
  \bibinfo{author}{\bibfnamefont{J.-J.} \bibnamefont{Su}}, and
  \bibinfo{author}{\bibfnamefont{A.~H.} \bibnamefont{MacDonald}},
  \bibinfo{year}{2020}{\natexlab{a}}, \bibinfo{journal}{Physical Review
  Letters} \textbf{\bibinfo{volume}{125}}(\bibinfo{number}{22}),
  \bibinfo{pages}{227702}.

\bibitem[{\citenamefont{Zhu}
  \emph{et~al.}(2020{\natexlab{b}})\citenamefont{Zhu, Bai, Hong, Geng, Jiang,
  Liu, Li, Shi, Wang, Li} \emph{et~al.}}]{zhu2020investigating}
\bibinfo{author}{\bibnamefont{Zhu}, \bibfnamefont{K.}},
  \bibinfo{author}{\bibfnamefont{Y.}~\bibnamefont{Bai}},
  \bibinfo{author}{\bibfnamefont{X.}~\bibnamefont{Hong}},
  \bibinfo{author}{\bibfnamefont{Z.}~\bibnamefont{Geng}},
  \bibinfo{author}{\bibfnamefont{Y.}~\bibnamefont{Jiang}},
  \bibinfo{author}{\bibfnamefont{R.}~\bibnamefont{Liu}},
  \bibinfo{author}{\bibfnamefont{Y.}~\bibnamefont{Li}},
  \bibinfo{author}{\bibfnamefont{M.}~\bibnamefont{Shi}},
  \bibinfo{author}{\bibfnamefont{L.}~\bibnamefont{Wang}},
  \bibinfo{author}{\bibfnamefont{W.}~\bibnamefont{Li}}, \emph{et~al.},
  \bibinfo{year}{2020}{\natexlab{b}}, \bibinfo{journal}{Journal of Physics:
  Condensed Matter} \textbf{\bibinfo{volume}{32}}(\bibinfo{number}{47}),
  \bibinfo{pages}{475002}.

\bibitem[{\citenamefont{Zhu} \emph{et~al.}(2019)\citenamefont{Zhu, Tu, Tong,
  and Yao}}]{zhu2019gate}
\bibinfo{author}{\bibnamefont{Zhu}, \bibfnamefont{Q.}},
  \bibinfo{author}{\bibfnamefont{M.~W.-Y.} \bibnamefont{Tu}},
  \bibinfo{author}{\bibfnamefont{Q.}~\bibnamefont{Tong}}, and
  \bibinfo{author}{\bibfnamefont{W.}~\bibnamefont{Yao}}, \bibinfo{year}{2019},
  \bibinfo{journal}{Science Advances}
  \textbf{\bibinfo{volume}{5}}(\bibinfo{number}{1}), \bibinfo{pages}{eaau6120}.

\bibitem[{\citenamefont{Zhu} \emph{et~al.}(2021)\citenamefont{Zhu, Bishop,
  Zhou, Zhu, O’Hara, Baker, Cheng, Walko, Repicky, Liu}
  \emph{et~al.}}]{zhu2021synthesis}
\bibinfo{author}{\bibnamefont{Zhu}, \bibfnamefont{T.}},
  \bibinfo{author}{\bibfnamefont{A.~J.} \bibnamefont{Bishop}},
  \bibinfo{author}{\bibfnamefont{T.}~\bibnamefont{Zhou}},
  \bibinfo{author}{\bibfnamefont{M.}~\bibnamefont{Zhu}},
  \bibinfo{author}{\bibfnamefont{D.~J.} \bibnamefont{O’Hara}},
  \bibinfo{author}{\bibfnamefont{A.~A.} \bibnamefont{Baker}},
  \bibinfo{author}{\bibfnamefont{S.}~\bibnamefont{Cheng}},
  \bibinfo{author}{\bibfnamefont{R.~C.} \bibnamefont{Walko}},
  \bibinfo{author}{\bibfnamefont{J.~J.} \bibnamefont{Repicky}},
  \bibinfo{author}{\bibfnamefont{T.}~\bibnamefont{Liu}}, \emph{et~al.},
  \bibinfo{year}{2021}, \bibinfo{journal}{Nano Letters} .

\bibitem[{\citenamefont{Zou} \emph{et~al.}(2020)\citenamefont{Zou, Zhan, Zheng,
  Wu, Fan, and Wang}}]{zou2020intrinsic}
\bibinfo{author}{\bibnamefont{Zou}, \bibfnamefont{R.}},
  \bibinfo{author}{\bibfnamefont{F.}~\bibnamefont{Zhan}},
  \bibinfo{author}{\bibfnamefont{B.}~\bibnamefont{Zheng}},
  \bibinfo{author}{\bibfnamefont{X.}~\bibnamefont{Wu}},
  \bibinfo{author}{\bibfnamefont{J.}~\bibnamefont{Fan}}, and
  \bibinfo{author}{\bibfnamefont{R.}~\bibnamefont{Wang}}, \bibinfo{year}{2020},
  \bibinfo{journal}{Physical Review B}
  \textbf{\bibinfo{volume}{101}}(\bibinfo{number}{16}),
  \bibinfo{pages}{161108}.

\end{thebibliography}

\end{document}